\documentclass[a4paper,12pt]{article}

\usepackage{amsmath,amsthm,amsbsy,amssymb,amsfonts,graphicx,mathrsfs,bm,accents,cite,color,float,fancybox}
\usepackage{hyperref}
\usepackage[all]{xy}

\makeatletter
\catcode`\@=11
\@addtoreset{equation}{section}

\makeatother

\makeatletter

\@addtoreset{table}{section}
\makeatother

\renewcommand{\thefootnote}{\arabic{footnote}}

\usepackage{tikz,tikz-3dplot,pgfplots}
\usetikzlibrary{arrows,calc,decorations,decorations.markings,intersections,positioning,snakes,3d,matrix,patterns}
\definecolor{darkslategray}{HTML}{2F4F4F}
\newcommand{\BN}[2]{\renewcommand{\arraystretch}{.75}\begin{tabular}{c}{#1}\cr{#2}\end{tabular}}
\definecolor{darkcyan}{HTML}{008B8B}
\makeatletter

\@addtoreset{figure}{section}
\makeatother

\makeatletter
\def\tbcaption{\def\@captype{table}\caption}
\def\figcaption{\def\@captype{figure}\caption}
\makeatother

\newcommand{\Exp}[1]{\operatorname{e}^{#1}}
\newcommand{\abs}[1]{\lvert {#1} \rvert}
\newcommand{\ket}[1]{\lvert {#1} \rangle}
\newcommand{\rmd}{{\mathrm{d}}}
\newcommand{\nn}{\nonumber}
\newcommand{\Lie}{\pounds}
\newcommand{\lp}{l_p}
\newcommand{\gs}{g_s}
\newcommand{\ls}{l_s}
\newcommand{\ii}{i}

\newcommand{\cE}{\mathcal E}\newcommand{\cF}{\mathcal F}
\newcommand{\cH}{\mathcal H}

\newcommand{\cK}{\mathcal K}\newcommand{\cL}{\mathcal L}
\newcommand{\cM}{\mathcal M}
\newcommand{\cO}{\mathcal O}
\newcommand{\cR}{\mathcal R}
\newcommand{\cT}{\mathcal T}

\newcommand{\sfd}{\mathsf{d}}
\newcommand{\sfy}{\mathsf{y}}
\newcommand{\sfF}{\mathsf{F}}
\newcommand{\sfI}{\mathsf{I}}
\newcommand{\sfL}{\mathsf{L}}
\newcommand{\sfM}{\mathsf{M}}
\newcommand{\sfPhi}{\mathsf{\Phi}}

\newcommand{\rmT}{\mathrm{T}}

\newcommand{\SL}{\mathrm{SL}}
\newcommand{\OO}{\mathrm{O}}

\newcommand{\Def}[1]{\textcolor{purple}{#1}}
\newcommand{\DoW}[1]{\textcolor{blue}{#1}}
\newcommand{\SpF}[1]{\textcolor{darkcyan}{#1}}
\newcommand{\colorDef}{purple}
\newcommand{\colorDoW}{blue}
\newcommand{\colorSpF}{darkcyan}

\newcommand{\rmM}{\text{\tiny M}}
\newcommand{\rmD}{\text{\tiny D}}
\newcommand{\rmS}{\text{\tiny S}}
\newcommand{\rmE}{\text{\tiny E}}

\newcommand{\onebb}{\mbox{1}\hspace{-0.25em}\mbox{l}}
\newcommand{\bdelta}{{\boldsymbol\delta}}

\newcommand{\sla}[1]{\setbox0=\hbox{$#1$} 
\dimen0=\wd0 \setbox1=\hbox{/} \dimen1=\wd1 
\ifdim\dimen0>\dimen1 \rlap{\hbox to \dimen0{\hfil/\hfil}} #1 
\else\rlap{\hbox to \dimen1{\hfil$#1$\hfil}} / \fi}

\allowdisplaybreaks[3]

\begin{document}

\begin{titlepage}
\renewcommand{\thefootnote}{\fnsymbol{footnote}}

\vspace*{-1cm} 
\begin{flushright}
YITP-18-51
\end{flushright}

\vspace*{1.0cm}

\begin{center}
\Large\textbf{Weaving the Exotic Web}
\end{center}

\vspace{1.0cm}

\centerline{
{\large Jos\'e J.~Fern\'andez-Melgarejo$^{a,b}$}%
\footnote{E-mail address: \texttt{josejuan@yukawa.kyoto-u.ac.jp}},
{\large Tetsuji Kimura$^{c}$}%
\footnote{E-mail address: \texttt{kimura.tetsuji@nihon-u.ac.jp}}, 
and {\large Yuho Sakatani$^{d}$}%
\footnote{E-mail address: \texttt{yuho@koto.kpu-m.ac.jp}}
}

\vspace{0.2cm}

\begin{center}
${}^a${\it Center for Gravitational Physics, Yukawa Institute for Theoretical Physics,}\\
{\it Kyoto University, Kyoto 606-8502, Japan}

\vspace*{1mm}

${}^b${\it Departamento de F\'isica, Universidad de Murcia,}\\
{\it Campus de Espinardo, 30100 Murcia, Spain}

\vspace*{1mm}

${}^c${\it Research Institute of Science and Technology, College of Science and Technology,} \\
{\it Nihon University, 1-8-14 Kanda Surugadai, Chiyoda-ku, Tokyo 101-8308, Japan}

\vspace*{1mm}

${}^d${\it Department of Physics, Kyoto Prefectural University of Medicine,}\\
{\it Kyoto 606-0823, Japan}
\end{center}

\vspace*{1mm}

\begin{abstract}
String and M-theory contain a family of branes forming $U$-duality multiplets. In particular, standard branes with codimension higher than or equal to two, can be explicitly found as supergravity solutions. However, whether domain-wall branes and space-filling branes can be found as supergravity solutions is still unclear. In this paper, we firstly provide a full list of exotic branes in type II string theory or M-theory compactified to three or higher dimensions. We show how to systematically obtain backgrounds of exotic domain-wall branes and space-filling branes as solutions of the double field theory or the exceptional field theory. Such solutions explicitly depend on the winding coordinates and cannot be given as solutions of the conventional supergravity theories. However, as the domain-wall solutions depend linearly on the winding coordinates, we describe them as solutions of deformed supergravities such as the Romans massive IIA supergravity or lower-dimensional gauged supergravities. We establish explicit relations among the domain-wall branes, the mixed-symmetry potentials, the locally non-geometric fluxes, and deformed supergravities. 
\end{abstract}
\thispagestyle{empty}
\end{titlepage}

\tableofcontents

\setcounter{footnote}{0}

\section{Introduction and summary}

\subsection{Background}

The ten-dimensional type II superstring theories contain a rich variety of extended objects such as the D-branes and the NS5-brane. 
The tension of a D$p$-brane is proportional to $\gs^{-1}$ ($\gs$: string coupling constant) and that of the NS5-brane is proportional to $\gs^{-2}$\,. 

It is conjectured that string theories are related by discrete non-perturbative dualities. 
When we compactify M/string-theory to lower dimensions, the $U$-duality group is enlarged and can relate objects that were not related in higher dimensions. 
That is to say, it can occur that, by a duality transformation, an extended object is mapped to a ``non-geometric'' one, being the latter an object that is not a solution of the higher-dimensional supergravity theories. 
This is a consequence of the geometric formulation of supergravity theories: the transition functions that are needed to ``glue'' the patches of the manifold on which the theory is defined do not contain the $U$-duality group transformations. 
Moreover, if we compactify the theory, the tension of the dualized extended object can change and be proportional to $\gs^{\alpha}$ with $\alpha\leq -3$. 
These objects are known as the \emph{exotic branes} \cite{Elitzur:1997zn,Blau:1997du,Hull:1997kb,Obers:1997kk,Obers:1998fb,Eyras:1999at,LozanoTellechea:2000mc}, and in this paper we are going to revisit various aspects of them. 

Conceptually, exotic branes should not be considered that exotic: because they are obtained by duality transformations, their role is as important as the standard branes.
Usually, the charges of these solutions are determined by the non-trivial monodromy that appears when we go around them. 
For example, let us consider type II supergravity on a $T^2$ with an NS5 brane extended along the external directions. 
The resulting background obtained after performing two $T$-dualizations is the so-called $5_2^2$-brane background. 
This background exhibits a non-trivial monodromy when going around the brane, which is not captured by the symmetries of the supergravity theories. 
The flux\footnote{We will consider fluxes as some field strengths with indices in the internal directions that have a non-trivial background value. Global treatments are omitted.} induced by this background, the so-called $Q$-flux, is the field strength of an antisymmetric object $\beta^{mn}$, which is related by $T$-duality to the Kalb--Ramond $B$-field. 
That is to say, the $5_2^2$ background is the source of the $Q$-flux, as it is magnetically coupled to the potential $\beta^{mn}$.

Just like the D$p$-brane is electrically coupled to the Ramond--Ramond (R--R) $(p+1)$-form potential, in general, exotic branes are electrically coupled to mixed-symmetry potentials. 
In a series of works \cite{Bergshoeff:2010xc,Bergshoeff:2011zk,Bergshoeff:2011mh,Bergshoeff:2011ee,Bergshoeff:2011se,Bergshoeff:2012ex,Bergshoeff:2012pm,Bergshoeff:2016ncb,Lombardo:2016swq,Bergshoeff:2017gpw}, it has been shown that there exists a one-to-one correspondence between exotic branes and some mixed-symmetry potentials that are defined in ten dimensions. 
A classification of mixed-symmetry potentials (and thus of exotic branes) has been done by considering different arguments. 
Firstly, the $E_{11}$ conjecture \cite{West:2001as,Schnakenburg:2001he,Kleinschmidt:2003mf,Riccioni:2006az} (see \cite{Tumanov:2015yjd,Tumanov:2016abm,West:2018lfn} for recent studies) allows one to predict the spectra and degeneracy of all possible mixed-symmetry potentials of any multiplet at any dimension. 
This prediction is based on the analysis of roots and weights of the $U$-duality group at any dimension \cite{West:2003fc,Kleinschmidt:2003jf,West:2004kb,West:2004iz,Cook:2008bi,Kleinschmidt:2011vu}. 
Lately, the so-called \emph{wrapping rules} were formulated \cite{Bergshoeff:2011ee,Bergshoeff:2012jb,Bergshoeff:2013spa,Pradisi:2014fqa,Bergshoeff:2017gpw}. 
This set of rules allows one to construct a set of mixed-symmetry potentials, depending on the type of $T$- and $S$-duality transformations that one performs.
This approach is in full agreement with the predictions given by the $E_{11}$ decomposition method.


Despite the study of the mixed-symmetry potentials allows to elucidate the spectra of exotic branes for any dimensions, we still lack a method to generate these backgrounds. 
Because the geometric isometries of supergravity are not enough to cover generic $U$-duality transformations, we would require a theory in which dualities are true symmetries. 
For example, we can consider extended field theories, such as Double Field Theory (DFT) \cite{Siegel:1993xq,Siegel:1993th,Siegel:1993bj,Hull:2009mi,Hull:2009zb,Hohm:2010jy,Hohm:2010pp,Jeon:2010rw,Hohm:2010xe,Jeon:2011cn,Hohm:2011zr,Hohm:2011dv,Jeon:2011vx,Hohm:2011si,Hohm:2011nu,Jeon:2011sq,Jeon:2012kd,Jeon:2012hp} and Exceptional Field Theory (EFT) \cite{West:2000ga,West:2001as,Hillmann:2009pp,Berman:2010is,Berman:2011jh,Berman:2012vc,West:2012qz,Hohm:2013pua,Hohm:2013vpa,Hohm:2013uia,Hohm:2014fxa,Hohm:2015xna,Abzalov:2015ega,Musaev:2015ces,Berman:2015rcc}, which are manifestly $T$-duality- and $U$-duality-symmetric theories, respectively. 

In the formulations of DFT and EFT, some additional coordinates have been added in order to realize a manifest duality symmetry. 
In this case, the usual spacetime coordinates and the dual coordinates, known as the \emph{winding coordinates}, are on the same footing. 
Because dualities are isometries of these theories, we should be able to realize the exotic branes that are obtained upon a chain of dualities as solutions within this theory. 
To consistently formulate DFT/EFT, we demand a constraint on the dependence of the fields on the full set of coordinates. 
The so-called \emph{section condition} (SC) imposes some restrictions among the physical and the winding coordinates, in such a way that when fields depend on all the physical coordinates, the usual supergravity theory is recovered. 
Nevertheless, there exist other solutions to the SC which imply that the fields can depend on the winding coordinates. 
In particular, it is known that solutions of the SC that allow the fields to depend on the winding coordinates reproduce the Romans massive type IIA supergravity \cite{Romans:1985tz} (see a derivation from DFT \cite{Hohm:2011cp} and a modified EFT called XFT \cite{Ciceri:2016dmd}) or the type IIB generalized supergravity equations (GSE) \cite{Arutyunov:2015mqj,Wulff:2016tju} (see a derivation from (a modified) DFT \cite{Sakatani:2016fvh,Sakamoto:2017wor} and EFT \cite{Baguet:2016prz}). 

Because DFT/EFT are duality-symmetric theories, they should be the appropriate scenarios to describe the set of exotic branes that can be obtained by duality transformations (see \cite{Bakhmatov:2017les} for a recent study on exotic branes in EFT).


The choice of a solution of the SC in which fields can depend on winding coordinates implies the existence of isometry directions along some physical coordinates. 
Then, we can think of these theories as effective lower-dimensional theories such that, when uplifted to ten/eleven-dimensional supergravities, they exhibit the isometry directions. 
Such isometry directions are determined by a set of Killing vectors, which become crucial in the description of these massive or deformed supergravities \cite{Bergshoeff:1996ui,Meessen:1998qm}. 
The first formulations of these theories were prior to the DFT/EFT formulations. 
The case of the GSE \cite{Arutyunov:2015mqj,Wulff:2016tju} and its derivation from DFT \cite{Sakatani:2016fvh,Sakamoto:2017wor} or EFT \cite{Baguet:2016prz} is one of the most recent examples that have been worked out in the literature. 

As mentioned above, because of their isometry directions, these deformed supergravities can be understood as effective lower-dimensional theories with massive deformations. 
Such deformations can be studied systematically:
using the embedding tensor formalism \cite{Cordaro:1998tx,deWit:2002vt} and constructing the tensor hierarchy of a theory, one can scan all the possible deformations of a particular lower-dimensional supergravity. 
Then, a dictionary between the fluxes associated to these deformed supergravities and the embedding tensor is estimated. 

It is the purpose of this paper to establish a systematic way of studying the exotic branes and their expected-to-be one-to-one related objects. 
Based on the above arguments, we guess that the distinct formulations of exotic branes, mixed-symmetry potentials and the massive supergravities are closely related. 
We would like to fill the gaps among these three approaches and establish precise mechanisms to show their equivalence. 

In this paper, we firstly generate the full web of exotic branes by applying $U$-duality transformations to standard branes. 
We only consider a subgroup of the $U$-duality group, which consists of the $T$- and $S$-duality transformations. 
As we could expect from the finiteness of the $U$-duality group for $d\geq 3$, we have obtained a finite set of exotic states, which have been classified into different orbits. 

After fully determining the web of branes, we find a systematic way to generate the exotic-brane backgrounds as solutions of duality-symmetric theories, namely DFT or EFT. 
Being a $U$-duality-symmetric theory, EFT is the appropriate framework where to describe these backgrounds. 
Starting from a fully geometric brane background, we should be able to perform $T$- and $S$-duality transformations to generate the dualized backgrounds. 
To do so, we need to understand how duality transformations act on both the fields that enter the EFT and on the winding coordinates. 
That is to say, firstly, we rewrite the usual $T$- and $S$-duality rules in terms of the supergravity fields that appear in the M-theory and type IIB parameterizations of EFT. 
Secondly, we apply the duality transformations on the generalized coordinates. 
Because we start from geometric solutions that correspond to the standard branes, the sections (i.e.~solutions of the SC) for the obtained solution are $T$- and $S$-duality-related to the geometric section. 

The dictionary between the supergravity fields and the dual (or the non-geometric) fields in DFT/EFT allows us to calculate the non-geometric fluxes. 
We find a relation between the non-geometric fluxes and the mixed-symmetry potentials obtained from the $E_{11}$ decomposition and the wrapping rules. 

Finally, we find a mechanism to systematically obtain ten/eleven-dimensional deformed supergravities that exhibit some isometry directions. 
The number of isometry directions depends on the specific solution of the SC, which will pick the non-physical coordinates that the fields can depend on. 
In general, we would like to engineer a systematic way to generate deformed supergravities that contain each of the exotic domain-wall branes as a solution. 
For instance, a relation between the domain-wall solutions and deformed supergravities has been suggested in \cite{Bergshoeff:2012pm}. 

\subsection{Main results}

In this subsection we summarize the results that we have obtained. 

In Section \ref{sec:Full-duality-web}, by brute force application of the $S$- and $T$-duality transformations, we have generated the full web of supersymmetric branes for each $p$-brane multiplet at any dimension $d\ge3$. 
In this classification, we have distinguished the defect, domain-wall and space-filling brane types (which have codimension 2, 1, and 0, respectively) from the standard branes. 
In \eqref{eq:all-M-branes}, \eqref{eq:all-IIA-branes}, and \eqref{eq:all-IIB-branes}, we have shown the spectrum of all the M-theory branes, type IIA branes, and type IIB branes. 
In Figures \ref{fig:web01}--\ref{fig:web20}, we have generated the web of type II branes and shown the $T$-duality and $S$-duality chains of transformations that relate them.
At any dimension $d\geq 3$, we have obtained the spectra of exotic branes for any $p$-brane multiplet together with their degeneracies, which are given in Appendix \ref{app:multiplets}. 

In Section \ref{sec:exotic-DFT}, by utilizing the manifest $\OO(d,d)$ $T$-duality symmetry of DFT, we have obtained some known domain-wall solutions, the D8 solution and the $5^3_2$ solution (also known as the $R$-brane solution). 
In order to obtain the backgrounds of the full web of branes, $T$-duality is not enough. 
That is to say, to generate the whole $T$-duality orbits of Figures \ref{fig:web01}--\ref{fig:web20} which only contain domain-wall and space-filling branes, we additionally need $S$-duality transformations. 
By $S$-dualizing some elements of the orbits containing standard and defect branes, we generate the orbits spanned by domain-wall and space-filling branes. 

Then, in Section \ref{sec:exotic-EFT}, by making use of $S$- and $T$-duality transformations, we have obtained the full web of branes as solutions of EFT. 
We have shown that to obtain the exotic branes as EFT solutions, we have to systematically apply $S$- and $T$-duality transformations on both the EFT generalized metric and the set of coordinates defined in EFT. 
Unlike the well-known defect-brane solutions, the obtained solutions include the explicit linear winding-coordinate dependence, while satisfying the SC.

In terms of the dual fields in EFT, we have calculated all the non-geometric fluxes for each exotic brane in the web. 
Similar to the fact that the defect branes can be regarded as the magnetic sources of the globally non-geometric fluxes, the domain-wall branes are identified as the magnetic sources of the locally non-geometric $R$-fluxes. 
We have proposed suitable definitions of the $R$-fluxes in the $E_{8(8)}$ EFT that transform covariantly under $U$-duality transformations. 
Then, we have shown that the domain-wall-brane solutions in EFT contain constant $R$-fluxes. 

We have clarified the relation between the non-geometric fluxes associated to each brane and the mixed-symmetry potentials predicted in the literature. 
In particular, we have shown that the non-geometric fluxes in EFT are dual to the field strengths of the mixed-symmetry potentials. 
To do so, we have to extend the electric-magnetic duality transformation of the mixed-symmetry potentials that was conjectured in DFT \cite{Bergshoeff:2015cba} to the EFT formulation, for both the M-theory/IIB parameterizations. 
The electric-magnetic duality transformation in EFT involves the dual spacetime metric, as it occurs in the DFT case.

Finally, in Section \ref{sec:deformed-sugra}, we have discussed various deformed supergravity theories, which generalize the Romans massive IIA supergravity. 
As mentioned earlier, these theories can enjoy one or more isometry directions, each of them characterized by a Killing vector. 
For a given exotic brane, we have provided a prescription to identify the lower-dimensional deformed supergravity theory that realizes that background. 
While standard and defect branes do not exhibit any dependence on winding coordinates, this is not the case for the domain-walls and the space-filling branes. 
The winding-coordinate dependence of the domain-wall solutions are transmuted into the $R$-fluxes (or the gaugings), which characterize the deformations of the supergravities, and the domain-wall solutions in the deformed supergravities are independent of the winding coordinates. 
That is to say, the dependence on the winding coordinate is encoded in the deformation parameter of the corresponding supergravities.
We have reproduced several known domain-wall solutions in certain deformed supergravities, which include known solutions in \cite{Meessen:1998qm}.

In summary, in this paper, we have explicitly established one-to-one mappings among several topics,
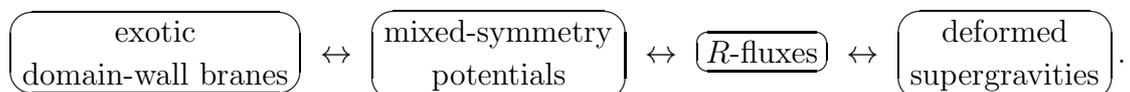
\begin{align*}
 \ovalbox{\normalsize$\genfrac{}{}{0pt}{0}{\text{exotic}}{\text{domain-wall branes}}$} 
 \ \leftrightarrow\ 
 \ovalbox{\normalsize$\genfrac{}{}{0pt}{0}{\text{mixed-symmetry}}{\text{potentials}}$}\ \leftrightarrow\ 
 \ovalbox{\normalsize$R$-fluxes}\ \leftrightarrow\ 
 \ovalbox{$\normalsize\genfrac{}{}{0pt}{0}{\text{deformed}}{\text{supergravities}}$}\,. 
\end{align*}
Nevertheless, several question remain unclear. 
Let us elaborate on them. 

\subsection{Future directions}

Let us comment on several open questions that have not been addressed in this paper. 

In this work, we have concentrated on the branes which are connected through Weyl reflections (which is a part of $U$-duality transformations). 
For the disconnected ``missing states,'' we have only translated their Dynkin labels to the mass formula and classified them into several families of missing sates connected under Weyl reflections. 
At the present stage, the physical properties of the missing states are totally unclear. 
In the literature \cite{Cook:2009ri,Houart:2009ya,Kleinschmidt:2011vu,Cook:2011ir}, the missing states are discussed as certain bound states of the ``elementary'' branes (i.e.~branes connected to the standard branes via Weyl reflections). 
Indeed, among the familiar triplet of seven branes in type IIB theory \cite{Meessen:1998qm,Bergshoeff:2005ac,Bergshoeff:2006jj,Bergshoeff:2007aa}, only two of them correspond to ``elementary'' branes, D7 and $7_3$\,. 
The remaining one is disconnected to these two and will be a member of the missing states, $7_2$\,. 
The supergravity solution for the missing seven brane is discussed in \cite{Bergshoeff:2007aa}. 
A more detailed study of the supergravity solution and its $U$-duality rotations will be an interesting future direction. 
It will also be important to study the supersymmetry of the exotic backgrounds. 
In Section \ref{sec:susy} we have only obtained the projection rule via $U$-duality rotations. 
More detailed studies, such as the explicit finding of the Killing spinor equations, will be important.
Other than single-brane solutions, by applying our $U$-duality rules in EFT, we can also rotate an arbitrary solution, such as a D$p$--D$q$ solution, and obtain various multiple-brane solutions that may include exotic branes. 

It is also interesting to investigate the physical meaning of the huge number of the exotic space-filling branes. 
In the case of the D9-brane and the $9_4$-brane in type IIB theory, their existence was closely related to type I or heterotic theory. 
It is interesting to study the (possibly non-covariant or lower-dimensional) theories that are associated with other exotic space-filling branes. 

According to the $E_{11}$ program, the existence of many mixed-symmetry potentials has been conjectured, which transform under certain representations of the $U$-duality group. 
However, it has not been discussed much about how to describe the mixed-symmetry potentials in the context of EFT. 
Only the mixed-symmetry potential $C_{i_1\cdots i_8,\,j}$ in M-theory or $D_{m_1\cdots m_7,\,n}$ in type II theory can appear in $E_{8(8)}$ EFT, but other potentials do not enter the generalized metric. 

From this paper's perspective, the generalized metric of EFT contains the dual fields, such as $\beta$, $\gamma$, and $\Omega$, whereas the Hodge duals (which are taken with respect to a combination of the ``dual'' metrics that parameterize the generalized metric) of their winding derivatives give the dual field strengths. 
Then, the mixed-symmetry potentials are defined as the potentials for the dual field strengths. 
In fact, the mixed-symmetry potentials can appear in a more direct manner.
In EFT, there exists an external 1-form $\mathfrak{A}_\mu^{I_1}$ that transforms in the particle multiplet, and a 2-form $\mathfrak{B}_{\mu\mu}^{I_2}$ that transforms in the string multiplet, and a 3-form $\mathfrak{C}_{\mu\mu\rho}^{I_3}$ that transforms in the membrane multiplet, and so on. 
As we have explicitly shown in this paper, these $p$-brane multiplets include all of the exotic branes and these $(p+1)$-form fields should be composed of the mixed-symmetry potentials. 
The explicit parameterizations of these $(p+1)$-form fields will be important. 
More importantly, there must be constraints for the derivative of the $(p+1)$-form fields and the derivative of the generalized metric, corresponding to the electric-magnetic duality between the field strength of the mixed-symmetry potentials and the $R$-fluxes. 
Such duality will correspond to the \emph{exotic duality} \cite{Bergshoeff:2016gub} recently discussed in DFT, and it is important to identify the duality relation in EFT. 

As a different direction, it would be interesting to determine the worldvolume actions for exotic branes and show that the Wess--Zumino term indeed contains the mixed-symmetry potential (see \cite{Bergshoeff:2010xc} for a related work). 

Let us comment on the $R$-fluxes. 
Despite in this work we have given some heuristic definitions of the locally non-geometric $R$-fluxes, we should provide more systematic definitions similar to \cite{Blair:2014zba,Gunaydin:2016axc,Lust:2017bwq}. 
Here, we have concentrated on the $R$-fluxes associated with the ``elementary'' domain-walls, but since the mixed-symmetry potentials for the missing states are also proposed, by performing the electric-magnetic duality, we may also obtain the $R$-fluxes associated with the missing states. 
Only after introducing such $R$-fluxes, we can obtain a $U$-duality multiplet of fluxes. 

It is also interesting to study the fluxes associated with the space-filling branes. 
In terms of the mixed-symmetry potentials in type II theory, mixed-symmetry potentials with a set of ten antisymmetric indices, such as $E_{10,7-p,2}$ have been proposed. 
Naively, by introducing their field strengths and performing the electric-magnetic duality, one may obtain the corresponding fluxes. 
However, proper definitions of the field strengths are not clear at present. 
Some hints may be found by studying the EFT solutions of the space-filling branes in more detail. 

In this paper, we have made clear how to obtain the action or the equations of motion of various deformed supergravities from EFT. 
According to the SC in EFT, the deformed supergravities are effectively defined in lower-dimensional spacetime, as some of the winding coordinates are used to provide the constant fluxes or deformation parameters. 
It would be interesting to establish a systematic relation between the exotic branes or, equivalently, their associated lower-dimensional deformed supergravities and the gaugings in the language of the embedding tensor. 
Some work in this direction has been recently done for spacetime-filling branes in \cite{Dibitetto:2018wvc}. 
In this paper, among all of the domain-wall branes contained in the $U$-duality multiplets, we have only considered the ``elementary'' domain-walls. 
In that case, the SC is not violated. 
On the other hand, in \cite{Dibitetto:2012rk}, gaugings that break the SC have been found. 
In particular, non-geometric fluxes that are in different orbits from the standard fluxes are introduced. 
It would be relevant to find a connection between these fluxes and the missing states in the $U$-duality multiplets. 

\subsection{Plan of the paper}

The paper is organized as follows. 
In Section \ref{sec:review} we briefly review the notation for exotic branes. 
In Section \ref{sec:Full-duality-web}, we review the relation between exotic branes and the weights of the $U$-duality group. 
We then construct the full web of ``elementary'' exotic branes and give the duality transformations that relate them. 
In Section \ref{sec:exotic-DFT}, we review how to obtain some of the known domain-wall backgrounds as solutions of DFT. 
In Section \ref{sec:exotic-EFT}, we explain how to obtain all the ``elementary'' exotic branes as solutions of EFT. 
The definitions of the non-geometric $R$-fluxes are also provided. 
In Section \ref{sec:deformed-sugra}, we review some deformed/massive supergravity theories and show that they can be obtained upon solving the SC of DFT/EFT in such a way that winding coordinates are allowed. 

We also provide several appendices. 
Appendix \ref{app:conventions} provides the notation used along this work. 
In Appendix \ref{app:EFT-parameterization}, we review parameterizations of the generalized metric in $E_{n(n)}$ EFT ($n\leq 7$). 
Appendix \ref{app:multiplets} shows the various spectra of $p$-brane multiplets for diverse dimensions. 
Finally, Appendix \ref{app:counting-mixed-symmetry} shows the relation between the exotic branes that we have obtained and the mixed-symmetries potentials that are coupled to. 

\medskip

\noindent\textbf{Addendum.} Upon publication of version 1 of this work, we learned that some aspects of Section \ref{sec:Full-duality-web} were also constructed by another group \cite{Berman:2018okd}.

\section{A brief review of duality rules and exotic branes}
\label{sec:review}

In this section, we provide a brief review of exotic branes in type II string theories and M-theory toroidally compactified to $d$-dimensions. 

\subsection{Type II branes}

In type II string theory, by denoting the radius of the torus along the $x^i$-direction as $R_i$, the mass of a fundamental string (denoted as F1) wrapped along the $x^i$-direction is given by
\begin{align}
 M_{\text{F1}(i)} = \frac{1}{2\pi\ls^2}\times (2\pi R_i) = \frac{R_i}{\ls^2} \qquad \bigl(\ls\equiv\sqrt{\alpha'}\bigr)\,.
\label{eq:mass-F1}
\end{align}
By using the familiar $T$- and $S$-duality transformation rules,
\begin{align}
\begin{split}
 \text{$T$-duality}:&\quad R_i \to \ls^2/R_i \,, \qquad \gs \to \gs\, \ls/R_i \,,\qquad \ls \to \ls \,,
\\
 \text{$S$-duality}:&\quad \gs \to 1/\gs \,, \qquad \ls \to \gs^{1/2}\,\ls\,,
\end{split}
\label{eq:T-S-rule}
\end{align}
we can see how the mass \eqref{eq:mass-F1} is transformed under duality transformations. 
For example, if we perform a $T$-duality along the $x^i$-direction, the mass \eqref{eq:mass-F1} becomes
\begin{align}
 M_{\text{P}(i)} = \frac{1}{R_i} \,,
\end{align}
which is interpreted as a mass of the pp-wave or the Kaluza--Klein (KK) momentum (denoted as P). 
If we instead perform an $S$-duality, the mass \eqref{eq:mass-F1} becomes
\begin{align}
 M_{\text{D1}(i)} = \frac{R_i}{\gs\ls^2} = \gs^{-1}\frac{1}{2\pi\ls^2}\times (2\pi R_i)\,,
\end{align}
which is interpreted as a mass of the D1-brane wrapped along the $x^i$-direction. 
By repeating duality transformations, we obtain masses of various branes. 
It is then useful to employ the notation of \cite{Obers:1998fb} (see also \cite{deBoer:2010ud,deBoer:2012ma}), which allows us to characterize various branes by their masses. 
If an object wrapped along $x^{n_1},\dotsc,x^{n_b}$-directions has a mass,
\begin{align}
 M = \frac{1}{\gs^{n}\,\ls} \Bigl(\frac{R_{n_1}\cdots R_{n_b}}{\ls^b}\Bigr) \Bigl(\frac{R_{m_1}\cdots R_{m_{c_2}}}{\ls^{c_2}}\Bigr)^2\cdots \Bigl(\frac{R_{p_1}\cdots R_{p_{c_s}}}{\ls^{c_s}}\Bigr)^{s}\,,
\label{eq:typeII-brane-mass}
\end{align}
we denote the brane as
\begin{align}
 b^{(c_s,\dotsc,c_2)}_{n}(n_1\cdots n_b,\,m_1\cdots m_{c_2},\cdots,\, p_1\cdots p_{c_s})\,,
\end{align}
or simply call it a $b^{(c_s,\dotsc,c_2)}_{n}$-brane.\footnote{We denote $b^{(0,c_{s-1},\dotsc,c_2)}_{n}$ by $b^{(c_{s-1},\dotsc,c_2)}_{n}$\,, whereas $b^{(c)}_{n}$ and $b^{(0)}_{n}$ are respectively denoted as $b^{c}_n$ and $b_n$\,.} 
With this notation, for example, the usual D$p$-brane and the NS5-brane are denoted as the $p_1$-brane and the $5_2$-brane. 

\subsection{M-theory branes}

We can uplift the mass of type IIA branes to the mass of M-theory branes by using the usual relation connecting 11D and 10D,
\begin{align}
 R_{\rmM} = \gs \ls\,,\qquad \lp = \gs^{1/3}\,\ls \,,
\label{eq:11D-10D}
\end{align}
where $R_{\rmM}$ represents the radius of the M-theory circle. 
After the uplift, M-theory branes generally have masses of the form
\begin{align}
 M = \frac{1}{\lp} \Bigl(\frac{R_{i_1}\cdots R_{i_b}}{\lp^b}\Bigr) \Bigl(\frac{R_{j_1}\cdots R_{j_{c_2}}}{\lp^{c_2}}\Bigr)^2\cdots \Bigl(\frac{R_{k_1}\cdots R_{k_{c_s}}}{\lp^{c_s}}\Bigr)^s\,. 
\label{eq:M-brane-mass}
\end{align}
We then denote the brane as
\begin{align}
 b^{(c_s,\dotsc,c_2)}_{n}(i_1\cdots i_b,\, j_1\cdots j_{c_2},\cdots,\, k_1\cdots k_{c_s})\,,
\end{align}
where $n\equiv 1+b+2\,c_2+3\,c_3+\cdots+s\,c_s$ represents the power of the Planck length in the denominator. 
In the literature, $n$ is omitted, but here we keep it since it is a good measure of the exoticism, similar to the power of $\gs$ in type II theory. 

In terms of M-theory, the transformation rule \eqref{eq:T-S-rule} can be nicely summarized as \cite{Aharony:1996wp,Elitzur:1997zn}
\begin{align}
 U_{i,j,k}:\quad 
 R_i \to \frac{\lp^3}{R_j R_k}\,,\quad
 R_j \to \frac{\lp^3}{R_k R_i}\,,\quad
 R_k \to \frac{\lp^3}{R_i R_j}\,,\quad
 \lp \to \frac{\lp^2}{(R_i R_j R_k)^{1/3}}\,.
\label{eq:U-dual}
\end{align}
It is noted that the inverse of the Newton constant in $d$-dimensions,
\begin{align}
 l_d \equiv \frac{R_1\cdots R_{n}}{\lp^9}\qquad (n\equiv 11-d)\,,
\end{align}
is invariant under the $U_{i,j,k}$ for arbitrary choices of three directions $\{i,j,k\}$ in the torus $T^n$\,. 

\subsection{Brane tension}

The $b^{(c_s,\dotsc,c_2)}_{n}$-brane in type II/M-theory can extend along the external spacetime up to $b$ number of spatial dimensions, although the indices in the second slots (i.e.~$j_1,\dotsc,j_{c_2}$) or later should be internal ones. 
Namely, we can consider an external $p$-brane,
\begin{align}
\begin{split}
 \text{Type II}:\quad
 &b^{(c_s,\dotsc,c_2)}_{n}(\mu_1\cdots \mu_p\, n_1\cdots n_{b-p},\,m_1\cdots m_{c_2},\cdots,\, \ell_1\cdots \ell_{c_s})\,,
\\
 \text{M-theory}:\quad
 &b^{(c_s,\dotsc,c_2)}_{n}(\mu_1\cdots \mu_p\, i_1\cdots i_{b-p},\,j_1\cdots j_{c_2},\cdots,\, k_1\cdots k_{c_s})\,,
\end{split}
\end{align}
where $\mu_i$ represents the external directions. 
Corresponding to the masses \eqref{eq:typeII-brane-mass} and \eqref{eq:M-brane-mass}, the tensions of the $p$-brane in type II/M-theory are given by
\begin{align}
\begin{split}
 \text{Type II}:\quad
  &T_p = \frac{1}{\gs^{n}\,\ls\,(2\pi l_s)^p} \Bigl(\frac{R_{n_1}\cdots R_{n_{b-p}}}{\ls^{b-p}}\Bigr) \Bigl(\frac{R_{m_1}\cdots R_{m_{c_2}}}{\ls^{c_2}}\Bigr)^2\cdots \Bigl(\frac{R_{\ell_1}\cdots R_{\ell_{c_s}}}{\ls^{c_s}}\Bigr)^s\,,
\\
 \text{M-theory}:\quad 
  &T_p = \frac{1}{\lp\,(2\pi\lp)^p} \Bigl(\frac{R_{i_1}\cdots R_{i_{b-p}}}{\lp^{b-p}}\Bigr) \Bigl(\frac{R_{j_1}\cdots R_{j_{c_2}}}{\lp^{c_2}}\Bigr)^2\cdots \Bigl(\frac{R_{k_1}\cdots R_{k_{c_s}}}{\lp^{c_s}}\Bigr)^s\,. 
\end{split}
\label{eq:tension}
\end{align}
This formula becomes necessary in the next section, when we associate various tensions appearing in the $p$-brane multiplets to branes in type II/M-theory. 

\medskip

As we review in Section \ref{sec:Weyl-reflection}, $T$- and $S$-duality in type II theory, or the duality transformation $U_{i,j,k}$ in M-theory can be regarded as the Weyl reflection associated with the simple roots of the $E_{n(n)}$ group, which is the $U$-duality group of the string/M-theory. 
In the next section, we provide a full list of branes obtained by the $U$-duality transformations. 

\section{Full duality web for $d\geq 3$}
\label{sec:Full-duality-web}

In this section, we provide the full duality web for string/M-theory compactified to $d$-dimensions with $d\geq 3$. 

\subsection{Duality rotations as Weyl reflections}
\label{sec:Weyl-reflection}

Before showing the duality web, here we explain that the chain of $T$- and $S$-duality can be regarded as Weyl reflections by closely following the discussion of \cite{Elitzur:1997zn,Obers:1998fb}. 

\subsubsection{Setup}

Let us rewrite the brane tension \eqref{eq:tension} as
\begin{align}
 \cT_p &\equiv (2\pi)^p\,T_p \equiv \lp^{3 v^0} (R_{i_1}\cdots R_{i_{b-p}}) (R_{j_1}\cdots R_{j_{c_2}})^2\cdots (R_{k_1}\cdots R_{k_{c_s}})^s
\nn\\
 &\equiv \Exp{v^0\,x_0+v^i\,x_i} \qquad 
 \bigl(\Exp{x_0} \equiv \lp^{3}\,,\quad \Exp{x_i} \equiv R_i \bigr)\,,
\label{eq:tension2}
\end{align}
which has the mass dimension
\begin{align}
 -\bigl(3\,v^0 + v^1 + \cdots + v^n\bigr) \equiv 1 + p \,,
\end{align}
and reduces to the mass \eqref{eq:M-brane-mass} when $p=0$. 
We also define a vector $v\equiv v^\mu\,e_\mu$ ($\mu=0,1,\dotsc,n$) by using a basis $e_\mu$ of $(n+1)$-dimensional vector space endowed with an inner product,
\begin{align}
 e_\mu \cdot e_\nu = \eta_{\mu\nu}\,, \qquad (\eta_{\mu\nu})\equiv \text{diag}(-1,1,\dotsc,1) \,. 
\end{align}
Then, a particular $U$-duality transformation $U_{1,2,3}$ of \eqref{eq:U-dual} can be realized as a reflection,
\begin{align}
 v \ \to \ v - 2\, \frac{v\cdot \alpha_n}{\alpha_n\cdot \alpha_n} \,\alpha_n \,,\qquad 
 \alpha_n \equiv e_0 - (e_{1} + e_{2} + e_{3}) \,. 
\end{align}
A general $U$-duality transformation \eqref{eq:U-dual} can be realized by combining $U_{1,2,3}$ and particular $U$-duality transformations $P_{i}$: $R_i \leftrightarrow R_{i+1}$ ($i=1,\dotsc,n-1$),
\footnote{In type II theory, by denoting a chain of $T$-dualities along the $x^{m_1},\dotsc,x^{m_n}$-direction by $T_{m_1\cdots m_n}$\,, a chain of dualities $T_m\,S\,T_{mn}\,S\,T_{nm}\,S\,T_{m}$\,, corresponds to a permutation $R_{m}\leftrightarrow R_{n}$ keeping $\gs$ and $\ls$ invariant. 
Therefore, an exchange $R_{m}\leftrightarrow R_{n}$ in 11D can be also realized as a combination of $U_{i,j,k}$\,. 
Furthermore, the 11D uplift of $T_m\,S\,T_m$ corresponds to a permutation $R_m\leftrightarrow R_{\rmM}$\,. 
Therefore, the $U$-duality \eqref{eq:U-dual} contains all possible permutations $R_i\leftrightarrow R_j$ in 11D.
} 
and $P_{i}$ can be also realized as a reflection,
\begin{align}
 v \ \to \ v - 2\, \frac{v\cdot \alpha_i}{\alpha_i\cdot \alpha_i} \,\alpha_i \,,\qquad
 \alpha_i \equiv e_{i} - e_{i+1} \qquad (i=1,\dotsc,n-1)\,. 
\end{align}
Since the inner products $a_{ij}\equiv \alpha_i\cdot \alpha_j$ among the vectors $\alpha_i$ $(i=1,\dotsc,n)$ take the form
\begin{align}
 (a_{ij}) = {\scriptsize\begin{pmatrix}
 2 & -1 & & & & &0\\
 -1&2&-1& & & &0 \\
  &-1&2&-1& & &-1 \\
  & & -1 & \ddots & \ddots & & 0 \\
  & & & \ddots & \ddots & -1 & \vdots \\
  & & & &-1&2&0 \\
 0&0&-1&0&\cdots&0&2 \end{pmatrix}} \quad\ \leftrightarrow \quad\ 
 \vcenter{{\small\xygraph{!{<0cm,0cm>;<0.8cm,0cm>:<0cm,0.8cm>::}
    *\cir<4pt>{}([]!{+(0,-.4)} {\alpha_1})-[r]
    *\cir<4pt>{}([]!{+(0,-.4)} {\alpha_2})-[r]
    *\cir<4pt>{}([]!{+(0,-.4)} {\alpha_3})
   (-[u] *\cir<4pt>{}([]!{+(.5,0)}{\alpha_{n}}),
    -[r] *\cir<4pt>{}([]!{+(0,-.4)}{\alpha_4})-[r] \cdots ([]!{+(0,-.3)} {})-[r]*\cir<4pt>{}([]!{+(0,-.4)}{\alpha_{n-1}}))}}}\,,
\end{align}
we can regard $\alpha_i$ as the simple roots of the $E_{n(n)}$ $U$-duality group and transformations \eqref{eq:U-dual} as the Weyl reflections. 
Corresponding to the invariance of the $d$-dimensional Newton constant (or $l_{d}=R_1\cdots R_n/\lp^9$), a vector
\begin{align}
 \delta \equiv e_1+\cdots +e_n -3\,e_0\qquad (\delta\cdot\delta=n-9)\,,
\end{align}
is invariant under the reflections associated with the simple roots $\alpha_i$\,. 
This shows that, for an arbitrary vector $v=v^\mu\,e_\mu$\,, a quantity $3\,v^0 + v^1+\cdots+v^d\,(=p)$ is invariant under the Weyl reflections. 

Now, we introduce the fundamental weights $\lambda^i$ associated with the simple roots (satisfying $\lambda^i \cdot \alpha_j=\delta^i_j$) as follows:
\begin{align}
\begin{split}
 &\lambda^1\equiv e_1 - e_0\,, \qquad
 \lambda^2\equiv e_1 + e_2 - 2\,e_0\,, \qquad
 \lambda^3\equiv e_1 + e_2 + e_3 - 3\,e_0\,, \quad \dotsc, 
\\
 &\lambda^{n-1}\equiv e_1+\cdots +e_{n-1} - 3\,e_0\,, \qquad
 \lambda^n\equiv - e_0 \,.
\end{split}
\end{align}
Since the vector $\delta$ is orthogonal to all of the simple roots, there is an ambiguity in the choice of $\lambda^i$\,; $\lambda^i\sim \lambda^i+c^i\,\delta$ ($c^i$: constant). 
We can determine the constants $c^i$ by requiring $\alpha_i=a_{ij}\,\lambda^j$\,, but the $\delta$-direction is irrelevant for our purpose, and we can mod out the direction from the $(n+1)$-dimensional space spanned by $e_\mu$\,. 
With the above choice, we have
\begin{align}
 \alpha_i=a_{ij}\,\lambda^j\quad (i\neq n-1)\,,\qquad 
 \alpha_{n-1}= a_{(n-1)j}\,\lambda^j - \delta \sim a_{(n-1)j}\,\lambda^j\,. 
\end{align}

\subsubsection{$p$-brane multiplet}

According to the relation \eqref{eq:tension2} between the tension and the vector, the tension associated with a fundamental weight $\lambda^1$ is $\cT_1=R_1/\lp^3$\,. 
This is the tension of a string in the external $d$-dimensional spacetime. 
More concretely, this string can be interpreted as an M2-brane wrapped along the internal $x^1$-direction. 
By acting $E_{n(n)}$ $U$-duality transformations, we obtain the $U$-duality multiplet, known as the \emph{string multiplet}, that is associated with the fundamental weight $\lambda^1$\,. 

The tension associated with $\lambda^2$ is $\cT_3=R_1R_2/\lp^6$, which corresponds to the tension of a 3-brane in the external $d$-dimensional spacetime. 
In terms of M-theory states, it is an M5-brane wrapped along the internal $x^1$ and $x^2$-directions. 
Performing $E_{n(n)}$ transformations, we obtain the \emph{3-brane multiplet}. 
Note that in order to consider the 3-brane multiplet, the dimension $d$ needs to satisfy $d\geq 4$\,. 

Similarly, the tension associated with $\lambda^3$ is $\cT'_5=R_1R_2R_3/\lp^9$ and it makes a certain 5-brane multiplet. 
If there is an ``M8-brane'' that has a tension $\frac{1}{\lp(2\pi\lp)^8}$\,, the tension can be interpreted as the tension of an M8-brane wrapped along the internal $x^1,\,x^2$, and $x^3$-directions. 
However, the existence of such object is not clearly understood. 
The tensions associated with $\lambda^i$ ($i=4,\dotsc,n-2$) do not have a clear interpretation either. 

The tension associated with $\lambda^{n-1}=\delta -e_n \sim -e_n$ is $\cT_0=1/R_n$\,. 
This can be regarded as the mass of the pp-wave, and the corresponding multiplet is called the \emph{particle multiplet}. 

Finally, the tension associated with $\lambda^{n}$ is $\cT_2=\lp^{-3}$ which is nothing but the tension of the M2-brane. 
The corresponding multiplet is known as the \emph{membrane multiplet}. 
This completes the fundamental representations of the $E_{n(n)}$ $U$-duality group. 

We can also consider the \emph{4-brane multiplet} by considering a tension of the M5-brane wrapped along the $x^1$-direction, $\cT_4=R_1/\lp^6$\,. 
This corresponds to a weight $\lambda^{(4)}=e_1-2\,e_0 = \lambda_1 + \lambda_n$\,. 
Thus, the 4-brane multiplet is the representation labelled by the Dynkin label $[1,0,\dotsc,0,1]$\,. 
Similarly, associated with the tension of the M5-brane $T^{(\text{\tiny M5})}_5=1/\lp^6$ that corresponds to $\lambda^{(\text{\tiny M5})}=2\,\lambda_n$, the representation for the \emph{5-brane multiplet} is labelled by the Dynkin label $[0,0,\dotsc,0,2]$. 
There is another 5-brane multiplet associated with the KKM with the Taub--NUT direction given by the $x^1$-direction and wrapping along the $x^2$-direction. 
The tension is given by $T^{(\text{\tiny KKM})}_5=R_1^2R_2/\lp^9$ and it corresponds to $\lambda^{(\text{\tiny KKM})}=\lambda_1 + \lambda_2$\,. 
Namely, the second 5-brane multiplet has the Dynkin label $[1,1,0,\dotsc,0]$. 
Higher $p$-brane multiplets can also be constructed similarly. 
We can summarize this subsection with Table \ref{tab:Dynkin-label}.
\begin{table}[tb]
 \begin{center}
 \begin{tabular}{|c|c|c|c|}\hline
 $p$ & Dynkin label & Tension & M-theory brane \\ \hline\hline
 $0$ & $[0,0,0,0,\dotsc,0,1,0]$ & $\cT_0=1/R_n$ & P \\\hline
 $1$ & $[1,0,0,0,\dotsc,0,0,0]$ & $\cT_1=R_1/\lp^3$ & M2 \\\hline
 $2$ & $[0,0,0,0,\dotsc,0,0,1]$ & $\cT_2=1/\lp^{3}$ & M2 \\\hline
 $3$ & $[0,1,0,0,\dotsc,0,0,0]$ & $\cT_3=R_1R_2/\lp^6$ & M5 \\\hline
 $4$ & $[1,0,0,0,\dotsc,0,0,1]$ & $\cT_4=R_1/\lp^6$ & M5 \\\hline
 $5$ & $[1,1,0,0,\dotsc,0,0,0]$ & $\cT_5=R_1^2R_2/\lp^9$ & KKM \\\hline
 $5$ & $[0,0,0,0,\dotsc,0,0,2]$ & $\cT_5=1/\lp^6$ & M5 \\\hline
 $5$ & $[0,0,1,0,\dotsc,0,0,0]$ & $\cT_5=R_1R_2R_3/\lp^9$ & ? \\\hline
\end{tabular}
\end{center}
\caption{Dynkin labels of the $p$-brane multiplets, and the tension associated with the highest weight vector and the associated M-theory brane.}
\label{tab:Dynkin-label}
\end{table}

\subsubsection{Example: string multiplet in $E_{6(6)}$}

As an example, let us consider a string multiplet in M-theory compactified on $T^6$\,, where the $U$-duality group is $E_{6(6)}$\,. 
We start from the highest weight vector $[1,0,0,0,0,0]$ that corresponds to a $2_3$-brane wrapped along the $x^1$-direction. 
In order to indicate that the $2_3$-brane behaves as a string in the external five-dimensional spacetime, we denote it as $2_3(\cdot 1)$-brane, where the dot ``$\ \cdot \ $'' corresponds to one external dimension. 
As described in Table \ref{tab:E66-string}, by subtracting the simple roots, we can obtain the weight diagram for the 27-dimensional string multiplet. 
\begin{table}
 \begin{center}
\begin{tikzpicture}[baseline=(A.center), scale=0.7, every node/.style={scale=0.7}]
\matrix (A) [matrix of math nodes, row sep=7mm, column sep=3mm]
{ 
& & \node (100000) {[1,0,0,0,0,0]}; \\
& & \node (-110000) {[-1,1,0,0,0,0]}; \\
& & \node (0-11000) {[0,-1,1,0,0,0]}; \\
& & \node (00-1101) {[0,0,-1,1,0,0,1]}; \\
& \node (000-111) {[0,0,0,-1,1,1]}; & & \node (00010-1) {[0,0,0,1,0,-1]}; \\
\node (0000-11) {[0,0,0,0,-1,1]}; & & \node (001-11-1) {[0,0,1,-1,1,-1]}; \\
& \node (0010-1-1) {[0,0,1,0,-1,-1]}; & & \node (01-1010) {[0,1,-1,0,1,0]}; \\
& & \node (01-11-10) {[0,1,-1,1,-1,0]}; & & \node (1-10010) {[1,-1,0,0,1,0]}; \\
& \node (010-100) {[0,1,0,-1,0,0]}; & & \node (1-101-10) {[1,-1,0,1,-1,0]}; & & \node (-100010) {[-1,0,0,0,1,0]}; \\ 
& & \node (1-11-100) {[1,-1,1,-1,0,0]}; & & \node (-1001-10) {[-1,0,0,1,-1,0]}; \\
& \node (10-1001) {[1,0,-1,0,0,1]}; & & \node (-101-100) {[-1,0,1,-1,0,0]}; \\ 
\node (10000-1) {[1,0,0,0,0,-1]}; & & \node (-11-1001) {[-1,1,-1,0,0,1]}; \\
& \node (-11000-1) {[-1,1,0,0,0,-1]}; & & \node (0-10001) {[0,-1,0,0,0,1]}; \\ 
& & \node (0-1100-1) {[0,-1,1,0,0,-1]}; \\
& & \node (00-1100) {[0,0,-1,1,0,0]}; \\
& & \node (000-110) {[0,0,0,-1,1,0]}; \\
& & \node (0000-10) {[0,0,0,0,-1,0]}; \\
};
\draw[-latex] (100000.south) -- (-110000.north) node[midway,right=1.5mm] {$\alpha_1$};
\draw[-latex] (-110000.south) -- (0-11000.north) node[midway,right=1.5mm] {$\alpha_2$};
\draw[-latex] (0-11000.south) -- (00-1101.north) node[midway,right=1.5mm] {$\alpha_3$};
\draw[-latex] ([xshift=-6mm]00-1101.south) -- (000-111.north) node[midway,left=5mm] {$\alpha_4$};
\draw[-latex] ([xshift=6mm]00-1101.south) -- (00010-1.north) node[midway,right=5mm] {$\alpha_6$};
\draw[-latex] ([xshift=-6mm]000-111.south) -- (0000-11.north) node[midway,left=5mm] {$\alpha_5$};
\draw[-latex] ([xshift=6mm]000-111.south) -- ([xshift=-6mm]001-11-1.north) node[midway,right=5mm] {$\alpha_3$};
\draw[-latex] ([xshift=0mm]00010-1.south) -- ([xshift=6mm]001-11-1.north) node[midway,left=5mm] {$\alpha_4$};
\draw[-latex] ([xshift=0mm]0000-11.south) -- ([xshift=-6mm]0010-1-1.north) node[midway,right=5mm,yshift=0mm] {$\alpha_6$};
\draw[-latex] ([xshift=-6mm]001-11-1.south) -- ([xshift=6mm]0010-1-1.north) node[midway,left=5mm] {$\alpha_5$};
\draw[-latex] ([xshift=6mm]001-11-1.south) -- ([xshift=-6mm]01-1010.north) node[midway,right=5mm,yshift=0mm] {$\alpha_3$};
\draw[-latex] ([xshift=0mm]0010-1-1.south) -- ([xshift=-6mm]01-11-10.north) node[midway,right=5mm,yshift=0mm] {$\alpha_3$};
\draw[-latex] ([xshift=-6mm]01-1010.south) -- ([xshift=6mm]01-11-10.north) node[midway,left=5mm] {$\alpha_5$};
\draw[-latex] ([xshift=6mm]01-1010.south) -- ([xshift=-6mm]1-10010.north) node[midway,right=5mm,yshift=0mm] {$\alpha_2$};
\draw[-latex] ([xshift=-6mm]01-11-10.south) -- ([xshift=0mm]010-100.north) node[midway,left=5mm] {$\alpha_4$};
\draw[-latex] ([xshift=6mm]01-11-10.south) -- ([xshift=-6mm]1-101-10.north) node[midway,right=5mm,yshift=0mm] {$\alpha_2$};
\draw[-latex] ([xshift=-6mm]1-10010.south) -- ([xshift=6mm]1-101-10.north) node[midway,left=5mm] {$\alpha_5$};
\draw[-latex] ([xshift=6mm]1-10010.south) -- ([xshift=0mm]-100010.north) node[midway,right=5mm,yshift=0mm] {$\alpha_1$};
\draw[-latex] ([xshift=0mm]010-100.south) -- ([xshift=-6mm]1-11-100.north) node[midway,right=5mm,yshift=0mm] {$\alpha_2$};
\draw[-latex] ([xshift=-6mm]1-101-10.south) -- ([xshift=6mm]1-11-100.north) node[midway,left=5mm] {$\alpha_4$};
\draw[-latex] ([xshift=6mm]1-101-10.south) -- ([xshift=-6mm]-1001-10.north) node[midway,right=5mm,yshift=0mm] {$\alpha_1$};
\draw[-latex] ([xshift=0mm]-100010.south) -- ([xshift=6mm]-1001-10.north) node[midway,left=5mm] {$\alpha_5$};
\draw[-latex] ([xshift=-6mm]1-11-100.south) -- ([xshift=0mm]10-1001.north) node[midway,left=5mm] {$\alpha_3$};
\draw[-latex] ([xshift=6mm]1-11-100.south) -- ([xshift=-6mm]-101-100.north) node[midway,right=5mm,yshift=0mm] {$\alpha_1$};
\draw[-latex] ([xshift=0mm]-1001-10.south) -- ([xshift=6mm]-101-100.north) node[midway,left=5mm] {$\alpha_4$};
\draw[-latex] ([xshift=-6mm]10-1001.south) -- ([xshift=0mm]10000-1.north) node[midway,left=5mm] {$\alpha_6$};
\draw[-latex] ([xshift=6mm]10-1001.south) -- ([xshift=-6mm]-11-1001.north) node[midway,right=5mm,yshift=0mm] {$\alpha_1$};
\draw[-latex] ([xshift=0mm]-101-100.south) -- ([xshift=6mm]-11-1001.north) node[midway,left=5mm] {$\alpha_3$};
\draw[-latex] ([xshift=0mm]10000-1.south) -- ([xshift=-6mm]-11000-1.north) node[midway,right=5mm,yshift=0mm] {$\alpha_1$};
\draw[-latex] ([xshift=-6mm]-11-1001.south) -- ([xshift=6mm]-11000-1.north) node[midway,left=5mm] {$\alpha_6$};
\draw[-latex] ([xshift=6mm]-11-1001.south) -- ([xshift=0mm]0-10001.north) node[midway,right=5mm,yshift=0mm] {$\alpha_2$};
\draw[-latex] ([xshift=0mm]-11000-1.south) -- ([xshift=-6mm]0-1100-1.north) node[midway,right=5mm,yshift=0mm] {$\alpha_2$};
\draw[-latex] ([xshift=0mm]0-10001.south) -- ([xshift=6mm]0-1100-1.north) node[midway,left=5mm] {$\alpha_6$};
\draw[-latex] (0-1100-1.south) -- (00-1100.north) node[midway,right=1.5mm] {$\alpha_3$};
\draw[-latex] (00-1100.south) -- (000-110.north) node[midway,right=1.5mm] {$\alpha_4$};
\draw[-latex] (000-110.south) -- (0000-10.north) node[midway,right=1.5mm] {$\alpha_5$};
\node[red,right] at (100000.east) {$2_3(\cdot 1)$};
\node[red,right] at (-110000.east) {$2_3(\cdot 2)$};
\node[red,right] at (0-11000.east) {$2_3(\cdot 3)$};
\node[red,right] at (00-1101.east) {$2_3(\cdot 4)$};
\node[red,right] at (000-111.east) {$2_3(\cdot 5)$};
\node[red,right] at (00010-1.east) {$5_6(\cdot 1234)$};
\node[red,right] at (0000-11.east) {$2_3(\cdot 6)$};
\node[red,right] at (001-11-1.east) {$5_6(\cdot 1235)$};
\node[red,right] at (0010-1-1.east) {$5_6(\cdot 1236)$};
\node[red,right] at (01-1010.east) {$5_6(\cdot 1245)$};
\node[red,right] at (01-11-10.east) {$5_6(\cdot 1246)$};
\node[red,right] at (1-10010.east) {$5_6(\cdot 1345)$};
\node[red,right] at (010-100.east) {$5_6(\cdot 1256)$};
\node[red,right] at (1-101-10.east) {$5_6(\cdot 1346)$};
\node[red,right] at (-100010.east) {$5_6(\cdot 2345)$};
\node[red,right] at (1-11-100.east) {$5_6(\cdot 1356)$};
\node[red,right] at (-1001-10.east) {$5_6(\cdot 2346)$};
\node[red,right] at (10-1001.east) {$5_6(\cdot 1456)$};
\node[red,right] at (-101-100.east) {$5_6(\cdot 2356)$};
\node[red,right] at (10000-1.east) {$6^1_9(\cdot 23456,1)$};
\node[red,right] at (-11-1001.east) {$5_6(\cdot 2456)$};
\node[red,right] at (-11000-1.east) {$6^1_9(\cdot 13456,2)$};
\node[red,right] at (0-10001.east) {$5_6(\cdot 3456)$};
\node[red,right] at (0-1100-1.east) {$6^1_9(\cdot 12456,3)$};
\node[red,right] at (00-1100.east) {$6^1_9(\cdot 12356,4)$};
\node[red,right] at (000-110.east) {$6^1_9(\cdot 12346,5)$};
\node[red,right] at (0000-10.east) {$6^1_9(\cdot 12345,6)$};
\end{tikzpicture}
\end{center}
\caption{The weight diagram for the string multiplet in $E_{6(6)}$ case.}
\label{tab:E66-string}
\end{table}

Here, let us briefly explain how to make the identification between the Dynkin labels and the brane charges. 
When we subtract a simple root $\alpha_{n} = e_0-(e_1+e_2+e_3)$\,, the brane tension is multiplied by $R_1R_2R_3/\lp^3$\,. 
At the same time, the Dynkin label is reduced by $[a_{n1},\dotsc,a_{nn}]$ corresponding to $\alpha_{n} = a_{nj}\,\lambda^j$\,. 
Similarly, when we subtract $\alpha_k=e_k-e_{k+1}=a_{ki}\,\lambda^i$ ($k\neq n-1$), the brane tension is multiplied by $R_k/R_{k+1}$ and the Dynkin label is reduced by $[a_{k1},\dotsc,a_{kn}]$\,. 
On the other hand, when we subtract $\alpha_{n-1} = e_{n-1}-e_{n} = a_{(n-1)j}\,\lambda^j - \delta$, the brane tension is multiplied by $R_{n-1}/R_{n}$\,. 
In this case, the Dynkin label is reduced by $[a_{(n-1)1},\dotsc,a_{(n-1)n}]$ and the information about $\delta$, which corresponds to the inverse of the $d$-dimensional Newton constant $l_d$, is lost. 
Accordingly, when we try to reproduce the tension from the Dynkin label, we should introduce $l_d$ appropriately. 
For example, the Dynkin label $[0,-1,1,0,0,-1]$ corresponds to
\begin{align}
 -\lambda_2+\lambda_3-\lambda_6\quad \leftrightarrow\quad R_3 \,. 
\end{align}
In order to make the mass dimension the same as that of the string tension, we multiply it by $l_5$ and obtain
\begin{align}
 \cT_1 = l_5\times R_3 = \frac{R_1R_2R_4R_5R_6R_3^2}{\lp^9}\,,
\end{align}
By using the convention \eqref{eq:tension}, this is interpreted as the tension of the $6^1_9(\cdot 12456,3)$-brane. 
From a similar consideration, we can find the identifications between the Dynkin labels and branes shown in Table \ref{tab:E66-string}. 

From Table \ref{tab:E66-string}, we can summarize the detailed number of degeneracy as
\begin{align}
 2_3~(\text{\bf 6})\,,\qquad 
 5_6~(\text{\bf 15})\,,\qquad 
 6_9^1~(\text{\bf 6})\,. 
\end{align}
which is consistent with Table \ref{tab:BPS-1-brane}. 
It is noted that, in this case, all of the states correspond to weight vectors with the same length, and the 27 states are connected via the Weyl reflections, or the $U$-duality transformations \eqref{eq:U-dual}. 

\subsection{Web of supersymmetric branes}

\begin{table}
 \begin{center}
{\scriptsize
 \begin{tabular}{|@{\,}c@{\,}|@{\,}c@{\,}|@{\,}c@{\,}|@{\,}c@{\,}|@{\,}c@{\,}|@{\,}c@{\,}|@{\,}c@{\,}|@{\,}c@{\,}|@{\,}c@{\,}|@{\,}c@{\,}|}\hline
 $d\backslash p$ & 0 & 1 & 2 & 3 & 4 & 5 & 6 & 7 & 8 \\ \hline\hline
 9 &    3 &    2 &     1 &    1 &   2 & $2+1$ &  \Def{$2\subset 3$} & \DoW{$2\subset 3$} & \SpF{$2\subset 4$} \\ \hline\hline
 8 &    6 &    3 &     2 &    3 &   6 & \Def{$6+2\subset 8\oplus 3$} & \DoW{$6\subset 12$} & \SpF{$6\subset 15$} &   \\ \hline\hline
 7 &   10 &    5 &     5 &   10 & \Def{$20\subset 24$} & \DoW{$20+5\subset 40\oplus 15$} & \SpF{$20\subset 70$} &   &   \\ \hline\hline
 6 &   16 &   10 & 16 & \Def{$40\subset 45$} & \DoW{$80\subset 144$} & \SpF{$80+16\subset 320\oplus 126$} &    &   &   \\ \hline\hline
 5 &   27 &   27 & \Def{$72\subset 78$} & \DoW{$216\subset 351$} & \SpF{$432\subset 1728$} &    &    &   &   \\ \hline\hline
 4 & 56 & \Def{$126\subset 133$} & \DoW{$576\subset 912$} & \SpF{$2016\subset 8645$} &     &    &    &   &   \\ \hline\hline
 3 & \Def{$240 \subset 248$} & \DoW{$2160\subset 3875$} & \SpF{$17280\subset 147250$} &      &     &    &    &   &   \\ \hline\hline
\end{tabular}
}
\end{center}
\caption{Dimensions of the $p$-brane multiplet and numbers of branes in the same Weyl orbit as the highest weight state. 
Only the 5-brane multiplet consists of two irreducible representations. 
This table was originally obtained in \cite{Kleinschmidt:2011vu} from a discussion based on the $E_{11}$.}
\label{tab:dim-multiplets}
\end{table}

Utilizing the duality transformation rule \eqref{eq:U-dual}, we can generate a chain of exotic branes in M-theory. 
Indeed, by brute force applications of duality \eqref{eq:U-dual} to the tensions of the standard branes, we obtain Tables \ref{tab:BPS-0-brane}--\ref{tab:BPS-8-brane}, which show the explicit brane charges and the degeneracies in each multiplet. 
By summing up the degeneracies of all branes, we obtain the size of the Weyl orbit in each $p$-brane multiplet in $d$-dimensions, as summarized in Table \ref{tab:dim-multiplets}. 
In M-theory compactified to $d$-dimensions with $d\geq 3$, all of the branes appearing in the Weyl orbit are summarized as follows (potentials that couple to the following branes are listed in \cite{Kleinschmidt:2011vu}):
\begin{align}
\begin{split}
 &\text{P}\,,\ 
  2_3\,,\ 
  5_6\,,\ 
  6_{9}^{1}\,,\ 
  \Def{5_{12}^{3}}\,,\ 
  \DoW{8_{12}^{(1,0)}}\,,\ 
  \Def{2_{15}^{6}}\,,\ 
  \DoW{5_{15}^{(1,3)}}\,,\ 
  \Def{0_{18}^{(1,7)}}\,,\ 
  \DoW{3_{18}^{(2,4)}}\,,\ 
  \SpF{6_{18}^{(3,1)}}\,,\ 
  \SpF{5_{18}^{(1,0,4)}}\,,
\\
 &\DoW{2_{21}^{(4,3)}}\,,\ 
  \DoW{1_{21}^{(1,1,6)}}\,,\ 
  \SpF{4_{21}^{(1,2,3)}}\,,\ 
  \DoW{2_{24}^{(7,0)}}\,,\ 
  \DoW{1_{24}^{(1,4,3)}}\,,\ 
  \SpF{4_{24}^{(1,5,0)}}\,,\ 
  \SpF{3_{24}^{(2,2,3)}}\,,\ 
  \SpF{2_{24}^{(1,0,2,5)}}\,,
\\
 &\DoW{1_{27}^{(2,5,1)}}\,,\ 
  \SpF{3_{27}^{(3,3,1)}}\,,\ 
  \SpF{2_{27}^{(4,0,4)}}\,,\ 
  \SpF{2_{27}^{(1,1,3,3)}}\,,\ 
  \DoW{1_{30}^{(4,4,0)}}\,,\ 
  \SpF{3_{30}^{(5,2,0)}}\,,\ 
  \SpF{2_{30}^{(1,3,2,2)}}\,,\ 
  \SpF{2_{30}^{(2,0,5,1)}}\,,\ 
  \SpF{2_{30}^{(1,0,0,7,0)}}\,,
\\
 &\DoW{1_{33}^{(7,1,0)}}\,,\ 
  \SpF{2_{33}^{(2,3,2,1)}}\,,\ 
  \SpF{2_{33}^{(1,0,3,4,0)}}\,,\ 
  \SpF{2_{36}^{(3,4,0,1)}}\,,\ 
  \SpF{2_{36}^{(4,1,3,0)}}\,,\ 
  \SpF{2_{36}^{(1,1,4,2,0)}}\,,\ 
  \SpF{2_{39}^{(1,3,3,1,0)}}\,,\ 
  \SpF{2_{39}^{(2,0,6,0,0)}}\,,
\\
 &\SpF{2_{42}^{(1,6,0,1,0)}}\,,\ 
  \SpF{2_{42}^{(2,3,3,0,0)}}\,,\ 
  \SpF{2_{45}^{(3,4,1,0,0)}}\,,\ 
  \SpF{2_{48}^{(5,3,0,0,0)}}\,,\ 
  \SpF{2_{51}^{(8,0,0,0,0)}}\,. 
\end{split}
\label{eq:all-M-branes}
\end{align}
In the literature, $2_3$, $5_6$, $6_{9}^{1}$, and $8_{12}^{(1,0)}$ are respectively called M2, M5, KKM, and M9-brane while the others do not have familiar common names. 
We consider a $b_{n}^{(c_s,\dotsc,c_2)}$-brane as a kind of $(b+c_2+\cdots+c_s)$-brane, and the codimension is given by $10-(b+c_2+\cdots+c_s)$. 
If the codimension of a brane is equal to 2, 1, or 0, we call it a \emph{defect brane}, a \emph{domain-wall brane}, or a \emph{space-filling brane}, respectively. 
These are also called the \emph{non-standard branes} while branes with codimension 3 or greater are called the \emph{standard branes}. 
For clarification, in \eqref{eq:all-M-branes}, we have colored the defect branes, the domain-wall branes, and the space-filling branes in \colorDef, \colorDoW, and \colorSpF, respectively. 

As we can see from Table \ref{tab:dim-multiplets}, in each dimension $d$, there is a symmetry between the dimensions of the $p$-brane multiplet and $(d-4-p)$ for $p\leq d-4$\,. 
The representation for $p=d-3$ is always the adjoint representation of the $E_{n(n)}$ group.

In the previous subsection, we have shown the Dynkin labels only for $p$-brane multiplets with $p\leq 5$\,. 
In $d\geq 7$\,, we also have the $p$-brane multiplet with $p=6,7,8$\,. 
For the $6$-brane multiplet, the highest weight corresponds to a $6^1_9$-brane with the Taub--NUT direction given by the $x^1$-direction. 
The tension is $\cT_6=R_1^2/\lp^9$ and it corresponds to $2\,\lambda_1 + \lambda_n$\,, whose Dynkin label is $[2,0,0,1]$ in $d=7$ and $[2,0,1]$ in $d=8$\,. 
For the $7$-brane multiplet, the highest weight corresponds to a $8_{12}^{(1,0)}$-brane with the special isometry direction given by the $x^1$-direction and wrapped along the $x^2$-direction. 
The tension is $\cT_7=R_1^3R_2/\lp^{12}$\,, corresponding to $2\,\lambda_1 + \lambda_2$ and $[2,1,0]$ in $d=8$. 

We can also consider the type II branes by using the 11D/10D relation \eqref{eq:11D-10D}, and the type IIA branes associated with all of the ``elementary'' M-branes are obtained in Table \ref{tab:M-IIA-map}. 
\begin{table}
\begin{align*}
\tiny
\xymatrix@C=1pt@R=0pt{
 \quad\text{M-branes}\quad & \quad\text{IIA branes}\quad & \quad\text{M-branes}\quad & \quad\text{IIA branes}\quad &
 \quad\text{M-branes}\quad & \quad\text{IIA branes}\quad & \quad\text{M-branes}\quad & \quad\text{IIA branes}\quad \\
 \text{P} \ar[r] \ar@<-0.5ex>[dr] & \text{P} & 
 2_{3} \ar[r] \ar@<-0.5ex>[dr] & 1_0 & 
 5_{6} \ar[r] \ar@<-0.5ex>[dr] & 4_1 & 
 6^{1}_{9} \ar[r] \ar@/_/@<-0.5ex>[dr] \ar@/_/@<-1ex>[ddr] & 6_1 
\\
  & 0_1 &
  & 2_1 &
  & 5_2 &
  & 5^1_2 
\\
  & & 
  & & 
  & & 
  & 6^1_3 
\\
 5^{3}_{12} \ar[r] \ar@/_/@<-0.5ex>[dr] \ar@/_/@<-1ex>[ddr] & 5^2_2 & 
 8^{(1, 0)}_{12} \ar[r] \ar@/_/@<-0.5ex>[dr] \ar@/_/@<-1ex>[ddr] & 8_1 & 
 2^{6}_{15} \ar[r] \ar@/_/@<-0.5ex>[dr] \ar@/_/@<-1ex>[ddr] & 2^5_3 & 
 5^{(1, 3)}_{15} \ar[r] \ar@/_/@<-0.5ex>[dr] \ar@/_/@<-1ex>[ddr] \ar@/_/@<-1.5ex>[dddr] & 5_2^3 
\\
  & 4^3_3 &
  & 7^{(1, 0)}_3 &
  & 1^6_4 &
  & 5_3^{(1, 2)}
\\
  & 5^{3}_4 & 
  & 8^{(1, 0)}_4 & 
  & 2^6_5 & 
  & 4_4^{(1, 3)} 
\\
  & & 
  & & 
  & & 
  & 5_5^{(1, 3)}
\\
 0^{(1, 7)}_{18} \ar[r] \ar@<-0.5ex>[dr] & 0^7_3 & 
 3^{(2, 4)}_{18} \ar[r] \ar@/_/@<-0.5ex>[dr] \ar@/_/@<-1ex>[ddr] \ar@/_/@<-1.5ex>[dddr] & 3^{(1, 4)}_3 & 
 5^{(1, 0, 4)}_{18} \ar[r] \ar@/_/@<-0.5ex>[dr] \ar@/_/@<-1ex>[ddr] & 5^{4}_2 & 
 6^{(3, 1)}_{18} \ar[r] \ar@/_/@<-0.5ex>[dr] \ar@/_/@<-1ex>[ddr] & 6^{(2, 1)}_3
\\
  & 0^{(1, 6)}_4 &
  & 3^{(2, 3)}_4 &
  & 5^{(1, 0, 3)}_4 &
  & 6^{(3, 0)}_4
\\
 & & 
  & 2^{(2, 4)}_5 & 
  & 4^{(1, 0, 4)}_5 & 
  & 5^{(3, 1)}_5 & 
\\
 & & 
  & 3^{(2, 4)}_6 &
 & & 
 & 
\\
 2^{(4, 3)}_{21} \ar[r] \ar@/_/@<-0.5ex>[dr] \ar@/_/@<-1ex>[ddr] \ar@/_/@<-1.5ex>[dddr] & 2^{(3, 3)}_{4} & 
 1^{(1, 1, 6)}_{21} \ar[r] \ar@/_/@<-0.5ex>[dr] \ar@/_/@<-1ex>[ddr] & 1^{(1, 6)}_{3} & 
 4^{(1, 2, 3)}_{21} \ar[r] \ar@/_/@<-0.5ex>[dr] \ar@/_/@<-1ex>[ddr] \ar@/_/@<-1.5ex>[dddr] & 4^{(2, 3)}_{3} & 
 2^{(7, 0)}_{24} \ar[r] \ar@/_/@<-0.5ex>[dr] \ar@/_/@<-1ex>[ddr] & 2^{(6, 0)}_{5}
\\
  & 2^{(4, 2)}_{5} &
  & 1^{(1, 0, 6)}_{4} &
  & 4^{(1, 1, 3)}_{4} &
  & 1^{(7, 0)}_{7}
\\
  & 1^{(4, 3)}_{6} &
  & 1^{(1, 1, 5)}_{5} &
  & 4^{(1, 2, 2)}_{5} &
  & 2^{(7, 0)}_{8} &
\\
  & 2^{(4, 3)}_{7} &
 & & 
  & 3^{(1, 2, 3)}_{6} &
 & 
\\
 1^{(1, 4, 3)}_{24} \ar[r] \ar@/_/@<-0.5ex>[dr] \ar@/_/@<-1ex>[ddr] & 1^{(4, 3)}_{4} &
 4^{(1, 5, 0)}_{24} \ar[r] \ar@/_/@<-0.5ex>[dr] \ar@/_/@<-1ex>[ddr] & 4^{(5, 0)}_{4} & 
 3^{(2, 2, 3)}_{24} \ar[r] \ar@/_/@<-0.5ex>[dr] \ar@/_/@<-1ex>[ddr] \ar@/_/@<-1.5ex>[dddr] & 3^{(1, 2, 3)}_{4} & 
 2^{(1, 0, 2, 5)}_{24} \ar[r] \ar@/_/@<-0.5ex>[dr] \ar@/_/@<-1ex>[ddr] & 2^{(2, 5)}_{3} 
\\
  & 1^{(1, 3, 3)}_{5} &
  & 4^{(1, 4, 0)}_{5} &
  & 3^{(2, 1, 3)}_{5} &
  & 2^{(1, 0, 1, 5)}_{5} &
\\
  & 1^{(1, 4, 2)}_{6} &
  & 3^{(1, 5, 0)}_{7} &
  & 3^{(2, 2, 2)}_{6} &
  & 2^{(1, 0, 2, 4)}_{6} &
\\
 & & 
 & & 
  & 2^{(2, 2, 3)}_{7} &
 & 
\\
 1^{(2, 5, 1)}_{27} \ar[r] \ar@/_/@<-0.5ex>[dr] \ar@/_/@<-1ex>[ddr] & 1^{(1, 5, 1)}_{5} & 
 3^{(3, 3, 1)}_{27} \ar[r] \ar@/_/@<-0.5ex>[dr] \ar@/_/@<-1ex>[ddr] \ar@/_/@<-1.5ex>[dddr] & 3^{(2, 3, 1)}_{5} & 
 2^{(4, 0, 4)}_{27} \ar[r] \ar@<-0.5ex>[dr] & 2^{(3, 0, 4)}_{5} & 
 2^{(1, 1, 3, 3)}_{27} \ar[r] \ar@/_/@<-0.5ex>[dr] \ar@/_/@<-1ex>[ddr] \ar@/_/@<-1.5ex>[dddr] & 2^{(1, 3, 3)}_{4} 
\\
  & 1^{(2, 4, 1)}_{6} &
  & 3^{(3, 2, 1)}_{6} &
  & 2^{(4, 0, 3)}_{7} &
  & 2^{(1, 0, 3, 3)}_{5}
\\
  & 1^{(2, 5, 0)}_{7} &
  & 3^{(3, 3, 0)}_{7} &
 & & 
  & 2^{(1, 1, 2, 3)}_{6} 
\\
 & & 
  & 2^{(3, 3, 1)}_{8} &
 & & 
  & 2^{(1, 1, 3, 2)}_{7}
\\
 1^{(4, 4, 0)}_{30} \ar[r] \ar@<-0.5ex>[dr] & 1^{(3, 4, 0)}_{6} & 
 3^{(5, 2, 0)}_{30} \ar[r] \ar@/_/@<-0.5ex>[dr] \ar@/_/@<-1ex>[ddr] & 3^{(4, 2, 0)}_{6} & 
 2^{(1, 3, 2, 2)}_{30} \ar[r] \ar@/_/@<-0.5ex>[dr] \ar@/_/@<-1ex>[ddr] \ar@/_/@<-1.5ex>[dddr] & 2^{(3, 2, 2)}_{5} & 
 2^{(2, 0, 5, 1)}_{30} \ar[r] \ar@/_/@<-0.5ex>[dr] \ar@/_/@<-1ex>[ddr] & 2^{(1, 0, 5, 1)}_{5} & 
\\
  & 1^{(4, 3, 0)}_{7} &
  & 3^{(5, 1, 0)}_{7} &
  & 2^{(1, 2, 2, 2)}_{6} &
  & 2^{(2, 0, 4, 1)}_{7} 
\\
  & & 
  & 2^{(5, 2, 0)}_{9} &
  & 2^{(1, 3, 1, 2)}_{7} &
  & 2^{(2, 0, 5, 0)}_{8}
\\
  & &
 & & 
  & 2^{(1, 3, 2, 1)}_{8} &
 & 
\\
 2^{(1, 0, 0, 7, 0)}_{30} \ar[r] \ar@<-0.5ex>[dr] & 2^{(7, 0)}_{4} & 
 2^{(2, 3, 2, 1)}_{33} \ar[r] \ar@/_/@<-0.5ex>[dr] \ar@/_/@<-1ex>[ddr] \ar@/_/@<-1.5ex>[dddr] & 2^{(1, 3, 2, 1)}_{6} & 
 2^{(1, 0, 3, 4, 0)}_{33} \ar[r] \ar@/_/@<-0.5ex>[dr] \ar@/_/@<-1ex>[ddr] & 2^{(3, 4, 0)}_{5} & 
 2^{(3, 4, 0, 1)}_{36} \ar[r] \ar@/_/@<-0.5ex>[dr] \ar@/_/@<-1ex>[ddr] & 2^{(2, 4, 0, 1)}_{7}
\\
  & 2^{(1, 0, 0, 6, 0)}_{7} &
  & 2^{(2, 2, 2, 1)}_{7} &
  & 2^{(1, 0, 2, 4, 0)}_{7} &
  & 2^{(3, 3, 0, 1)}_{8} 
\\
 1^{(7, 1, 0)}_{33} \ar[r] \ar@<-0.5ex>[dr] & 1^{(6, 1, 0)}_{7} & 
  & 2^{(2, 3, 1, 1)}_{8} &
  & 2^{(1, 0, 3, 3, 0)}_{8} &
  & 2^{(3, 4, 0, 0)}_{10} 
\\
 & 1^{(7, 0, 0)}_{8} & 
  & 2^{(2, 3, 2, 0)}_{9} &
 & & 
 & 
\\
 2^{(4, 1, 3, 0)}_{36} \ar[r] \ar@/_/@<-0.5ex>[dr] \ar@/_/@<-1ex>[ddr] & 2^{(3, 1, 3, 0)}_{7} & 
 2^{(1, 1, 4, 2, 0)}_{36} \ar[r] \ar@/_/@<-0.5ex>[dr] \ar@/_/@<-1ex>[ddr] \ar@/_/@<-1.5ex>[dddr] & 2^{(1, 4, 2, 0)}_{6} & 
 2^{(1, 3, 3, 1, 0)}_{39} \ar[r] \ar@/_/@<-0.5ex>[dr] \ar@/_/@<-1ex>[ddr] \ar@/_/@<-1.5ex>[dddr] & 2^{(3, 3, 1, 0)}_{7} & 
 2^{(2, 0, 6, 0, 0)}_{39} \ar[r] \ar@<-0.5ex>[dr] & 2^{(1, 0, 6, 0, 0)}_{7} &
\\
  & 2^{(4, 0, 3, 0)}_{8} & 
  & 2^{(1, 0, 4, 2, 0)}_{7} & 
  & 2^{(1, 2, 3, 1, 0)}_{8} & 
  & 2^{(2, 0, 5, 0, 0)}_{9} & 
\\
  & 2^{(4, 1, 2, 0)}_{9} & 
  & 2^{(1, 1, 3, 2, 0)}_{8} & 
  & 2^{(1, 3, 2, 1, 0)}_{9} & 
 & & 
\\
 & & 
  & 2^{(1, 1, 4, 1, 0)}_{9} & 
  & 2^{(1, 3, 3, 0, 0)}_{10} & 
 & & 
\\
 2^{(1, 6, 0, 1, 0)}_{42} \ar[r] \ar@/_/@<-0.5ex>[dr] \ar@/_/@<-1ex>[ddr] & 2^{(6, 0, 1, 0)}_{8} & 
 2^{(2, 3, 3, 0, 0)}_{42} \ar[r] \ar@/_/@<-0.5ex>[dr] \ar@/_/@<-1ex>[ddr] & 2^{(1, 3, 3, 0, 0)}_{8} & 
 2^{(3, 4, 1, 0, 0)}_{45} \ar[r] \ar@/_/@<-0.5ex>[dr] \ar@/_/@<-1ex>[ddr] & 2^{(2, 4, 1, 0, 0)}_{9} & 
 2^{(5, 3, 0, 0, 0)}_{48} \ar[r] \ar@<-0.5ex>[dr] & 2^{(4, 3, 0, 0, 0)}_{10} 
\\
  & 2^{(1, 5, 0, 1, 0)}_{9} & 
  & 2^{(2, 2, 3, 0, 0)}_{9} & 
  & 2^{(3, 3, 1, 0, 0)}_{10} & 
  & 2^{(5, 2, 0, 0, 0)}_{11} 
\\
  & 2^{(1, 6, 0, 0, 0)}_{11} & 
  & 2^{(2, 3, 2, 0, 0)}_{10} & 
  & 2^{(3, 4, 0, 0, 0)}_{11} & 
 2^{(8, 0, 0, 0, 0)}_{51} \ar[r] & 2^{(7, 0, 0, 0, 0)}_{11} 
}
\end{align*}
\vspace{0mm}
\caption{A map between branes in M-theory and type IIA theory.}
\label{tab:M-IIA-map}
\end{table}
It may be more convenient to summarize a list of the type IIA branes as follows, where defect branes, domain-wall branes, and space-filling branes are colored in the same way as the M-theory branes:
\begin{align}
 &0_1\,,\ 
 \text{P}\,,\ 
 0_1\,,\ 
 2_1\,,\ 
 4_1\,,\ 
 6_1\,,\ 
 \DoW{8_1}\,,\ 
 5_2\,,\ 
 5_2^1\,,\ 
 \Def{5_{2}^{2}}\,,\ 
 \DoW{5_{2}^{3}}\,,\ 
 \SpF{5_{2}^{4}}\,,\ 
 \Def{6_{3}^{1}}\,,\ 
 \Def{4_{3}^{3}}\,,\ 
 \Def{2_{3}^{5}}\,,\ 
 \Def{0_{3}^{7}}\,,\ 
 \DoW{7_{3}^{(1,0)}}\,,\ 
 \DoW{5_{3}^{(1,2)}}\,, 
\nn\\
 &
 \DoW{3_{3}^{(1,4)}}\,,\ 
 \DoW{1_{3}^{(1,6)}}\,,\ 
 \SpF{6_{3}^{(2,1)}}\,,\ 
 \SpF{4_{3}^{(2,3)}}\,,\ 
 \SpF{2_{3}^{(2,5)}}\,,\ 
 \Def{1_{4}^{6}}\,,\ 
 \Def{0_{4}^{(1,6)}}\,,\ 
 \DoW{1_{4}^{(1,0,6)}}\,,\ 
 \DoW{5_{4}^{3}}\,,\ 
 \DoW{4_{4}^{(1,3)}}\,,\ 
 \DoW{3_{4}^{(2,3)}}\,,\ 
 \DoW{2_{4}^{(3,3)}}\,,\ 
 \DoW{1_{4}^{(4,3)}}\,, 
\nn\\
 &\SpF{5_{4}^{(1,0,3)}}\,,\ 
 \SpF{4_{4}^{(1,1,3)}}\,,\ 
 \SpF{3_{4}^{(1,2,3)}}\,,\ 
 \SpF{2_{4}^{(1,3,3)}}\,,\ 
 \SpF{8_{4}^{(1,0)}}\,,\ 
 \SpF{6_{4}^{(3,0)}}\,,\ 
 \SpF{4_{4}^{(5,0)}}\,,\ 
 \SpF{2_{4}^{(7,0)}}\,,\ 
 \DoW{2_{5}^{6}}\,,\ 
 \DoW{2_{5}^{(2,4)}}\,,\ 
 \DoW{2_{5}^{(4,2)}}\,,\ 
 \DoW{2_{5}^{(6,0)}}\,, 
\nn\\
 &\DoW{1_{5}^{(1,1,5)}}\,,\ 
 \DoW{1_{5}^{(1,3,3)}}\,,\ 
 \DoW{1_{5}^{(1,5,1)}}\,,\ 
 \SpF{2_{5}^{(1,0,1,5)}}\,,\ 
 \SpF{2_{5}^{(1,0,3,3)}}\,,\ 
 \SpF{2_{5}^{(1,0,5,1)}}\,,\ 
 \SpF{5_{5}^{(1,3)}}\,,\ 
 \SpF{5_{5}^{(3,1)}}\,,\ 
 \SpF{4_{5}^{(1,0,4)}}\,,\ 
 \SpF{4_{5}^{(1,2,2)}}\,,
\nn\\
 &\SpF{4_{5}^{(1,4,0)}}\,,\ 
 \SpF{3_{5}^{(2,1,3)}}\,,\ 
 \SpF{3_{5}^{(2,3,1)}}\,,\ 
 \SpF{2_{5}^{(3,0,4)}}\,,\ 
 \SpF{2_{5}^{(3,2,2)}}\,,\ 
 \SpF{2_{5}^{(3,4,0)}}\,,\ 
 \DoW{1_{6}^{(4,3)}}\,,\ 
 \DoW{1_{6}^{(1,4,2)}}\,,\ 
 \DoW{1_{6}^{(2,4,1)}}\,,\ 
 \DoW{1_{6}^{(3,4,0)}}\,,
\nn\\
 &\SpF{3_{6}^{(2,4)}}\,,\ 
 \SpF{3_{6}^{(1,2,3)}}\,,\ 
 \SpF{3_{6}^{(2,2,2)}}\,,\ 
 \SpF{3_{6}^{(3,2,1)}}\,,\ 
 \SpF{3_{6}^{(4,2,0)}}\,,\ 
 \SpF{2_{6}^{(1,0,2,4)}}\,,\ 
 \SpF{2_{6}^{(1,1,2,3)}}\,,\ 
 \SpF{2_{6}^{(1,2,2,2)}}\,,\ 
 \SpF{2_{6}^{(1,3,2,1)}}\,,\ 
 \SpF{2_{6}^{(1,4,2,0)}}\,, 
\nn\\
 & \DoW{1_{7}^{(7,0)}}\,,\ 
 \DoW{1_{7}^{(2,5,0)}}\,,\ 
 \DoW{1_{7}^{(4,3,0)}}\,,\ 
 \DoW{1_{7}^{(6,1,0)}}\,,\ 
 \SpF{3_{7}^{(1,5,0)}}\,,\ 
 \SpF{3_{7}^{(3,3,0)}}\,,\ 
 \SpF{3_{7}^{(5,1,0)}}\,,\ 
 \SpF{2_{7}^{(1,0,0,6,0)}}\,,\ 
 \SpF{2_{7}^{(1,0,2,4,0)}}\,,\ 
 \SpF{2_{7}^{(1,0,4,2,0)}}\,,
\nn\\
 &\SpF{2_{7}^{(1,0,6,0,0)}}\,,\ 
 \SpF{2_{7}^{(4,3)}}\,,\ 
 \SpF{2_{7}^{(2,2,3)}}\,,\ 
 \SpF{2_{7}^{(4,0,3)}}\,,\ 
 \SpF{2_{7}^{(1,1,3,2)}}\,,\ 
 \SpF{2_{7}^{(1,3,1,2)}}\,,\ 
 \SpF{2_{7}^{(2,0,4,1)}}\,,\ 
 \SpF{2_{7}^{(2,2,2,1)}}\,,\ 
 \SpF{2_{7}^{(2,4,0,1)}}\,, 
\nn\\
 & 
 \SpF{2_{7}^{(3,1,3,0)}}\,,\ 
 \SpF{2_{7}^{(3,3,1,0)}}\,,\ 
 \DoW{1_{8}^{(7,0,0)}}\,,\ 
 \SpF{2_{8}^{(7,0)}}\,,\ 
 \SpF{2_{8}^{(2,0,5,0)}}\,,\ 
 \SpF{2_{8}^{(4,0,3,0)}}\,,\ 
 \SpF{2_{8}^{(6,0,1,0)}}\,,\ 
 \SpF{2_{8}^{(3,3,1)}}\,,\ 
 \SpF{2_{8}^{(1,3,2,1)}}\,,
\nn\\
 & 
 \SpF{2_{8}^{(2,3,1,1)}}\,,\ 
 \SpF{2_{8}^{(3,3,0,1)}}\,,\ 
 \SpF{2_{8}^{(1,0,3,3,0)}}\,,\ 
 \SpF{2_{8}^{(1,1,3,2,0)}}\,,\ 
 \SpF{2_{8}^{(1,2,3,1,0)}}\,,\ 
 \SpF{2_{8}^{(1,3,3,0,0)}}\,,\ 
 \SpF{2_{9}^{(5,2,0)}}\,,\ 
 \SpF{2_{9}^{(2,3,2,0)}}\,,\ 
 \SpF{2_{9}^{(4,1,2,0)}}\,,
\nn\\
 & 
 \SpF{2_{9}^{(1,1,4,1,0)}}\,,\ 
 \SpF{2_{9}^{(1,3,2,1,0)}}\,,\ 
 \SpF{2_{9}^{(1,5,0,1,0)}}\,,\ 
 \SpF{2_{9}^{(2,0,5,0,0)}}\,,\ 
 \SpF{2_{9}^{(2,2,3,0,0)}}\,,\ 
 \SpF{2_{9}^{(2,4,1,0,0)}}\,,\ 
 \SpF{2_{10}^{(3,4,0,0)}}\,,
 \SpF{2_{10}^{(1,3,3,0,0)}}\,, 
\nn\\
 & 
 \SpF{2_{10}^{(2,3,2,0,0)}}\,,\ 
 \SpF{2_{10}^{(3,3,1,0,0)}}\,,\ 
 \SpF{2_{10}^{(4,3,0,0,0)}}\,,\ 
 \SpF{2_{11}^{(1,6,0,0,0)}}\,,\ 
 \SpF{2_{11}^{(3,4,0,0,0)}}\,,\ 
 \SpF{2_{11}^{(5,2,0,0,0)}}\,,\ 
 \SpF{2_{11}^{(7,0,0,0,0)}}\,.
\label{eq:all-IIA-branes}
\end{align}
Here, $1_0$, $p_1$, $5_2$, and $5_2^1$ respectively represent the standard F1, D$p$, NS5, and KKM, while $p^{7-p}_3$ are known as the higher KK branes denoted as D$p_{7-p}$ \cite{LozanoTellechea:2000mc}. 
In addition, $7_{3}^{(1,0)}$ is known as the KK8A-brane in \cite{Meessen:1998qm}. 
As one can clearly see, in dimensions $d\geq 3$\,, there exist the type IIA branes with tensions proportional to $\gs^{\alpha}$ with $-11\leq \alpha\leq 0$\,.

In order to obtain all of the ``elementary'' type IIB branes, we act a $T$-duality to each of the type IIA branes. 
Since a $T$-duality does not change the power of $\gs$\,, the type IIB branes also have tensions proportional to $\gs^{\alpha}$ with $-11\leq \alpha\leq 0$\,. 
A list of all of the ``elementary'' type IIB branes is as follows:
\begin{align}
 &1_0\,,\ 
 \text{P}\,,\ 
 1_1\,,\ 
 3_1\,,\ 
 5_1\,,\ 
 \Def{7_1}\,,\ 
 \SpF{9_1}\,,\ 
 5_2\,,\ 
 5_2^1\,,\ 
 \Def{5_{2}^{2}}\,,\ 
 \DoW{5_{2}^{3}}\,,\ 
 \SpF{5_{2}^{4}}\,,\ 
 \Def{7_{3}}\,,\ 
 \Def{5_{3}^{2}}\,,\ 
 \Def{3_{3}^{4}}\,,\ 
 \Def{1_{3}^{6}}\,,\ 
 \DoW{6_{3}^{(1,1)}}\,,\ 
 \DoW{4_{3}^{(1,3)}}\,,\ 
 \DoW{2_{3}^{(1,5)}}\,,
\nn\\
 &\SpF{7_{3}^{(2,0)}}\,,\ 
 \SpF{5_{3}^{(2,2)}}\,,\ 
 \SpF{3_{3}^{(2,4)}}\,,\ 
 \Def{1_{4}^{6}}\,,\ 
 \Def{0_{4}^{(1,6)}}\,,\ 
 \DoW{1_{4}^{(1,0,6)}}\,,\ 
 \DoW{5_{4}^{3}}\,,\ 
 \DoW{4_{4}^{(1,3)}}\,,\ 
 \DoW{3_{4}^{(2,3)}}\,,\ 
 \DoW{2_{4}^{(3,3)}}\,,\ 
 \DoW{1_{4}^{(4,3)}}\,,\ 
 \SpF{5_{4}^{(1,0,3)}}\,,\ 
 \SpF{4_{4}^{(1,1,3)}}\,,
\nn\\
 &
 \SpF{3_{4}^{(1,2,3)}}\,,\ 
 \SpF{2_{4}^{(1,3,3)}}\,,\ 
 \SpF{9_{4}}\,,\ 
 \SpF{7_{4}^{(2,0)}}\,,\ 
 \SpF{5_{4}^{(4,0)}}\,,\ 
 \SpF{3_{4}^{(6,0)}}\,,\ 
 \DoW{2_{5}^{(1,5)}}\,,\ 
 \DoW{2_{5}^{(3,3)}}\,,\ 
 \DoW{2_{5}^{(5,1)}}\,,\ 
 \DoW{1_{5}^{(1,0,6)}}\,,\ 
 \DoW{1_{5}^{(1,2,4)}}\,,\ 
 \DoW{1_{5}^{(1,4,2)}}\,,\ 
 \DoW{1_{5}^{(1,6,0)}}\,,
\nn\\
 &
 \SpF{2_{5}^{(1,0,0,6)}}\,,\ 
 \SpF{2_{5}^{(1,0,2,4)}}\,,\ 
 \SpF{2_{5}^{(1,0,4,2)}}\,,\ 
 \SpF{2_{5}^{(1,0,6,0)}}\,,\ 
 \SpF{5_{5}^{4}}\,,\ 
 \SpF{5_{5}^{(2,2)}}\,,\ 
 \SpF{5_{5}^{(4,0)}}\,,\ 
 \SpF{4_{5}^{(1,1,3)}}\,,\ 
 \SpF{4_{5}^{(1,3,1)}}\,,\ 
 \SpF{3_{5}^{(2,0,4)}}\,,\ 
 \SpF{3_{5}^{(2,2,2)}}\,,
\nn\\
 &
 \SpF{3_{5}^{(2,4,0)}}\,,\ 
 \SpF{2_{5}^{(3,1,3)}}\,,\ 
 \SpF{2_{5}^{(3,3,1)}}\,,\ 
 \DoW{1_{6}^{(4,3)}}\,,\ 
 \DoW{1_{6}^{(1,4,2)}}\,,\ 
 \DoW{1_{6}^{(2,4,1)}}\,,\ 
 \DoW{1_{6}^{(3,4,0)}}\,,\ 
 \SpF{3_{6}^{(2,4)}}\,,\ 
 \SpF{3_{6}^{(1,2,3)}}\,,\ 
 \SpF{3_{6}^{(2,2,2)}}\,,\ 
 \SpF{3_{6}^{(3,2,1)}}\,,\ 
 \SpF{3_{6}^{(4,2,0)}}\,,
\nn\\
 &
 \SpF{2_{6}^{(1,0,2,4)}}\,,\ 
 \SpF{2_{6}^{(1,1,2,3)}}\,,\ 
 \SpF{2_{6}^{(1,2,2,2)}}\,,\ 
 \SpF{2_{6}^{(1,3,2,1)}}\,,\ 
 \SpF{2_{6}^{(1,4,2,0)}}\,,\ 
 \DoW{1_{7}^{(1,6,0)}}\,,\ 
 \DoW{1_{7}^{(3,4,0)}}\,,\ 
 \DoW{1_{7}^{(5,2,0)}}\,,\ 
 \DoW{1_{7}^{(7,0,0)}}\,,\ 
 \SpF{3_{7}^{(6,0)}}\,,\ 
 \SpF{3_{7}^{(2,4,0)}}\,,
\nn\\
 &
 \SpF{3_{7}^{(4,2,0)}}\,,\ 
 \SpF{3_{7}^{(6,0,0)}}\,,\ 
 \SpF{2_{7}^{(1,0,1,5,0)}}\,,\ 
 \SpF{2_{7}^{(1,0,3,3,0)}}\,,\ 
 \SpF{2_{7}^{(1,0,5,1,0)}}\,,\ 
 \SpF{2_{7}^{(1,3,3)}}\,,\ 
 \SpF{2_{7}^{(3,1,3)}}\,,\ 
 \SpF{2_{7}^{(1,0,4,2)}}\,,\ 
 \SpF{2_{7}^{(1,2,2,2)}}\,,\ 
 \SpF{2_{7}^{(1,4,0,2)}}\,,
\nn\\
 &
 \SpF{2_{7}^{(2,1,3,1)}}\,,\ 
 \SpF{2_{7}^{(2,3,1,1)}}\,,\ 
 \SpF{2_{7}^{(3,0,4,0)}}\,,\ 
 \SpF{2_{7}^{(3,2,2,0)}}\,,\ 
 \SpF{2_{7}^{(3,4,0,0)}}\,,\ 
 \DoW{1_{8}^{(7,0,0)}}\,,\ 
 \SpF{2_{8}^{(1,0,6,0)}}\,,\ 
 \SpF{2_{8}^{(3,0,4,0)}}\,,\ 
 \SpF{2_{8}^{(5,0,2,0)}}\,,\ 
 \SpF{2_{8}^{(7,0,0,0)}}\,,
\nn\\
 &
 \SpF{2_{8}^{(3,3,1)}}\,,\ 
 \SpF{2_{8}^{(1,3,2,1)}}\,,\ 
 \SpF{2_{8}^{(2,3,1,1)}}\,,\ 
 \SpF{2_{8}^{(3,3,0,1)}}\,,\ 
 \SpF{2_{8}^{(1,0,3,3,0)}}\,,\ 
 \SpF{2_{8}^{(1,1,3,2,0)}}\,,\ 
 \SpF{2_{8}^{(1,2,3,1,0)}}\,,\ 
 \SpF{2_{8}^{(1,3,3,0,0)}}\,,\ 
 \SpF{2_{9}^{(1,4,2,0)}}\,,
\nn\\
 &
 \SpF{2_{9}^{(3,2,2,0)}}\,,\ 
 \SpF{2_{9}^{(5,0,2,0)}}\,,\ 
 \SpF{2_{9}^{(1,0,5,1,0)}}\,,\ 
 \SpF{2_{9}^{(1,2,3,1,0)}}\,,\ 
 \SpF{2_{9}^{(1,4,1,1,0)}}\,,\ 
 \SpF{2_{9}^{(2,1,4,0,0)}}\,,\ 
 \SpF{2_{9}^{(2,3,2,0,0)}}\,,\ 
 \SpF{2_{9}^{(2,5,0,0,0)}}\,,\ 
 \SpF{2_{10}^{(3,4,0,0)}}\,,
\nn\\
 &
 \SpF{2_{10}^{(1,3,3,0,0)}}\,,\ 
 \SpF{2_{10}^{(2,3,2,0,0)}}\,,\ 
 \SpF{2_{10}^{(3,3,1,0,0)}}\,,\ 
 \SpF{2_{10}^{(4,3,0,0,0)}}\,,\ 
 \SpF{2_{11}^{(7,0,0,0)}}\,,\ 
 \SpF{2_{11}^{(2,5,0,0,0)}}\,,\ 
 \SpF{2_{11}^{(4,3,0,0,0)}}\,,\ 
 \SpF{2_{11}^{(6,1,0,0,0)}}\,.
\label{eq:all-IIB-branes}
\end{align}

We can also summarize the $T$-duality web between the type IIA branes (upper) and the type IIB branes (lower) as in Figures \ref{fig:web01}--\ref{fig:web20}:
\begin{center}
\setlength{\abovecaptionskip}{-3mm}
\setlength{\belowcaptionskip}{4mm}
\scalebox{.7}{
}%
\figcaption{$T$-duality chain of F(undamental)-branes.} 
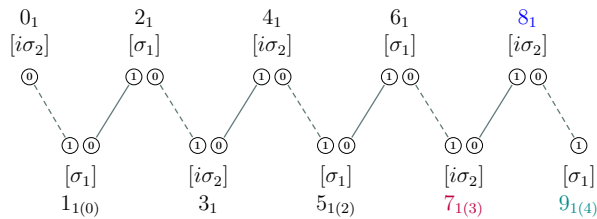
\label{fig:web01}

\scalebox{0.7}{%
}
\figcaption{$T$-duality chain of D(irichlet)-branes.} \label{fig:web02}

\scalebox{0.7}{%
}
\figcaption{$T$-duality chain of S(olitonic)-branes.} \label{fig:web03}

\scalebox{0.7}{%
}
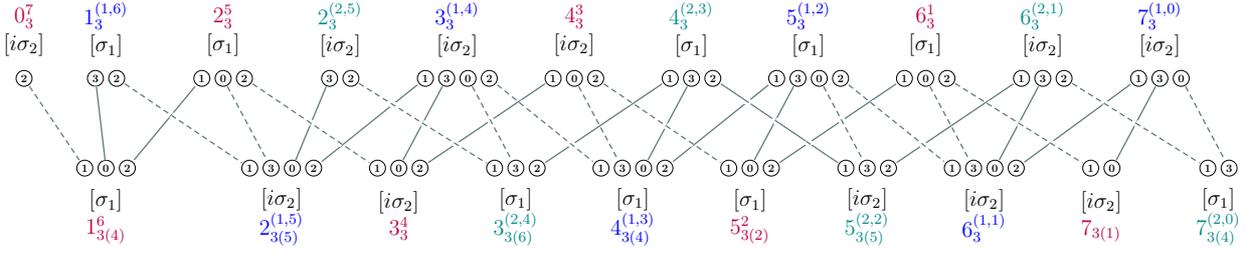
\figcaption{$T$-duality chain of the E(xotic)-branes.} \label{fig:web04}

\scalebox{0.7}{%
}
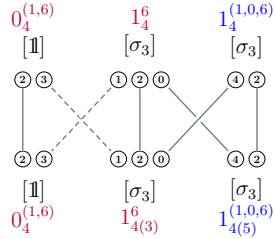
\figcaption{$T$-duality chain of the E$^{(4;6)}$-branes.} \label{fig:web05}

\scalebox{0.7}{%
}
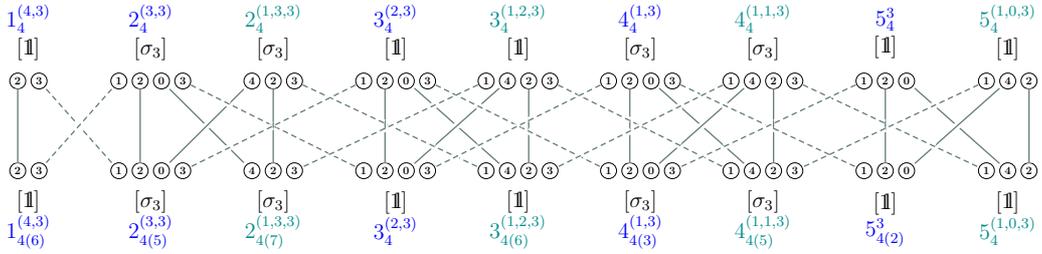
\figcaption{$T$-duality chain of the E$^{(4;3)}$-branes.} \label{fig:web06}

\scalebox{0.7}{%
}
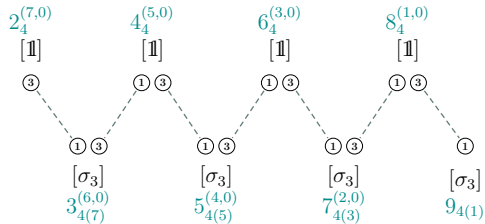
\figcaption{$T$-duality chain of the E$^{(4;0)}$-branes.} \label{fig:web07}

\scalebox{0.7}{%
}
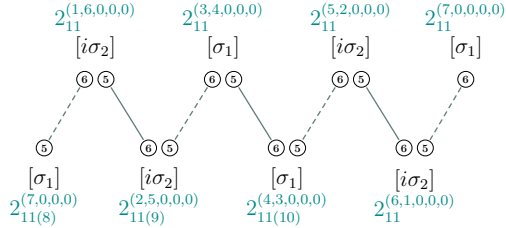
\figcaption{$T$-duality chain of the E$^{(11;7)}$-branes.} \label{fig:web20}
\end{center}
Here, for the $S$-duality non-singlets in the type IIB side, we have appended the subscript with round brackets. 
For example, $2^{(7,0,0,0)}_{11(8)}$ in Figure \ref{fig:web20} represents the $2^{(7,0,0,0)}_{11}$-brane, and also denotes that its $S$-dual partner is the $2^{(7,0,0,0)}_{8}$-brane. 
The characters in the squared brackets are not important here, and will be explained in Section \ref{sec:susy}. 
Each (solid or dashed) line corresponds to a $T$-duality and the circled numbers have the following meaning. 
For example, the $5^3_2$-brane in Figure \ref{fig:web03} has three types of direction along which we can perform $T$-duality; (i) directions along which the mass does not depend on the radii, which we call {\raise0.2ex\hbox{\textcircled{\scriptsize{$\bm{0}$}}}}, (ii) directions, denoted as {\raise0.2ex\hbox{\textcircled{\scriptsize{$\bm{1}$}}}}, along which the mass linearly depends on the radii, (iii) three directions denoted as {\raise0.2ex\hbox{\textcircled{\scriptsize{$\bm{2}$}}}} along which the mass quadratically depends on the radii. 
If we perform a $T$-duality along the {\raise0.2ex\hbox{\textcircled{\scriptsize{$\bm{0}$}}}} direction, we obtain the $5^4_2$-brane, while if we perform a $T$-duality along the {\raise0.2ex\hbox{\textcircled{\scriptsize{$\bm{1}$}}}} direction, we obtain the $5^3_2$-brane, and along the {\raise0.2ex\hbox{\textcircled{\scriptsize{$\bm{2}$}}}} direction, we obtain the $5^2_2$-brane. 
Namely, the circled number can be understood as the power of the radius dependence along which the $T$-duality is performed. 
The meaning between the solid or dashed line, which is not important here, is that each line connected to even/odd number in the type IIB side is a solid/dashed line. 

\subsection{Web of the missing states}
\label{sec:missing}

In the previous subsection, we have only considered the branes that are connected to the standard branes via $T$- and $S$-duality transformations, i.e.~the Weyl reflections \eqref{eq:U-dual}. 
However, as we can clearly see from Table \ref{tab:dim-multiplets}, if we consider the non-standard branes (i.e.~colored branes with codimension 2 or less), these are not enough to make up the whole $U$-duality multiplet. 
We need to introduce additional states, which we call \emph{missing states} for obvious reason. 

The existence of the missing states was originally noted in \cite{Elitzur:1997zn}, and they were later discussed for example in \cite{Cook:2009ri,Houart:2009ya,Kleinschmidt:2011vu}. 
Properties of such missing states are not clearly understood, and they may not be supersymmetric states as conjectured in \cite{Kleinschmidt:2011vu}.
Here, we only compute the tensions of these states by simply extrapolating the correspondence between tensions and Dynkin labels discussed in the previous sections to arbitrary weight vectors (see \cite{Cook:2008bi} for a similar work in the context of $E_{11}$). 

\subsubsection{Example: 4-brane multiplet in $E_{4(4)}$}

Let us start with a simple example, a 4-brane multiplet in M-theory compactified on $T^4$. 
In this case, Table \ref{tab:dim-multiplets} shows that the number of supersymmetric branes is 20, although the dimension of the 4-brane multiplet is 24. 
Thus, there are four missing states. 
In order to identify the missing states, let us consider the weight diagram for the 4-brane multiplet given in Table \ref{tab:4-brane-T4}. 
\begin{table}
\begin{center}
\begin{tikzpicture}[baseline=(A.center), scale=0.6, every node/.style={scale=0.6}]
\matrix (A) [matrix of math nodes, row sep=10mm, column sep=3mm]
{ & & & \node (1001) {[1,0,0,1]}; & \\ 
& & \node (-1101) {[-1,1,0,1]}; & & \node (101-1) {[1,0,1,-1]}; \\
& \node (0-111) {[0,-1,1,1]}; & & \node (-111-1) {[-1,1,1,-1]}; & & \node (11-10) {[1,1,-1,0]}; \\
\node (00-12) {[0,0,-1,2]}; & & \node (0-12-1) {[0,-1,2,-1]}; & & \node (-12-10) {[-1,2,-1,0]}; & & \node (2-100) {[2,-1,0,0]}; \\
\node (0000ll) {[0,0,0,0]}; & & \node (0000lr) {[0,0,0,0]}; & & \node (0000rl) {[0,0,0,0]}; & & \node (0000rr) {[0,0,0,0]}; \\
\node (001-2) {[0,0,1,-2]}; & & \node (01-21) {[0,1,-2,1]}; & & \node (1-210) {[1,-2,1,0]}; & & \node (-2100) {[-2,1,0,0]}; \\
& \node (01-1-1) {[0,1,-1,-1]}; & & \node (1-1-11) {[1,-1,-1,1]}; & & \node (-1-110) {[-1,-1,1,0]}; \\
& & \node (1-10-1) {[1,-1,0,-1]}; & & \node (-10-11) {[-1,0,-1,1]}; \\
& & & \node (-100-1) {[-1,0,0,-1]}; & \\ 
};
\draw[draw=blue, rounded corners=2.5mm] ([xshift=-27mm,yshift=-2mm]0000ll.south west) rectangle ([xshift=27mm,yshift=4.5mm]0000rr.north east);
\node[below=.8mm] (A-5-4.center) {\fcolorbox{blue}{white}{``missing states''}};
\draw[-latex] ([xshift=-5mm]1001.south) -- (-1101.north) node[midway, left=3mm] {$\alpha_1$};
\draw[-latex] ([xshift=5mm]1001.south) -- (101-1.north) node[midway, right=2mm] {$\alpha_4$};
\draw[-latex] ([xshift=-5mm]-1101.south) -- (0-111.north) node[midway, left=3mm] {$\alpha_2$};
\draw[-latex] ([xshift=5mm]-1101.south) -- ([xshift=-6mm]-111-1.north) node[midway, left=1.5mm] {$\alpha_4$};
\draw[-latex] ([xshift=-5mm]101-1.south) -- ([xshift=6mm]-111-1.north) node[midway, left=1.5mm] {$\alpha_1$};
\draw[-latex] ([xshift=5mm]101-1.south) -- (11-10.north) node[midway, right=2mm] {$\alpha_3$};
\draw[-latex] ([xshift=-5mm]0-111.south) -- (00-12.north) node[midway, left=3mm] {$\alpha_3$};
\draw[-latex] ([xshift=5mm]0-111.south) -- ([xshift=-6mm]0-12-1.north) node[midway, left=1.5mm] {$\alpha_4$};
\draw[-latex] ([xshift=-5mm]-111-1.south) -- ([xshift=6mm]0-12-1.north) node[midway, left=1.5mm] {$\alpha_2$};
\draw[-latex] ([xshift=5mm]-111-1.south) -- ([xshift=-6mm]-12-10.north) node[midway, left=1.5mm] {$\alpha_3$};
\draw[-latex] ([xshift=-5mm]11-10.south) -- ([xshift=6mm]-12-10.north) node[midway, left=1.5mm] {$\alpha_1$};
\draw[-latex] ([xshift=5mm]11-10.south) -- (2-100.north) node[midway, right=2mm] {$\alpha_2$};
\draw[-latex] (00-12.south) -- (0000ll.north) node[midway,left=2mm] {$\alpha_4$};
\draw[-latex] (0-12-1.south) -- (0000lr.north) node[midway,left=2mm] {$\alpha_3$};
\draw[-latex] (-12-10.south) -- (0000rl.north) node[midway,right=2mm] {$\alpha_2$};
\draw[-latex] (2-100.south) -- (0000rr.north) node[midway,right=2mm] {$\alpha_1$};
\draw[-latex] (0000ll.south) -- (001-2.north) node[midway,left=2mm] {$\alpha_4$};
\draw[-latex] (0000lr.south) -- (01-21.north) node[midway,left=2mm] {$\alpha_3$};
\draw[-latex] (0000rl.south) -- (1-210.north) node[midway,right=2mm] {$\alpha_2$};
\draw[-latex] (0000rr.south) -- (-2100.north) node[midway,right=2mm] {$\alpha_1$};
\draw[-latex] (001-2.south) -- ([xshift=-6mm]01-1-1.north) node[midway, left=1.5mm] {$\alpha_3$};
\draw[-latex] ([xshift=-5mm]01-21.south) -- ([xshift=6mm]01-1-1.north) node[midway, left=1.5mm] {$\alpha_4$};
\draw[-latex] ([xshift=5mm]01-21.south) -- ([xshift=-6mm]1-1-11.north) node[midway, left=1.5mm] {$\alpha_2$};
\draw[-latex] ([xshift=-5mm]1-210.south) -- ([xshift=6mm]1-1-11.north) node[midway, left=1.5mm] {$\alpha_3$};
\draw[-latex] ([xshift=5mm]1-210.south) -- ([xshift=-6mm]-1-110.north) node[midway, left=1.5mm] {$\alpha_1$};
\draw[-latex] (-2100.south) -- ([xshift=6mm]-1-110.north) node[midway, right=1.5mm] {$\alpha_2$};
\draw[-latex] (01-1-1.south) -- ([xshift=-6mm]1-10-1.north) node[midway, left=1.5mm] {$\alpha_2$};
\draw[-latex] ([xshift=-5mm]1-1-11.south) -- ([xshift=6mm]1-10-1.north) node[midway, left=1.5mm] {$\alpha_4$};
\draw[-latex] ([xshift=5mm]1-1-11.south) -- ([xshift=-6mm]-10-11.north) node[midway, left=1.5mm] {$\alpha_1$};
\draw[-latex] (-1-110.south) -- ([xshift=6mm]-10-11.north) node[midway, right=1.5mm] {$\alpha_3$};
\draw[-latex] (1-10-1.south) -- ([xshift=-6mm]-100-1.north) node[midway, left=1.5mm] {$\alpha_1$};
\draw[-latex] (-10-11.south) -- ([xshift=6mm]-100-1.north) node[midway, right=1.5mm] {$\alpha_4$};
\node[red,left] at (1001.west) {$5_6(\cdot\cdot\cdot\cdot1)$};
\node[red,left] at (-1101.west) {$5_6(\cdot\cdot\cdot\cdot2)$};
\node[red,right] at (101-1.east) {$6^1_9(\cdot\cdot\cdot\cdot23,1)$};
\node[red,left] at (0-111.west) {$5_6(\cdot\cdot\cdot\cdot3)$};
\node[red,right] at (-111-1.east) {$6^1_9(\cdot\cdot\cdot\cdot13,2)$};
\node[red,right] at (11-10.east) {$6^1_9(\cdot\cdot\cdot\cdot24,1)$};
\node[red,left] at (00-12.west) {$5_6(\cdot\cdot\cdot\cdot4)$};
\node[red,left] at (0-12-1.west) {$6^1_9(\cdot\cdot\cdot\cdot12,3)$};
\node[red,right] at (-12-10.east) {$6^1_9(\cdot\cdot\cdot\cdot14,2)$};
\node[red,right] at (2-100.east) {$6^1_9(\cdot\cdot\cdot\cdot34,1)$};
\node[red,left] at (0000ll.west) {$8_9(\cdot\cdot\cdot\cdot1234)$};
\node[red,left] at (0000lr.west) {$8_9(\cdot\cdot\cdot\cdot1234)$};
\node[red,right] at (0000rl.east) {$8_9(\cdot\cdot\cdot\cdot1234)$};
\node[red,right] at (0000rr.east) {$8_9(\cdot\cdot\cdot\cdot1234)$};
\node[red,left] at (001-2.west) {$5^3_{12}(\cdot\cdot\cdot\cdot4,123)$};
\node[red,left] at (01-21.west) {$6^1_9(\cdot\cdot\cdot\cdot12,4)$};
\node[red,right] at (1-210.east) {$6^1_9(\cdot\cdot\cdot\cdot14,3)$};
\node[red,right] at (-2100.east) {$6^1_9(\cdot\cdot\cdot\cdot34,2)$};
\node[red,left] at (01-1-1.west) {$5^3_{12}(\cdot\cdot\cdot\cdot3,124)$};
\node[red,right] at (1-1-11.east) {$6^1_9(\cdot\cdot\cdot\cdot13,4)$};
\node[red,right] at (-1-110.east) {$6^1_9(\cdot\cdot\cdot\cdot24,3)$};
\node[red,left] at (1-10-1.west) {$5^3_{12}(\cdot\cdot\cdot\cdot2,134)$};
\node[red,right] at (-10-11.east) {$6^1_9(\cdot\cdot\cdot\cdot23,4)$};
\node[red,left] at (-100-1.west) {$5^3_{12}(\cdot\cdot\cdot\cdot1,234)$};
\end{tikzpicture}
\end{center}
\caption{The weight diagram for the 4-brane multiplet in M-theory compactified on $T^4$.}
\label{tab:4-brane-T4}
\end{table}

Since the mass dimension of the tension $\cT_4$ is five, the four degenerate Dynkin labels $[0,0,0,0]$ correspond to
\begin{align}
 \cT_4 = l_7 = \frac{R_1R_2R_3R_4}{\lp^9} \,. 
\end{align}
By using the convention \eqref{eq:tension}, this should be understood as a tension of the $8_9(\cdot\cdot\cdot\cdot1234)$-brane, where the four dots $\cdot\cdot\cdot\cdot$ represent that the 8-brane is extended along certain four external spatial directions. 
This kind of 8-brane was predicted in \cite{Elitzur:1997zn} and called M8-brane in \cite{Hull:1997kb}, although its properties are unclear so far. 
We can just extrapolate their tensions. 
From the tension, we can find that these states are singlets under the Weyl reflections. 

\subsubsection{List of missing states}

Generalizing the above procedure, we can compute the tensions of missing states in all of the multiplets. 
In order to obtain a list of the weights for the exceptional groups $E_{6(6)}$, $E_{7(7)}$, and $E_{8(8)}$, it will be useful to use a computer program such as SimpLie \cite{SimpLie}. 
By transforming the Dynkin labels into the tensions, we obtain Tables \ref{tab:missing-1-brane}--\ref{tab:missing-7-brane}. 
The states contained in a single column have the weights with the same length, and we have checked that they are indeed in a single $U$-duality orbit of \eqref{eq:U-dual}. 

In terms of M-theory, the following states are contained in Tables \ref{tab:missing-1-brane}--\ref{tab:missing-7-brane}:
\begin{align*}
 &8_9\,,
 7_{12}^{2}\,,
 9_{12}^{1}\,,
 4_{15}^{5}\,,
 6_{15}^{4}\,,
 7_{15}^{(1,2)}\,,
 1_{18}^{8}\,,
 2_{18}^{(1,6)}\,,
 3_{18}^{7}\,,
 4_{18}^{(1,5)}\,,
 5_{18}^{(2,3)}\,,
 1_{21}^{(3,5)}\,,
 2_{21}^{(2,6)}\,,
 3_{21}^{(3,4)}\,,
 4_{21}^{(4,2)}\,,
 2_{21}^{(1,0,7)}\,,
 3_{21}^{(1,1,5)}\,,
\\
 &
 1_{24}^{(6,2)}\,,
 2_{24}^{(5,3)}\,,
 3_{24}^{(6,1)}\,,
 2_{24}^{(1,3,4)}\,,
 3_{24}^{(1,4,2)}\,,
 2_{24}^{(2,1,5)}\,,
 1_{27}^{(1,7,0)}\,,
 2_{27}^{(8,0)}\,,
 2_{27}^{(1,6,1)}\,,
 2_{27}^{(2,4,2)}\,,
 3_{27}^{(2,5,0)}\,,
 2_{27}^{(3,2,3)}\,,
 2_{27}^{(1,0,5,2)}\,,
\\
 &2_{30}^{(3,5,0)}\,,
 2_{30}^{(4,3,1)}\,,
 2_{30}^{(5,1,2)}\,,
 2_{30}^{(1,1,6,0)}\,,
 2_{30}^{(1,2,4,1)}\,,
 2_{33}^{(6,2,0)}\,,
 2_{33}^{(7,0,1)}\,,
 2_{33}^{(1,4,3,0)}\,,
 2_{33}^{(1,5,1,1)}\,,
 2_{33}^{(2,2,4,0)}\,,
\\
 &2_{36}^{(1,7,0,0)}\,,
 2_{36}^{(2,5,1,0)}\,,
 2_{36}^{(3,3,2,0)}\,,
 2_{36}^{(1,0,6,1,0)}\,,
 2_{39}^{(5,2,1,0)}\,,
 2_{39}^{(4,4,0,0)}\,,
 2_{39}^{(1,2,5,0,0)}\,,
 2_{42}^{(7,1,0,0)}\,,
 2_{42}^{(1,5,2,0,0)}\,,
 2_{45}^{(2,6,0,0,0)}\,.
\end{align*}
One can also make a list of type II branes appearing in Tables \ref{tab:missing-1-brane}--\ref{tab:missing-7-brane} and draw a duality web along these states. 

As we can see from Tables \ref{tab:missing-1-brane}--\ref{tab:missing-7-brane}, the missing states in the $p$-brane multiplet have degeneracies which depend on $p$\,. 
For example, the $8_9$-brane in the $p$-brane multiplet ($1\leq p\leq 6$) has degeneracy $(8-p)$\,, although for $p=6$ the degeneracy becomes 1. 
The $p$-dependence is non-trivial, but the degeneracy is independent of $d$ for all missing states. 
The missing states in higher $d$ can be obtained from the missing states in lower $d$ just by truncating the states that are disallowed by dimensionality. 

\subsection{Web of mixed-symmetry potentials}

The standard branes in type II theory couple to certain potentials in type II supergravity. 
For example, the F1 and the pp-wave (electrically) couple to the $B$-field $B_2$ and the graviphoton $A_1^m$\,, and D$p$-branes couple to the R--R potentials $C_{p+1}$\,. 
Following a series of works \cite{Bergshoeff:2010xc,Bergshoeff:2011zk,Bergshoeff:2011mh,Bergshoeff:2011ee,Bergshoeff:2011se,Bergshoeff:2012ex,Bergshoeff:2012pm,Bergshoeff:2016ncb,Lombardo:2016swq,Bergshoeff:2017gpw}, we call F1 and the pp-wave the \emph{F(undamental)-branes}. 
There are also the \emph{S(olitonic)-branes}, which consist of $5_2$-brane (NS5-brane), $5_2^1$-brane (KK monopole), $5^2_2$-brane, $5^3_2$-brane, and $5^4_2$-brane. 
This chain of 5-branes is recently studied well and the $5_2^n$-branes are known to couple to a set of mixed-symmetry potentials $D_{6+n,n}$\,, which represents a mixed-symmetry potential $D_{m_1\cdots m_{6+n},\,p_1\cdots p_n}$ where multiple indices separated by comma are totally antisymmetrized. 
For $n=0$ (NS5-brane), the 6-form potential $D_6$ is nothing but the magnetic dual of the $B$-field. 
For $n=1$ (KK monopole), the potential $D_{7,1}$ is the magnetic dual of the graviphoton, known as the dual graviton. 
The worldvolume action of the KK monopole including the Wess--Zumino term has been obtained in \cite{Eyras:1998hn}. 
For $n=2$, the potential $D_{8,2}$ is rather non-standard but it is known to be the magnetic dual of the so-called $\beta$-field. 
Its coupling to the $5^2_2$-brane has been determined in \cite{Chatzistavrakidis:2013jqa,Kimura:2014upa} (see also \cite{Andriot:2014uda,Sakatani:2014hba}). 
The M-theory uplift, the action for the $5^3_{12}$-brane was also studied in \cite{Kimura:2016anf}. 
Generalizations to $n=3$ and $4$ in the manifestly $T$-duality covariant approach have been achieved in \cite{Blair:2017hhy}. 

The $S$-dual of the $5^2_2$-brane is $5^2_3$-brane is a member of the \emph{E(xotic)-branes}. 
The E-brane $(p+n)_3^{(n,7-p-n)}$ couples to the mixed-symmetry potential $E_{8+n,7-p,n}$\,. 
In general, there is a conjectural relation between supersymmetric branes and the mixed-symmetry potentials:
\begin{align}
 \text{$b_n^{(c_s,\dotsc,c_2)}$-brane}\qquad\Leftrightarrow\qquad 
 \text{potential $E_{1+b+c_2+\cdots+c_s,\dotsc,c_{s-1}+c_s,c_s}$}\,.
\label{eq:brane-potential}
\end{align}
In the convention of \cite{Bergshoeff:2012ex}, depending on the power of the string coupling $n$, the potentials are denoted as $E$ ($n=3$), $F$ ($n=4$), $G$ ($n=5$), $H$ ($n=6$), $\dotsc$\,.
In this paper, since the integer $n$ runs up to $11$, we call them $E^{(n)}$ (i.e.~$E^{(4)}=F$, $E^{(5)}=G$, $E^{(6)}=H$, and so on) and denote the corresponding brane the $E^{(n)}$-brane. 
\begin{table}
\begin{center}
{\tiny
 \begin{tabular}{|c|c|c|}\hline
 $\alpha=0$ & $B_{1M}$ & F1/P (F-brane) \\ \hline
 A/B & \multicolumn{2}{|c|}{$
  1_0\,,\ 
  \text{P}
$} \\\hline\hline
 $\alpha=-1$ & $C_{p+1}$ & $p_1$-brane (D-brane) \\ \hline
 IIA & \multicolumn{2}{|c|}{$
  0_1\,,\ 
  2_1\,,\ 
  4_1\,,\ 
  6_1\,,\ 
  \DoW{8_1}
$} \\\hline
 IIB & \multicolumn{2}{|c|}{$
  1_1\,,\ 
  3_1\,,\ 
  5_1\,,\ 
  \Def{7_1}\,,\ 
  \SpF{9_1}
$} \\\hline\hline
 $\alpha=-2$ & $D_{6+n,n}$ & $5_2^{n}$-brane (S-brane) \\ \hline
 A/B & \multicolumn{2}{|c|}{$
  5_{2}\,,\ 
  5_{2}^{1}\,,\ 
  \Def{5_{2}^{2}}\,,\ 
  \DoW{5_{2}^{3}}\,,\ 
  \SpF{5_{2}^{4}}
$} \\\hline\hline
 $\alpha=-3$ & $E_{8+n,7-p,n}$ & $(p+n)_3^{(n,7-p-n)}$-brane (E-brane) \\ \hline
 IIA & \multicolumn{2}{|c|}{$
  \Def{6_{3}^{1}}\,,\ 
  \Def{4_{3}^{3}}\,,\ 
  \Def{2_{3}^{5}}\,,\ 
  \Def{0_{3}^{7}}\,,\ 
  \DoW{7_{3}^{(1,0)}}\,,\ 
  \DoW{5_{3}^{(1,2)}}\,,\ 
  \DoW{3_{3}^{(1,4)}}\,,\ 
  \DoW{1_{3}^{(1,6)}}\,,\ 
  \SpF{6_{3}^{(2,1)}}\,,\ 
  \SpF{4_{3}^{(2,3)}}\,,\ 
  \SpF{2_{3}^{(2,5)}}
$} \\\hline
 IIB & \multicolumn{2}{|c|}{$
  \Def{7_{3}}\,,\ 
  \Def{5_{3}^{2}}\,,\ 
  \Def{3_{3}^{4}}\,,\ 
  \Def{1_{3}^{6}}\,,\ 
  \DoW{6_{3}^{(1,1)}}\,,\ 
  \DoW{4_{3}^{(1,3)}}\,,\ 
  \DoW{2_{3}^{(1,5)}}\,,\ 
  \SpF{7_{3}^{(2,0)}}\,,\ 
  \SpF{5_{3}^{(2,2)}}\,,\ 
  \SpF{3_{3}^{(2,4)}}
$} \\\hline\hline
 $\alpha=-4$ & $E^{(4)}_{8+n,6+m+n,m+n,n}$ & $(1-m)_4^{(n,m,6)}$-brane (E$^{(4;6)}$-brane) \\ \hline
 A/B & \multicolumn{2}{|c|}{$
  \Def{1_{4}^{6}}\,,\ 
  \Def{0_{4}^{(1,6)}}\,,\ 
  \DoW{1_{4}^{(1,0,6)}}
$}
\\\hline\hline
 $\alpha=-4$ & $E^{(4)}_{9+n,3+m+n,m+n,n}$ & $(5-m)_4^{(n,m,3)}$-brane (E$^{(4;3)}$-brane) \\ \hline
 A/B & \multicolumn{2}{|c|}{$
  \DoW{5_{4}^{3}}\,,\ 
  \DoW{4_{4}^{(1,3)}}\,,\ 
  \DoW{3_{4}^{(2,3)}}\,,\ 
  \DoW{2_{4}^{(3,3)}}\,,\ 
  \DoW{1_{4}^{(4,3)}}\,,\ 
  \SpF{5_{4}^{(1,0,3)}}\,,\ 
  \SpF{4_{4}^{(1,1,3)}}\,,\ 
  \SpF{3_{4}^{(1,2,3)}}\,,\ 
  \SpF{2_{4}^{(1,3,3)}}
$}
\\\hline\hline
 $\alpha=-4$ & $E^{(4)}_{10,q,q}$ & $(9-q)_4^{(q,0)}$-brane (E$^{(4;0)}$-brane) \\ \hline
 IIA & \multicolumn{2}{|c|}{$
  \SpF{8_{4}^{(1,0)}}\,,\ 
  \SpF{6_{4}^{(3,0)}}\,,\ 
  \SpF{4_{4}^{(5,0)}}\,,\ 
  \SpF{2_{4}^{(7,0)}}
$} \\\hline
 IIB & \multicolumn{2}{|c|}{$
  \SpF{9_{4}}\,,\ 
  \SpF{7_{4}^{(2,0)}}\,,\ 
  \SpF{5_{4}^{(4,0)}}\,,\ 
  \SpF{3_{4}^{(6,0)}}
$}
\\\hline\hline
 $\alpha=-5$ & $E^{(5)}_{9+n,6+m,p,m,n}$ & $(2-m)_5^{(n,m-n,p-m,6+m-p)}$-brane (E$^{(5;6)}$-brane) \\ \hline
 IIA & \multicolumn{2}{|c|}{$
  \DoW{2_{5}^{6}}\,,\ 
  \DoW{2_{5}^{(2,4)}}\,,\ 
  \DoW{2_{5}^{(4,2)}}\,,\ 
  \DoW{2_{5}^{(6,0)}}\,,\ 
  \DoW{1_{5}^{(1,1,5)}}\,,\ 
  \DoW{1_{5}^{(1,3,3)}}\,,\ 
  \DoW{1_{5}^{(1,5,1)}}\,,\ 
  \SpF{2_{5}^{(1,0,1,5)}}\,,\ 
  \SpF{2_{5}^{(1,0,3,3)}}\,,\ 
  \SpF{2_{5}^{(1,0,5,1)}}
$} \\\hline
 IIB & \multicolumn{2}{|c|}{$
  \DoW{2_{5}^{(1,5)}}\,,\ 
  \DoW{2_{5}^{(3,3)}}\,,\ 
  \DoW{2_{5}^{(5,1)}}\,,\ 
  \DoW{1_{5}^{(1,0,6)}}\,,\ 
  \DoW{1_{5}^{(1,2,4)}}\,,\ 
  \DoW{1_{5}^{(1,4,2)}}\,,\ 
  \DoW{1_{5}^{(1,6,0)}}\,,\ 
  \SpF{2_{5}^{(1,0,0,6)}}\,,\ 
  \SpF{2_{5}^{(1,0,2,4)}}\,,\ 
  \SpF{2_{5}^{(1,0,4,2)}}\,,\ 
  \SpF{2_{5}^{(1,0,6,0)}}
$}
\\\hline\hline
 $\alpha=-5$ & $E^{(5)}_{10,4+n,q,n}$ & $(5-n)_5^{(n,q-n,4-q+n)}$-brane (E$^{(5;4)}$-brane) \\ \hline
 IIA & \multicolumn{2}{|c|}{$
  \SpF{5_{5}^{(1,3)}}\,,\ 
  \SpF{5_{5}^{(3,1)}}\,,\ 
  \SpF{4_{5}^{(1,0,4)}}\,,\ 
  \SpF{4_{5}^{(1,2,2)}}\,,\ 
  \SpF{4_{5}^{(1,4,0)}}\,,\ 
  \SpF{3_{5}^{(2,1,3)}}\,,\ 
  \SpF{3_{5}^{(2,3,1)}}\,,\ 
  \SpF{2_{5}^{(3,0,4)}}\,,\ 
  \SpF{2_{5}^{(3,2,2)}}\,,\ 
  \SpF{2_{5}^{(3,4,0)}}
$} \\\hline
 IIB & \multicolumn{2}{|c|}{$
  \SpF{5_{5}^{4}}\,,\ 
  \SpF{5_{5}^{(2,2)}}\,,\ 
  \SpF{5_{5}^{(4,0)}}\,,\ 
  \SpF{4_{5}^{(1,1,3)}}\,,\ 
  \SpF{4_{5}^{(1,3,1)}}\,,\ 
  \SpF{3_{5}^{(2,0,4)}}\,,\ 
  \SpF{3_{5}^{(2,2,2)}}\,,\ 
  \SpF{3_{5}^{(2,4,0)}}\,,\ 
  \SpF{2_{5}^{(3,1,3)}}\,,\ 
  \SpF{2_{5}^{(3,3,1)}}
$}
\\\hline\hline
 $\alpha=-6$ & $E^{(6)}_{9,7,4+n,n}$ & $1_6^{(n,4,3-n)}$-brane (E$^{(6;4)}$-brane) \\ \hline
 A/B & \multicolumn{2}{|c|}{$
  \DoW{1_{6}^{(4,3)}}\,,\ 
  \DoW{1_{6}^{(1,4,2)}}\,,\ 
  \DoW{1_{6}^{(2,4,1)}}\,,\ 
  \DoW{1_{6}^{(3,4,0)}}
$}
\\\hline\hline
 $\alpha=-6$ & $E^{(6)}_{10,6+n,2+m+n,m+n,n}$ & $(3-n)_6^{(n,m,2,4-m)}$-brane (E$^{(7;2)}$-brane) \\ \hline
 A/B & \multicolumn{2}{|c|}{$
  \SpF{3_{6}^{(2,4)}}\,,\ 
  \SpF{3_{6}^{(1,2,3)}}\,,\ 
  \SpF{3_{6}^{(2,2,2)}}\,,\ 
  \SpF{3_{6}^{(3,2,1)}}\,,\ 
  \SpF{3_{6}^{(4,2,0)}}\,,\ 
  \SpF{2_{6}^{(1,0,2,4)}}\,,\ 
  \SpF{2_{6}^{(1,1,2,3)}}\,,\ 
  \SpF{2_{6}^{(1,2,2,2)}}\,,\ 
  \SpF{2_{6}^{(1,3,2,1)}}\,,\ 
  \SpF{2_{6}^{(1,4,2,0)}}
$}
\\\hline\hline
 $\alpha=-7$ & $E^{(7)}_{9,7,7,p}$ & $1_7^{(p,7-p,0)}$-brane (E$^{(7;7)}$-brane) \\ \hline
 IIA & \multicolumn{2}{|c|}{$
  \DoW{1_{7}^{(7,0)}}\,,\ 
  \DoW{1_{7}^{(2,5,0)}}\,,\ 
  \DoW{1_{7}^{(4,3,0)}}\,,\ 
  \DoW{1_{7}^{(6,1,0)}}
$} \\\hline
 IIB & \multicolumn{2}{|c|}{$
  \DoW{1_{7}^{(1,6,0)}}\,,\ 
  \DoW{1_{7}^{(3,4,0)}}\,,\ 
  \DoW{1_{7}^{(5,2,0)}}\,,\ 
  \DoW{1_{7}^{(7,0,0)}}
$}
\\\hline\hline
 $\alpha=-7$ & $E^{(7)}_{10,6+n,6+n,q,n,n}$ & $(3-n)_7^{(n,0,q-n,6+n-q,0)}$-brane (E$^{(7;6)}$-brane) \\ \hline
 IIA & \multicolumn{2}{|c|}{$
  \SpF{3_{7}^{(1,5,0)}}\,,\ 
  \SpF{3_{7}^{(3,3,0)}}\,,\ 
  \SpF{3_{7}^{(5,1,0)}}\,,\ 
  \SpF{2_{7}^{(1,0,0,6,0)}}\,,\ 
  \SpF{2_{7}^{(1,0,2,4,0)}}\,,\ 
  \SpF{2_{7}^{(1,0,4,2,0)}}\,,\ 
  \SpF{2_{7}^{(1,0,6,0,0)}}
$} \\\hline
 IIB & \multicolumn{2}{|c|}{$
  \SpF{3_{7}^{(6,0)}}\,,\ 
  \SpF{3_{7}^{(2,4,0)}}\,,\ 
  \SpF{3_{7}^{(4,2,0)}}\,,\ 
  \SpF{3_{7}^{(6,0,0)}}\,,\ 
  \SpF{2_{7}^{(1,0,1,5,0)}}\,,\ 
  \SpF{2_{7}^{(1,0,3,3,0)}}\,,\ 
  \SpF{2_{7}^{(1,0,5,1,0)}}
$}
\\\hline\hline
 $\alpha=-7$ & $E^{(7)}_{10,7,4+n,p,n}$ & $2_7^{(n,p-n,4-p+n,3-n)}$-brane (E$^{(7;4)}$-brane) \\ \hline
 IIA & \multicolumn{2}{|c|}{$
  \SpF{2_{7}^{(4,3)}}\,,\ 
  \SpF{2_{7}^{(2,2,3)}}\,,\ 
  \SpF{2_{7}^{(4,0,3)}}\,,\ 
  \SpF{2_{7}^{(1,1,3,2)}}\,,\ 
  \SpF{2_{7}^{(1,3,1,2)}}\,,\ 
  \SpF{2_{7}^{(2,0,4,1)}}\,,\ 
  \SpF{2_{7}^{(2,2,2,1)}}\,,\ 
  \SpF{2_{7}^{(2,4,0,1)}}\,,\ 
  \SpF{2_{7}^{(3,1,3,0)}}\,,\ 
  \SpF{2_{7}^{(3,3,1,0)}}
$} \\\hline
 IIB & \multicolumn{2}{|c|}{$
  \SpF{2_{7}^{(1,3,3)}}\,,\ 
  \SpF{2_{7}^{(3,1,3)}}\,,\ 
  \SpF{2_{7}^{(1,0,4,2)}}\,,\ 
  \SpF{2_{7}^{(1,2,2,2)}}\,,\ 
  \SpF{2_{7}^{(1,4,0,2)}}\,,\ 
  \SpF{2_{7}^{(2,1,3,1)}}\,,\ 
  \SpF{2_{7}^{(2,3,1,1)}}\,,\ 
  \SpF{2_{7}^{(3,0,4,0)}}\,,\ 
  \SpF{2_{7}^{(3,2,2,0)}}\,,\ 
  \SpF{2_{7}^{(3,4,0,0)}}
$}
\\\hline\hline
 $\alpha=-8$ & $E^{(8)}_{9,7,7,7}$ & $1_8^{(7,0,0)}$-brane (E$^{(8;7)}$-brane) \\ \hline
 A/B & \multicolumn{2}{|c|}{$\DoW{1_{8}^{(7,0,0)}}$}
\\\hline\hline
 $\alpha=-8$ & $E^{(8)}_{10,7,7,p,p}$ & $2_8^{(p,0,7-p,0)}$-brane (E$^{(8;0)}$-brane) \\ \hline
 IIA & \multicolumn{2}{|c|}{$
  \SpF{2_{8}^{(7,0)}}\,,\ 
  \SpF{2_{8}^{(2,0,5,0)}}\,,\ 
  \SpF{2_{8}^{(4,0,3,0)}}\,,\ 
  \SpF{2_{8}^{(6,0,1,0)}}
$} \\\hline
 IIB & \multicolumn{2}{|c|}{$
  \SpF{2_{8}^{(1,0,6,0)}}\,,\ 
  \SpF{2_{8}^{(3,0,4,0)}}\,,\ 
  \SpF{2_{8}^{(5,0,2,0)}}\,,\ 
  \SpF{2_{8}^{(7,0,0,0)}}
$}
\\\hline\hline
 $\alpha=-8$ & $E^{(8)}_{10,7,6+n,3+m+n,m+n,n}$ & $2_8^{(n,m,3,3-m,1-n)}$-brane (E$^{(8;3)}$-brane) \\ \hline
 A/B & \multicolumn{2}{|c|}{$
  \SpF{2_{8}^{(3,3,1)}}\,,\ 
  \SpF{2_{8}^{(1,3,2,1)}}\,,\ 
  \SpF{2_{8}^{(2,3,1,1)}}\,,\ 
  \SpF{2_{8}^{(3,3,0,1)}}\,,\ 
  \SpF{2_{8}^{(1,0,3,3,0)}}\,,\ 
  \SpF{2_{8}^{(1,1,3,2,0)}}\,,\ 
  \SpF{2_{8}^{(1,2,3,1,0)}}\,,\ 
  \SpF{2_{8}^{(1,3,3,0,0)}}
$}
\\\hline\hline
 $\alpha=-9$ & $E^{(9)}_{10,7,7,5+n,p,n}$ & $2_9^{(n,p-n,5-p+n,2-n,0)}$-brane (E$^{(9;5)}$-brane) \\ \hline
 IIA & \multicolumn{2}{|c|}{$
  \SpF{2_{9}^{(5,2,0)}}\,,\ 
  \SpF{2_{9}^{(2,3,2,0)}}\,,\ 
  \SpF{2_{9}^{(4,1,2,0)}}\,,\ 
  \SpF{2_{9}^{(1,1,4,1,0)}}\,,\ 
  \SpF{2_{9}^{(1,3,2,1,0)}}\,,\ 
  \SpF{2_{9}^{(1,5,0,1,0)}}\,,\ 
  \SpF{2_{9}^{(2,0,5,0,0)}}\,,\ 
  \SpF{2_{9}^{(2,2,3,0,0)}}\,,\ 
  \SpF{2_{9}^{(2,4,1,0,0)}}
$} \\\hline
 IIB & \multicolumn{2}{|c|}{$
  \SpF{2_{9}^{(1,4,2,0)}}\,,\ 
  \SpF{2_{9}^{(3,2,2,0)}}\,,\ 
  \SpF{2_{9}^{(5,0,2,0)}}\,,\ 
  \SpF{2_{9}^{(1,0,5,1,0)}}\,,\ 
  \SpF{2_{9}^{(1,2,3,1,0)}}\,,\ 
  \SpF{2_{9}^{(1,4,1,1,0)}}\,,\ 
  \SpF{2_{9}^{(2,1,4,0,0)}}\,,\ 
  \SpF{2_{9}^{(2,3,2,0,0)}}\,,\ 
  \SpF{2_{9}^{(2,5,0,0,0)}}
$}
\\\hline\hline
 $\alpha=-10$ & $E^{(10)}_{10,7,7,7,3+n,n}$ & $2_{10}^{(n,3,4-n,0,0)}$-brane (E$^{(10;3)}$-brane) \\ \hline
 A/B & \multicolumn{2}{|c|}{$
  \SpF{2_{10}^{(3,4,0,0)}}\,,\ 
  \SpF{2_{10}^{(1,3,3,0,0)}}\,,\ 
  \SpF{2_{10}^{(2,3,2,0,0)}}\,,\ 
  \SpF{2_{10}^{(3,3,1,0,0)}}\,,\ 
  \SpF{2_{10}^{(4,3,0,0,0)}}
$}
\\\hline\hline
 $\alpha=-11$ & $E^{(11)}_{10,7,7,7,7,q}$ & $2_{11}^{(q,7-q,0,0,0)}$-brane (E$^{(11;7)}$-brane) \\ \hline
 IIA & \multicolumn{2}{|c|}{$
  \SpF{2_{11}^{(1,6,0,0,0)}}\,,\ 
  \SpF{2_{11}^{(3,4,0,0,0)}}\,,\ 
  \SpF{2_{11}^{(5,2,0,0,0)}}\,,\ 
  \SpF{2_{11}^{(7,0,0,0,0)}}
$} \\\hline
 IIB & \multicolumn{2}{|c|}{$
  \SpF{2_{11}^{(7,0,0,0)}}\,,\ 
  \SpF{2_{11}^{(2,5,0,0,0)}}\,,\ 
  \SpF{2_{11}^{(4,3,0,0,0)}}\,,\ 
  \SpF{2_{11}^{(6,1,0,0,0)}}
$}
\\\hline
\end{tabular}
}
\end{center}
\caption{Exotic branes with the tension proportional to $\gs^{\alpha}$ ($\alpha\leq -3$) and the corresponding mixed-symmetry potentials in type II string theories. 
Here, $n$ and $m$ are non-negative integers while $p$ ($q$) runs over non-negative even/odd (odd/even) numbers in type IIA/IIB theory.}
\label{tab:mixed-symmetry-list}
\end{table}
In fact, as we can see from Table \ref{tab:mixed-symmetry-list}, for example, there are three families of $E^{(4)}$-branes, and we distinguish them by introducing additional integers as $E^{(4;6)}$, $E^{(4;3)}$, and $E^{(4;0)}$\,. 
The general rule for the second integer is very simple; 
an exotic brane $b_{2n}^{(c_s,\,\cdots,\,c_2)}$ is a member of the $E^{(2n;c_n)}$-brane and an exotic brane $b_{2n+1}^{(c_s,\,\cdots,\,c_2)}$ is a member of the $E^{(2n+1;c_n+c_{n+1})}$-brane.

The first set of indices in the mixed-symmetry potential $E^{(n)}_{m_1\cdots m_{1+b+c_2+\cdots+c_s,\cdots}}$ corresponds to the worldvolume directions of the brane, and the directions after the first comma correspond to the isometry directions, namely the internal toroidal directions. 
As it is suggested from the relation \eqref{eq:brane-potential}, in order to relate the supersymmetric branes to the mixed-symmetry potentials, a set of indices delimited by commas has to be a subset of the set of indices sitting to the left. 
For example, $E_{0\mu\nu\rho 1234,34}$ couples to the exotic $5^2_3(\mu\nu\rho 12,34)$-brane while there is no supersymmetric brane which couples to $E_{0\mu\nu\rho 1234,45}$. 
By considering this argument, there is a one-to-one correspondence between supersymmetric branes and mixed-symmetry potentials. 
The explicit counting of the number of mixed-symmetry potentials in each dimension is summarized in Appendix \ref{app:counting-mixed-symmetry}. 

\section{Exotic-brane solutions in DFT}
\label{sec:exotic-DFT}

In this section, we explain how to construct the supergravity solutions for the variety of exotic branes discussed in the previous section. 
If we consider only the standard branes or the defect branes, we can (at least locally) write down the solutions satisfying the standard supergravity equations of motion. 
However, as we discuss in this section, for domain-wall branes or space-filling branes, we need to employ the manifestly duality-covariant formulations of supergravity, such as the DFT or EFT. 
This section is devoted to descriptions of exotic-brane solutions in DFT while the descriptions in EFT are discussed in Section \ref{sec:exotic-EFT}. 

Solutions of the domain-wall branes or the space-filling branes can be obtained from the standard-brane solutions or the defect-brane solutions by performing duality transformations. 
Since the standard branes or the defect branes are contained only in the $T$-duality webs given in Figures \ref{fig:web01}--\ref{fig:web05}, in DFT, we can at most construct brane solutions contained in Figures \ref{fig:web01}--\ref{fig:web05}. 
For branes contained in Figures \ref{fig:web06}--\ref{fig:web20}, we need to perform $S$-duality as well, and we need EFT. 
In this section, we consider only two examples, the D8-brane solution and the $5^3_2$-brane solution, and consider other solutions in Section \ref{sec:exotic-EFT}. 

\subsection{D7-brane solution}

Let us begin with the standard D7(1234567)-brane solution,
\begin{align}
 \rmd s^2 = \tau_2^{-1/2} \,\bigl(\rmd x^2_{01\cdots 7} + \tau_2 \,\rmd x_{89}^2\bigr)\,,\qquad 
 \Exp{-2\Phi}= \tau_2^{2} \,, \qquad
 \ket{A} = \bigl(\tau_1 - \tau_2^{-1}\,\gamma^{0 \cdots 7}\bigr)\ket{0} \,,
\label{eq:D7-soln}
\end{align}
where $\tau(x^8,\,x^9) \equiv \tau_1 +\ii\,\tau_2$ is given by
\begin{align}
\begin{split}
 &\tau(x^8,\,x^9) \equiv \frac{\ii\,\sigma}{2\pi}\,\ln(r_c/z) = \sigma\,\Bigl[\frac{\theta}{2\pi}+\ii\,\ln\Bigl(\frac{r_c}{r}\Bigr)\Bigr]\qquad \bigl(z\equiv x^8+\ii\,x^9 \equiv r \Exp{\ii\theta}\bigr) 
\\
 &\bigl[\sigma:\text{constant representing the number of D7-branes}\,,\quad r_c(>0):\text{cut-off parameter}\bigr]\,.
\end{split}
\end{align}
Here, we have introduced a shorthand notation,
\begin{align}
 \rmd x_{0m_1\cdots m_p}^2 \equiv -(\rmd x^0)^2 + \rmd x_{m_1\cdots m_p}^2 \,,\qquad 
 \rmd x_{m_1\cdots m_p}^2 \equiv \sum_{k=1}^p (\rmd x^{m_k})^2 \,. 
\end{align}

We now consider a domain-wall solution, the D8 solution. 
Since the D7 solution has codimension two, we need to implement the standard smearing procedure, which changes the function $\tau(x^8,\,x^9)$ keeping the expression \eqref{eq:D7-soln} intact. 
The resulting functions after the smearing are
\begin{align}
 \tau_1\equiv m\, x^8\,,\qquad \tau_2\equiv h_0 + m\,\abs{x^9} \,,
\end{align}
where $h_0$ and $m$ are constants. 
In the case of the D$p$-brane solution with $p\leq 6$ that depends on the transverse $(9-p)$ coordinates, the standard smearing procedure produces an additional isometry direction, and by performing a $T$-duality in the isometry direction, we can obtain the D$(p+1)$-brane solution. 
However, in the case of $p=7$, the smeared solution still depends on the two coordinates $x^8$ and $x^9$ and we cannot perform the usual $T$-duality to obtain the D$8$-brane solution. 
This is a new feature of the domain-wall solution. 

\subsection{A quick review of DFT}

In order to perform a formal $T$-duality, we utilize the DFT on a 20-dimensional doubled spacetime with the generalized coordinates $(x^M)=(x^m,\,\tilde{x}_m)$\,.\footnote{For our purpose, it is not necessary to double the time direction, but just for notational simplicity, we double all of the directions.} 
In DFT, all of the bosonic fields are packaged into the generalized metric $\cH_{MN}(x)$\,, the $T$-duality-invariant dilaton $d(x)$\,, and the $\OO(10,10)$ spinor $\ket{A}$\,. 
In the framework of DFT, we can consider formal $T$-dualities even in the absence of isometries. 
Indeed, without assuming the existence of any isometries, the equations of motion are transformed covariantly under a $T$-duality along the $x^i$-direction,
\begin{align}
\begin{split}
 &\cH_{MN}\ \to \ (\Lambda^{\rmT})_M{}^P\,\cH_{PQ}\,\Lambda^Q{}_N\,,\qquad d\to d\,, \qquad \ket{A}\ \to \ \gamma^{11}\,(\gamma_i-\gamma^i)\,\ket{A}\,,
\\
 &x^M\ \to \ (\Lambda^{-1})^M{}_N\,x^N\,,\qquad (\Lambda^M{}_N)\equiv {\footnotesize\begin{pmatrix} \bm{1} -e_i & e_i \\ e_i & \bm{1} -e_i \end{pmatrix}}, 
\end{split}
\label{eq:T-dual-DFT}
\end{align}
where we have defined a matrix $e_i\equiv \text{diag}(0,\dotsc,0,\overset{i}{1},0,\dotsc,0)$\,. 

The relation between the DFT fields and the usual supergravity fields is as follows. 
The generalized metric and the dilaton can be parameterized as\footnote{In our convention, the sign of the $B$-field is opposite to the conventional DFT.}
\begin{align}
 (\cH_{MN}) = \begin{pmatrix} (g-B\,g^{-1}\,B)_{mn} & -B_{mp}\,g^{pn} \\ g^{np}\,B_{pn} & g^{mn} \end{pmatrix} , \qquad 
 \Exp{-2d} = \sqrt{-g}\,\Exp{-2\Phi}\,. 
\end{align}
On the other hand, the $\OO(10,10)$ spinor $\ket{A}$ is defined on the Clifford vacuum $\ket{0}$ $\bigl($satisfying $\gamma_m\,\ket{0}=0\bigr)$ as
\begin{align}
 \ket{A} \equiv \sum_p \frac{1}{p!}\,A_{m_1\cdots m_p} \, \gamma^{m_1\cdots m_p}\,\ket{0}\,. 
\end{align} 
Here, the gamma matrices $(\gamma_M)\equiv (\gamma_m,\,\gamma^m)$ are defined by
\begin{align}
 \{\gamma_M\,,\gamma_N\} = \eta_{MN}\,,\qquad (\eta_{MN})\equiv \begin{pmatrix} 0 & \delta_m^n \\ \delta^m_n & 0 \end{pmatrix} ,
\end{align}
and $\gamma^{m_1\cdots m_p}\equiv \gamma^{[m_1}\cdots\gamma^{m_p]}$\,. 
Note that the $\OO(10,10)$ metric $\eta_{MN}$ and its inverse $\eta^{MN}$ are used to raise or lower the indices $M,N,\ldots$\,. 
The coefficients of the $\OO(10,10)$ spinor $A_{m_1\cdots m_p}$ are identified as the (curved) components of the R--R potential in type II supergravity, and they are related to another definition of the R--R potential $C_{m_1\cdots m_p}$ as $A=\Exp{-B_2\wedge}C$ ($A,\,C:$ polyform). 
In type IIA/IIB theory, only the R--R odd/even-form potentials are included, and thus $\ket{A}$ is defined to satisfy
\begin{align}
 \gamma^{11}\,\ket{A} = \mp \ket{A} \qquad (\text{IIA/IIB})\,. 
\end{align}
Here, $\gamma^{11}$ is defined as $\gamma^{11}\equiv (-1)^{N_f}$\,, where $N_f\equiv\gamma^m\,\gamma_m$ counts the number of gamma matrices. 
The field strength is defined as
\begin{align}
 \ket{F} \equiv \sla{\partial}\,\ket{A}\qquad \bigl(\sla{\partial}\equiv \gamma^M\,\partial_M\bigr)\,. 
\label{eq:DFT-RR-FS}
\end{align}

Unlike the standard supergravity fields, the DFT fields can depend on the generalized coordinates $x^M$ but the consistency condition, namely the SC,
\begin{align}
 \eta^{MN}\,\partial_M \otimes \partial_N =0 \,,
\end{align}
requires that the DFT fields cannot depend on more than ten coordinates out of twenty. 
If we keep the dependence on the standard coordinates $x^m$\,, for example, the field strength \eqref{eq:DFT-RR-FS} is reduced to the usual one,
\begin{align}
 \ket{F} = \sum_p \frac{1}{(p+1)!}\,F_{m_1\cdots m_{p+1}} \, \gamma^{m_1\cdots m_{p+1}}\,\ket{0}\,,\qquad 
 F_{m_1\cdots m_{p+1}}\equiv (p+1)\,\partial_{[m_1}A_{m_2\cdots m_{p+1}]}\,. 
\end{align}
In this paper, we consider different choices of coordinates where supergravity fields depend on some of the winding coordinates $\tilde{x}_m$\,. 

\subsection{D8-brane solution}

By using the above setup, let us construct the D8-brane solution in DFT. 
We start from the smeared D7 solution \eqref{eq:D7-soln}, and perform the formal $T$-duality \eqref{eq:T-dual-DFT} along the $x^8$-direction.
We then obtain
\begin{align}
 \rmd s^2 = \tau_2^{-1/2} \,\bigl(\rmd x^2_{01\cdots 8} + \tau_2 \,\rmd x_{9}^2\bigr)\,,\qquad 
 \Exp{-2\Phi}= \tau_2^{5/2} \,, \qquad
 \ket{A} = \bigl(\tau_1\,\gamma^8 - \tau_2^{-1}\,\gamma^{0\cdots 8}\bigr)\ket{0} \,. 
\label{eq:D8-soln}
\end{align}
Since the formal $T$-duality changes the coordinates $x^8\ \leftrightarrow\ \tilde{x}_8$\,, the $\tau_1$ here has the linear winding-coordinate dependence; $\tau_1=m\,\tilde{x}_8$\,. 
In fact, this is precisely the D8 solution in DFT \cite{Hohm:2011cp} (which corresponds to the familiar D8 solution \cite{Bergshoeff:1996ui}). 
The field strength becomes
\begin{align}
 \ket{F} = \sla{\partial} \ket{A}
 = \bigl(m + \partial_9\tau_2^{-1}\,\gamma^{0 \cdots 9}\bigr)\ket{0}\,,
\end{align}
which means that the background has the constant 0-form and the dual 10-form field strengths
\begin{align}
 F_0 = m\,,\qquad F_0= *_{10}F_{10} \,. 
\end{align}
The relation to the Romans massive IIA supergravity \cite{Romans:1985tz} is discussed in Section \ref{sec:deformed-sugra}. 

\subsection{$5^3_2$-brane solution}
\label{sec:532-DFT}

There is another known domain-wall solution in DFT, the $5^3_2$-brane solution, also known as the $R$-brane solution \cite{Hassler:2013wsa,Bakhmatov:2016kfn,Kimura:2018hph}. 

Let us start with the background of the (smeared) exotic $5^2_2(12345,67)$-brane \cite{Meessen:1998qm},
\begin{align}
\begin{split}
 &\rmd s^2 = \rmd x^2_{01\cdots 5} + \frac{\tau_2}{\abs{\tau}^2} \, \rmd x^2_{67} + \tau_2 \,\rmd x_{89}^2\,,\qquad 
  \Exp{-2\Phi}= \frac{\abs{\tau}^2}{\tau_2} \,, 
\\
 &B_2 = -\frac{\tau_1}{\abs{\tau}^2}\,\rmd x^6\wedge\rmd x^7 \,,\qquad
  D_6 = - \frac{\abs{\tau}^2}{\tau_2} \rmd x^0\wedge \cdots \wedge \rmd x^5 \,,
\end{split}
\end{align}
where $\tau_1= m\, x^8$ and $\tau_2= h_0+m\,\abs{x^9}$\,. 
Here, $D_6$ is the potential of the dual field strength $H_7\equiv \rmd D_6+\cdots$ (where the ellipses denote non-linear terms depending on type IIA or IIB) satisfying $H_7=\Exp{-2\Phi}*_{10}H_3$\,. 
Again by performing a formal $T$-duality along the $x^8$-direction, we obtain the background of the $5^3_2(12345,678)$-brane,
\begin{align}
 \rmd s^2 = \rmd x^2_{01\cdots 5} + \frac{\tau_2}{\abs{\tau}^2} \, \rmd x^2_{67} + \tau_2^{-1} \,\rmd x_{8}^2 + \tau_2 \,\rmd x_{9}^2\,,\quad 
 \Exp{-2\Phi}= \abs{\tau}^2 \,, \quad
 B_2 = -\frac{\tau_1}{\abs{\tau}^2}\,\rmd x^6\wedge\rmd x^7 \,,
\label{eq:532-soln-conv}
\end{align}
where $\tau_1= m\, \tilde{x}_8$ and $\tau_2= h_0+m\,\abs{x^9}$\,. 
The DFT fields associated with these fields $(g_{mn},\,B_{mn},\,\Phi)$ satisfy the equations of motion of DFT as it is expected, as the formal $T$-duality always maps a solution to a solution.

In fact, in order to describe the $5^2_2$ or the $5^3_2$ backgrounds, it is more convenient to introduce the dual supergravity fields $(\tilde{g}_{mn},\,\beta^{mn},\,\tilde{\phi})$ suggested in \cite{Duff:1989tf,Andriot:2011uh}. 
They are defined through
\begin{align}
\begin{split}
 &(\cH_{MN}) = \begin{pmatrix} (g-B\,g^{-1}\,B)_{mn} & -B_{mp}\,g^{pn} \\ g^{np}\,B_{pn} & g^{mn} \end{pmatrix} 
            = \begin{pmatrix} \tilde{g}_{mn} & -\tilde{g}_{mp}\,\beta^{pn} \\ \beta^{np}\,\tilde{g}_{pn} & (\tilde{g}^{-1}-\beta\,\tilde{g}\,\beta)^{mn} \end{pmatrix} ,
\\
 & \Exp{-2d} = \sqrt{-g}\,\Exp{-2\Phi} = \sqrt{-\tilde{g}}\,\Exp{-2\tilde{\phi}} \,,
\end{split}
\label{eq:conv-dual-DFT}
\end{align}
and can be regarded as redefinitions of the supergravity fields. 
More explicitly, we obtain\footnote{Let us note that this map can be singular in certain backgrounds, for example, when $E_{mn}$ is not invertible.}
\begin{align}
\begin{split}
 &\tilde{g}_{mn} = E_{mp}\,E_{nq}\,g^{pq}\,,\qquad 
 \beta^{mn} = E^{mp}\,E^{nq}\,B_{pq}\,,\qquad 
 \Exp{-2\tilde{\phi}} \equiv \frac{\det (g_{mn})}{\det (E_{mn})}\,\Exp{-2\Phi}\,,
\\
 &E_{mn}\equiv g_{mn}+B_{mn}\,,\qquad 
 E^{mn}\equiv (E^{-1})^{mn} = \tilde{g}^{mn} -\beta^{mn}\,. 
\end{split}
\end{align}

From the relation, we can determine the dual parameterization for the $5^2_2$ background as \cite{Hassler:2013wsa}
\footnote{Here, we have dropped an unimportant minus sign in front of $m$ by a redefinition of $m$\,. 
Similarly, in the following computation, such minus sign can appear during a course of duality transformations, but we will always absorb the sign into $m$ for simplicity.}
\begin{align}
 \rmd \tilde{s}^2 = \rmd x^2_{01\cdots 5} + \tau_2^{-1} \, \rmd x^2_{67} + \tau_2 \,\rmd x_{89}^2\,,\qquad 
 \Exp{-2\tilde{\phi}}= \tau_2 \,, \qquad
 \beta^{67} = m\,x^8 \,,
\label{eq:522-soln}
\end{align}
and the $5^3_2$ background as
\begin{align}
 \rmd \tilde{s}^2 = \rmd x^2_{01\cdots 5} + \tau_2^{-1} \, \rmd x^2_{678} + \tau_2 \,\rmd x_{9}^2\,,\qquad 
 \Exp{-2\tilde{\phi}}= \tau_2^2 \,, \qquad
 \beta^{67} = m\,\tilde{x}_8 \,.
\label{eq:532-soln}
\end{align}

In the dual description, the winding-coordinate dependence in the $5^3_2$ background is contained only in the $\beta$-field. 
Moreover, its dependence is only linear similar to the D8 background. 
In fact, as we show in the next section, in all of the ``elementary'' domain-wall solutions, the winding-coordinate dependence appears only in a certain gauge field linearly. 

\paragraph{Non-geometric fluxes and mixed-symmetry potentials\\}

In the dual parameterization, we can define the so-called non-geometric $Q$-flux \cite{Marchesano:2007vw,Wecht:2007wu,Halmagyi:2008dr,Grana:2008yw} as 
\begin{align}
 Q_1^{pq}\equiv Q_m{}^{pq}\,\rmd x^m \equiv \rmd \beta^{pq} \,. 
\end{align}
The non-geometricity of the $5^2_2$-brane (or the $Q$-brane) background was pointed out in \cite{deBoer:2010ud}, and as shown in \cite{Hassler:2013wsa,Geissbuhler:2013uka}, the $5^2_2(12345,67)$ background has a constant $Q$-flux,
\begin{align}
 Q_{\bar8}{}^{\bar6\bar7} = m \,.
\end{align}
Here and hereafter, in order to avoid confusion, we may add bars on integers, like $\bar6$ and $\bar7$, indicating that these integers are associated with certain spacetime directions.

The low-energy effective Lagrangian for the non-geometric $Q$-flux was obtained in \cite{Andriot:2011uh} as
\begin{align}
 \cL \sim \sqrt{-\tilde{g}}\Exp{-2\tilde{\phi}}\Bigl(\tilde{R} + 4\,\abs{\rmd \tilde{\phi}}^2 - \frac{1}{2}\,\abs{Q}^2 \Bigr) \,,
\end{align}
where $\abs{Q}^2\equiv \tilde{g}^{mn}\,\tilde{g}_{pq,\,rs}\,Q_m{}^{pq}\,Q_n{}^{rs}$\,, and we have used $\tilde{g}_{p_1\cdots p_n,\,q_1\cdots q_n}\equiv \tilde{g}_{p_1r_1}\cdots\tilde{g}_{p_nr_n}\,\delta^{r_1\cdots r_n}_{q_1\cdots q_n}$\, and  $\delta^{r_1\cdots r_n}_{q_1\cdots q_n}\equiv \delta^{[r_1}_{[q_1}\cdots \delta^{r_n]}_{q_n]}$\,, and $\tilde{R}$ is the Ricci scalar associated with $\tilde{g}_{mn}$\,. 
The equation of motion for the $\beta$-field takes the form
\begin{align}
 \partial_m \bigl(\Exp{-2\tilde{\phi}}\sqrt{-\tilde{g}}\,\tilde{g}^{mn}\,\tilde{g}_{pq,\,rs}\, Q_n{}^{rs} \bigr) = 0 \,,
\end{align}
and this suggests to introduce the dual field strength as \cite{Sakatani:2014hba}
\begin{align}
 Q_{9,2} \equiv \Exp{-2\tilde{\phi}} \tilde{g}_{pq,\,rs}\, \tilde{*}_{10}Q_1^{pq} \otimes \rmd x^r\wedge\rmd x^s\,.
\end{align}
Here, the subscript ``$9,2$'' represents that the field strength is the mixed-symmetry tensor with 9 antisymmetric indices and 2 antisymmetric indices, and the Hodge star operator $\tilde{*}_{10}$ is associated with the dual metric $\tilde{g}_{mn}$\,. 
By introducing the associated potential $Q_{9,2}\equiv \rmd D_{8,2}$\,, we can find a connection between the non-geometric $Q$-flux and the mixed-symmetry potential $D_{8,2}$ introduced in a series of works \cite{Bergshoeff:2010xc,Bergshoeff:2011zk,Bergshoeff:2011mh,Bergshoeff:2011ee,Bergshoeff:2011se,Bergshoeff:2012ex,Bergshoeff:2012pm,Bergshoeff:2016ncb,Lombardo:2016swq,Bergshoeff:2017gpw} (see \cite{Bergshoeff:2011se,Bergshoeff:2015cba} for a similar Hodge duality between $Q$-flux and the mixed-symmetry potential $D_{8,2}$). 
In the $5^2_2(12345,67)$ background, we obtain
\begin{align}
 Q_{9,\,\bar6\bar7} = \rmd \bigl(-m\,\tau_2^{-1}\,\rmd x^0\wedge\cdots\wedge \rmd x^7\bigr) \,,\qquad 
 D_{\bar0\bar1\bar2\bar3\bar4\bar5\bar6\bar7,\,\bar6\bar7} = -m\,\tau_2^{-1} \,. 
\end{align}

As discussed in \cite{Hassler:2013wsa}, by $T$-dualizing the $Q$-brane background, we can obtain the background of the $R$-brane, which is nothing but the $5^3_2$-brane. 
By defining the non-geometric $R$-flux,
\begin{align}
 R^{mnp} \equiv 3\,\tilde{\partial}^{[m} \beta^{np]} \,, 
\end{align}
we can show that the $5^3_2$ background contains a constant $R$-flux,
\begin{align}
 R^{\bar8\bar6\bar7} \equiv 3\,\tilde{\partial}^{[8} \beta^{67]} = m \,. 
\end{align}
The $R$-flux is sometimes called the \emph{locally non-geometric flux}. 
In \cite{Andriot:2012wx}, the effective Lagrangian for the $R$-flux was derived from the DFT Lagrangian as (see also \cite{Blumenhagen:2012nk,Blumenhagen:2012nt,Blumenhagen:2013aia,Asakawa:2015jza,Kaneko:2016eqp})
\begin{align}
 \cL \sim \sqrt{-\tilde{g}}\Exp{-2\tilde{\phi}}\Bigl(\tilde{R} + 4\,\abs{\rmd \tilde{\phi}}^2 - \frac{1}{2}\,\abs{R}^2 \Bigr) \,,
\end{align}
where $\abs{R}^2\equiv \frac{1}{3!}\,\tilde{g}_{m_1m_2m_3,\,n_1n_2n_3}\,R^{m_1m_2m_3}\,R^{n_1n_2n_3}$\,. 
This again suggests to define the dual field strength as
\begin{align}
 R_{10,\,m_1m_2m_3} \equiv \Exp{-2\tilde{\phi}} \tilde{g}_{m_1m_2m_3,\,n_1n_2n_3}\, \tilde{*}_{10}R^{n_1n_2n_3} \,. 
\end{align}
By defining the corresponding potential $R_{10,3} \equiv \rmd D_{9,3}$\,, we obtain
\begin{align}
 R_{10,\,\bar6\bar7\bar8} = \rmd \bigl(-m\, \tau_2^{-1}\,\rmd x^0\wedge\cdots\wedge \rmd x^8\bigr) \,,\qquad
 D_{\bar0\bar1\bar2\bar3\bar4\bar5\bar6\bar7\bar8,\,\bar6\bar7\bar8} = -m\,\tau_2^{-1} \,,
\end{align}
in the $5^3_2(12345,678)$ background. 
A similar duality relation between the mixed-symmetry potential $D_{9,3}$ and the $R$-flux was recently discussed in \cite{Bergshoeff:2015cba}. 
As it has been discussed there, $D_{9,3}$ is the $T$-dual of $D_{8,2}$\,,
\begin{align}
 D_{a_1\cdots a_8,\,b_1b_2} \ \overset{T_z}{\longleftrightarrow} \ D_{a_1\cdots a_8y,\,b_1b_2y}\qquad 
 \bigl(y\not\in \{a_1,\dotsc,a_8\}\,,\quad \{b_1,b_2\}\in \{a_1,\dotsc,a_8\}\bigr)\,. 
\end{align}
By observing the explicit form of the $D_{8,2}$ in the $5^2_2$ background and the $D_{9,3}$ in the $5^3_2$ background, our result is consistent with the above $T$-duality rule. 

According to the above relation between $Q$- and $R$-fluxes and the mixed-symmetry potentials, we can summarize the famous $T$-duality chain \cite{Wecht:2007wu} as follows:
\begin{align}
 \begin{pmatrix}
 5_2 \\ H_{xyz} \\ D_6
 \end{pmatrix}
 \ \overset{T_z}{\longleftrightarrow} \ 
 \begin{pmatrix}
 5_2^1 \\ f_{xy}{}^z \\ D_{6z,\,z}
 \end{pmatrix}
 \ \overset{T_y}{\longleftrightarrow} \ 
 \begin{pmatrix}
 5_2^2 \\ Q_x{}^{yz} \\ D_{6yz,\,yz}
 \end{pmatrix}
 \ \overset{T_x}{\longleftrightarrow} \ 
 \begin{pmatrix}
 5_2^3 \\ R^{xyz} \\ D_{6xyz,\,xyz}
 \end{pmatrix}\,.
\end{align}

Let us make an additional comment on the $5^3_2(12345,678)$ solution \eqref{eq:532-soln}. 
It does not have a symmetry between the $\{x^6,\,x^7\}$ and $x^8$-directions although there is no particular difference between these three internal directions $\{x^6,\,x^7,\,x^8\}$. 
However, this is just a matter of a gauge choice, and there is no asymmetry at the level of the field strength $R^{\bar8\bar6\bar7}=3\,\tilde{\partial}^{[\bar8}\beta^{\bar6\bar7]}$ since the indices are totally antisymmetrized. 
In the next section, we encounter many locally non-geometric backgrounds, for which proper definitions of the locally non-geometric $R$-fluxes are not known. 
In such cases, we provide heuristic definitions of the $R$-fluxes by considering the symmetry of exotic branes and providing an appropriate antisymmetrization, like $R^{mnp}=3\,\tilde{\partial}^{[m}\beta^{np]}$\,. 

\section{All exotic-brane solutions in EFT}
\label{sec:exotic-EFT}

In this section, we give a prescription to construct all of the elementary exotic-brane solutions in EFT. 
After introducing duality transformations in Sections \ref{sec:duality-rotations-in-EFT} and \ref{sec:dual-param}, in Sections \ref{sec:first-dw} to \ref{sec:exotic-sol-ii} we construct all the domain-wall solutions contained in \eqref{eq:all-M-branes}, \eqref{eq:all-IIA-branes}, and \eqref{eq:all-IIB-branes} as well as their associated $R$-fluxes and mixed-symmetry potentials. 
In Section \ref{sec:exotic-sol-m}, we uplift these domain-wall solutions to M-theory and obtain various domain-wall solutions in M-theory. 
In Section \ref{sec:exotic-spacefilling}, we show the validity of our method to construct any exotic-brane background and give some space-filling branes solutions as examples. 
Finally, in Section \ref{sec:susy} we discuss the projection conditions for Killing spinors.

\subsection{Duality rotations in EFT}
\label{sec:duality-rotations-in-EFT}

Type II string theory compactified on an $(n-1)$-torus has the $E_{n(n)}$ $U$-duality symmetry, which contains the $\OO(n-1,n-1)$ $T$-duality symmetry as a subgroup. 
The $E_{n(n)}$ EFT is a generalization of DFT that manifests the $U$-duality symmetry in supergravity \cite{West:2000ga,West:2001as,Hillmann:2009pp,Berman:2010is,Berman:2011jh,Berman:2012vc,West:2012qz,Hohm:2013pua,Hohm:2013vpa,Hohm:2013uia,Hohm:2014fxa,Hohm:2015xna,Abzalov:2015ega,Musaev:2015ces,Berman:2015rcc}. 
Similar to DFT, it is defined on an extended spacetime with the generalized coordinates $x^I$ associated with the branes in the particle multiplet of the $E_{n(n)}$ group. 
In particular, when we consider M-theory/$T^n$, the set of coordinates $x^I$ is parameterized as
\begin{align}
 (x_{\text{\tiny (M)}}^I) = (x^i ,\, y_{i_1i_2} ,\, y_{i_1\cdots i_5} ,\, y_{i_1\cdots i_7,\,j} ,\, y_{i_1\cdots i_8,\,j_1j_2j_3} ,\, y_{i_1\cdots i_8,\,j_1\cdots j_6} ,\, y_{i_1\cdots i_8,\,j_1\cdots j_8,\,k}) \,,
\label{eq:M-coordinates}
\end{align}
where the indices $i,j,k$ run over the internal toroidal directions $i=d,\dotsc,11$ and the external spacetime has the usual coordinates $x^\mu$ ($\mu=0,\cdots,d-1$). 
Each of the generalized coordinates corresponds to that of M-theory branes such as P, M2, M5, KKM etc., and the total number is equal to the dimension of the particle multiplet of the $E_{n(n)}$ group. 
On the other hand, when we consider type IIB theory/$T^{n-1}$, we can parameterize the same generalized coordinates as
\begin{align}
\begin{split}
 (x^I_{\text{\tiny (IIB)}}) = &(x^m,\, \sfy^\alpha_m ,\, \sfy_{m_1m_2m_3} ,\, \sfy^\alpha_{m_1\cdots m_5} ,\, \sfy_{m_1\cdots m_6,\,n} ,\, \sfy^{\alpha\beta}_{m_1\cdots m_7} , 
\\
 &\, \sfy^\alpha_{m_1\cdots m_7,\,n_1n_2} ,\, \sfy_{m_1\cdots m_7,\,n_1\cdots n_4} ,\, \sfy^\alpha_{m_1\cdots m_7,\,n_1\cdots n_6} ,\, \sfy_{m_1\cdots m_7,\,n_1\cdots n_7,\,p}) \,,
\end{split}
\end{align}
where $m,n,p=d,\dotsc,10$\,. 
In the type IIB parameterization, all of the coordinates are associated with the type IIB branes, such as P, F1/D1, D3, NS5/D5 etc., and the index $\alpha$ represents the $\SL(2)$ $S$-duality doublet. 
The winding-coordinates $\sfy^{\alpha\beta}_{m_1\cdots m_7}=\sfy^{(\alpha\beta)}_{m_1\cdots m_7}$ correspond to the triplet of the 7-branes.

In EFT, the supergravity fields are contained in the generalized metric $\cM_{IJ}$ and additional fields which contain the external $d$-dimensional indices $\mu$. 
For simplicity, we here concentrate on the generalized metric $\cM_{IJ}$\,. 
Corresponding to the two parameterizations of $x^I$, we can parameterize the generalized metric $\cM_{IJ}$ in two ways; in terms of the bosonic fields in 11D supergravity (\emph{M-theory parameterization}) and the bosonic fields in type IIB supergravity (\emph{type IIB parameterization}). 
We refer to Appendix \ref{app:EFT-parameterization} for a more detailed study of these parameterizations. 
As determined in \cite{Sakatani:2017nfr} for the cases $E_{n(n)}$ EFT ($n\leq 7$), the two parameterizations can be related by a linear map,
\begin{align}
 \cM_{IJ}^{\text{\tiny (IIB)}} = (S^\rmT)_I{}^K\,\cM^{\text{\tiny (M)}}_{KL}\,S^L{}_J \,,\qquad 
 x^I_{\text{\tiny (IIB)}} = (S^{-1})^I{}_J \,x_{\text{\tiny (M)}}^J\,.
\label{eq:linear-map}
\end{align}
Here, the $S^I{}_J$ is a constant matrix and under this transformation, the equations of motion of EFT (prior to choosing a particular solution of the SC) are transformed covariantly. 
If we rewrite the fields in 11D supergravity in terms of those in type IIA supergravity, by comparing both sides in \eqref{eq:linear-map}, we find the standard $T$-duality rules between type IIA and type IIB supergravity \cite{Sakatani:2017nfr}. 
Therefore, we can do $T$-duality transformations throughout the linear map \eqref{eq:linear-map}, as the matrix $S^I{}_J$ contains the information of the $T$-duality direction. 
On the other hand, the $S$-duality rule is rather trivial. 
In the type IIB parameterization, all of the generalized coordinates and the supergravity fields are $\SL(2)$ tensors, and the indices $\alpha$ and $\beta$ are rotated by a matrix $\Lambda^{\alpha}{}_{\beta}\equiv \bigl(\begin{smallmatrix} 0 & 1\\ -1 & 0\end{smallmatrix}\bigr)$ as usual. 

\subsection{Dual parameterization in the whole bosonic sector}
\label{sec:dual-param}

In the case of DFT, the conventional fields and the dual fields are related through the expression \eqref{eq:conv-dual-DFT}. 
Here, we briefly explain how to generalize the relation \eqref{eq:conv-dual-DFT} to EFT. 

As we already explained, the generalized metric $\cM_{IJ}$ in EFT can be parameterized by the bosonic fields in type IIB supergravity, which we call $\cM_{IJ}^{\text{\tiny(IIB)}}$\,.
We can also parameterize the same generalized metric in terms of the dual fields in type IIB supergravity such as $(\tilde{g},\,\tilde{\phi},\,\beta^{mn},\,\gamma^{m_1\cdots m_p},\,\cdots)$ \cite{Blair:2014zba,Lee:2016qwn,Sakatani:2017nfr}. 
This is called the \emph{non-geometric parameterization} since the dual fields are related to the non-geometric fluxes, and we call the generalized metric $\cM_{IJ}^{\text{\tiny(IIB, non-geometric)}}$\,. 
Similar to \eqref{eq:conv-dual-DFT}, by comparing the two parameterizations as
\begin{align}
 \cM_{IJ}^{\text{\tiny(IIB)}} = \cM_{IJ}^{\text{\tiny(IIB, non-geometric)}} \,,
\label{eq:IIB-parameterization-map1}
\end{align}
we can, in principle, determine the dual fields in terms of the conventional fields. 

Since the generalized metric contains only the supergravity fields with internal (toroidal) components, for the metric with external indices $g_{\mu\nu}$ and $g_{\mu m}$\,, we need a more elaborated recipe. 
For our purposes, it is enough to know the transformation rule for the components $g_{\mu\nu}$. 
By truncating other external fields, the duality relation becomes \cite{Lee:2016qwn}
\begin{align}
 \bigl(\det g^{\rmE}_{mn}\bigr)^{\frac{1}{d-2}}\,g^{\rmE}_{\mu\nu} = \bigl(\det\tilde{g}^{\rmE}_{mn}\bigr)^{\frac{1}{d-2}}\,\tilde{g}^{\rmE}_{\mu\nu}\,,
\label{eq:IIB-parameterization-map2}
\end{align}
where $g^{\rmE}_{mn}\equiv\Exp{-\frac{1}{2}\,\Phi}g_{mn}$ and $\tilde{g}^{\rmE}_{mn}\equiv\Exp{-\frac{1}{2}\,\tilde{\phi}}\tilde{g}_{mn}$ are internal components of the Einstein-frame metric that are contained in $\cM_{IJ}$\,. 
We can compute the external components of the Einstein-frame dual metric $\tilde{g}^{\rmE}_{\mu\nu}$ and the string-frame metric is obtained as $\tilde{g}_{\mu\nu}\equiv\Exp{\frac{1}{2}\,\tilde{\phi}}\tilde{g}^{\rmE}_{\mu\nu}$\,.

\paragraph*{Reorganization of the generalized coordinates\\}

In order to simplify the $T$-duality rule, we here consider the following redefinitions of the generalized coordinates in type II theory. 
The winding coordinates for P and F1 (that appear also in DFT) are defined as
\begin{align}
 \big(x^m,\,\tilde{x}_m\bigr) \equiv 
 \begin{cases}
  \bigl(x^m,\, y_{m\rmM}\bigr) & (\text{IIA})\,, \\
  \bigl(x^m,\, \sfy^1_m\bigr) & (\text{IIB}) \,,
 \end{cases}
\end{align}
in terms of the generalized coordinates in the M-theory/type IIB parameterizations. 
Here, $\rmM$ denotes the M-theory direction. 
The winding coordinates for the D-branes $y^{\rmD}_{m_1\cdots m_p}$ ($p=0,\dotsc,7$) and the solitonic branes $5_2^n$ ($n=0,1,2$) are denoted as
\begin{align}
 &(y^{\rmD}_{m_1\cdots m_p}) \equiv 
 \bigl(\underbrace{-x^\rmM}_{\text{D0}},\,\underbrace{-\sfy^2_m}_{\text{D1}},\,\underbrace{y_{m_1m_2}}_{\text{D2}},\, \underbrace{\sfy_{m_1m_2m_3}}_{\text{D3}},\,\underbrace{y_{m_1\cdots m_4\rmM}}_{\text{D4}},\, \underbrace{\sfy^1_{m_1\cdots m_5}}_{\text{D5}},\,\underbrace{y_{m_1\cdots m_6\rmM,\rmM}}_{\text{D6}},\,\underbrace{\sfy^{11}_{m_1\cdots m_7}}_{\text{D7}}\bigr)\,,
\\
 &\bigl(y^{\rmS}_{m_1\cdots m_5n_1\cdots n_n,\,n_1\cdots n_n}\bigr) \equiv 
  \begin{cases}
   \bigl(\underbrace{y_{m_1\cdots m_5}}_{5_2},\, \underbrace{y_{m_1\cdots m_6\rmM,\,n}}_{5^1_2},\, \underbrace{y_{m_1\cdots m_7\rmM,\,n_1n_2\rmM}}_{5^2_2}\bigr) & (\text{IIA})\,,\\
   \bigl(\underbrace{-\sfy^2_{m_1\cdots m_5}}_{5_2},\, \underbrace{\sfy_{m_1\cdots m_6,\,n}}_{5^1_2},\, \underbrace{\sfy^1_{m_1\cdots m_7,\,n_1n_2}}_{5^2_2}\bigr) & (\text{IIB})\,. 
\end{cases}
\end{align}
The winding coordinates for the exotic $p_3^{7-p}$-branes ($p=0,\dotsc,7$) and the $(1^6_4,\,0^{(1,6)}_4)$-branes are called
\begin{align}
 \begin{split}
 &(y^{\rmE}_{m_1\cdots m_7,\,n_1\cdots n_{7-p}})
 \equiv \bigl(\underbrace{y_{m_1\cdots m_7\rmM,\,n_1\cdots n_7\rmM,\,\rmM}}_{0^7_3},\,
 \underbrace{\sfy^1_{m_1\cdots m_7,\,n_1\cdots n_6}}_{1^6_3},\,
 \underbrace{y_{m_1\cdots m_7\rmM,\,n_1\cdots n_5\rmM}}_{2^5_3},\,
 \underbrace{\sfy_{m_1\cdots m_7,\,n_1\cdots n_4}}_{3^4_3},\,
\\
 &\qquad\qquad\qquad\qquad\quad \underbrace{y_{m_1\cdots m_7\rmM,\,n_1n_2n_3}}_{4^3_3},\,
 \underbrace{-\sfy^2_{m_1\cdots m_7,\,n_1n_2}}_{5^2_3},\,
 \underbrace{y_{m_1\cdots m_7,\,s}}_{6^1_3},\,
 \underbrace{\sfy^{22}_{m_1\cdots m_7}}_{7_3}\bigr)\,.
\end{split}
\\
 &\bigl(\underbrace{\tilde{x}_{m_1\cdots m_7,\,n_1\cdots n_6}}_{1^6_4},\, \underbrace{\tilde{x}_{m_1\cdots m_7,\,n_1\cdots n_7,\,p}}_{0^{(1,6)}_4}\bigr)\equiv
 \begin{cases}
  \bigl(y_{m_1\cdots m_7\rmM,\,n_1\cdots n_6},\,y_{m_1\cdots m_7\rmM,\,n_1\cdots n_7\rmM,\,p}\bigr) & (\text{IIA})\,,\\
  \bigl(-\sfy^2_{m_1\cdots m_7,\,n_1\cdots n_6},\, \sfy_{m_1\cdots m_7,\,n_1\cdots n_7,\,p}\bigr) & (\text{IIB})\,. 
 \end{cases}
\end{align}

The remaining eight coordinates in the $E_{8(8)}$ exceptional space, 
\begin{align}
 &y_{m_1\cdots m_6\rmM,\,n}\quad \bigl(n\not\in\{m_1\cdots m_6\}\bigr)\,,\quad y_{m_1\cdots m_7\rmM}\qquad (\text{IIA})\,, 
\\
 &\sfy_{m_1\cdots m_6,\,n}\quad \bigl(n\not\in\{m_1\cdots m_6\}\bigr)\,,\quad \sfy^{12}_{m_1\cdots m_7}\qquad (\text{IIB})\,, 
\end{align}
correspond to the eight missing states that are not connected to other branes under $T$- and $S$-dualities (see Section \ref{sec:missing}). 
In the following discussion, we do not consider these coordinates any more since these do not appear in our solutions. 

In summary, when we consider type II theory, we use the following set of generalized coordinates:
\begin{align}
\begin{split}
 (x^I)=\bigl\{&x^m,\,\tilde{x}_m,\,y^{\rmD}_{m_1\cdots m_{p}},\,y^{\rmS}_{m_1\cdots m_5},\,y^{\rmS}_{m_1\cdots m_5n,\, n},\,y^{\rmS}_{m_1\cdots m_5n_1n_2,\,n_1n_2}, 
\\
 &y^{\rmE}_{m_1\cdots m_7,\,n_1\cdots n_{7-p}},\, \tilde{x}_{m_1\cdots m_7,\,n_1\cdots n_6},\, \tilde{x}_{m_1\cdots m_7,\,n_1\cdots n_7,\,p},\,\text{\footnotesize{(missing states)}}\bigr\}\,,
\end{split}
\end{align}
where $p$ is an even/odd number in type IIA/IIB. 
On the other hand, when we consider M-theory, we employ the original parameterization \eqref{eq:M-coordinates}. 

\paragraph*{$T$-duality rule\\}

With the above parameterization, the linear map \eqref{eq:linear-map} (a $T$-duality along the $y$-direction) between the generalized coordinates can be summarized as follows:\footnote{A sign flip may happen in the maps for winding coordinates associated with $5^1_2\leftrightarrow 5^2_2$, $1^6_3\leftrightarrow 2^5_3$, $3^4_3\leftrightarrow 4^3_3$, $5^2_3\leftrightarrow 6^1_3$, $1^6_4\leftrightarrow 1^6_4$, and $1^6_4\,(\text{IIA})\leftrightarrow 0^{(1,6)}_4\,(\text{IIB})$\,, since the linear map for the $E_{8(8)}$ case is determined only up to sign in \cite{Sakatani:2017nfr}. 
However, even if a minus sign can appear, it does not affect the following computations for obtaining exotic-brane solutions in EFT since the minus sign can be absorbed into a free parameter $m$\,.}
\begin{align}
 &x^a\ \leftrightarrow\ x^a\,,\qquad 
 \tilde{x}_a\ \leftrightarrow\ \tilde{x}_a\,,\qquad 
 x^y\ \leftrightarrow\ \tilde{x}_y\,, \qquad
 y^{\rmD}_{a_1\cdots a_p}\ \leftrightarrow \ y^{\rmD}_{a_1\cdots a_py}\,, 
\nn\\
 &y^{\rmS}_{a_1\cdots a_4b_1\cdots b_ny,\,b_1\cdots b_n}\ \leftrightarrow \ y^{\rmS}_{a_1\cdots a_4b_1\cdots b_ny,\,b_1\cdots b_n}\,,\qquad 
 y^{\rmS}_{a_1\cdots a_5b_1\cdots b_n,\,b_1\cdots b_n}\ \leftrightarrow \ y^{\rmS}_{a_1\cdots a_5b_1\cdots b_ny,\,b_1\cdots b_ny}\,,
\nn\\
 &y^{\rmE}_{a_1\cdots a_pb_1\cdots b_{7-p}y,\,b_1\cdots b_{7-p}}\ \leftrightarrow \ y^{\rmE}_{a_1\cdots a_pb_1\cdots b_{7-p}y,\,b_1\cdots b_{7-p}y}\,,\qquad 
 \tilde{x}_{ab_1\cdots b_5y,\,b_1\cdots b_5y}\ \leftrightarrow \ \tilde{x}_{ib_1\cdots b_5y,\,b_1\cdots b_5y}\,,
\nn\\
 &\tilde{x}_{a_1\cdots a_6y,\,a_1\cdots a_6}\ \leftrightarrow \ \tilde{x}_{a_1\cdots a_6y,\,a_1\cdots a_6y,\,y}\,,\qquad 
  \tilde{x}_{a_1\cdots a_6y,\,a_1\cdots a_6y,\,a_6}\ \leftrightarrow \ \tilde{x}_{a_1\cdots a_6y,\,a_1\cdots a_6y,\,a_6}\,,
\label{eq:EFT-T-dual1}
\end{align}
where $a$ and $b$ run over the internal directions other than $y$\,. 

In the following, we utilize the dual parameterization of the generalized metric. 
The dual fields in M-theory and type IIB theory can be summarized as
\begin{align}
 \text{M-theory}:\quad &\bigl\{\tilde{G}_{ij},\,\Omega^{i_1i_2i_3},\,\Omega^{i_1\cdots i_6},\,\Omega^{i_1\cdots i_8,\,j}\bigr\}\,,
\\
 \text{type IIB}:\quad &\bigl\{\tilde{g}_{mn},\, \tilde{\phi},\,\gamma,\,\beta^{mn},\, \gamma^{mn},\,\gamma^{m_1\cdots m_4},\,\beta^{m_1\cdots m_6},\,\gamma^{m_1\cdots m_6},\,\beta^{m_1\cdots m_7,\,n} \bigr\}\,.
\end{align}
If we decompose the dual fields in M-theory as
\begin{align}
\begin{split}
 \tilde{G}^{mn} &= \Exp{-\frac{2}{3}\,\tilde{\phi}_{\text{\tiny(A)}}}\,\tilde{g}_{\text{\tiny(A)}}^{mn}+\Exp{\frac{4}{3}\,\tilde{\phi}_{\text{\tiny(A)}}}\,\gamma_{\text{\tiny(A)}}^m\, \gamma_{\text{\tiny(A)}}^n \,,\quad 
 \tilde{G}^{m\rmM} = -\Exp{\frac{4}{3}\,\tilde{\phi}_{\text{\tiny(A)}}}\, \gamma_{\text{\tiny(A)}}^m \,,
\\
 \tilde{G}^{\rmM\rmM} &= \Exp{\frac{4}{3}\,\tilde{\phi}_{\text{\tiny(A)}}}\,,\quad
 \Omega^{mn \rmM} = \beta_{\text{\tiny(A)}}^{mn} \,,\quad 
 \Omega^{mnp} = \gamma_{\text{\tiny(A)}}^{mnp} \,,
\\
 \Omega^{m_1\cdots m_5\rmM} &= \gamma_{\text{\tiny(A)}}^{m_1\cdots m_5} - 5\,\gamma_{\text{\tiny(A)}}^{[m_1m_2m_3}\,\beta_{\text{\tiny(A)}}^{m_4m_5]} \,,\quad 
 \Omega^{m_1\cdots m_6} = \beta_{\text{\tiny(A)}}^{m_1\cdots m_6} \,,
\end{split}
\end{align}
the linear map \eqref{eq:linear-map} reproduces the following $T$-duality transformation rules \cite{Sakatani:2017nfr}:
\begin{align}
\begin{split}
 \tilde{g}_{\text{\tiny(A)}}^{ab} &= \tilde{g}^{ab} - \frac{\tilde{g}^{a y}\,\tilde{g}^{b y}-\beta^{a y}\,\beta^{b y}}{\tilde{g}^{yy}}\,,\quad 
 \tilde{g}_{\text{\tiny(A)}}^{a y}= \frac{\beta^{a y}}{\tilde{g}^{yy}}\,,\quad 
 \tilde{g}_{\text{\tiny(A)}}^{yy}=\frac{1}{\tilde{g}^{yy}}\,,
\\
 \beta_{\text{\tiny(A)}}^{ab} &= \beta^{ab} + \frac{\beta^{a y}\,\tilde{g}^{b y}-\tilde{g}^{a y}\,\beta^{b y}}{\tilde{g}^{yy}}\,,\quad 
 \beta_{\text{\tiny(A)}}^{a y} = \frac{\tilde{g}^{a y}}{\tilde{g}^{yy}} \,, \quad
 \Exp{-2\tilde{\phi}_{\text{\tiny(A)}}}= \frac{\Exp{-2\tilde{\phi}}}{\tilde{g}^{yy}} \,,
\\
 \gamma_{\text{\tiny(A)}}^{a_1\cdots a_{n-1}y}&= \gamma^{a_1\cdots a_{n-1}} - (n-1)\,\frac{\gamma^{[a_1\cdots a_{n-2}|y|}\,\tilde{g}^{a_{n-1}]y}}{\tilde{g}^{yy}}\,,
\\
 \gamma_{\text{\tiny(A)}}^{a_1\cdots a_n} &= \gamma^{a_1\cdots a_n y} + n\, \gamma^{[a_1\cdots a_{n-1}}\, \beta^{a_n]y} + n\,(n-1)\,\frac{\gamma^{[a_1\cdots a_{n-2}|y|}\, \beta^{a_{n-1}|y|}\,\tilde{g}^{a_n] y}}{\tilde{g}^{yy}}\,,
\\
 \beta_{\text{\tiny(A)}}^{a_1\cdots a_5 y}&= \beta^{a_1\cdots a_5 y} + 5\,\eta^{[a_1\cdots a_4}\,\gamma^{a_5]y} + 5\,\eta^{[a_1a_2a_3|y|}\,\gamma^{a_4a_5]} 
 - \frac{45}{2}\, \gamma^{[a_1a_2}\,\beta^{a_3a_4}\,\gamma^{a_5]y} 
\\
 &- \frac{15}{2}\, \gamma^{[a_1a_2}\,\gamma^{a_3a_4}\,\beta^{a_5]y} + \frac{10\,\eta^{[a_1\cdots a_3|y|}\,\gamma^{a_4|y|}\tilde{g}^{a_5]y}}{\tilde{g}^{yy}} - \frac{15\, \gamma^{[a_1a_2}\, \beta^{a_3|y|}\,\gamma^{a_4|y|}\,\tilde{g}^{a_5]y}}{\tilde{g}^{yy}}
\,, 
\end{split}
\label{eq:EFT-T-dual2}
\end{align}
where $\eta^{m_1\cdots m_4} \equiv \gamma^{m_1\cdots m_4} + 3\,\beta^{[m_1m_2}\, \gamma^{m_3m_4]}$\,. 
In the following, for simplicity, we drop the subscript ${\text{\tiny(A)}}$ for the type IIA fields. 

In summary, the $T$-duality transformation \eqref{eq:EFT-T-dual1} and \eqref{eq:EFT-T-dual2} maps a solution of EFT to another solution of EFT. 
Although \eqref{eq:EFT-T-dual2} is the result for the $E_{7(7)}$ EFT \cite{Sakatani:2017nfr}, if we consider the $E_{8(8)}$ EFT, there appears an additional dual field $\beta^{m_1\cdots m_7,\,n}$\,. 
Its duality rule has not been determined yet in the context of EFT, but in the following discussion, it is enough to employ the $T$-duality rule\footnote{Our purpose is to study supergravity solutions of the exotic branes. In these brane solutions, only a single potential has a non-vanishing value and we can ignore non-linear terms.}
\begin{align}
 \beta^{a_1\cdots a_6} \ \overset{T_y}{\longleftrightarrow} \ \beta^{a_1\cdots a_6y,\,y} + \text{(irrelevant non-linear terms)}\,,
\end{align}
which is the dual of the transformation rule $D_{6} \ \overset{T_y}{\longleftrightarrow} \ D_{6y,y}+ \text{(irrelevant non-linear terms)}$ \cite{Eyras:1998hn}. 

\paragraph*{$S$-duality rule\\}

The $S$-duality transformation rules are as follows:
\begin{align}
\begin{split}
 &\tilde{g}'_{mn} = \sqrt{\Exp{-2\tilde{\phi}}+\gamma^2}\,\tilde{g}_{mn}\,,\quad 
 \Exp{-\tilde{\phi}'}=\frac{\Exp{-\tilde{\phi}}}{\Exp{-2\tilde{\phi}}+\gamma^2}\,,\quad 
 \gamma' = -\frac{\gamma}{\Exp{-2\tilde{\phi}}+\gamma^2}\,, 
\\
 &\beta'^{mn} = -\gamma^{mn}\,,\quad \gamma'^{mn} = \beta^{mn} \,,\quad 
  \gamma'^{m_1\cdots m_4} = \gamma^{m_1\cdots m_4} + 6\,\beta^{[m_1m_2}\,\gamma^{m_3m_4]} \,, 
\\
 &\gamma'^{m_1\cdots m_6} = -\beta^{m_1\cdots m_6} + 45\,\gamma^{[m_1m_2}\,\gamma^{m_3m_4}\,\beta^{m_5m_6]}\,,
\\
 &\beta'^{m_1\cdots m_6} = \gamma^{m_1\cdots m_6} - 45\,\beta^{[m_1m_2}\,\beta^{m_3m_4}\,\gamma^{m_5m_6]}\,.
\end{split}
\end{align}
The $S$-duality rule for $\beta^{m_1\cdots m_7,\,n}$ will be $\beta'^{m_1\cdots m_7,\,n}=\beta^{m_1\cdots m_7,\,n}+\text{(non-linear terms)}$\,. 
The $S$-duality transformation also rotates the generalized coordinates as
\begin{align}
\begin{split}
 &\tilde{x}'_m = - y^{\rmD}_m\,,\quad 
 y'^{\rmD}_m = \tilde{x}_m\,,\quad 
 y'^{\rmD}_{m_1\cdots m_5} = - y^{\rmS}_{m_1\cdots m_5}\,, 
\\ 
 &y'^{\rmS}_{m_1\cdots m_5} = y^{\rmD}_{m_1\cdots m_5}\,,\quad 
 y'^{\rmD}_{m_1\cdots m_7} = y^{\rmE}_{m_1\cdots m_7}\,,\quad 
 y'^{\rmE}_{m_1\cdots m_7} = y^{\rmD}_{m_1\cdots m_7}\,,
\\
 &y'^{\rmS}_{m_1\cdots m_7,\,n_1n_2} = -y^{\rmE}_{m_1\cdots m_7,\,n_1n_2}\,,\quad 
 y'^{\rmE}_{m_1\cdots m_7,\,n_1n_2} = y^{\rmS}_{m_1\cdots m_7,\,n_1n_2}\,,
\\
 &y'^{\rmE}_{m_1\cdots m_7,\,n_1\cdots n_6} = -\tilde{x}_{m_1\cdots m_7,\,n_1\cdots n_6}\,,\quad \tilde{x}'_{m_1\cdots m_7,\,n_1\cdots n_6} = y^{\rmE}_{m_1\cdots m_7,\,n_1\cdots n_6}\,, 
\end{split}
\end{align}
while keeping other coordinates invariant.

\subsection{First two examples of domain-wall solutions in EFT}
\label{sec:first-dw}

Before considering all of the ``elementary'' domain-wall solutions, let us begin with two simple examples. 

\subsubsection{$p_3^{(1,7-p)}$-brane background}

We start with the smeared exotic $p_3^{7-p}(1\cdots p,p+1\cdots 7)$-brane solution,
\begin{align}
\begin{split}
 &\rmd s^2 = \frac{\abs{\tau}}{\tau_2^{1/2}}\, \bigl(\rmd x^2_{01\cdots p}+\tau_2\,\rmd x_{89}^2\bigr) + \frac{\tau_2^{1/2}}{\abs{\tau}}\, \rmd x^2_{(p+1)\cdots 7} \,, 
\\
 &\Exp{-2\Phi} = \biggl(\frac{\abs{\tau}^2}{\tau_2}\biggr)^{\frac{3-p}{2}} \,, \qquad C_{(p+1) \cdots 7} = -\frac{\tau_1}{\abs{\tau}^2} \,,
\end{split}
\label{eq:p(7-p)3-soln-conv}
\end{align}
where $\tau_1= m\,x^8$\,, $\tau_2= h_0+m\,\abs{x^9}$\,, $\rmd x^2_{(7+1)\cdots 7}\equiv 0$\,, and $C_{(7+1) \cdots 7}$ represents the R--R 0-form $C_0$\,. 
By performing the $T$-duality along the $x^8$-direction, we obtain the $p_3^{(1,7-p)}(1\cdots p,p+1\cdots 7,8)$-brane solution
\begin{align}
\begin{split}
 &\rmd s^2 = \frac{\abs{\tau}}{\tau_2^{1/2}}\, \bigl(\rmd x^2_{01\cdots p}+\tau_2\,\rmd x_{9}^2\bigr)
  + \frac{\tau_2^{1/2}}{\abs{\tau}}\, \rmd x^2_{(p+1)\cdots 7}
  + \frac{\tau_2^{-1/2}}{\abs{\tau}}\, \rmd x_8^2\,,
\\
 &\Exp{-2\Phi}= \frac{\abs{\tau}^{4-p}}{\tau_2^{\frac{2-p}{2}}} \,,\qquad
  C_{(p+1) \cdots 78} = -\frac{\tau_1}{\abs{\tau}^2} \,. 
\end{split}
\label{eq:p1(7-p)3-soln-conv}
\end{align}
where $\tau_1= m\, \tilde{x}_8$ and $\tau_2= h_0+m\,\abs{x^9}$\,. 

\paragraph*{Dual parameterization for the $7_3$-brane solution\\}

For the $7_3$ background, the non-vanishing fields are the $(g_{\mu\nu},\,g_{mn},\,\Phi,\,C_0)$\,. 
In this case, \eqref{eq:IIB-parameterization-map1} and \eqref{eq:IIB-parameterization-map2} are reduced to
\begin{align}
\begin{split}
 &g^{\rmE}_{mn} = \tilde{g}^{\rmE}_{mn}\,,\qquad 
 \bigl(m_{\alpha\beta}\bigr) = \Exp{\Phi}\,\begin{pmatrix}
 \Exp{-2\Phi} + (C_0)^2 & C_0 \\
 C_0 & 1
 \end{pmatrix} = \Exp{\tilde{\phi}}\,\begin{pmatrix}
 1 & \gamma \\
 \gamma & \Exp{-2\tilde{\phi}} + \gamma^2
 \end{pmatrix} \,,
\\
 &\bigl(\det g^{\rmE}_{mn}\bigr)^{\frac{1}{d-2}}\,g^{\rmE}_{\mu\nu}= \bigl(\det\tilde{g}^{\rmE}_{mn}\bigr)^{\frac{1}{d-2}}\,\tilde{g}^{\rmE}_{\mu\nu}\,,
\end{split}
\end{align}
and we obtain the dual parameterization for the $7_3$-brane solution,
\begin{align}
 \rmd \tilde{s}^2 = \tau_2^{1/2}\,\bigl(\rmd x^2_{01\cdots 7}+\tau_2\,\rmd x_{89}^2\bigr) \,, \qquad 
 \Exp{-2\tilde{\phi}} = \tau_2^{-2} \,,\qquad \gamma = m\,x^8 \,. 
\end{align}

\paragraph*{Dual parameterization for other $p$-brane solutions\\}

We can similarly obtain the dual parameterizations for other $p$-brane solutions in \eqref{eq:p(7-p)3-soln-conv} and \eqref{eq:p1(7-p)3-soln-conv}. 
However, in general, a direct comparison of the generalized metrics is very complicated. 

For simplicity, we instead use the $T$-duality rules \eqref{eq:EFT-T-dual1} and \eqref{eq:EFT-T-dual2}. 
Then, we can easily obtain the dual parameterization of the $p_3^{7-p}$ background. 
The $p_3^{(1,7-p)}$ background is also obtained by further performing a $T$-duality along the $x^8$-direction. 
The results are as follows:
\begin{align}
\begin{split}
 &\bm{\underline{p_3^{7-p}(1\cdots p,p+1\cdots 7):}}
\\
 &\rmd \tilde{s}^2 = \tau_2^{1/2}\,\bigl(\rmd x^2_{01\cdots p}+\tau_2\,\rmd x_{89}^2\bigr) + \tau_2^{-1/2}\, \rmd x^2_{(p+1)\cdots 7}\,, \quad 
 \Exp{-2\tilde{\phi}} = \tau_2^{\frac{3-p}{2}} \,,\quad \gamma^{(p+1)\cdots 7} = m\,x^8 \,,
\end{split}
\label{eq:p(7-p)3-soln}
\\
\begin{split}
 &\bm{\underline{p_3^{(1,7-p)}(1\cdots p,p+1\cdots 7,8):}}
\\
 &\rmd \tilde{s}^2 = \tau_2^{1/2}\,\bigl(\rmd x^2_{01\cdots p}+\tau_2\,\rmd x_{9}^2\bigr) + \tau_2^{-1/2}\, \rmd x^2_{(p+1)\cdots 7} + \tau_2^{-3/2}\,\rmd x_8^2\,, 
\\
 &\Exp{-2\tilde{\phi}} = \tau_2^{\frac{6-p}{2}} \,,\qquad \gamma^{(p+1)\cdots 8} = m\,\tilde{x}_8 \,. 
\end{split}
\label{eq:p1(7-p)3-soln}
\end{align}
In the $p_3^{(1,7-p)}$ solution, similar to the $5^3_2$ solution \eqref{eq:532-soln}, the winding-coordinate dependence is appearing only in the $\gamma$-field linearly. 

\paragraph{Non-geometric flux and mixed-symmetry potentials\\}

As discussed in \cite{Sakatani:2014hba}, backgrounds of the exotic $p_3^{7-p}$-branes are the magnetic sources of the non-geometric $P$-fluxes. 
The non-geometric $P$-fluxes were introduced in \cite{Aldazabal:2006up,Aldazabal:2008zza,Aldazabal:2010ef} and in particular, a $P$-flux $P_m{}^{pq}$ is $S$-dual of the $Q$-flux. 
They are roughly defined as (see \cite{Lee:2016qwn} for more details)
\begin{align}
 P_1^{n_1\cdots n_{7-p}} \equiv P_m{}^{n_1\cdots n_{7-p}}\,\rmd x^m \equiv \rmd \gamma^{n_1\cdots n_{7-p}}\,,
\end{align}
and in the $p_3^{7-p}(1\cdots p,p+1\cdots 7)$ background, it takes the form
\begin{align}
 P_{\bar8}{}^{\overline{p+1}\cdots\bar7} = m \,.
\end{align}

According to \cite{Sakatani:2014hba,Lee:2016qwn}, the effective Lagrangian for the $P$-fluxes has the form
\begin{align}
 \cL \sim - \frac{1}{2}\,\sqrt{-\tilde{g}}\Exp{-4\tilde{\phi}}\!\!\!\! \sum_{p=\text{even/odd}}\!\! \abs{P_1^{7-p}}^2 \qquad \text{(IIA/IIB)}\,,
\end{align}
where $\abs{P_1^{7-p}}^2 \equiv \tilde{g}^{pq}\,\tilde{g}_{m_1\cdots m_{7-p},\,n_1\cdots n_{7-p}}\,P_p{}^{m_1\cdots m_{7-p}}\,P_q{}^{n_1\cdots n_{7-p}}$\,. 
The equations of motion suggest the dual field strength of the form \cite{Sakatani:2014hba}
\begin{align}
 P_{9,7-p} \equiv \Exp{-4\tilde{\phi}} \tilde{g}_{n_1\cdots n_{7-p},\,q_1\cdots q_{7-p}}\, \tilde{*}_{10}P_1^{n_1\cdots n_{7-p}} \otimes \rmd x^{q_1}\wedge\cdots\wedge\rmd x^{q_{7-p}}\,,
\label{eq:Hodge-E-P-flux}
\end{align}
and its potential may be defined as $P_{9,7-p}\equiv \rmd E_{8,7-p}$\,.
These mixed-symmetry potentials were discussed in \cite{Bergshoeff:2011ee} and the Hodge duality similar to \eqref{eq:Hodge-E-P-flux} was discussed in \cite{Bergshoeff:2011se,Bergshoeff:2015cba}. 
In the $p_3^{7-p}(1\cdots p,p+1\cdots 7)$ solution, we obtain
\begin{align}
 E_{\bar0\bar1\bar2\bar3\bar4\bar5\bar6\bar7,\,\overline{p+1}\cdots\bar7} = -m\,\tau_2^{-1} \,. 
\end{align}

Similarly, in the $p_3^{(1,7-p)}(1\cdots p,p+1\cdots 7,8)$ background, the derivative of the $\gamma$-field gives a locally non-geometric flux introduced in \cite{Bergshoeff:2015cba}, 
\begin{align}
 R^{m_1\cdots m_{7-p},\,m} \equiv \tilde{\partial}^{m} \gamma^{m_1\cdots m_{7-p}} \,. 
\end{align}
In the $p_3^{(1,7-p)}(1\cdots p,p+1\cdots 7,8)$ solution, we obtain a constant flux,
\begin{align}
 R^{\overline{p+1}\cdots \bar8,\,\bar8} \equiv \tilde{\partial}^{\bar8} \gamma^{\overline{p+1}\cdots\bar 8} = m \,. 
\end{align}
The dual field strengths may be defined as
\begin{align}
 R_{10,8-p,1} \equiv \Exp{-4\tilde{\phi}} \tilde{g}_{m_1\cdots m_{8-p},\,n_1\cdots n_{8-p}}\,\tilde{g}_{rs}\, \tilde{*}_{10}R^{m_1\cdots m_{8-p},\,r} \otimes \rmd x^{n_1}\wedge \cdots \wedge\rmd x^{n_{8-p}}\otimes \rmd x^r \,,
\end{align}
which should be derived from DFT/EFT but here we introduced them heuristically. 
If we also introduce the potential as $R_{10,8-p,1} \equiv \rmd E_{9,8-p,1}$\,, we obtain
\begin{align}
 E_{\bar0\bar1\bar2\bar3\bar4\bar5\bar6\bar7\bar8,\,\overline{p}\cdots\bar8,\bar8} = -m\,\tau_2^{-1} \,,
\label{eq:E-potential-simple}
\end{align}
in the $p_3^{(1,7-p)}(1\cdots p,p+1\cdots 7,8)$ background, suggesting the $T$-duality rule
\begin{align}
 E_{a_1\cdots a_8,\,b_1\cdots b_{7-p}} \ \overset{T_y}{\longleftrightarrow} \ E_{a_1\cdots a_8y,\,b_1\cdots b_{7-p}y,\,y}\qquad 
 \bigl(y\not\in \{a_1,\dotsc,a_8\}\,,\quad \{b_1,\,b_2\}\in \{a_1,\dotsc,a_8\}\bigr)\,,
\end{align}
which is consistent with the result of \cite{Lombardo:2016swq}. 

\subsubsection{$1_4^{(1,0,6)}$-brane background}

As the second example, we consider the $1_4^{(1,0,6)}$-brane background. 
It can be obtained from the smeared $1_4^6(1,234567)$-brane background:
\begin{align}
\begin{split}
 &\rmd s^2= \frac{\abs{\tau}^2}{\tau_2} \, \bigl(\rmd x_{01}^2 + \tau_2\,\rmd x_{89}\bigr)+ \rmd x^2_{2\cdots 7} \,, \quad 
  \Exp{-2\Phi} = \frac{\tau_2}{\abs{\tau}^2} \,, 
\\
 &B_2 = -\frac{\abs{\tau}^2}{\tau_2}\,\rmd x^0 \wedge \rmd x^1 \,,\qquad
  D_6 = -\frac{\tau_1}{\abs{\tau}^2}\,\rmd x^2 \wedge \cdots\wedge \rmd x^7 \,. 
\end{split}
\end{align}
By performing a formal $T$-duality along the $x^8$-direction, we obtain the $1_4^{(1,0,6)}(1,234567,,8)$ background,
\begin{align}
 \rmd s^2= \frac{\abs{\tau}^2}{\tau_2} \, \bigl(\rmd x_{01}^2 + \tau_2\,\rmd x_9\bigr)+ \rmd x^2_{2\cdots 8} \,, \quad 
 \Exp{-2\Phi} = \frac{\tau_2}{\abs{\tau}} \,, \qquad
 B_2 = -\frac{\abs{\tau}^2}{\tau_2}\,\rmd x^0 \wedge \rmd x^1 \,,
\end{align}
where $\tau$ is $\tau_1= m\, \tilde{x}_8$ and $\tau_2= h_0+ m\,\abs{x^9}$\,. 
Since there are no R--R fields, we can easily check that this is a solution of DFT. 

The dual parameterization again can be obtained by comparing two parameterizations of the generalized metric in EFT, but in order to obtain the dual parameterization for the $1_4^6$ solution, it is easier to $S$-dualize the $1_3^6$ solution \eqref{eq:p(7-p)3-soln}. 
The dual parameterization for the $1_4^{(1,0,6)}$ solution can be obtained by further performing a $T$-duality. 
The results are as follows:
\begin{align}
\begin{split}
 &\bm{\underline{1_4^6(1,234567):}}
\\
 &\rmd \tilde{s}^2 = \tau_2 \,\bigl(\rmd x^2_{01}+\tau_2\,\rmd x_{89}^2\bigr) + \rmd x^2_{2\cdots 7}\,, \quad 
 \Exp{-2\tilde{\phi}} = \tau_2^{-1} \,,\quad \beta^{2\cdots 7} = m\,x^8 \,,
\end{split}
\label{eq:164-soln}
\\
\begin{split}
 &\bm{\underline{1_4^{(1,0,6)}(1,234567,,8):}}
\\
 &\rmd \tilde{s}^2 = \tau_2 \,\bigl(\rmd x^2_{01}+\tau_2\,\rmd x_{9}^2\bigr) + \tau_2^{-1}\,\rmd x_8^2 + \rmd x^2_{2\cdots 7}\,, \quad 
 \Exp{-2\tilde{\phi}} = \tau_2 \,,\quad \beta^{2\cdots 8,\,8} = m\,\tilde{x}^8 \,. 
\end{split}
\label{eq:11064-soln}
\end{align}

\paragraph{Non-geometric fluxes and mixed-symmetry potentials\\}

We can again consider a definition of a non-geometric flux \cite{Sakatani:2014hba,Lee:2016qwn}
\begin{align}
 Q_1^{n_1\cdots n_6}\equiv Q_m{}^{n_1\cdots n_6}\,\rmd x^m \equiv \rmd \beta^{n_1\cdots n_6} \,.
\end{align}
In the $1_4^6(1,234567)$ background, we obtain
\begin{align}
 Q_{\bar8}{}^{\bar2\cdots\bar7} = m \,.
\end{align}
In this case, the effective Lagrangian becomes \cite{Sakatani:2014hba,Lee:2016qwn}
\begin{align}
 \cL \sim - \frac{1}{2}\,\sqrt{-\tilde{g}}\Exp{-6\tilde{\phi}} \abs{Q_1^6}^2 \,,
\end{align}
where $\abs{Q_1^{6}}^2 \equiv \tilde{g}^{pq}\,\tilde{g}_{m_1\cdots m_6,\,n_1\cdots n_6}\,Q_p{}^{m_1\cdots m_6}\,Q_q{}^{n_1\cdots n_6}$ and the dual field strength is defined as \cite{Sakatani:2014hba}
\begin{align}
 Q_{9,6}\equiv \Exp{-6\tilde{\phi}} \tilde{g}_{m_1\cdots m_6,\,n_1 \cdots n_6}\, \tilde{*}_{10} Q_1^{m_1\cdots m_6} \otimes \rmd x^{n_1}\wedge\cdots\wedge\rmd x^{n_6}\,.
\end{align}
This kind of Hodge duality has been also suggested in \cite{Bergshoeff:2011se}. 
The mixed-symmetry potential may be defined through $Q_{9,6}\equiv \rmd E^{(4)}_{8,6}$\,, and in the $1_4^6(1,234567)$ background, we obtain
\begin{align}
 E^{(4)}_{\bar0\cdots\bar7,\,\bar2\cdots\bar7} = -m\,\tau_2^{-1} \,. 
\end{align}

Similarly, in the $1_4^{(1,0,6)}(1,234567,,8)$ background, a new locally non-geometric flux may be defined as
\begin{align}
 R^{\bar2\cdots \bar8,\,\bar8,\,\bar8} \equiv \tilde{\partial}^{\bar8} \gamma^{\bar2\cdots\bar 8,\,\bar 8} = m \,. 
\end{align}
If we define the dual field strength as
\begin{align}
\begin{split}
 R_{10,7,1,1} &\equiv \Exp{-6\tilde{\phi}} \tilde{g}_{m_1\cdots m_7,\,n_1\cdots n_{7}}\,\tilde{g}_{pr}\,\tilde{g}_{qs}\, \tilde{*}_{10}R^{m_1\cdots m_{7},\,p,\,q}\\
 &\quad \otimes \rmd x^{n_1}\wedge \cdots \wedge\rmd x^{n_{7}}\otimes \rmd x^r\otimes \rmd x^s \,,
\end{split}
\end{align}
and also define the potential through $R_{10,7,1,1} \equiv \rmd E^{(4)}_{9,7,1,1}$\,, we obtain
\begin{align}
 E^{(4)}_{\bar0\cdots\bar8,\bar2\cdots\bar8,\,\bar8,\,\bar8} = -m\,\tau_2^{-1} \,. 
\end{align}

\subsubsection{A short summary}

Up to here, we have discussed the defect-brane solutions
\begin{align}
 \text{D7}\ \eqref{eq:D7-soln},\quad 5^2_2\ \eqref{eq:522-soln},\quad p_3^{7-p}\ \eqref{eq:p(7-p)3-soln},\quad 1_4^{(1,0,6)}\ \eqref{eq:164-soln}\,,
\end{align}
and the domain-wall-brane solutions
\begin{align}
 \text{D8}\ \eqref{eq:D8-soln},\quad 5^3_2\ \eqref{eq:532-soln},\quad p_3^{(1,7-p)}\ \eqref{eq:p1(7-p)3-soln},\quad 1_4^{(1,0,6)}\ \eqref{eq:11064-soln}\,,
\end{align}
in DFT or EFT. 
The D7 and D8-branes are rather exceptional, but for other branes, the results can be summarized as follows. 
The defect branes are the magnetic sources of the 1-form fluxes $\{\bm{Q}_1^A\}\equiv\{Q_1^2,\,P_1^{7-p},\,Q_1^6\}$, which are known as the \emph{globally non-geometric fluxes}. 
The electric potentials are 8-forms $\{\bm{E}_{8,A}\}\equiv\{D_{8,2},\,E_{8,7-p},\,E^{(4)}_{8,6}\}$\,, whose field strengths are related to the magnetic fluxes, schematically written as
\begin{align}
 \bm{E}_{8,A} = \Exp{2(1-n)\,\tilde{\phi}}\, (\tilde{g} \cdots\tilde{g})\,\tilde{*}_{10}\bm{Q}_1^A\,,
\end{align}
where $n$ represents the power of $\gs$ in the tension of the exotic brane $T\sim \gs^{-n}$ and $(\tilde{g} \cdots\tilde{g})$ denotes that all of the upper indices in $\bm{Q}_1^A$ are lowered with the dual metric $\tilde{g}_{mn}$\,. 
On the other hand, the domain-wall branes are the magnetic sources of the locally non-geometric $R$-fluxes $\{\bm{R}^{\{A\}}\}\equiv\{R^3,\,R^{8-p,1},\,R^{7,1,1}\}$. 
The electric potentials are 9-forms $\{\bm{E}_{9,\{A\}}\}\equiv\{D_{9,3},\,E_{9,8-p,1},\,E^{(4)}_{9,7,1,1}\}$\,, whose field strengths are related to the magnetic fluxes as
\begin{align}
 \bm{E}_{9,\{A\}} = \Exp{2(1-n)\,\tilde{\phi}}\, (\tilde{g} \cdots\tilde{g})\,\tilde{*}_{10}\bm{R}^{\{A\}}\,. 
\end{align}
This suggests that there is a one-to-one correspondence between domain-wall branes, the 9-form mixed-symmetry potentials, and the $R$-fluxes. 
Further, the set of indices $\{A\}$ in the $R$-fluxes can be found from the set of indices $\{A\}$ in the mixed-symmetry potentials, which are consistent with the general rule \eqref{eq:brane-potential}. 
In fact, this appears to be a general structure as we see below. 

In the following, we will firstly introduce a generalization of the locally non-geometric $R$-fluxes, and then show that the domain-wall solutions in EFT have a constant $R$-flux. 

\subsection{Locally non-geometric fluxes}

As we have already discussed, a domain-wall brane, say the $b^{(c_s,\dotsc,c_2)}_n$-brane, is the magnetic source of the non-geometric flux with a set of antisymmetrized indices, $R^{c_2+\cdots+c_s,\dotsc,c_{s-1}+c_{s},c_s}$, which is a $U$-duality version of the familiar $R$-flux (see \cite{Blair:2014zba,Gunaydin:2016axc,Lust:2017bwq} for definitions of locally non-geometric fluxes in M-theory compactified on up to the 7-torus). 
In this subsection, we obtain a generalization of the locally non-geometric fluxes for the case of $E_{8(8)}$ EFT, both in terms of type II theories and M-theory. 

Rather than the systematic analysis similar to \cite{Blair:2014zba,Gunaydin:2016axc,Lust:2017bwq}, we here take a heuristic approach to define the locally non-geometric fluxes. 
Our starting point is the familiar $R$-flux, $R^{m_1m_2m_3} \equiv 3\,\tilde{\partial}^{[m_1} \beta^{m_2m_3]}$\,, whose magnetic source is the $5^3_2$-brane. 
Under $S$-duality transformation, the $5^3_2$-brane is mapped to the $5^3_4$-brane, and at the same time, the $R$-flux is mapped to another flux, $R^{m_1m_2m_3} \equiv 3\,\partial_{\rmD}^{[m_1} \gamma_{\vphantom{\rmD}}^{m_2m_3]}$\,. 
By using the transformation rules given in Section \ref{sec:duality-rotations-in-EFT}, we can further perform the $T$- and $S$-duality transformations, and obtain the $U$-dual counterpart of the locally non-geometric $R$-fluxes. 
At certain points, we need to prescribe appropriate antisymmetrizations such that the $R$-fluxes have the expected index structure. 
By repeating the procedures, we can find appropriate definitions of the $R$-fluxes that transform covariantly under the $T$- and $S$-duality transformations. 
For simplicity, we only consider the linear dependence on the potentials and ignore non-linear terms of the form $\beta^{\cdots}\,\partial_{\cdots}\,\gamma^{\cdots}$\,, but it is enough to study backgrounds of exotic brane, where only a single potential is non-vanishing.

\subsubsection{Locally non-geometric fluxes in type IIA theory/$T^7$}

The obtained $R$-fluxes in type IIA theory and the corresponding domain-wall branes can be summarized as follows:
\begin{align}
\begin{split}
 &\underline{R_{(2)}^{3}\ \leftrightarrow\ \text{$5^3_2$-brane:}}
\\
 &R_{(2)}^{m_1m_2m_3} \equiv 3\,\tilde{\partial}^{[m_1} \beta^{m_2m_3]} \,,
\end{split}
\\
\begin{split}
 &\underline{R_{(3)}^{7,1}\ \leftrightarrow\ \text{$1^{(1,6)}_3$-brane:}}
\\
 &R_{(3)}^{m_1\cdots m_7,\,n} \equiv 
 \tilde{\partial}^n\,\gamma^{m_1\cdots m_7} - 7\,\partial_{\rmD}^{[m_1\cdots m_6}\beta^{m_7]n} \,,
\end{split}
\\
\begin{split}
 &\underline{R_{(3)}^{5,1}\ \leftrightarrow\ \text{$3^{(1,4)}_3$-brane:}}
\\
 &R_{(3)}^{m_1\cdots m_5,\,n} \equiv 
 \tilde{\partial}^n\,\gamma^{m_1\cdots m_5} - 5\,\partial_{\rmD}^{[m_1\cdots m_4}\beta^{m_5]n} 
 - \partial_{\rmD}^{m_1\cdots m_5q} A^n_q \,,
\end{split}
\\
\begin{split}
 &\underline{R_{(3)}^{3,1}\ \leftrightarrow\ \text{$5^{(1,2)}_3$-brane:}}
\\
 &R_{(3)}^{m_1m_2m_3,\,n} \equiv 
 \tilde{\partial}^n\,\gamma^{m_1m_2m_3} - 3\,\partial_{\rmD}^{[m_1m_2}\beta^{m_3]n} 
 - \partial_{\rmD}^{m_1m_2m_3q} A^n_q \,,
\end{split}
\\
\begin{split}
 &\underline{R_{(3)}^{1,1}\ \leftrightarrow\ \text{$7^{(1,0)}_3$-brane:}}
\\
 &R_{(3)}^{m,\,n} \equiv 
 \tilde{\partial}^n\,\gamma^{m} - \partial_{\rmD}^{mq} A^n_q \,,
\end{split}
\\
\begin{split}
 &\underline{R_{(4)}^{7,1,1}\ \leftrightarrow\ \text{$1^{(1,0,6)}_4$-brane:}}
\\
 &R_{(4)}^{m_1\cdots m_7,\,n,\,p} \equiv
 \tilde{\partial}^{n} \beta^{m_1\cdots m_7,\,p}
 -7\,\partial_{\rmS}^{[m_1\cdots m_6|,\,p}\beta^{|m_7]n} 
 + \partial_{\rmS}^{m_1\cdots m_7,\,nq}A^p_q \,,
\end{split}
\\
\begin{split}
 &\underline{R_{(4)}^{3}\ \leftrightarrow\ \text{$5^3_4$-brane:}}
\\
 &R_{(4)}^{m_1m_2m_3} \equiv
 \partial_{\rmD} \gamma^{m_1m_2m_3} 
 - 3\,\partial_{\rmD}^{[m_1m_2}\,\gamma^{m_3]} \,,
\end{split}
\\
\begin{split}
 &\underline{R_{(4)}^{4,1}\ \leftrightarrow\ \text{$4^{(1,3)}_4$-brane:}}
\\
 &R_{(4)}^{m_1\cdots m_4,\,n} \equiv
 \frac{4!}{2!\,2!}\,\partial_{\rmD}^{[m_1m_2} \gamma^{m_3m_4]n} 
 - \partial_{\rmD}^{m_1\cdots m_4}\,\gamma^{n} 
 + \partial_{\rmS}^{m_1\cdots m_4q}A^{n}_q \,,
\end{split}
\\
 &\underline{R_{(4)}^{5,2}\ \leftrightarrow\ \text{$3^{(2,3)}_4$-brane:}}
\\
 &R_{(4)}^{m_1\cdots m_5,\,n_1n_2} \equiv
 \partial_{\rmD}^{n_1n_2} \gamma^{m_1\cdots m_5} 
 - 5\,\partial_{\rmD}^{[m_1\cdots m_4}\,\gamma^{m_5]n_1n_2} 
 + \partial_{\rmS}^{m_1\cdots m_5}\,\beta^{n_1n_2} 
 + 2\,\partial_{\rmS}^{m_1\cdots m_5q,\,[n_1}A^{n_2]}_q \,,
\nn\\
\begin{split}
 &\underline{R_{(4)}^{6,3}\ \leftrightarrow\ \text{$2^{(3,3)}_4$-brane:}}
\\
 &R_{(4)}^{m_1\cdots m_6,\,n_1n_2n_3} \equiv
 \frac{6!}{4!\,2!}\,\partial_{\rmD}^{[m_1\cdots m_4} \gamma^{m_5m_6]n_1n_2n_3} 
 - \partial_{\rmD}^{m_1\cdots m_6}\,\gamma^{n_1n_2n_3} 
 + 3\,\partial_{\rmS}^{m_1\cdots m_6,\,[n_1}\,\beta^{n_2n_3]} 
\\
 &\qquad\qquad\qquad\quad +3\, \partial_{\rmS}^{m_1\cdots m_6q,\,[n_1n_2}A^{n_3]}_q \,,
\end{split}
\\
\begin{split}
 &\underline{R_{(4)}^{7,4}\ \leftrightarrow\ \text{$1^{(4,3)}_4$-brane:}}
\\
 &R_{(4)}^{m_1\cdots m_7,\,n_1\cdots n_4} \equiv
 \partial_{\rmD}^{n_1\cdots n_4} \gamma^{m_1\cdots m_7}
 -7\,\partial_{\rmD}^{[m_1\cdots m_6} \gamma^{m_7]n_1\cdots n_4} 
 + \frac{4!}{2!\,2!}\, \partial_{\rmS}^{m_1\cdots m_7,\,[n_1n_2}\,\beta^{n_3n_4]} \,,
\end{split}
\\
\begin{split}
 &\underline{R_{(5)}^{6}\ \leftrightarrow\ \text{$2^{6}_5$-brane:}}
\\
 &R_{(5)}^{m_1\cdots m_6} \equiv
  \partial_{\rmD} \beta^{m_1\cdots m_6} + 6\,\partial_{\rmS}^{[m_1\cdots m_5}\,\gamma^{m_6]} \,,
\end{split}
\\
 &\underline{R_{(5)}^{6,2}\ \leftrightarrow\ \text{$2^{(2,4)}_5$-brane:}}
\\
 &R_{(5)}^{m_1\cdots m_6,\,n_1n_2} \equiv
  \partial_{\rmD}^{n_1n_2} \beta^{m_1\cdots m_6} + 2\,\partial_{\rmS}^{m_1\cdots m_6,\,[n_1}\,\gamma^{n_2]} 
 + 6\,\partial_{\rmS}^{[m_1\cdots m_5}\,\gamma^{m_6]n_1n_2} 
 -2\, \partial_{\rmE}^{m_1\cdots m_6q,\,[n_1}A^{n_2]}_q \,,
\nn\\
\begin{split}
 &\underline{R_{(5)}^{6,4}\ \leftrightarrow\ \text{$2^{(4,2)}_5$-brane:}}
\\
 &R_{(5)}^{m_1\cdots m_6,\,n_1\cdots n_4} \equiv
  \partial_{\rmD}^{n_1\cdots n_4} \beta^{m_1\cdots m_6} + 4\,\partial_{\rmS}^{m_1\cdots m_6,\,[n_1}\,\gamma^{n_2n_3n_4]} 
 + 6\,\partial_{\rmS}^{[m_1\cdots m_5}\,\gamma^{m_6]n_1\cdots n_4} 
\\
 &\qquad\qquad\qquad\ -4\, \partial_{\rmE}^{m_1\cdots m_6q,\,[n_1n_2n_3}A^{n_4]}_q \,,
\end{split}
\\
\begin{split}
 &\underline{R_{(5)}^{6,6}\ \leftrightarrow\ \text{$2^{(6,0)}_5$-brane:}}
\\
 &R_{(5)}^{m_1\cdots m_6,\,n_1\cdots n_6} \equiv
  \partial_{\rmD}^{n_1\cdots n_6} \beta^{m_1\cdots m_6} + 6\,\partial_{\rmS}^{m_1\cdots m_6,\,[n_1}\,\gamma^{n_2\cdots n_6]} 
  -6\, \partial_{\rmE}^{m_1\cdots m_6q,\,[n_1\cdots n_5}A^{n_6]}_q \,,
\end{split}
\\
\begin{split}
 &\underline{R_{(5)}^{7,2,1}\ \leftrightarrow\ \text{$1^{(1,1,5)}_5$-brane:}}
\\
 &R_{(5)}^{m_1\cdots m_7,\,n_1n_2,\,p} \equiv
  \partial_{\rmD}^{n_1n_2} \beta^{m_1\cdots m_7,\,p} 
  -2\,\partial_{\rmS}^{m_1\cdots m_7,\,p[n_1}\,\gamma^{n_2]} 
  -7\,\partial_{\rmS}^{[m_1\cdots m_6|,\,p}\,\gamma^{|m_7]n_1n_2} 
\\
 &\qquad\qquad\qquad\quad -2\,\partial_{\rmE}^{m_1\cdots m_7,\,[n_1}\,\beta^{n_2]p} 
  -\partial_{\rmE}^{m_1\cdots m_7,\,n_1n_2q}\, A_q^{p} \,,
\end{split}
\\
\begin{split}
 &\underline{R_{(5)}^{7,4,1}\ \leftrightarrow\ \text{$1^{(1,3,3)}_5$-brane:}}
\\
 &R_{(5)}^{m_1\cdots m_7,\,n_1\cdots n_4,\,p} \equiv
  \partial_{\rmD}^{n_1\cdots n_4} \beta^{m_1\cdots m_7,\,p} 
  -4\,\partial_{\rmS}^{m_1\cdots m_7,\,p[n_1}\,\gamma^{n_2n_3n_4]} 
  -7\,\partial_{\rmS}^{[m_1\cdots m_6|,\,p}\,\gamma^{|m_7]n_1\cdots n_4} 
\\
 &\qquad\qquad\qquad\qquad -4\,\partial_{\rmE}^{m_1\cdots m_7,\,[n_1n_2n_3}\,\beta^{n_4]p} 
  -\partial_{\rmE}^{m_1\cdots m_7,\,n_1\cdots n_4q}\, A_q^{p} \,,
\end{split}
\\
\begin{split}
 &\underline{R_{(5)}^{7,6,1}\ \leftrightarrow\ \text{$1^{(1,5,1)}_5$-brane:}}
\\
 &R_{(5)}^{m_1\cdots m_7,\,n_1\cdots n_6,\,p} \equiv
  \partial_{\rmD}^{n_1\cdots n_6} \beta^{m_1\cdots m_7,\,p} 
  -6\,\partial_{\rmS}^{m_1\cdots m_7,\,p[n_1}\,\gamma^{n_2\cdots n_6]} 
  -7\,\partial_{\rmS}^{[m_1\cdots m_6|,\,p}\,\gamma^{|m_7]n_1\cdots n_6} 
\\
 &\qquad\qquad\qquad\qquad -6\,\partial_{\rmE}^{m_1\cdots m_7,\,[n_1\cdots n_5}\,\beta^{n_6]p} 
  -\partial_{\rmE}^{m_1\cdots m_7,\,n_1\cdots n_6q}\, A_q^{p} \,,
\end{split}
\\
 &\underline{R_{(6)}^{7,4}\ \leftrightarrow\ \text{$1^{(4,3)}_6$-brane:}}
\\
 &R_{(6)}^{m_1\cdots m_7,\,n_1\cdots n_4} \equiv
   \frac{7!}{5!\,2!}\,\partial_{\rmS}^{[m_1\cdots m_5} \beta^{m_6m_7]n_1\cdots n_4} 
  +4\,\partial_{\rmE}^{m_1\cdots m_7,\,[n_1}\,\gamma^{n_2n_3n_4]} 
  -4\,\partial_{\rmE}^{m_1\cdots m_7,\,[n_1n_2n_3}\,\gamma^{n_4]} \,,
\nn\\
\begin{split}
 &\underline{R_{(6)}^{7,5,1}\ \leftrightarrow\ \text{$1^{(1,4,2)}_6$-brane:}}
\\
 &R_{(6)}^{m_1\cdots m_7,\,n_1\cdots n_5,\,p} \equiv
   7\,\partial_{\rmS}^{[m_1\cdots m_6|,\,p} \beta^{|m_7]n_1\cdots n_5} 
  - \partial_{\rmS}^{n_1\cdots n_5}\,\beta^{m_1\cdots m_7,\,p} 
  - \tilde{\partial}^{m_1\cdots m_7,\,n_1\cdots n_5q}\,A_q^{p} 
\\
 &\qquad\qquad\qquad\quad 
  + \partial_{\rmE}^{m_1\cdots m_7,\,p}\,\gamma^{n_1\cdots n_5}
  - \frac{5!}{2!\,3!}\,\partial_{\rmE}^{m_1\cdots m_7,\,p[n_1n_2}\,\gamma^{n_3n_4n_5]}
  + \tilde{\partial}^{m_1\cdots m_7,\,n_1\cdots n_5}\gamma^p\,,
\end{split}
\\
\begin{split}
 &\underline{R_{(6)}^{7,6,2}\ \leftrightarrow\ \text{$1^{(2,4,1)}_6$-brane:}}
\\
 &R_{(6)}^{m_1\cdots m_7,\,n_1\cdots n_6,\,p_1p_2} \equiv
  \partial_{\rmS}^{m_1\cdots m_7,\,p_1p_2} \beta^{n_1\cdots n_6} 
  -2\,\partial_{\rmS}^{n_1\cdots n_6,\,[p_1|} \beta^{m_1\cdots m_7,\,|p_2]} 
\\
 &\qquad\qquad\qquad\qquad\ 
  +\tilde{\partial}^{m_1\cdots m_7,\,n_1\cdots n_6}\beta^{p_1p_2}
  +2\,\tilde{\partial}^{m_1\cdots m_7,\,n_1\cdots n_6q,\,[p_1}A_q^{p_2]} 
\\
 &\qquad\qquad\qquad\qquad\ 
 +6\,\partial_{\rmE}^{m_1\cdots m_7,\,p_1p_2[n_1}\gamma^{n_2\cdots n_6]}
 -\frac{6!}{3!\,3!}\,\partial_{\rmE}^{m_1\cdots m_7,\,p_1p_2[n_1n_2n_3}\gamma^{n_4n_5n_6]}\,,
\end{split}
\\
\begin{split}
 &\underline{R_{(6)}^{7,7,3}\ \leftrightarrow\ \text{$1^{(3,4,0)}_6$-brane:}}
\\
 &R_{(6)}^{m_1\cdots m_7,\,n_1\cdots n_7,\,p_1p_2p_3} \equiv
   3\,\partial_{\rmS}^{m_1\cdots m_7,\,[p_1p_2|} \beta^{n_1\cdots n_7,\,|p_3]} 
  +3\,\tilde{\partial}^{n_1\cdots n_7,\,m_1\cdots m_7,\,[p_1}\beta^{p_2p_3]}
\\
 &\qquad\qquad\qquad\qquad\quad\ 
  -\partial_{\rmE}^{m_1\cdots m_7,\,p_1p_2p_3}\,\gamma^{n_1\cdots n_7} 
  + \frac{7!}{2!\,5!}\,\partial_{\rmE}^{m_1\cdots m_7,\,p_1p_2p_3[n_1n_2}\,\gamma^{n_3\cdots n_7]}
\\
 &\qquad\qquad\qquad\qquad\quad\ 
  -\partial_{\rmE}^{n_1\cdots n_7,\,m_1\cdots m_7,\,n_1\cdots n_7}\,\gamma^{p_1p_2p_3} \,,
\end{split}
\\
\begin{split}
 &\underline{R_{(7)}^{7,7}\ \leftrightarrow\ \text{$1^{(7,0)}_7$-brane:}}
\\
 &R_{(7)}^{m_1\cdots m_7,\,n_1\cdots n_7} \equiv 7\, \partial_{\rmE}^{n_1\cdots n_7,\,[m_1} \beta^{m_2\cdots m_7]} + 7\,\tilde{\partial}^{m_1\cdots m_7,\,[n_1\cdots n_6} \gamma^{n_7]} \,,
\end{split}
\\
\begin{split}
 &\underline{R_{(7)}^{7,7,2}\ \leftrightarrow\ \text{$1^{(2,5,0)}_7$-brane:}}
\\
 &R_{(7)}^{m_1\cdots m_7,\,n_1\cdots n_7,\,p_1p_2} \equiv
  -2\,\partial_{\rmE}^{n_1\cdots n_7,\,[p_1|} \beta^{m_1\cdots m_7,\,|p_2]} 
  +7\,\partial_{\rmE}^{n_1\cdots n_7,\,p_1p_2[m_1}\,\beta^{m_2\cdots m_7]} 
\\
 &\qquad\qquad\qquad\qquad +7\,\tilde{\partial}^{m_1\cdots m_7,\,[n_1\cdots n_6}\,\gamma^{n_7]p_1p_2}
  +2\,\tilde{\partial}^{m_1\cdots m_7,\,n_1\cdots n_7,\,[p_1}\gamma^{p_2]}\,,
\end{split}
\\
\begin{split}
 &\underline{R_{(7)}^{7,7,4}\ \leftrightarrow\ \text{$1^{(4,3,0)}_7$-brane:}}
\\
 &R_{(7)}^{m_1\cdots m_7,\,n_1\cdots n_7,\,p_1\cdots p_4} \equiv
  -4\,\partial_{\rmE}^{n_1\cdots n_7,\,[p_1p_2p_3|} \beta^{m_1\cdots m_7,\,|p_4]} 
  +7\,\partial_{\rmE}^{n_1\cdots n_7,\,p_1\cdots p_4[m_1}\,\beta^{m_2\cdots m_7]} 
\\
 &\qquad\qquad\qquad\qquad\quad\ +7\,\tilde{\partial}^{m_1\cdots m_7,\,[n_1\cdots n_6}\,\gamma^{n_7]p_1\cdots p_4}
  +4\,\tilde{\partial}^{m_1\cdots m_7,\,n_1\cdots n_7,\,[p_1}\gamma^{p_2p_3p_4]}\,,
\end{split}
\\
\begin{split}
 &\underline{R_{(7)}^{7,7,6}\ \leftrightarrow\ \text{$1^{(6,1,0)}_7$-brane:}}
\\
 &R_{(7)}^{m_1\cdots m_7,\,n_1\cdots n_7,\,p_1\cdots p_6} \equiv
  -6\,\partial_{\rmE}^{n_1\cdots n_7,\,[p_1\cdots p_5|} \beta^{m_1\cdots m_7,\,|p_6]} 
  +7\,\partial_{\rmE}^{n_1\cdots n_7,\,p_1\cdots p_6[m_1}\,\beta^{m_2\cdots m_7]} 
\\
 &\qquad\qquad\qquad\qquad\quad\ +7\,\tilde{\partial}^{m_1\cdots m_7,\,[n_1\cdots n_6}\,\gamma^{n_7]p_1\cdots p_6}
  +6\,\tilde{\partial}^{m_1\cdots m_7,\,n_1\cdots n_7,\,[p_1}\gamma^{p_2\cdots p_6]}\,,
\end{split}
\\
\begin{split}
 &\underline{R_{(8)}^{7,7,7}\ \leftrightarrow\ \text{$1^{(7,0,0)}_8$-brane:}}
\\
 &R_{(8)}^{m_1\cdots m_7,\,n_1\cdots n_7,\,p_1\cdots p_7} \equiv 
  7\,\tilde{\partial}^{n_1\cdots n_7,\,[p_1\cdots p_6|} \beta^{m_1\cdots m_7,\,|p_7]} 
  +7\,\tilde{\partial}^{m_1\cdots m_7,\,n_1\cdots n_7,\,[p_1} \beta^{p_2\cdots p_7]} \,.
\end{split}
\end{align}
Here, the vector field $A^m_n\equiv \tilde{g}^{mn}/\tilde{g}^{nn}$ is the graviphoton, which is $T$-dual of the $\beta$-field; $A^m_y\ \overset{T_y}{\leftrightarrow}\ \beta^{my}$\,. 
We have attached the subscript $(n)$ to the $R$-flux that is associated with the exotic brane $b^{(c_s,\dotsc,c_2)}_{n}$\,. 

In the type IIA case, there is another famous domain-wall brane, the D8-brane. 
As we have already discussed, it is the magnetic source of the R--R 0-form flux $F_0=\tilde{\partial}^m A_m$\,. 
Only in this example, the conventional supergravity field contains the winding-coordinate dependence. 

By the construction, indices of the obtained $R$-flux $R_{(n)}^{m_1,\dotsc,m_i,\,n_1,\dotsc,n_j,\,p_1,\dotsc,p_k}$ satisfy
\begin{align}
 \{m_1,\dotsc,m_i\}\supset\{n_1,\dotsc,n_j\}\supset\{p_1,\dotsc,p_k\}\,.
\label{eq:index-rule}
\end{align}
Then, for example, a combination $2\,A^{\cdots,\,n[m_1}\,B^{m_2]}$ is equal to $-A^{\cdots,\,m_1m_2}\,B^{n}$ since
\begin{align}
 3\,A^{\cdots,\,[m_1m_2}\,B^{n]} =0 \,,
\label{eq:Schouten-like-identity}
\end{align}
which follows from $\{m_1,\,m_2\}\supset\{n\}$\,. 
Similarly, $R_{(n)}^{m_1\cdots m_p,\,n_1\cdots n_p,\,\cdots}$ is equal to $R_{(n)}^{m_1\cdots m_p,\,m_1\cdots m_p,\,\cdots}$\,. 
The above expressions for the $R$-fluxes are correct only up to such identities, and in order to obtain the precise definitions of the $R$-fluxes, other approaches such as \cite{Blair:2014zba,Gunaydin:2016axc,Lust:2017bwq} will be necessary. 
In such approaches, the constraints \eqref{eq:index-rule} may be relaxed. 

\subsubsection{Locally non-geometric fluxes in type IIB theory/$T^7$}

In type IIB theory, the $R$-fluxes and the corresponding domain-wall branes are as follows:
\begin{align}
\begin{split}
 &\underline{R_{(2)}^{3}\ \leftrightarrow\ \text{$5^3_2$-brane:}}
\\
 &R_{(2)}^{m_1m_2m_3} \equiv 3\,\tilde{\partial}^{[m_1} \beta^{m_2m_3]} \,, 
\end{split}
\\
\begin{split}
 &\underline{R_{(3)}^{6,1}\ \leftrightarrow\ \text{$2^{(1,5)}_3$-brane:}}
\\
 &R_{(3)}^{m_1\cdots m_6,\,n} \equiv 
 \tilde{\partial}^n\,\gamma^{m_1\cdots m_6} + 6\,\partial_{\rmD}^{[m_1\cdots m_5}\beta^{m_6]n} 
 + \partial_{\rmD}^{m_1\cdots m_6q} A^n_q \,,
\end{split}
\\
\begin{split}
 &\underline{R_{(3)}^{4,1}\ \leftrightarrow\ \text{$4^{(1,3)}_3$-brane:}}
\\
 &R_{(3)}^{m_1\cdots m_4,\,n} \equiv 
 \tilde{\partial}^n\,\gamma^{m_1\cdots m_4} + 4\,\partial_{\rmD}^{[m_1m_2m_3}\beta^{m_4]n} 
 + \partial_{\rmD}^{m_1\cdots m_4q} A^n_q \,,
\end{split}
\\
\begin{split}
 &\underline{R_{(3)}^{2,1}\ \leftrightarrow\ \text{$6^{(1,1)}_3$-brane:}}
\\
 &R_{(3)}^{m_1m_2,\,n} \equiv 
 \tilde{\partial}^n\,\gamma^{m_1m_2} + 2\,\partial_{\rmD}^{[m_1}\beta^{m_2]n} 
 + \partial_{\rmD}^{m_1m_2q} A^n_q \,,
\end{split}
\\
\begin{split}
 &\underline{R_{(4)}^{7,1,1}\ \leftrightarrow\ \text{$1^{(1,0,6)}_4$-brane:}}
\\
 &R_{(4)}^{m_1\cdots m_7,\,n,\,p} \equiv
 \tilde{\partial}^{n} \beta^{m_1\cdots m_7,\,p}
 -7\,\partial_{\rmS}^{[m_1\cdots m_6|,\,p}\beta^{|m_7]n} 
 + \partial_{\rmS}^{m_1\cdots m_7,\,nq}A^p_q \,,
\end{split}
\\
\begin{split}
 &\underline{R_{(4)}^{3}\ \leftrightarrow\ \text{$5^3_4$-brane:}}
\\
 &R_{(4)}^{m_1m_2m_3} \equiv
 3\,\partial_{\rmD}^{[m_1} \gamma^{m_2m_3]} - \partial_{\rmD}^{m_1m_2m_3}\,\gamma \,,
\end{split}
\\
\begin{split}
 &\underline{R_{(4)}^{4,1}\ \leftrightarrow\ \text{$4^{(1,3)}_4$-brane:}}
\\
 &R_{(4)}^{m_1\cdots m_4,\,n} \equiv
 -4\,\partial_{\rmD}^{[m_1} \gamma^{m_2m_3m_4]n} 
 +4\, \partial_{\rmD}^{[m_1m_2m_3}\,\gamma^{m_4]n} 
 + \partial_{\rmS}^{m_1\cdots m_4q}A^{n}_q \,,
\end{split}
\\
 &\underline{R_{(4)}^{5,2}\ \leftrightarrow\ \text{$3^{(2,3)}_4$-brane:}}
\\
 &R_{(4)}^{m_1\cdots m_5,\,n_1n_2} \equiv
 \frac{5!}{3!\,2!}\,\partial_{\rmD}^{[m_1m_2m_3} \gamma^{m_4m_5]n_1n_2} 
 - \partial_{\rmD}^{m_1\cdots m_5}\,\gamma^{n_1n_2} 
 + \partial_{\rmS}^{m_1\cdots m_5}\,\beta^{n_1n_2} 
 + 2\,\partial_{\rmS}^{m_1\cdots m_5q,\,[n_1}A^{n_2]}_q \,,
\nn\\
\begin{split}
 &\underline{R_{(4)}^{6,3}\ \leftrightarrow\ \text{$2^{(3,3)}_4$-brane:}}
\\
 &R_{(4)}^{m_1\cdots m_6,\,n_1n_2n_3} \equiv
 \partial_{\rmD}^{n_1n_2n_3} \gamma^{m_1\cdots m_6} 
 + 6\,\partial_{\rmD}^{[m_1\cdots m_5}\,\gamma^{m_6]n_1n_2n_3} 
 + 3\,\partial_{\rmS}^{m_1\cdots m_6,\,[n_1}\,\beta^{n_2n_3]} 
\\
 &\qquad\qquad\qquad\quad +3\, \partial_{\rmS}^{m_1\cdots m_6q,\,[n_1n_2}A^{n_3]}_q \,,
\end{split}
\\
 &\underline{R_{(4)}^{7,4}\ \leftrightarrow\ \text{$1^{(4,3)}_4$-brane:}}
\\
 &R_{(4)}^{m_1\cdots m_7,\,n_1\cdots n_4} \equiv
 \frac{7!}{5!\,2!}\,\partial_{\rmD}^{[m_1\cdots m_5} \gamma^{m_6m_7]n_1\cdots n_4} 
 - \partial_{\rmD}^{m_1\cdots m_7}\,\gamma^{n_1\cdots n_4} 
 + \frac{4!}{2!\,2!}\, \partial_{\rmS}^{m_1\cdots m_7,\,[n_1n_2}\,\beta^{n_3n_4]} \,,
\nn\\
\begin{split}
 &\underline{R_{(5)}^{6,1}\ \leftrightarrow\ \text{$2^{(1,5)}_5$-brane:}}
\\
 &R_{(5)}^{m_1\cdots m_6,\,n} \equiv
 \partial_{\rmD}^n \beta^{m_1\cdots m_6} + \partial_{\rmS}^{m_1\cdots m_6,\,n}\,\gamma 
 + 6\,\partial_{\rmS}^{[m_1\cdots m_5}\,\gamma^{m_6]n} 
 + \partial_{\rmE}^{m_1\cdots m_6q}A^n_q \,,
\end{split}
\\
\begin{split}
 &\underline{R_{(5)}^{6,3}\ \leftrightarrow\ \text{$2^{(3,3)}_5$-brane:}}
\\
 &R_{(5)}^{m_1\cdots m_6,\,n_1n_2n_3} \equiv
 \partial_{\rmD}^{n_1n_2n_3} \beta^{m_1\cdots m_6} + 3\,\partial_{\rmS}^{m_1\cdots m_6,\,[n_1}\,\gamma^{n_2n_3]} 
 + 6\,\partial_{\rmS}^{[m_1\cdots m_5}\,\gamma^{m_6]n_1n_2n_3} 
\\
 &\qquad\qquad\qquad\quad +3\, \partial_{\rmE}^{m_1\cdots m_6q,\,[n_1n_2}A^{n_3]}_q \,,
\end{split}
\\
\begin{split}
 &\underline{R_{(5)}^{6,5}\ \leftrightarrow\ \text{$2^{(5,1)}_5$-brane:}}
\\
 &R_{(5)}^{m_1\cdots m_6,\,n_1\cdots n_5} \equiv
 \partial_{\rmD}^{n_1\cdots n_5} \beta^{m_1\cdots m_6} + 5\,\partial_{\rmS}^{m_1\cdots m_6,\,[n_1}\,\gamma^{n_2\cdots n_5]} 
 + 6\,\partial_{\rmS}^{[m_1\cdots m_5}\,\gamma^{m_6]n_1\cdots n_5} 
\\
 &\qquad\qquad\qquad +5\, \partial_{\rmE}^{m_1\cdots m_6q,\,[n_1\cdots n_4}A^{n_5]}_q \,,
\end{split}
\\
\begin{split}
 &\underline{R_{(5)}^{7,1,1}\ \leftrightarrow\ \text{$1^{(1,0,6)}_5$-brane:}}
\\
 &R_{(5)}^{m_1\cdots m_7,\,n,\,p} \equiv
  \partial_{\rmD}^{n} \beta^{m_1\cdots m_7,\,p} 
  -7\,\partial_{\rmS}^{[m_1\cdots m_6|,\,p}\,\gamma^{|m_7]n} 
  +\partial_{\rmE}^{m_1\cdots m_7,\,nq}\, A_q^{p} \,,
\end{split}
\\
\begin{split}
 &\underline{R_{(5)}^{7,3,1}\ \leftrightarrow\ \text{$1^{(1,2,4)}_5$-brane:}}
\\
 &R_{(5)}^{m_1\cdots m_7,\,n_1n_2n_3,\,p} \equiv
  \partial_{\rmD}^{n_1n_2n_3} \beta^{m_1\cdots m_7,\,p} 
  -3\,\partial_{\rmS}^{m_1\cdots m_7,\,p[n_1}\,\gamma^{n_2n_3]} 
  -7\,\partial_{\rmS}^{[m_1\cdots m_6|,\,p}\,\gamma^{|m_7]n_1n_2n_3} 
\\
 &\qquad\qquad\qquad\qquad +3\,\partial_{\rmE}^{m_1\cdots m_7,\,[n_1n_2}\,\beta^{n_3]p} 
  +\partial_{\rmE}^{m_1\cdots m_7,\,n_1n_2n_3q}\, A_q^{p} \,,
\end{split}
\\
\begin{split}
 &\underline{R_{(5)}^{7,5,1}\ \leftrightarrow\ \text{$1^{(1,4,2)}_5$-brane:}}
\\
 &R_{(5)}^{m_1\cdots m_7,\,n_1\cdots n_5,\,p} \equiv
  \partial_{\rmD}^{n_1\cdots n_5} \beta^{m_1\cdots m_7,\,p} 
  -5\,\partial_{\rmS}^{m_1\cdots m_7,\,p[n_1}\,\gamma^{n_2\cdots n_5]} 
  -7\,\partial_{\rmS}^{[m_1\cdots m_6|,\,p}\,\gamma^{|m_7]n_1\cdots n_5} 
\\
 &\qquad\qquad\qquad\qquad +5\,\partial_{\rmE}^{m_1\cdots m_7,\,[n_1\cdots n_4}\,\beta^{n_5]p} 
  +\partial_{\rmE}^{m_1\cdots m_7,\,n_1\cdots n_5q}\, A_q^{p} \,,
\end{split}
\\
\begin{split}
 &\underline{R_{(5)}^{7,7,1}\ \leftrightarrow\ \text{$1^{(1,6,0)}_5$-brane:}}
\\
 &R_{(5)}^{m_1\cdots m_7,\,n_1\cdots n_7,\,p} \equiv
  \partial_{\rmD}^{n_1\cdots n_7} \beta^{m_1\cdots m_7,\,p} 
  -7\,\partial_{\rmS}^{m_1\cdots m_7,\,p[n_1}\,\gamma^{n_2\cdots n_7]} 
  + 7\,\partial_{\rmE}^{m_1\cdots m_7,\,[n_1\cdots n_6}\,\beta^{n_7]p} \,,
\end{split}
\\
\begin{split}
 &\underline{R_{(6)}^{7,4}\ \leftrightarrow\ \text{$1^{(4,3)}_6$-brane:}}
\\
 &R_{(6)}^{m_1\cdots m_7,\,n_1\cdots n_4} \equiv
   \frac{7!}{5!\,2!}\,\partial_{\rmS}^{[m_1\cdots m_5} \beta^{m_6m_7]n_1\cdots n_4} 
  -\partial_{\rmE}^{m_1\cdots m_7}\,\gamma^{n_1\cdots n_4} 
\\
 &\qquad\qquad\qquad\quad 
  +\frac{4!}{2!\,2!}\,\partial_{\rmE}^{m_1\cdots m_7,\,[n_1n_2}\,\gamma^{n_3n_4]} 
  -\partial_{\rmE}^{m_1\cdots m_7,\,n_1\cdots n_4}\,\gamma \,,
\end{split}
\\
\begin{split}
 &\underline{R_{(6)}^{7,5,1}\ \leftrightarrow\ \text{$1^{(1,4,2)}_6$-brane:}}
\\
 &R_{(6)}^{m_1\cdots m_7,\,n_1\cdots n_5,\,p} \equiv
   7\,\partial_{\rmS}^{[m_1\cdots m_6|,\,p} \beta^{|m_7]n_1\cdots n_5} 
  - \partial_{\rmS}^{n_1\cdots n_5}\,\beta^{m_1\cdots m_7,\,p} 
  - \tilde{\partial}^{m_1\cdots m_7,\,n_1\cdots n_5q}\,A_q^{p} 
\\
 &\qquad\qquad\qquad\qquad 
  + 5\,\partial_{\rmE}^{m_1\cdots m_7,\,p[n_1}\,\gamma^{n_2\cdots n_5]}
  - \frac{5!}{3!\,2!}\,\partial_{\rmE}^{m_1\cdots m_7,\,p[n_1n_2n_3}\,\gamma^{n_4n_5]} \,,
\end{split}
\\
\begin{split}
 &\underline{R_{(6)}^{7,6,2}\ \leftrightarrow\ \text{$1^{(2,4,1)}_6$-brane:}}
\\
 &R_{(6)}^{m_1\cdots m_7,\,n_1\cdots n_6,\,p_1p_2} \equiv
  \partial_{\rmS}^{m_1\cdots m_7,\,p_1p_2} \beta^{n_1\cdots n_6} 
  -2\,\partial_{\rmS}^{n_1\cdots n_6,\,[p_1|} \beta^{m_1\cdots m_7,\,|p_2]} 
\\
 &\qquad\qquad\quad 
  +\tilde{\partial}^{m_1\cdots m_7,\,n_1\cdots n_6}\beta^{p_1p_2}
  +2\,\tilde{\partial}^{m_1\cdots m_7,\,n_1\cdots n_6q,\,[p_1}A_q^{p_2]} 
\\
 &\qquad\qquad\quad 
  -\partial_{\rmE}^{m_1\cdots m_7,\,p_1p_2}\gamma^{n_1\cdots n_6}
  +\frac{6!}{2!\,4!}\,\partial_{\rmE}^{m_1\cdots m_7,\,p_1p_2[n_1n_2}\gamma^{n_3\cdots n_6]}
  -\partial_{\rmE}^{m_1\cdots m_7,\,n_1\cdots n_7}\gamma^{p_1p_2}\,,
\end{split}
\\
\begin{split}
 &\underline{R_{(6)}^{7,7,3}\ \leftrightarrow\ \text{$1^{(3,4,0)}_6$-brane:}}
\\
 &R_{(6)}^{m_1\cdots m_7,\,n_1\cdots n_7,\,p_1p_2p_3} \equiv
   3\,\partial_{\rmS}^{m_1\cdots m_7,\,[p_1p_2|} \beta^{n_1\cdots n_7,\,|p_3]} 
  +3\,\tilde{\partial}^{n_1\cdots n_7,\,m_1\cdots m_7,\,[p_1}\beta^{p_2p_3]}
\\
 &\qquad\qquad\qquad\qquad\quad\ -7\,\partial_{\rmE}^{m_1\cdots m_7,\,p_1p_2p_3[n_1}\,\gamma^{n_2\cdots n_7]}
  +7\,\partial_{\rmE}^{n_1\cdots n_7,\,[m_1\cdots m_6}\,\gamma^{m_7]p_1p_2p_3} \,,
\end{split}
\\
\begin{split}
 &\underline{R_{(7)}^{7,7,1}\ \leftrightarrow\ \text{$1^{(1,6,0)}_7$-brane:}}
\\
 &R_{(7)}^{m_1\cdots m_7,\,n_1\cdots n_7,\,p} \equiv
  \partial_{\rmE}^{n_1\cdots n_7} \beta^{m_1\cdots m_7,\,p} 
  -7\,\partial_{\rmE}^{n_1\cdots n_7,\,p[m_1}\,\beta^{m_2\cdots m_7]} 
\\
 &\qquad\qquad\qquad\qquad + 7\,\tilde{\partial}^{m_1\cdots m_7,\,[n_1\cdots n_6}\,\gamma^{n_7]p}
  + \tilde{\partial}^{m_1\cdots m_7,\,n_1\cdots n_7,\,p}\gamma\,,
\end{split}
\\
\begin{split}
 &\underline{R_{(7)}^{7,7,3}\ \leftrightarrow\ \text{$1^{(3,4,0)}_7$-brane:}}
\\
 &R_{(7)}^{m_1\cdots m_7,\,n_1\cdots n_7,\,p_1p_2p_3} \equiv
   3\,\partial_{\rmE}^{n_1\cdots n_7,\,[p_1p_2|} \beta^{m_1\cdots m_7,\,|p_3]} 
  -7\,\partial_{\rmE}^{n_1\cdots n_7,\,p_1p_2p_3[m_1}\,\beta^{m_2\cdots m_7]} 
\\
 &\qquad\qquad\qquad\qquad\quad\ +7\,\tilde{\partial}^{m_1\cdots m_7,\,[n_1\cdots n_6}\,\gamma^{n_7]p_1p_2p_3}
  +3\,\tilde{\partial}^{m_1\cdots m_7,\,n_1\cdots n_7,\,[p_1}\gamma^{p_2p_3]}\,,
\end{split}
\\
\begin{split}
 &\underline{R_{(7)}^{7,7,5}\ \leftrightarrow\ \text{$1^{(5,2,0)}_7$-brane:}}
\\
 &R_{(7)}^{m_1\cdots m_7,\,n_1\cdots n_7,\,p_1\cdots p_5} \equiv
   5\,\partial_{\rmE}^{n_1\cdots n_7,\,[p_1\cdots p_4|} \beta^{m_1\cdots m_7,\,|p_5]} 
  -7\,\partial_{\rmE}^{n_1\cdots n_7,\,p_1\cdots p_5[m_1}\,\beta^{m_2\cdots m_7]} 
\\
 &\qquad\qquad\qquad\qquad\quad\ +7\,\tilde{\partial}^{m_1\cdots m_7,\,[n_1\cdots n_6}\,\gamma^{n_7]p_1\cdots p_5}
  +5\,\tilde{\partial}^{m_1\cdots m_7,\,n_1\cdots n_7,\,[p_1}\gamma^{p_2\cdots p_5]}\,,
\end{split}
\\
\begin{split}
 &\underline{R_{(7)}^{7,7,7}\ \leftrightarrow\ \text{$1^{(7,0,0)}_7$-brane:}}
\\
 &R_{(7)}^{m_1\cdots m_7,\,n_1\cdots n_7,\,p_1\cdots p_7} \equiv
   7\,\partial_{\rmE}^{n_1\cdots n_7,\,[p_1\cdots p_6|} \beta^{m_1\cdots m_7,\,|p_7]} 
  +7\,\tilde{\partial}^{m_1\cdots m_7,\,n_1\cdots n_7,\,[p_1}\gamma^{p_2\cdots p_7]}\,,
\end{split}
\\
\begin{split}
 &\underline{R_{(8)}^{7,7,7}\ \leftrightarrow\ \text{$1^{(7,0,0)}_8$-brane:}}
\\
 &R_{(8)}^{m_1\cdots m_7,\,n_1\cdots n_7,\,p_1\cdots p_7} \equiv 
  7\,\tilde{\partial}^{n_1\cdots n_7,\,[p_1\cdots p_6|} \beta^{m_1\cdots m_7,\,|p_7]} 
  +7\,\tilde{\partial}^{m_1\cdots m_7,\,n_1\cdots n_7,\,[p_1} \beta^{p_2\cdots p_7]} \,.
\end{split}
\end{align}
For an exotic brane that is self-dual under the $S$-duality, we can check that the associated $R$-flux also behaves as a singlet. 
Under the $S$-duality, the scalar $\gamma$ is mapped to the R--R 0-form $-C_0$\,, but since $C_0$ is non-linear in terms of the dual fields, we have just truncated $C_0$ in the above expressions. 

\subsubsection{Locally non-geometric fluxes in M-theory/$T^8$}

We can easily uplift the $R$-fluxes obtained in type IIA theory to M-theory. 
The results are as follows:
\begin{align}
\begin{split}
 &\underline{R^{1,1}\ \leftrightarrow\ \text{$8^{(1,0)}_{12}$-brane:}}
\\
 &R^{i,\,j} \equiv \partial^{ki}A^{j}_k \,,
\end{split}
\\
\begin{split}
 &\underline{R^{4,1}\ \leftrightarrow\ \text{$5^{(1,3)}_{15}$-brane:}}
\\
 &R^{i_1\cdots i_4,\,j} \equiv
 \frac{4!}{2!\,2!}\,\partial^{[i_1i_2} \Omega^{i_3i_4]j} + \partial^{i_1\cdots i_4k}A^{j}_k \,,
\end{split}
\\
\begin{split}
 &\underline{R^{6,2}\ \leftrightarrow\ \text{$3^{(2,4)}_{18}$-brane:}}
\\
 &R^{i_1\cdots i_6,\,j_1j_2} \equiv
 \partial^{j_1j_2} \Omega^{i_1\cdots i_6} + 6\,\partial^{[i_1\cdots i_5}\,\Omega^{i_6]j_1j_2} -2\, \partial^{i_1\cdots i_6k,\,[j_1}A^{j_2]}_k \,,
\end{split}
\\
\begin{split}
 &\underline{R^{7,4}\ \leftrightarrow\ \text{$2^{(4,3)}_{21}$-brane:}}
\\
 &R^{i_1\cdots i_7,\,j_1\cdots j_4} \equiv
 \frac{7!}{5!\,2!}\,\partial^{[i_1\cdots i_5} \Omega^{i_6i_7]j_1\cdots j_4} + 4\,\partial^{i_1\cdots i_7,\,[j_1}\,\Omega^{j_2j_3j_4]} 
 +4\, \partial^{i_1\cdots i_7l,\,[j_1j_2j_3}A^{j_4]}_l \,,
\end{split}
\\
\begin{split}
 &\underline{R^{8,2,1}\ \leftrightarrow\ \text{$1^{(1,1,6)}_{21}$-brane:}}
\\
 &R^{i_1\cdots i_8,\,j_1j_2,\,k} \equiv
  \partial^{j_1j_2} \Omega^{i_1\cdots i_8,\,k} 
  +8\,\partial^{[i_1\cdots i_7|,\,k}\,\Omega^{|i_8]j_1j_2} 
  -\partial^{i_1\cdots i_8,\,j_1j_2l}\,A_l^{k} \,,
\end{split}
\\
\begin{split}
 &\underline{R^{7,7}\ \leftrightarrow\ \text{$2^{(7,0)}_{24}$-brane:}}
\\
 &R^{i_1\cdots i_7,\,j_1\cdots j_7} \equiv 7\, \partial^{j_1\cdots j_7,\,[i_1} \Omega^{i_2\cdots i_7]} - 7\, \partial^{i_1\cdots i_7k,\,[j_1\cdots j_6} A_k^{j_7]} \,,
\end{split}
\\
\begin{split}
 &\underline{R^{8,5,1}\ \leftrightarrow\ \text{$1^{(1,4,3)}_{24}$-brane:}}
\\
 &R^{i_1\cdots i_8,\,j_1\cdots j_5,\,k} \equiv 
 \partial^{j_1\cdots j_5} \Omega^{i_2\cdots i_8,\,k}
  +\frac{5!}{2!\,3!}\,\partial^{i_1\cdots i_8,\,k[j_1j_2} \Omega^{j_3j_4j_5]} 
\\
 &\qquad\qquad\qquad\quad +8\,\partial^{[i_1\cdots i_7|,\,k} \Omega^{|i_8]j_1\cdots j_5} 
  +\partial^{i_1\cdots i_8,\,j_1\cdots j_5l} A_l^k \,,
\end{split}
\\
\begin{split}
 &\underline{R^{8,7,2}\ \leftrightarrow\ \text{$1^{(2,5,1)}_{27}$-brane:}}
\\
 &R^{i_1\cdots i_8,\,j_1\cdots j_7,\,k_1k_2} \equiv 
  7\,\partial^{i_1\cdots i_8,\,k_1k_2[j_1} \Omega^{j_2\cdots j_7]}
  -2\,\partial^{j_1\cdots j_7,\,[k_1|} \Omega^{i_1\cdots i_8,\,|k_2]}
\\
 &\qquad\qquad\qquad\qquad +7\,\partial^{i_1\cdots i_8,\,[j_1\cdots j_6} \Omega^{j_7]k_1k_2}
  -2\,\partial^{i_1\cdots i_8,\,j_1\cdots j_7l,\,[k_1} A_l^{k_2]} \,,
\end{split}
\\
\begin{split}
 &\underline{R^{8,8,4}\ \leftrightarrow\ \text{$1^{(4,4,0)}_{30}$-brane:}}
\\
 &R^{i_1\cdots i_8,\,j_1\cdots j_8,\,k_1\cdots k_4} \equiv 
  -4\,\partial^{j_1\cdots j_8,\,[k_1k_2k_3|} \Omega^{i_1\cdots i_8,\,|k_4]}
\\
 &\qquad\qquad\qquad\qquad\ 
  +\frac{8!}{2!\,4!}\,\partial^{j_1\cdots j_8,\,k_1\cdots k_4[i_1i_2} \Omega^{i_3\cdots i_8]}
  +4\,\partial^{i_1\cdots i_8,\,j_1\cdots j_8,\,[k_1} \Omega^{k_2k_3k_4]}\,,
\end{split}
\\
\begin{split}
 &\underline{R^{8,8,7}\ \leftrightarrow\ \text{$1^{(7,1,0)}_{33}$-brane:}}
\\
 &R^{i_1\cdots i_8,\,j_1\cdots j_8,\,k_1\cdots k_7} \equiv 
   7\,\partial^{j_1\cdots j_8,\,[k_1\cdots k_6|} \Omega^{i_1\cdots i_8,\,|k_7]}
  +7\,\partial^{i_1\cdots i_8,\,j_1\cdots j_8,\,[k_1} \Omega^{k_2\cdots k_7]}\,.
\end{split}
\end{align}
Note that the $A_{\rmM}^m$ is equal to the $-\gamma^m$ in type IIA theory while $A^{\rmM}_m$ is a complicated non-linear expression that will be related to R--R 1-form $C_m$. 

By using the identities such as \eqref{eq:Schouten-like-identity}, the fluxes $R^{4,1}$, $R^{7,4}$, and $R^{7,7}$ appear to be consistent with the locally non-geometric fluxes $R^{i,\,jklm}$, $R^{ijkl}$, and $R$ of \cite{Lust:2017bwq}, respectively. 
The flux $R^{6,2}$ also may be related to $R^{ij}{}_k$ and $R^i$\,.

\subsection{Mixed-symmetry potentials in EFT}

In the previous subsection, we have introduced various $R$-fluxes on a heuristic basis. 
Similar to the $R$-fluxes in DFT, we here consider the introduction of the dual field strength to the $R$-flux in type II theory, $R_{(n)}^{m_1\cdots m_{a_1},\,\dotsc,\,p_1\cdots p_{a_s}}$\,. 
As we check in the next subsection, if we define the dual field strength and its potential as
\begin{align}
\begin{split}
 R^{(n)}_{10,a_1,\dotsc,a_s} &\equiv \Exp{2(1-n)\,\tilde{\phi}} \tilde{g}_{m_1\cdots m_{a_1},\,n_1\cdots n_{a_1}}\cdots\tilde{g}_{p_1\cdots p_{a_s},\,q_1\cdots q_{a_s}}\, \tilde{*}_{10} R_{(n)}^{m_1\cdots m_{a_1},\,\dotsc,\,p_1\cdots p_{a_s}} 
\\
 &\quad \otimes \rmd x^{n_1\cdots n_{a_1}}\otimes \cdots \otimes \rmd x^{q_1\cdots q_{a_s}} \equiv \rmd E^{(n)}_{9,a_1,\dotsc,a_s} \,,
\end{split}
\end{align}
the mixed-symmetry potential $E^{(n)}_{9,a_1,\dotsc,a_s}$ in the exotic domain-wall background always has the simple form $E^{(n)}_{\bar0\cdots\bar8,\cdots,\cdots} = -m\,\tau_2^{-1}$ similar to \eqref{eq:E-potential-simple}. 
This implies that under duality transformations, the structure of the domain-wall background is not essentially changed and only the name of the mixed-symmetry potential $E^{(n)}_{9,a_1,\dotsc,a_s}$ are changed. 
The transformation rule (at the linearized level) is perfectly consistent with the rule \cite{Lombardo:2016swq,Bergshoeff:2017gpw},
\begin{align}
 E^{(n)}_{\underbrace{{\tiny\cdots y,\cdots y,\cdots y}}_{p}}\quad \overset{T_y}{\leftrightarrow}\quad E^{(n)}_{\underbrace{\scriptsize\cdots y,\cdots y,\cdots y}_{n-p}} \,. 
\end{align}
We thus expect that the dual potentials $E^{(n)}_{9,a_1,\dotsc,a_s}$ defined above are precisely the $U$-dual extensions of the familiar mixed-symmetry potentials, such as $D_{9,3}$ (i.e.~the electric dual potential for the usual $R$-flux), which electrically couples to the exotic domain-wall. 

In terms of M-theory, we can similarly define the dual field strength and the mixed potentials as
\begin{align}
\begin{split}
 R_{11,a_1,\dotsc,a_s} &\equiv \tilde{G}_{i_1\cdots i_{a_1},\,j_1\cdots j_{a_1}}\cdots\tilde{G}_{k_1\cdots k_{a_s},\,l_1\cdots l_{a_s}}\, \tilde{*}_{11} R^{i_1\cdots i_{a_1},\,\dotsc,\,k_1\cdots k_{a_s}} 
\\
 &\quad \otimes \rmd x^{j_1\cdots j_{a_1}}\otimes \cdots \otimes \rmd x^{l_1\cdots l_{a_s}} \equiv \rmd E_{10,a_1,\dotsc,a_s} \,. 
\end{split}
\label{eq:dual-potential-M}
\end{align}
The mixed-symmetry potential again has the same form $E_{\bar0\cdots\bar9,\cdots,\cdots} = -m\,\tau_2^{-1}$ in the exotic domain-wall background. 

In this manner, the locally non-geometric fluxes and the mixed-symmetry potentials are in one-to-one correspondence, and moreover, they are associated with the domain-wall branes. 

Note that the conjectured electric-magnetic duality relation does not have a manifestly duality symmetric form. 
It will be an important task to manifest the covariance similar to the approach of \cite{Bergshoeff:2016gub}. 

\subsection{Exotic-brane solutions in type II theory}
\label{sec:exotic-sol-ii}

Utilizing the technique of the duality rotations in EFT, we here provide a full list of the type II domain-wall solutions in EFT. 
The structure of the solutions is quite similar to the domain-wall solutions discussed above, and only a certain gauge field contains a winding-coordinate dependence. 
Similar to the domain-wall solutions in DFT, we can check that the non-vanishing component of the $R$-flux $R_{(n)}^{a_1,\dotsc,a_s}$ and the mixed-symmetry potential $E^{(n)}_{9,a_1,\dotsc,a_s}$ always have the same form,
\begin{align}
 R_{(n)}^{\cdots,\cdots,\cdots} = m \,, \qquad 
 E^{(n)}_{\bar0\cdots\bar8,\cdots,\cdots,\cdots} = -m\,\tau_2^{-1}\,. 
\end{align}

\subsubsection{E$^{(4;3)}$-branes}

By considering $S$-dual of the $5^3_2$-brane or the $4^{(1,3)}_3$-brane, we obtain the backgrounds of a $T$-duality family, E$^{(4;3)}$-branes. 
The explicit forms of the dual fields and the locally non-geometric fluxes are as follows:
\begin{align}
\begin{split}
 &\underline{\bm{5^3_4(12345,678):} \qquad \bigl\{\,R_{(4)}^{3},\,E^{(4)}_{9,3}\,\bigr\}}
\\
 &\rmd \tilde{s}^2 = \tau_2\, \rmd x^2_{01\cdots 5} + \rmd x^2_{678} + \tau_2^2 \,\rmd x_{9}^2 \,,\qquad 
 \Exp{-2\tilde{\phi}}= \tau_2^{-2} \,,
\\
 &\begin{cases}
 \gamma^{\bar6\bar7\bar8} = m\, y^{\rmD}\,,\quad R_{(4)}^{\bar6\bar7\bar8} = \partial_{\rmD} \gamma^{\bar6\bar7\bar8}=m & \text{(IIA)}
\\
 \gamma^{\bar6\bar7} = m\,y^{\rmD}_8\,,\quad R_{(4)}^{\bar6\bar7\bar8} = 3\, \partial_{\rmD}^{[\bar8} \gamma^{\bar6\bar7]}=m & \text{(IIB)}
 \end{cases} \,,
\end{split}
\\
\begin{split}
 &\underline{\bm{4^{(1,3)}_4(1234,678,5):} \qquad \bigl\{\,R_{(4)}^{4,1},\,E^{(4)}_{9,4,1}\,\bigr\}}
\\
 &\rmd \tilde{s}^2 = \tau_2\, \rmd x^2_{01\cdots 4} + \tau_2^{-1}\,\rmd x_5^2 + \rmd x^2_{678} + \tau_2^2 \,\rmd x_{9}^2 \,,\qquad 
  \Exp{-2\tilde{\phi}}= \tau_2^{-1} \,,
\\
 &\begin{cases}
 \gamma^{\bar5\bar6\bar7} = m\, y^{\rmD}_{58}\,,\quad R_{(4)}^{\bar5\bar6\bar7\bar8,\bar5} = \frac{4!}{2!\,2!}\, \partial_{\rmD}^{[\bar5\bar6} \gamma^{\bar7\bar8]\bar5}=m & \text{(IIA)}
\\
 \gamma^{\bar5\bar6\bar7\bar8} = m\, y^{\rmD}_{5}\,,\quad R_{(4)}^{\bar5\bar6\bar7\bar8,\bar5} = 4\,\partial_{\rmD}^{[\bar5|} \gamma^{\bar5|\bar6\bar7\bar8]}=m & \text{(IIB)}
 \end{cases} \,,
\end{split}
\\
\begin{split}
 &\underline{\bm{3^{(2,3)}_4(123,678,45):} \qquad \bigl\{\,R_{(4)}^{5,2},\,E^{(4)}_{9,5,2}\,\bigr\}}
\\
 &\rmd \tilde{s}^2 = \tau_2\, \rmd x^2_{0123} + \tau_2^{-1}\,\rmd x_{45}^2 + \rmd x^2_{678} + \tau_2^2 \,\rmd x_{9}^2 \,, \qquad 
  \Exp{-2\tilde{\phi}}= 1 \,,
\\
 &
 \begin{cases}
 \gamma^{\bar4\bar5\bar6\bar7\bar8} = m\, y^{\rmD}_{45}\,,\quad R_{(4)}^{\bar4\bar5\bar6\bar7\bar8,\bar4\bar5} = \partial_{\rmD}^{\bar4\bar5} \gamma^{\bar4\bar5\bar6\bar7\bar8}=m & \text{(IIA)}
\\
 \gamma^{\bar4\bar5\bar6\bar7} = m\, y^{\rmD}_{458}\,,\quad R_{(4)}^{\bar4\bar5\bar6\bar7\bar8,\bar4\bar5} = \frac{5!}{3!\,2!}\, \partial_{\rmD}^{[\bar4\bar5\bar8} \gamma^{\bar6\bar7]\bar4\bar5}=m & \text{(IIB)}
 \end{cases} \,,
\end{split}
\\
\begin{split}
 &\underline{\bm{2^{(3,3)}_4(12,678,345):} \qquad \bigl\{\,R_{(4)}^{6,3},\,E^{(4)}_{9,6,3}\,\bigr\}}
\\
 &\rmd \tilde{s}^2 = \tau_2\, \rmd x^2_{012} + \tau_2^{-1}\,\rmd x_{345}^2 + \rmd x^2_{678} + \tau_2^2 \,\rmd x_{9}^2 \,,\qquad
 \Exp{-2\tilde{\phi}}= \tau_2 \,, 
\\
 &\begin{cases}
 \gamma^{\bar3\cdots \bar7} = m\, y^{\rmD}_{3458}\,,\quad R_{(4)}^{\bar3\bar4\bar5\bar6\bar7\bar8,\bar3\bar4\bar5} = \frac{6!}{4!\,2!}\, \partial_{\rmD}^{[\bar3\bar4\bar5\bar8} \gamma^{\bar6\bar7]\bar3\bar4\bar5}=m & \text{(IIA)}
\\
 \gamma^{\bar3\cdots \bar7\bar8} = m\, y^{\rmD}_{345}\,,\quad R_{(4)}^{\bar3\bar4\bar5\bar6\bar7\bar8,\bar3\bar4\bar5} = \partial_{\rmD}^{\bar3\bar4\bar5} \gamma^{\bar3\bar4\bar5\bar6\bar7\bar8}=m & \text{(IIB)}
 \end{cases} \,,
\end{split}
\\
\begin{split}
 &\underline{\bm{1^{(4,3)}_4(1,678,2345):} \qquad \bigl\{\,R_{(4)}^{7,4},\,E^{(4)}_{9,7,4}\,\bigr\}}
\\
 &\rmd \tilde{s}^2 = \tau_2\, \rmd x^2_{01} + \tau_2^{-1}\,\rmd x_{2345}^2 + \rmd x^2_{678} + \tau_2^2 \,\rmd x_{9}^2 \,, \qquad 
 \Exp{-2\tilde{\phi}}= \tau_2^2 \,, 
\\
 &\begin{cases}
 \gamma^{\bar2\cdots \bar7\bar8} = m\, y^{\rmD}_{2345}\,,\quad R_{(4)}^{\bar2\bar3\bar4\bar5\bar6\bar7\bar8,\bar2\bar3\bar4\bar5} = \partial_{\rmD}^{\bar2\bar3\bar4\bar5} \gamma^{\bar2\bar3\bar4\bar5\bar6\bar7\bar8}=m & \text{(IIA)}
\\
  \gamma^{\bar2\cdots \bar7} = m\, y^{\rmD}_{23458}\,,\quad R_{(4)}^{\bar2\bar3\bar4\bar5\bar6\bar7\bar8,\bar2\bar3\bar4\bar5} = \frac{7!}{5!\,2!}\, \partial_{\rmD}^{[\bar2\bar3\bar4\bar5\bar8} \gamma^{\bar6\bar7]\bar2\bar3\bar4\bar5} = m & \text{(IIB)}
 \end{cases} \,.
\end{split}
\end{align}

\subsubsection{E$^{(5;6)}$-branes}

Similarly, by performing the $S$-duality in the $2^{(1,5)}_3$ or the $2^{(3,3)}_4$ solution, we obtain a $T$-duality chain of the E$^{(5;6)}$-branes. 
The dual fields and the locally non-geometric fluxes are as follows:
\begin{align}
\begin{split}
 &\underline{\bm{2^{6}_5(12,345678):} \qquad \bigl\{\,R_{(5)}^{6},\,E^{(5)}_{9,6}\,\bigr\}}
\\
 &\rmd \tilde{s}^2 = \tau_2^{3/2}\, \bigl(\rmd x^2_{012} + \tau_2\,\rmd x_{9}^2\bigr) + \tau_2^{1/2}\,\rmd x^2_{345678} \,,\qquad 
 \Exp{-2\tilde{\phi}}= \tau_2^{-5/2} \,, 
\\
 &\beta^{\bar3\cdots \bar8} = m\, y^{\rmD}\,,\qquad R_{(5)}^{\bar3\cdots \bar8} = \partial_{\rmD} \beta^{\bar3\cdots \bar8} = m \,,
\end{split}
\\
\begin{split}
 &\underline{\bm{2^{(1,5)}_5(12,45678,3):} \qquad \bigl\{\,R_{(5)}^{6,1},\,E^{(5)}_{9,6,1}\,\bigr\}}
\\
 &\rmd \tilde{s}^2 = \tau_2^{3/2}\, \bigl(\rmd x^2_{012} + \tau_2\,\rmd x_{9}^2\bigr) + \tau_2^{-1/2}\,\rmd x_{3}^2 + \tau_2^{1/2}\,\rmd x^2_{45678} \,,\qquad 
  \Exp{-2\tilde{\phi}}= \tau_2^{-2} \,, 
\\
 &\beta^{\bar3\cdots \bar8} = m\, y^{\rmD}_{3}\,,\qquad R_{(5)}^{\bar3\cdots \bar8,\bar3} = \partial_{\rmD}^{\bar3} \beta^{\bar3\cdots \bar8} = m \,,
\end{split}
\\
\begin{split}
 &\underline{\bm{2^{(2,4)}_5(12,5678,34):} \qquad \bigl\{\,R_{(5)}^{6,2},\,E^{(5)}_{9,6,2}\,\bigr\}}
\\
 &\rmd \tilde{s}^2 = \tau_2^{3/2}\, \rmd x^2_{012} + \tau_2^{-1/2}\,\rmd x_{34}^2 + \tau_2^{1/2}\,\rmd x^2_{5678} + \tau_2^{5/2} \,\rmd x_{9}^2 \,,\qquad 
 \Exp{-2\tilde{\phi}}= \tau_2^{-3/2} \,, 
\\
 &\beta^{\bar3\cdots \bar8} = m\, y^{\rmD}_{34} \,,\qquad R_{(5)}^{\bar3\cdots \bar8,\bar3\bar4} = \partial_{\rmD}^{\bar3\bar4} \beta^{\bar3\cdots \bar8} = m \,,
\end{split}
\\
\begin{split}
 &\underline{\bm{2^{(3,3)}_5(12,678,345):} \qquad \bigl\{\,R_{(5)}^{6,3},\,E^{(5)}_{9,6,3}\,\bigr\}}
\\
 &\rmd \tilde{s}^2 = \tau_2^{3/2}\, \rmd x^2_{012} + \tau_2^{-1/2}\,\rmd x_{345}^2 + \tau_2^{1/2}\,\rmd x^2_{678} + \tau_2^{5/2} \,\rmd x_{9}^2 \,,\qquad 
 \Exp{-2\tilde{\phi}}= \tau_2^{-1} \,, 
\\
 &\beta^{\bar3\cdots \bar8} = m\, y^{\rmD}_{345} \,,\qquad R_{(5)}^{\bar3\cdots \bar8,\bar3\bar4\bar5} = \partial_{\rmD}^{\bar3\bar4\bar5} \beta^{\bar3\cdots \bar8} = m\,,
\end{split}
\\
\begin{split}
 &\underline{\bm{2^{(4,2)}_5(12,78,3456):} \qquad \bigl\{\,R_{(5)}^{6,4},\,E^{(5)}_{9,6,4}\,\bigr\}}
\\
 &\rmd \tilde{s}^2 = \tau_2^{3/2}\, \bigl(\rmd x^2_{012} + \tau_2\,\rmd x_{9}^2\bigr) + \tau_2^{-1/2}\,\rmd x^2_{3456} + \tau_2^{1/2}\,\rmd x^2_{78} \,,\qquad 
 \Exp{-2\tilde{\phi}}= \tau_2^{-1/2} \,, 
\\
 &\beta^{\bar3\cdots \bar8} = m\, y^{\rmD}_{3456} \,,\qquad R_{(5)}^{\bar3\cdots \bar8,\bar3\cdots \bar6} = \partial_{\rmD}^{\bar3\cdots \bar6} \beta^{\bar3\cdots \bar8} = m\,,
\end{split}
\\
\begin{split}
 &\underline{\bm{2^{(5,1)}_5(12,8,34567):} \qquad \bigl\{\,R_{(5)}^{6,5},\,E^{(5)}_{9,6,5}\,\bigr\}}
\\
 &\rmd \tilde{s}^2 = \tau_2^{3/2}\, \bigl(\rmd x^2_{012} + \tau_2\,\rmd x_{9}^2\bigr) + \tau_2^{-1/2}\,\rmd x^2_{34567} + \tau_2^{1/2}\,\rmd x^2_{8} \,,\qquad 
 \Exp{-2\tilde{\phi}}= 1 \,, 
\\
 &\beta^{\bar3\cdots \bar8} = m\, y^{\rmD}_{34567} \,,\qquad R_{(5)}^{\bar3\cdots \bar8,\bar3\cdots \bar7} = \partial_{\rmD}^{\bar3\cdots \bar7} \beta^{\bar3\cdots \bar8} = m\,,
\end{split}
\\
\begin{split}
 &\underline{\bm{2^{(6,0)}_5(12,,345678):} \qquad \bigl\{\,R_{(5)}^{6,6},\,E^{(5)}_{9,6,6}\,\bigr\}}
\\
 &\rmd \tilde{s}^2 = \tau_2^{3/2}\, \bigl(\rmd x^2_{012} + \tau_2\,\rmd x_{9}^2\bigr) + \tau_2^{-1/2}\,\rmd x^2_{345678} \,,\qquad 
 \Exp{-2\tilde{\phi}}= \tau_2^{1/2} \,, 
\\
 &\beta^{\bar3\cdots \bar8} = m\, y^{\rmD}_{345678} \,,\qquad R_{(5)}^{\bar3\cdots \bar8,\bar3\cdots \bar8} = \partial_{\rmD}^{\bar3\cdots \bar8} \beta^{\bar3\cdots \bar8} = m\,,
\end{split}
\\
\begin{split}
 &\underline{\bm{1^{(1,0,6)}_5(1,345678,,2):} \qquad \bigl\{\,R_{(5)}^{7,1,1},\,E^{(5)}_{9,7,1,1}\,\bigr\}}
\\
 &\rmd \tilde{s}^2 = \tau_2^{3/2}\, \bigl(\rmd x^2_{01} + \tau_2\,\rmd x_{9}^2\bigr) + \tau_2^{-3/2}\,\rmd x_{2}^2 + \tau_2^{1/2}\,\rmd x^2_{345678} \,,\qquad 
 \Exp{-2\tilde{\phi}}= \tau_2^{-1} \,, 
\\
 &\beta^{\bar2\cdots \bar8,\bar2} = m\, y^{\rmD}_2 \,,\qquad R_{(5)}^{\bar2\cdots \bar8,\bar2,\bar2} = \partial_{\rmD}^{\bar2} \beta^{\bar2\cdots \bar8,\bar2} = m\,,
\end{split}
\\
\begin{split}
 &\underline{\bm{1^{(1,1,5)}_5(1,45678,3,2):} \qquad \bigl\{\,R_{(5)}^{7,2,1},\,E^{(5)}_{9,7,2,1}\,\bigr\}}
\\
 &\rmd \tilde{s}^2 = \tau_2^{3/2}\, \bigl(\rmd x^2_{01} + \tau_2\,\rmd x_{9}^2\bigr) + \tau_2^{-3/2}\,\rmd x_{2}^2 + \tau_2^{1/2}\,\rmd x^2_{45678} + \tau_2^{-1/2}\,\rmd x^2_{3} \,,\qquad 
 \Exp{-2\tilde{\phi}}= \tau_2^{-1/2} \,, 
\\
 &\beta^{\bar2\cdots \bar8,\bar2} = m\, y^{\rmD}_{23} \,,\qquad R_{(5)}^{\bar2\cdots \bar8,\bar2\bar3,\bar2} = \partial_{\rmD}^{\bar2\bar3} \beta^{\bar2\cdots \bar8,\bar2} = m\,,
\end{split}
\\
\begin{split}
 &\underline{\bm{1^{(1,2,4)}_5(1,5678,34,2):} \qquad \bigl\{\,R_{(5)}^{7,3,1},\,E^{(5)}_{9,7,3,1}\,\bigr\}}
\\
 &\rmd \tilde{s}^2 = \tau_2^{3/2}\, \bigl(\rmd x^2_{01} + \tau_2\,\rmd x_{9}^2\bigr) + \tau_2^{-3/2}\,\rmd x_{2}^2 + \tau_2^{1/2}\,\rmd x^2_{5678} + \tau_2^{-1/2}\,\rmd x^2_{34} \,,\qquad 
 \Exp{-2\tilde{\phi}}= 1 \,,
\\
 &\beta^{\bar2\cdots \bar8,\bar2} = m\, y^{\rmD}_{234} \,,\qquad R_{(5)}^{\bar2\cdots \bar8,\bar2\bar3\bar4,\bar2} = \partial_{\rmD}^{\bar2\bar3\bar4} \beta^{\bar2\cdots \bar8,\bar2} = m\,,
\end{split}
\\
\begin{split}
 &\underline{\bm{1^{(1,3,3)}_5(1,678,345,2):} \qquad \bigl\{\,R_{(5)}^{7,4,1},\,E^{(5)}_{9,7,4,1}\,\bigr\}}
\\
 &\rmd \tilde{s}^2 = \tau_2^{3/2}\, \bigl(\rmd x^2_{01} + \tau_2\,\rmd x_{9}^2\bigr) + \tau_2^{-3/2}\,\rmd x_{2}^2 + \tau_2^{1/2}\,\rmd x^2_{678} + \tau_2^{-1/2}\,\rmd x^2_{345} \,,\qquad 
 \Exp{-2\tilde{\phi}}= \tau_2^{1/2} \,, 
\\
 &\beta^{\bar2\cdots \bar8,\bar2} = m\, y^{\rmD}_{2345} \,,\qquad R_{(5)}^{\bar2\cdots \bar8,\bar2\cdots\bar5,\bar2} = \partial_{\rmD}^{\bar2\cdots\bar5} \beta^{\bar2\cdots \bar8,\bar2} = m\,,
\end{split}
\\
\begin{split}
 &\underline{\bm{1^{(1,4,2)}_5(1,78,3456,2):} \qquad \bigl\{\,R_{(5)}^{7,5,1},\,E^{(5)}_{9,7,5,1}\,\bigr\}}
\\
 &\rmd \tilde{s}^2 = \tau_2^{3/2}\, \bigl(\rmd x^2_{01} + \tau_2\,\rmd x_{9}^2\bigr) + \tau_2^{-3/2}\,\rmd x_{2}^2 + \tau_2^{1/2}\,\rmd x^2_{78} + \tau_2^{-1/2}\,\rmd x^2_{3456} \,,\qquad 
 \Exp{-2\tilde{\phi}}= \tau_2 \,, 
\\
 &\beta^{\bar2\cdots \bar8,\bar2} = m\, y^{\rmD}_{23456} \,,\qquad R_{(5)}^{\bar2\cdots \bar8,\bar2\cdots\bar6,\bar2} = \partial_{\rmD}^{\bar2\cdots\bar6} \beta^{\bar2\cdots \bar8,\bar2} = m\,,
\end{split}
\\
\begin{split}
 &\underline{\bm{1^{(1,5,1)}_5(1,8,34567,2):} \qquad \bigl\{\,R_{(5)}^{7,6,1},\,E^{(5)}_{9,7,6,1}\,\bigr\}}
\\
 &\rmd \tilde{s}^2 = \tau_2^{3/2}\, \bigl(\rmd x^2_{01} + \tau_2\,\rmd x_{9}^2\bigr) + \tau_2^{-3/2}\,\rmd x_{2}^2 + \tau_2^{1/2}\,\rmd x^2_{8} + \tau_2^{-1/2}\,\rmd x^2_{34567} \,,\qquad 
 \Exp{-2\tilde{\phi}}= \tau_2^{3/2} \,, 
\\
 &\beta^{\bar2\cdots \bar8,\bar2} = m\, y^{\rmD}_{234567} \,,\qquad R_{(5)}^{\bar2\cdots \bar8,\bar2\cdots\bar7,\bar2} = \partial_{\rmD}^{\bar2\cdots\bar7} \beta^{\bar2\cdots \bar8,\bar2} = m\,,
\end{split}
\\
\begin{split}
 &\underline{\bm{1^{(1,6,0)}_5(1,,345678,2):} \qquad \bigl\{\,R_{(5)}^{7,7,1},\,E^{(5)}_{9,7,7,1}\,\bigr\}}
\\
 &\rmd \tilde{s}^2 = \tau_2^{3/2}\, \bigl(\rmd x^2_{01} + \tau_2\,\rmd x_{9}^2\bigr) + \tau_2^{-3/2}\,\rmd x_{2}^2 + \tau_2^{-1/2}\,\rmd x^2_{345678} \,,\qquad 
 \Exp{-2\tilde{\phi}}= \tau_2^{2} \,, 
\\
 &\beta^{\bar2\cdots \bar8,\bar2} = m\, y^{\rmD}_{2345678} \,,\qquad R_{(5)}^{\bar2\cdots \bar8,\bar2\cdots\bar8,\bar2} = \partial_{\rmD}^{\bar2\cdots\bar8} \beta^{\bar2\cdots \bar8,\bar2} = m \,.
\end{split}
\end{align}

\subsubsection{E$^{(6;4)}$-branes}

We can repeat the duality transformations and obtain the following solutions,
\begin{align}
\begin{split}
 &\underline{\bm{1^{(4,3)}_6(1,678,2345):} \qquad \bigl\{\,R_{(6)}^{7,4},\,E^{(6)}_{9,7,4}\,\bigr\}}
\\
 &\rmd \tilde{s}^2 = \tau_2^2\, \rmd x^2_{01} + \rmd x_{2345}^2 + \tau_2\,\rmd x^2_{678} + \tau_2^3 \,\rmd x_{9}^2 \,, \qquad 
 \Exp{-2\tilde{\phi}}= \tau_2^{-2} \,, 
\\
 &\beta^{\bar2\cdots \bar7} = m\, y^{\rmS}_{2\cdots 58}\,,\quad R_{(6)}^{\bar2\cdots\bar8,\bar2\cdots\bar5} = \frac{7!}{5!\,2!}\, \partial_{\rmS}^{[\bar2\cdots\bar5\bar8} \beta^{\bar6\bar7]\bar2\cdots\bar5} =m \quad \text{(IIA/IIB)} \,,
\end{split}
\\
\begin{split}
 &\underline{\bm{1^{(1,4,2)}_6(1,78,2345,6):} \qquad \bigl\{\,R_{(6)}^{7,5,1},\,E^{(6)}_{9,7,5,1}\,\bigr\}}
\\
 &\rmd \tilde{s}^2 = \tau_2^2\, \rmd x^2_{01} + \rmd x_{2345}^2 + \tau_2^{-1}\,\rmd x^2_{6} + \tau_2\,\rmd x^2_{78} + \tau_2^3 \,\rmd x_{9}^2 \,, \qquad 
 \Exp{-2\tilde{\phi}}= \tau_2^{-1}\,, 
\\
 &\beta^{\bar2\cdots \bar7} = m\, y^{\rmS}_{2\cdots 68,6}\,,\quad R_{(6)}^{\bar2\cdots \bar8,\bar2\cdots\bar6,\bar6} = 7\, \partial_{\rmS}^{[\bar2\cdots\bar7|,\bar6} \beta^{|\bar8]\bar2\cdots\bar6} = m \quad \text{(IIA/IIB)} \,,
\end{split}
\\
\begin{split}
 &\underline{\bm{1^{(2,4,1)}_6(1,8,2345,67):} \qquad \bigl\{\,R_{(6)}^{7,6,2},\,E^{(6)}_{9,7,6,2}\,\bigr\}}
\\
 &\rmd \tilde{s}^2 = \tau_2^2\, \rmd x^2_{01} + \rmd x_{2345}^2 + \tau_2^{-1}\,\rmd x^2_{67} + \tau_2\,\rmd x^2_{8} + \tau_2^3 \,\rmd x_{9}^2 \,, \qquad 
 \Exp{-2\tilde{\phi}}= 1\,, 
\\
 &\beta^{\bar2\cdots \bar7} = m\, y^{\rmS}_{2\cdots 8,67}\,,\quad R_{(6)}^{\bar2\cdots\bar8,\bar2\cdots\bar7,\bar6\bar7} = \partial_{\rmS}^{\bar2\cdots\bar8,\bar6\bar7} \beta^{\bar2\cdots\bar7} =m \quad \text{(IIA/IIB)} \,,
\end{split}
\\
\begin{split}
 &\underline{\bm{1^{(3,4,0)}_6(1,,2345,678):} \qquad \bigl\{\,R_{(6)}^{7,7,3},\,E^{(6)}_{9,7,7,3}\,\bigr\}}
\\
 &\rmd \tilde{s}^2 = \tau_2^2\, \rmd x^2_{01} + \rmd x_{2345}^2 + \tau_2^{-1}\,\rmd x^2_{678} + \tau_2^3 \,\rmd x_{9}^2 \,, \qquad 
 \Exp{-2\tilde{\phi}}= \tau_2 \,, 
\\
 &\beta^{\bar2\cdots \bar8,\bar8} = m\, y^{\rmS}_{2\cdots 8,67}\,,\quad R_{(6)}^{\bar2\cdots\bar8,\bar2\cdots\bar8,\bar6\bar7\bar8} = 3\, \partial_{\rmS}^{\bar2\cdots\bar8,[\bar6\bar7|} \beta^{\bar2\cdots\bar8,|\bar8]} =m \quad \text{(IIA/IIB)} \,.
\end{split}
\end{align}
Note that the $1^{(2,4,1)}_6$-brane is self-dual under the $S$-duality transformation. 
Apparently, the above $1^{(2,4,1)}_6$ is not invariant under the $S$-duality transformation, but since the $R$-flux is invariant under the $S$-duality, the apparent non-invariance is due to a particular gauge choice. 
The $R$-flux, or the magnetic charge of the $1^{(2,4,1)}_6$-brane is invariant under the $S$-duality. 

\subsubsection{E$^{(7;7)}$-branes}

We can further obtain the following family of solutions:
\begin{align}
\begin{split}
 &\underline{\bm{1^{(7,0)}_7(1,,2345678):} \qquad \bigl\{\,R_{(7)}^{7,7},\,E^{(7)}_{9,7,7}\,\bigr\}}
\\
 &\rmd \tilde{s}^2 = \tau_2^{5/2}\, \bigl(\rmd x^2_{01} + \tau_2\,\rmd x_{9}^2\bigr) + \tau_2^{1/2}\,\rmd x^2_{2345678} \,,\qquad 
 \Exp{-2\tilde{\phi}}= \tau_2^{-5/2} \,, 
\\
 &\beta^{\bar3\cdots \bar8} = m\, y^{\rmE}_{2\cdots 8,2} \,,\qquad
 R_{(7)}^{\bar2\cdots \bar8,\bar2\cdots\bar8} = 7\, \partial_{\rmE}^{\bar2\cdots\bar8,[\bar2} \beta^{\bar3\cdots \bar8]} = m\,,
\end{split}
\\
\begin{split}
 &\underline{\bm{1^{(1,6,0)}_7(1,,345678,2):} \qquad \bigl\{\,R_{(7)}^{7,7,1},\,E^{(7)}_{9,7,7,1}\,\bigr\}}
\\
 &\rmd \tilde{s}^2 = \tau_2^{5/2}\, \bigl(\rmd x^2_{01} + \tau_2\,\rmd x_{9}^2\bigr) + \tau_2^{-1/2}\,\rmd x_{2}^2 + \tau_2^{1/2}\,\rmd x^2_{345678} \,,\qquad 
 \Exp{-2\tilde{\phi}}= \tau_2^{-2} \,, 
\\
 &\beta^{\bar2\cdots \bar8,\bar2} = m\, y^{\rmE}_{2\cdots 8} \,,\qquad
 R^{\bar2\cdots \bar8,\bar2\cdots\bar8,\bar2} = \partial_{\rmE}^{\bar2\cdots\bar8} \beta^{\bar2\cdots \bar8,\bar2} = m\,,
\end{split}
\\
\begin{split}
 &\underline{\bm{1^{(2,5,0)}_7(1,,45678,23):} \qquad \bigl\{\,R_{(7)}^{7,7,2},\,E^{(7)}_{9,7,7,2}\,\bigr\}}
\\
 &\rmd \tilde{s}^2 = \tau_2^{5/2}\, \bigl(\rmd x^2_{01} + \tau_2\,\rmd x_{9}^2\bigr) + \tau_2^{-1/2}\,\rmd x_{23}^2 + \tau_2^{1/2}\,\rmd x^2_{45678} \,,\qquad 
 \Exp{-2\tilde{\phi}}= \tau_2^{-3/2} \,, 
\\
 &\beta^{\bar2\cdots \bar8,\bar2} = m\, y^{\rmE}_{2\cdots 8,3} \,,\qquad
 R^{\bar2\cdots \bar8,\bar2\cdots\bar8,\bar2\bar3} = 2\, \partial_{\rmE}^{\bar2\cdots\bar8,[\bar3|} \beta^{\bar2\cdots \bar8,|\bar2]} = m\,,
\end{split}
\\
\begin{split}
 &\underline{\bm{1^{(3,4,0)}_7(1,,5678,234):} \qquad \bigl\{\,R_{(7)}^{7,7,3},\,E^{(7)}_{9,7,7,3}\,\bigr\}}
\\
 &\rmd \tilde{s}^2 = \tau_2^{5/2}\, \bigl(\rmd x^2_{01} + \tau_2\,\rmd x_{9}^2\bigr) + \tau_2^{-1/2}\,\rmd x_{234}^2 + \tau_2^{1/2}\,\rmd x^2_{5678} \,,\qquad 
 \Exp{-2\tilde{\phi}}= \tau_2^{-1} \,, 
\\
 &\beta^{\bar2\cdots \bar8,\bar2} = m\, y^{\rmE}_{2\cdots 8,34} \,,\qquad
 R_{(7)}^{\bar2\cdots \bar8,\bar2\cdots\bar8,\bar2\bar3\bar4} = 3\, \partial_{\rmE}^{\bar2\cdots\bar8,[\bar3\bar4|} \beta^{\bar2\cdots \bar8,|\bar2]} = m\,,
\end{split}
\\
\begin{split}
 &\underline{\bm{1^{(4,3,0)}_7(1,,678,2345):} \qquad \bigl\{\,R_{(7)}^{7,7,4},\,E^{(7)}_{9,7,7,4}\,\bigr\}}
\\
 &\rmd \tilde{s}^2 = \tau_2^{5/2}\, \bigl(\rmd x^2_{01} + \tau_2\,\rmd x_{9}^2\bigr) + \tau_2^{-1/2}\,\rmd x_{2345}^2 + \tau_2^{1/2}\,\rmd x^2_{678} \,,\qquad 
 \Exp{-2\tilde{\phi}}= \tau_2^{-1/2} \,, 
\\
 &\beta^{\bar2\cdots \bar8,\bar2} = m\, y^{\rmE}_{2\cdots 8,345} \,,\qquad
 R_{(7)}^{\bar2\cdots \bar8,\bar2\cdots\bar8,\bar2\cdots\bar5} = 4\, \partial_{\rmE}^{\bar2\cdots\bar8,[\bar3\bar4\bar5|} \beta^{\bar2\cdots \bar8,|\bar2]} = m\,,
\end{split}
\\
\begin{split}
 &\underline{\bm{1^{(5,2,0)}_7(1,,78,23456):} \qquad \bigl\{\,R_{(7)}^{7,7,5},\,E^{(7)}_{9,7,7,5}\,\bigr\}}
\\
 &\rmd \tilde{s}^2 = \tau_2^{5/2}\, \bigl(\rmd x^2_{01} + \tau_2\,\rmd x_{9}^2\bigr) + \tau_2^{-1/2}\,\rmd x_{23456}^2 + \tau_2^{1/2}\,\rmd x^2_{78} \,,\qquad 
 \Exp{-2\tilde{\phi}}= 1 \,, 
\\
 &\beta^{\bar2\cdots \bar8,\bar2} = m\, y^{\rmE}_{2\cdots 8,3456} \,,\qquad
 R_{(7)}^{\bar2\cdots \bar8,\bar2\cdots\bar8,\bar2\cdots\bar6} = 5\, \partial_{\rmE}^{\bar2\cdots\bar8,[\bar3\cdots\bar6|} \beta^{\bar2\cdots \bar8,|\bar2]} = m\,,
\end{split}
\\
\begin{split}
 &\underline{\bm{1^{(6,1,0)}_7(1,,8,234567):} \qquad \bigl\{\,R_{(7)}^{7,7,6},\,E^{(7)}_{9,7,7,6}\,\bigr\}}
\\
 &\rmd \tilde{s}^2 = \tau_2^{5/2}\, \bigl(\rmd x^2_{01} + \tau_2\,\rmd x_{9}^2\bigr) + \tau_2^{-1/2}\,\rmd x_{234567}^2 + \tau_2^{1/2}\,\rmd x^2_{8} \,,\qquad 
 \Exp{-2\tilde{\phi}}= \tau_2^{1/2} \,, 
\\
 &\beta^{\bar2\cdots \bar8,\bar2} = m\, y^{\rmE}_{2\cdots 8,3\cdots 7} \,,\qquad
 R_{(7)}^{\bar2\cdots \bar8,\bar2\cdots\bar8,\bar2\cdots\bar7} = 6\, \partial_{\rmE}^{\bar2\cdots\bar8,[\bar3\cdots\bar7|} \beta^{\bar2\cdots \bar8,|\bar2]} = m\,,
\end{split}
\\
\begin{split}
 &\underline{\bm{1^{(7,0,0)}_7(1,,,2345678):} \qquad \bigl\{\,R_{(7)}^{7,7,7},\,E^{(7)}_{9,7,7,7}\,\bigr\}}
\\
 &\rmd \tilde{s}^2 = \tau_2^{5/2}\, \bigl(\rmd x^2_{01} + \tau_2\,\rmd x_{9}^2\bigr) + \tau_2^{-1/2}\,\rmd x_{2345678}^2 \,,\qquad 
 \Exp{-2\tilde{\phi}}= \tau_2 \,, 
\\
 &\beta^{\bar2\cdots \bar8,\bar2} = m\, y^{\rmE}_{2\cdots 8,3\cdots 8} \,,\qquad 
 R_{(7)}^{\bar2\cdots \bar8,\bar2\cdots\bar8,\bar2\cdots\bar8} = 7\, \partial_{\rmE}^{\bar2\cdots\bar8,[\bar3\cdots\bar8|} \beta^{\bar2\cdots \bar8,|\bar2]} = m\,.
\end{split}
\end{align}

\subsubsection{E$^{(8;7)}$-branes}

Finally, by performing the $S$-duality in the $1^{(7,0,0)}_7$ background, we obtain
\begin{align}
\begin{split}
 &\underline{\bm{1^{(7,0,0)}_8(1,,,2345678):} \qquad \bigl\{\,R_{(8)}^{7,7,7},\,E^{(8)}_{9,7,7,7}\,\bigr\}}
\\
 &\rmd \tilde{s}^2 = \tau_2^{3}\, \bigl(\rmd x^2_{01} + \tau_2\,\rmd x_{9}^2\bigr) + \rmd x_{2345678}^2 \,,\qquad 
 \Exp{-2\tilde{\phi}}= \tau_2^{-1} \,, 
\\
&\beta^{\bar2\cdots \bar8,\bar2} = m\, \tilde{x}_{2345678,345678} \,,\qquad 
 R_{(8)}^{\bar2\cdots \bar8,\bar2\cdots\bar8,\bar2\cdots\bar8} = 7\,\tilde{\partial}^{\bar2\cdots\bar8,[\bar3\cdots\bar8|} \beta^{\bar2\cdots \bar8,|\bar2]} = m \quad \text{(IIA/IIB)} \,.
\end{split}
\end{align}

\subsection{Exotic-brane solutions in M-theory}
\label{sec:exotic-sol-m}

By uplifting the defect-brane solutions in type II theories to M-theory, we obtain the following defect-brane solutions:
\begin{align}
\begin{split}
 &\underline{\bm{5_{12}^{3}(12345,67z):} \qquad \bigl\{\,S_1^{3},\,E_{9,3}\,\bigr\}}
\\
 &\rmd \tilde{s}^2 = \tau_2^{1/3}\, \bigl(\rmd x^2_{012345} + \tau_2\,\rmd x_{89}^2\bigr) + \tau_2^{-2/3}\,\rmd x^2_{67z} \,,
\\
 &\Omega^{\bar6\bar7\bar z} = m\, x^8 \,,\qquad 
 S_{\bar8}{}^{\bar6\bar7\bar z} = m\,,
\end{split}
\\
\begin{split}
 &\underline{\bm{2^{6}_5(12,34567z):} \qquad \bigl\{\,S_1^{6},\,E_{9,6}\,\bigr\}}
\\
 &\rmd \tilde{s}^2 = \tau_2^{2/3}\, \bigl(\rmd x^2_{012} + \tau_2\,\rmd x_{89}^2\bigr) + \tau_2^{-1/3}\,\rmd x^2_{34567z} \,,
\\
 &\Omega^{\bar3\cdots\bar7\bar z} = m\,x^{8} \,, \qquad 
 S_{\bar8}{}^{\bar3\cdots\bar7\bar z} = m \,,
\end{split}
\\
\begin{split}
 &\underline{\bm{0_{18}^{(1,7)}(2\cdots7z,1):} \qquad \bigl\{\,S_1^{8,1},\,E_{9,8,1}\,\bigr\}}
\\
 &\rmd \tilde{s}^2 = \tau_2\, \bigl(\rmd x^2_{0} + \tau_2\,\rmd x_{89}^2\bigr) + \tau_2^{-1}\,\rmd x^2_{1} + \rmd x^2_{2\cdots7z} \,,
\\
 &\Omega^{1\cdots7z,1} = m\,x^8 \,, \qquad 
  S_{\bar8}{}^{1\cdots7z,1} = m \,,
\end{split}
\end{align}
where $S_i{}^{j_1\cdots j_p}\equiv \partial_i \Omega^{j_1\cdots j_p}$ and $S_i{}^{j_1\cdots j_8,\,k}\equiv \partial_i \Omega^{j_1\cdots j_8,\,k}$\,. 
The direction $z$ represents one of the internal ones that is not necessary to be the M-theory direction, which we denote M. 
If we define the dual field strengths,
\begin{align}
\begin{split}
 S_{10,\bar m_1\bar m_2\bar m_3} &\equiv \tilde{G}_{\bar m_1\bar m_2\bar m_3,\bar n_1\bar n_2\bar n_3}\,\tilde{*}_{11} S_{1}^{\bar n_1\bar n_2\bar n_3} \equiv \rmd E_{9,\bar m_1\bar m_2\bar m_3}\,,
\\
 S_{10,\bar m_1\cdots \bar m_6} &\equiv \tilde{G}_{\bar m_1\cdots \bar m_6,\bar n_1\cdots \bar n_6}\,\tilde{*}_{11} S_{1}^{\bar n_1\cdots \bar n_6} \equiv \rmd E_{9,\bar m_1\cdots \bar m_6}\,,
\\
 S_{10,\bar m_1\cdots \bar m_7,m} &\equiv \tilde{G}_{\bar m_1\cdots \bar m_7,\bar n_1\cdots \bar n_7}\,\tilde{G}_{\bar m\bar n}\,\tilde{*}_{11} S_{1}^{\bar n_1\cdots \bar n_7,\bar n} \equiv \rmd E_{9,\bar m_1\cdots \bar m_7,\bar m}\,,
\end{split}
\end{align}
we find that the non-vanishing component of the dual potentials again has the form
\begin{align}
 E_{\bar0\cdots\bar8,\cdots,\cdots,\cdots} = -m\,\tau_2^{-1}\,. 
\end{align}

Similarly, we can obtain all of the ``elementary'' domain-wall-brane solutions in M-theory as follows:
\begin{align}
\begin{split}
 &\underline{\bm{8_{12}^{(1,0)}(1234567z,,8):} \qquad \bigl\{\,R^{1,1},\,E_{10,1,1}\,\bigr\}}
\\
 &\rmd \tilde{s}^2 = \tau_2^{1/3}\, \bigl(\rmd x^2_{0\cdots7z} + \tau_2\,\rmd x_{9}^2\bigr) + \tau_2^{-5/3}\,\bigl(\rmd x^{8}+m\,y_{8z}\,\rmd x^z\bigr)^2\,,
\\
 &A^{\bar8}_{\bar z} \equiv \tilde{G}^{\bar8\bar z}/\tilde{G}^{\bar z\bar z}= -m\,y_{8z} \,,\qquad 
 R^{\bar8,\,\bar8} = \partial^{\bar z\bar8}A^{\bar8}_{\bar z} = m\,,
\end{split}
\label{eq:810M-soln}
\\
\begin{split}
 &\underline{\bm{5_{15}^{(1,3)}(1234z,678,5):} \qquad \bigl\{\,R^{4,1},\,E_{10,4,1}\,\bigr\}}
\\
 &\rmd \tilde{s}^2 = \tau_2^{2/3}\, \bigl(\rmd x^2_{01234z}+\tau_2\,\rmd x_{9}^2\bigr) + \tau_2^{-4/3}\,\rmd x^2_5 +\tau_2^{-1/3}\,\rmd x^2_{678} \,, 
\\
 &\Omega^{\bar5\bar6\bar7} = m\, y_{58} \,,\qquad R^{\bar5\cdots\bar8,\bar5}= \frac{4!}{2!\,2!}\,\partial^{[56}\Omega^{78]5}= m\,,
\end{split}
\\
\begin{split}
 &\underline{\bm{3_{18}^{(2,4)}(12z,5678,34):} \qquad \bigl\{\,R^{6,2},\,E_{10,6,2}\,\bigr\}}
\\
 &\rmd \tilde{s}^2 = \tau_2 \, \bigl(\rmd x^2_{012z} + \tau_2\,\rmd x_{9}^2\bigr) + \tau_2^{-1}\,\rmd x^2_{34} + \rmd x^2_{5\cdots8} \,, 
\\
 &\Omega^{3\cdots 8} = m\, y_{34} \,,\qquad 
  R^{\bar3\cdots\bar8,\bar3\bar4} = \partial^{\bar3\bar4}\Omega^{\bar3\cdots \bar8}= m\,,
\end{split}
\\
\begin{split}
 &\underline{\bm{2_{21}^{(4,3)}(12,78z,3456):} \qquad \bigl\{\,R^{7,4},\,E_{10,7,4}\,\bigr\}}
\\
 &\rmd \tilde{s}^2 = \tau_2^{4/3}\, \bigl(\rmd x^2_{012} + \tau_2\,\rmd x_{9}^2\bigr) + \tau_2^{-2/3}\,\rmd x^2_{3456} + \tau_2^{1/3}\,\rmd x^2_{78z} \,, 
\\
 &\Omega^{\bar3\cdots \bar8} = m\, y_{3456z} \,,\qquad R^{\bar3\cdots\bar8\bar z,\bar3\cdots\bar6}= \frac{7!}{5!\,2!}\,\partial^{[\bar3\cdots\bar7}\Omega^{\bar8\bar z]\bar3\cdots\bar6}= m\,,
\end{split}
\\
\begin{split}
 &\underline{\bm{1_{21}^{(1,1,6)}(1,45678z,3,2):} \qquad \bigl\{\,R^{8,2,1},\,E_{10,8,2,1}\,\bigr\}}
\\
 &\rmd \tilde{s}^2 =\tau_2^{4/3}\, \bigl(\rmd x^2_{01} + \tau_2\,\rmd x_{9}^2\bigr) + \tau_2^{-5/3}\,\rmd x^2_{2} + \tau_2^{-2/3}\,\rmd x^2_{3} + \tau_2^{1/3}\,\rmd x^2_{4\cdots8z} \,,
\\
 &\Omega^{\bar2\cdots\bar8\bar z,\bar2} = m\,y_{23} \,,\qquad R^{\bar2\cdots\bar8\bar z,\bar2\bar3,\bar2} = \partial^{\bar2\bar3}\Omega^{\bar2\cdots\bar8\bar z,\bar2}= m\,,
\end{split}
\\
\begin{split}
 &\underline{\bm{2_{24}^{(7,0)}(12,,345678z):} \qquad \bigl\{\,R^{7,7},\,E_{10,7,7}\,\bigr\}}
\\
 &\rmd \tilde{s}^2 = \tau_2^{5/3}\, \bigl(\rmd x^2_{012} + \tau_2\,\rmd x_{9}^2\bigr) + \tau_2^{-1/3}\,\rmd x^2_{345678z} \,, 
\\
 &\Omega^{\bar3\cdots\bar8}=m\,y_{3\cdots8z,z} \,,\qquad 
 R^{\bar3\cdots\bar8\bar z,\bar3\cdots\bar8\bar z}= 7\,\partial^{\bar3\cdots\bar8\bar z,[\bar z}\Omega^{\bar3\cdots\bar8]}= m\,,
\end{split}
\\
\begin{split}
 &\underline{\bm{1_{24}^{(1,4,3)}(1,678,345z,2):} \qquad \bigl\{\,R^{8,5,1},\,E_{10,8,5,1}\,\bigr\}}
\\
 &\rmd \tilde{s}^2 = \tau_2^{5/3}\, \bigl(\rmd x^2_{01} + \tau_2\,\rmd x_{9}^2\bigr) + \tau_2^{-4/3}\,\rmd x^2_{2} + \tau_2^{2/3}\,\rmd x^2_{678} + \tau_2^{-1/3}\,\rmd x^2_{345z} \,, 
\\
 &\Omega^{\bar2\cdots\bar8\bar z,\bar2} = m\,y_{2345z} \,,\qquad
 R^{\bar2\cdots\bar8\bar z,\bar2\bar3\bar4\bar5\bar z,\bar2}= \partial^{\bar2\bar3\bar4\bar5\bar z} \Omega^{\bar2\cdots\bar8\bar z,\bar2}= m\,,
\end{split}
\\
\begin{split}
 &\underline{\bm{1_{27}^{(2,5,1)}(1,8,34567,2z):} \qquad \bigl\{\,R^{8,7,2},\,E_{10,8,7,2}\,\bigr\}}
\\
 &\rmd \tilde{s}^2 = \tau_2^{2}\, \bigl(\rmd x^2_{01} + \tau_2\,\rmd x_{9}^2\bigr) + \tau_2^{-1}\,\rmd x^2_{2z} + \rmd x^2_{3\cdots7} + \tau_2\,\rmd x^2_{8} \,, 
\\
 &\Omega^{\bar2\cdots\bar8\bar z,\bar2} = m\,y_{2\cdots7z,z} \,,\qquad
 R^{\bar2\cdots\bar8\bar z,\bar2\cdots\bar7\bar z,\bar2\bar z}= 2\,\partial^{\bar2\cdots\bar7\bar z,[\bar z|} \Omega^{\bar2\cdots\bar8\bar z,|\bar2]}= m\,,
\end{split}
\\
\begin{split}
 &\underline{\bm{1_{30}^{(4,4,0)}(1,,2345,678z):} \qquad \bigl\{\,R^{8,8,4},\,E_{10,8,8,4}\,\bigr\}}
\\
 &\rmd \tilde{s}^2 = \tau_2^{7/3}\, \bigl(\rmd x^2_{01} + \tau_2\,\rmd x_{9}^2\bigr) + \tau_2^{1/3}\,\rmd x^2_{2345} + \tau_2^{-2/3}\,\rmd x^2_{678z} \,, 
\\
 &\Omega^{\bar2\cdots\bar8\bar z,\bar8}=m\,y_{2\cdots8z,67z} \,,\qquad 
 R^{\bar2\cdots\bar8\bar z,\bar2\cdots\bar8\bar z,\bar6\bar7\bar8\bar z}= 4\,\partial^{\bar2\cdots\bar8\bar z,[\bar6\bar7\bar z|}\Omega^{\bar2\cdots\bar8\bar z,|\bar8]}= m\,,
\end{split}
\\
\begin{split}
 &\underline{\bm{1_{33}^{(7,1,0)}(1,,8,234567z):} \qquad \bigl\{\,R^{8,8,7},\,E_{10,8,8,7}\,\bigr\}}
\\
 &\rmd \tilde{s}^2 = \tau_2^{8/3}\, \bigl(\rmd x^2_{01} + \tau_2\,\rmd x_{9}^2\bigr) + \tau_2^{-1/3}\,\rmd x^2_{2\cdots7z}+ \tau_2^{2/3}\,\rmd x^2_{8} \,, 
\\
 &\Omega^{\bar2\cdots\bar8\bar z,\bar2}=m\,y_{2\cdots8z,34567z} \,,\qquad 
 R^{\bar2\cdots\bar8\bar z,\bar2\cdots\bar8\bar z,\bar2\cdots\bar7\bar z} = 7\,\partial^{\bar2\cdots\bar8\bar z,[\bar3\cdots\bar7\bar z|}\Omega^{\bar2\cdots\bar8\bar z,|\bar2]} = m\,.
\end{split}
\end{align}
Again, by computing the dual mixed-symmetry potentials through \eqref{eq:dual-potential-M}, we obtain
\begin{align}
 E_{\bar0\cdots\bar9,\cdots,\cdots,\cdots} = -m\,\tau_2^{-1}\,. 
\end{align}

\subsection{Solutions for space-filling branes}
\label{sec:exotic-spacefilling}

We have completed the full list of the ``elementary'' domain-wall solutions. 
We can straightforwardly continue the duality rotations to obtain all of the space-filling branes given in Figures \ref{fig:web01}--\ref{fig:web20} or their M-theory extensions. 
Since there are too many space-filling branes in EFT, we will just show several examples and leave the construction of the other branes as future work.

Let us perform a formal $T$-duality along the $x^9$-direction to the D8-brane solution \eqref{eq:D8-soln}. 
We then obtain the D9-brane solution,
\begin{align}
 \rmd s^2 = \tau_2^{-\frac{1}{2}} \, \rmd x^2_{01\cdots 9} \,,\qquad 
 \Exp{-2\Phi}= \tau_2^{3} \,, \qquad 
 \ket{A} = \bigl(\tau_1\,\gamma^{89} - \tau_2^{-1}\,\gamma^{0 \cdots 9}\bigr)\ket{0} \,,
\end{align}
where $\tau_1=m\,\tilde{x}_8$ and $\tau_2= h_0+m\,\abs{\tilde{x}_9}$\,. 
The field strength becomes
\begin{align}
 \ket{F} = \gamma^M\,\partial_M \ket{A}
 = \bigl(m\,\gamma^9 + \tilde{\partial}^9\tau_2^{-1}\,\gamma^{0 \cdots 8}\bigr)\ket{0}\,,
\end{align}
and the background has the constant 1-form and 9-form field strengths
\begin{align}
 F_1 = m\,\rmd x^9\,,\qquad F_1= *_{10}F_{9} \,. 
\end{align}

Similarly, we can obtain the solutions of the $5^4_2$-brane as
\begin{align}
 \rmd s^2 = \rmd x^2_{01\cdots 5} + \frac{\tau_2}{\abs{\tau}^2} \, \rmd x^2_{67} + \tau_2^{-1} \,\rmd x_{89}^2 \,,\qquad 
 \Exp{-2\Phi}= \tau_2\,\abs{\tau}^2 \,, \qquad
 B_2 = -\frac{\tau_1}{\abs{\tau}^2}\,\rmd x^6\wedge\rmd x^7 \,,
\end{align}
where $\tau_1=m\,\tilde{x}_8$ and $\tau_2= h_0+m\,\abs{\tilde{x}_9}$\,. 
This is also a solution of DFT. 

As the last example, let us consider the $2^{(1,0,0,6)}_5$-brane appearing in Figure \ref{fig:web08}. 
By performing a $T$-duality along the $x^9$-direction in the $2^{6}_5(12,345678)$ solution, we obtain the $2^{(1,0,0,6)}_5(12,345678,9)$ solution,
\begin{align}
 \rmd \tilde{s}^2 = \tau_2^{3/2}\, \rmd x^2_{012} + \tau_2^{1/2}\,\rmd x^2_{345678} + \tau_2^{-5/2}\,\rmd x_{9}^2 \,,\qquad 
 \Exp{-2\tilde{\phi}}= 1 \,, \qquad \beta^{\bar3\cdots \bar9,\,\bar9} = m\, y_9^{\rmD}\,,
\end{align}
where $\tau_2= h_0+m\,\abs{\tilde{x}_9}$\,. 

We can easily construct the space-filling solutions, but their interpretation is not clear. 
For example, the D9-brane background is expected to be a flat spacetime, but the above solution contains non-trivial winding-coordinate dependence. 
Recently, a certain limit which removes the winding-coordinate dependence was discussed in \cite{Kimura:2018hph}, where the harmonic function becomes a constant. 
This may be useful to relate the above solutions to the conventional space-filling solutions such as the D9 solution. 
It is also not clear how to define the suitable fluxes in these backgrounds. 
In this paper, we will not address any further issues about the backgrounds of the space-filling branes. 

\subsection{Projection condition for Killing spinors}
\label{sec:susy}

As it is well known, actions of the standard type II branes, such as the D-branes, are invariant under the half of the spacetime supersymmetry, which is generated by the 32-component Majorana--Weyl Killing spinors $\varepsilon_1$ and $\varepsilon_2$ satisfying
\begin{align}
 \Gamma^{11}\,\varepsilon_1 = \varepsilon_1 \,,\qquad 
 \Gamma^{11}\,\varepsilon_2 = \mp \varepsilon_2 \quad (\text{IIA/IIB})\,.
\end{align}
The supersymmetries preseved by a type II brane can be characterized by a certain projection operator $\cO$ acting on the Killing spinors. 
Here, for convenience, we introduce the Weyl basis
\begin{align}
 \Gamma^{11}\equiv \begin{pmatrix} \bm{1} & \bm{0} \\ \bm{0} & - \bm{1} \end{pmatrix} \,,\qquad 
 \Gamma^{m}\equiv \begin{pmatrix} \bm{0} & \gamma^m \\ \gamma^m & \bm{0} \end{pmatrix} \,, 
\end{align}
and the 16-component Majorana--Weyl Killing spinors $\epsilon_1$ and $\epsilon_1$\,, and define matrices
\begin{align}
 \onebb \equiv \begin{pmatrix} 1 & 0 \\ 0 & 1 \end{pmatrix}\,,\quad 
 \sigma_1 \equiv \begin{pmatrix} 1 & 0 \\ 0 & 1 \end{pmatrix}\,,\quad 
 \sigma_2 \equiv \begin{pmatrix} 0 & -\ii \\ \ii & 0 \end{pmatrix}\,,\quad 
 \sigma_3 \equiv \begin{pmatrix} 1 & 0 \\ 0 & -1 \end{pmatrix}\,, 
\end{align}
which act on $\epsilon\equiv (\epsilon_1,\,\epsilon_2)^\rmT$\,. 
Further, we consider a probe type II brane in a flat target spacetime. 
Then, the projection condition for each type II brane is expressed as follows (see \cite{Ortin:2015hya} for a textbook):
\begin{align}
\begin{split}
 \text{P(1)}:&\quad \bigl(\bm{1}\mp\gamma^{01}\,\onebb\bigr)\,\epsilon =0\,, \qquad 
 \text{F1(1)}:\quad \bigl(\bm{1}\mp\gamma^{01}\,\sigma_3\bigr)\, \epsilon =0 \,,
\\
 \text{NS5(12345)}: & \quad \bigl(\bm{1}\mp\gamma^{012345}\,\cO_{\tiny\text{NS}5}\bigr)\,\epsilon =0 \qquad 
 \bigl[\cO_{\tiny\text{NS}5} \equiv \onebb\ (\text{IIA})\,,\quad \cO_{\tiny\text{NS}5} \equiv\sigma_3\ (\text{IIB})\bigr],
\\
\text{D$p$($1\cdots p$)}:& \quad \bigl(\bm{1}\mp\gamma^{01\cdots p}\,\cO_{\tiny\text{D}p}\bigr)\,\epsilon =0 \,, \quad 
 \cO_{\tiny\text{D}p} \equiv \left\{
\begin{array}{cl}
 \sigma_1 &:\ p = 1,2,5,6,9,
\\
 \ii\sigma_2 &:\ p = 3,4,7,8,
\end{array}
\right.
\end{split}
\end{align}
In our convention, under a $T$-duality transformation along the $y$-direction, the spinor $\varepsilon_1$ is invariant while $\varepsilon_2$ is transformed as $\varepsilon_2\ \to \ \Gamma^y\,\varepsilon_2$\,. 
The $S$-duality rule has been studied in \cite{Ortin:1994su,Imamura:1998gk} and it mixes $\epsilon_1$ and $\epsilon_2$ as
\begin{align}
 \epsilon \ \to \ S\,\epsilon\,,\qquad S\equiv \frac{1}{\sqrt{2}}\,\bigl(\onebb-\ii\,\sigma_2\bigr)\,. 
\end{align}
By using these rules, the projection conditions for many exotic branes were studied in \cite{Kimura:2016xzd} (see also \cite{LozanoTellechea:2000mc} for the conditions for the exotic defect-brane backgrounds in M-theory). 

Here, we extend the analysis of \cite{Kimura:2016xzd} to all of the ``elementary'' exotic branes. 
We will not show the detailed computation, but the result is very simple. 
The projection condition for an exotic brane that electrically couples to the mixed-symmetry potential $E^{(n)}_{m_1\cdots m_{a_1},\,\cdots,\,n_1\cdots n_{a_s}}$ is given by
\begin{align}
 \bigl(\bm{1}\mp \gamma^{m_1\cdots m_{a_1}}\cdots \gamma^{n_1\cdots n_{a_s}}\,\cO\bigr)\,\epsilon = 0\,,
\end{align}
where $\cO$ is a $2\times 2$ matrix attached to each brane in Figures \ref{fig:web01}--\ref{fig:web20}. 
Note that two $\cO$ connected with a dashed line in Figures \ref{fig:web01}--\ref{fig:web20} are different while those connected with a solid line are the same. 

We can easily uplift the type IIA results to M-theory. 
For a brane that electrically couples to a mixed-symmetry potential $E_{i_1\cdots i_{a_1},\cdots,\,j_1\cdots j_{a_s}}$\,, the projection rule becomes
\begin{align}
 \bigl(\bm{1}_{32}\mp \Gamma^{i_1\cdots i_{a_1}}\cdots \Gamma^{j_1\cdots j_{a_s}} \bigr)\,\varepsilon = 0\qquad (\varepsilon\equiv \varepsilon_1+\varepsilon_2)\,. 
\end{align}

The above discussion is about the supersymmetry preserved by probe-brane actions, but the same projection condition (with a small modification by the background supergravity fields) also will apply to the Killing spinors in the EFT solutions for the exotic branes. 
In this paper, we do not check the Killing spinor equations in our EFT solutions explicitly, and leave the detailed analysis for future work. 

\section{Exotic brane solutions in deformed supergravities}
\label{sec:deformed-sugra}

In the previous sections, we have constructed various exotic-brane solutions in DFT/EFT. 
Unlike the case of the standard branes or the defect branes, the obtained solutions explicitly depend on the dual winding coordinates. 
In this section, we explain that the winding-coordinate dependence in the domain-wall solutions can be removed, allowing us to go back to the standard description. 
The price to pay is the appearing of massive deformations, together with isometry directions in the supergravity theory. 

For example, in the literature, the D8-brane background \cite{Bergshoeff:1996ui} is known as the solution of the massive type IIA supergravity. 
In this example, the deformation parameter is nothing but the R--R 0-form potential $F_0$. 
Once we include the winding-coordinate dependence of the D8 solution \eqref{eq:D8-soln} into the mass parameter $F_0$, the solution \eqref{eq:D8-soln} without the R--R potential
\begin{align}
 \rmd s^2 = \tau_2^{-1/2} \,\bigl(\rmd x^2_{01\cdots 8} + \tau_2 \,\rmd x_{9}^2\bigr)\,,\qquad 
 \Exp{-2\Phi}= \tau_2^{5/2} \,, 
\end{align}
becomes the solution of the massive IIA supergravity. 
Since the winding-coordinate dependence has disappeared, we no longer need the DFT formulation. 

A similar story can also be applied to other domain-wall solutions we have obtained. 
In this section, we explain how the linear winding-coordinate dependence provides the equations of motion of the deformed supergravities and discuss the domain-wall solutions in such deformed supergravities. 
As particular examples, the known domain-wall solutions, KK8A, KK9M and ``Unknown (6,2,1)'' solutions in \cite{Meessen:1998qm,Eyras:1999at}, are reproduced. 

Despite in this section we provide several examples of deformed supergravities, we leave for a future work the problem of systematically relate exotic branes and gaugings.

\subsection{Generalized type II supergravity}

In order to get a feeling of the deformed supergravity, it is instructive to review the derivation of GSE \cite{Arutyunov:2015mqj,Wulff:2016tju} from DFT \cite{Sakatani:2016fvh,Sakamoto:2017wor} (see also \cite{Baguet:2016prz} for a derivation from EFT). 

\subsubsection{Bosonic sector of type II DFT}

The equations of motion of the type II DFT are given as
\begin{align}
 \cR_{MN} + \cE_{MN} = 0 \,,\qquad \cR = 0 \,, \qquad \sla{\partial} \, \cK\, \ket{F} = 0\,,
\label{eq:EOM-DFT}
\end{align}
where $\cR_{MN}$ and $\cR$ are the generalized Ricci tensor/scalar, and $\cK$ contains the information of $\cH_{MN}$\,, and the energy-momentum tensor $\cE_{MN}$ is defined in \cite{Hohm:2011zr,Hohm:2011dv} (see also for other conventions). 
If we parameterize the generalized metric $\cH_{MN}$ in terms of the conventional fields $(g_{mn},\,B_{mn})$, and remove the dependence on the winding coordinates $\tilde{x}_m$, the equations of motion of DFT reproduce those of the conventional supergravity. 

On the other hand, in order to derive the GSE from DFT, we suppose that the background admits an isometry. 
In this case, we can choose a set of ten-dimensional coordinates $(x^m)=(x^i,\,x^z)$ such that all fields are independent of $x^z$\,. 
Since the SC allows for one more coordinate dependence, let us introduce a linear $\tilde{x}_z$-dependence into the dilaton
\begin{align}
 d(x) = \sfd(x^i) + I^m\,\tilde{x}_m\,,\quad 
 \Phi(x) = \sfPhi(x^i) + I^m\,\tilde{x}_m \,,\quad 
 I^m\equiv c\,\delta_z^m \quad (c\,: \text{constant}) \,. 
\label{eq:dilaton-dual}
\end{align}
Here, we have decomposed the dilaton field into two parts, and in the following, we interpret the $x^i$-dependent fields $\sfd(x^i)$ and $\sfPhi(x^i)$ as ``physical'' dilatons whereas the winding-coordinate-dependent part $I^m$ is a (non-dynamical) Killing vector. 
Regarding the R--R field, since the field strength $F$ takes the form
\begin{align}
 F=\Exp{-\Phi}\Exp{-B_2\wedge}\hat{\cF} \,,
\end{align}
from \eqref{eq:dilaton-dual}, we suppose that the R--R fields have the following winding-coordinate dependence:
\begin{align}
 F(x) = \Exp{-I^m\,\tilde{x}_m} \sfF(x^i) \,. 
\end{align}
Since $I^m$ trivially satisfies the Killing properties
\begin{align}
 \Lie_I g_{mn}=0\,,\qquad \Lie_I B_{mn}=0\,,\qquad \Lie_I \sfPhi =0\,, \qquad \Lie_I \sfF =0\,,
\end{align}
we are essentially considering a nine-dimensional background. 

For the ``nine-dimensional'' supergravity fields $(g_{mn},\,B_{mn},\,\sfPhi,\,\sfF)$, the equations of motion of DFT take the following form:
\begin{align}
\begin{split}
 R_{mn}-\frac{1}{4}\,H_{mpq}\,H_n{}^{pq} + 2 D_m \partial_n \sfPhi + D_m U_n +D_n U_m 
 =&\ T_{mn} \,,
\\
 \frac{1}{2}\,D^k H_{kmn} - \partial_k\sfPhi\,H^k{}_{mn} - U^k\,H_{kmn} + D_m I_n - D_n I_m =&\ \cK_{mn} \,,
\\
 R + 4\,D^m \partial_m \sfPhi - 4\,\abs{\partial \sfPhi}^2 - \frac{1}{2}\,\abs{H_3}^2 - 4\,\bigl(I^m I_m+U^m U_m + 2\,U^m\,\partial_m \sfPhi - D_m U^m\bigr) =&\ 0\,,
\\
 \rmd \hat{\sfF} - H_3\wedge \hat{\sfF} =&\ 0 \,,
\end{split}
\end{align}
where $U_m \equiv B_{mn}\,I^n$\,, $\hat{\sfF} \equiv \Exp{B_2\wedge} \sfF$\,, and
\begin{align}
\begin{split}
 T_{mn} &\equiv \frac{1}{4} \Exp{2\Phi}\, \sum_p \biggl[ \frac{1}{(p-1)!}\, 
 \hat{\sfF}_{(m}{}^{k_1\cdots k_{p-1}} \hat{\sfF}_{n) k_1\cdots k_{p-1}} - \frac{1}{2}\, g_{mn}\,\abs{\hat{\sfF}_p}^2 \biggr] \,,
\\
 \cK_{mn}&\equiv \frac{1}{4}\Exp{2\Phi}\, \sum_p \frac{1}{(p-2)!}\, \hat{\sfF}_{k_1\cdots k_{p-2}}\, \hat{\sfF}_{mn}{}^{k_1\cdots k_{p-2}} \,. 
\end{split}
\end{align}
These are precisely the generalized type II supergravity equations of motion \cite{Arutyunov:2015mqj,Wulff:2016tju}. 
When the winding-coordinate dependence vanishes (i.e.~$I^m=0$), they have the same form as the usual supergravity equations of motion. 

In this manner, we can consider a slight modification of the supergravity equations of motion by assuming the existence of an isometry in the doubled space. 
Since DFT is defined well for arbitrary solutions of the SC, we can systematically determine the modifications of, for example, the gauge transformation and the duality transformation rules (see \cite{Arutyunov:2015mqj,Sakamoto:2017wor} for the $I$-modified $T$-duality transformation rule). 

\subsubsection{Another viewpoint in terms of the Scherk--Schwarz reduction}

As discussed in the addendum of \cite{Sakatani:2016fvh}, the ansatz for the dilaton \eqref{eq:dilaton-dual} can be understood as the Scherk--Schwarz ansatz in DFT \cite{Aldazabal:2011nj,Geissbuhler:2011mx,Grana:2012rr,Berman:2013cli,Cho:2015lha} (see also \cite{Baguet:2016prz} where the derivation of GSE from a Scherk--Schwarz compactification of EFT was originally discussed). 
An ansatz
\begin{align}
 \cH_{MN}(x,y) = (U^{\rmT})_M{}^K(y)\, \hat{\cH}_{KL}(x)\, U^L{}_N(y)\,,\quad 
 d(x,y) = \hat{d}(x) + \lambda(y) \,,
\label{eq:SS-ansatz}
\end{align}
generally introduces gaugings
\begin{align}
 f_{MNP} \equiv 3\,\eta_{S[M} \, (U^{-1})^Q{}_N\, (U^{-1})^R{}_{P]} \,\partial_Q U^S{}_R \,, \qquad 
 f_M \equiv \partial_N (U^{-1})^N{}_M -2\,(U^{-1})^N{}_M\,\partial_N \lambda\,, 
\end{align}
which are constrained to satisfy the consistency constraints such as
\begin{align}
 f^M{}_{NP}\,\partial_M = 0 \,,\qquad f^M\,\partial_M =0\,,
\label{consistency-GDFT}
\end{align}
which are closely related to the SC. 
For the ansatz \eqref{eq:dilaton-dual} and a constant twist matrix $U$\,, we obtain a constant flux $f_M$ which satisfies the consistency conditions. 
According to \cite{Baguet:2016prz}, this corresponds to a nine-dimensional deformed supergravity generated by the gauging of the trombone symmetry and a Cartan subgroup of $\SL(2)$ in type IIB \cite{Bergshoeff:2002nv,FernandezMelgarejo:2011wx}. 
In this sense, the introduction of the linear winding-coordinate dependence can be regarded as a systematic way to introduce constant gaugings satisfying the consistency constraints. 

\subsection{$5^3_2$ solution in a deformed type II supergravity}

Let us next consider the $5^3_2(12345,678)$ solution \eqref{eq:532-soln}. 
In this case, the linear winding-coordinate dependence is contained in the $\beta$-field $\beta^{67}=m\,\tilde{x}_8$\,. 
For generality, we introduce an arbitrary constant $c$ and decompose the $\beta$-field as
\begin{align}
 \beta = m\,\tilde{x}_8\,\partial_{6}\wedge \partial_7 = c\,\partial_{6}\wedge\partial_7 + \frac{1}{2}\,U^{mn}\,\partial_m\wedge \partial_n \,,\qquad U^{mn}=2\,\bigl(m\,\tilde{x}_8-c\bigr)\,\delta_{6}^{[m}\,\delta_7^{n]} \,. 
\end{align}
We then regard the $U^{mn}$ as a part of a $\beta$-twist matrix,
\begin{align}
 \cH_{MN} = \bigl(U^\rmT\,\hat{\cH}\,U\bigr)_{MN}\,,\qquad 
 U^M{}_N \equiv \begin{pmatrix} \delta^m_n & -U^{mn} \\ 0 & \delta_m^n \end{pmatrix} ,
\label{eq:532-twist}
\end{align}
in the sense of the Scherk--Schwarz ansatz \eqref{eq:SS-ansatz} (see \cite{Catal-Ozer:2017ycb} for a recent study on this twist). 

As we infer from their definition, the dual fields associated with the untwisted (or physical) generalized metric $\hat{\cH}_{MN}$ are
\begin{align}
 \rmd s^2 = \rmd x^2_{01\cdots 5} + \tau_2^{-1} \, \rmd x^2_{678} + \tau_2 \,\rmd x_{9}^2\,,\qquad 
 \Exp{-2\tilde{\phi}}= \tau_2^2 \,, \qquad
 \beta^{67} = c \,,
\end{align}
or equivalently, in terms of the conventional fields [see \eqref{eq:532-soln-conv}],
\begin{align}
\begin{split}
 &\rmd s^2 = \rmd x^2_{01\cdots 5} + \frac{\tau_2}{c^2+\tau_2^2} \, \rmd x^2_{67} + \tau_2^{-1} \,\rmd x_{8}^2 + \tau_2 \,\rmd x_{9}^2\,,
\\
 &\Exp{-2\Phi}= c^2+ \tau_2^2 \,, \qquad
 B_2 = -\frac{c}{c^2+\tau_2^2}\,\rmd x^6\wedge\rmd x^7 \,. 
\end{split}
\end{align}
Since the winding-coordinate dependence is absorbed into the twist matrix, the solution no longer depends on the winding coordinates. 
In particular, when we choose $c=0$\,, the solution is simplified as
\begin{align}
 \rmd s^2 = \rmd x^2_{01\cdots 5} + \tau_2^{-1} \,\rmd x_{678}^2 + \tau_2 \,\rmd x_{9}^2\,, \qquad
 \Exp{-2\Phi}= \tau_2^2 \,, 
\label{eq:532-soln-deformed}
\end{align}
where the asymmetry between $\{6,7\}$ and $8$ disappears. 

According to the gauged DFT \cite{Aldazabal:2011nj,Geissbuhler:2011mx,Grana:2012rr,Berman:2013cli,Cho:2015lha}, the twist matrix changes the NS--NS part of the DFT Lagrangian as
\begin{align}
 \cL_{\text{DFT}}&\equiv\Exp{-2d}\,\cR \quad \to \quad \cL_{\text{GDFT}}\equiv\Exp{-2d}\,\bigl(\hat{\cR} + \cR_f\bigr) \,,
\\
 \cR_f&\equiv -\frac{1}{2}\,f^{MNP}\,\hat{\cH}_{NQ}\,\hat{\cH}_{PR}\,\partial^Q \hat{\cH}_M{}^{R}
             -\frac{1}{12}\,f^{MNP}\,f^{QRS}\,\hat{\cH}_{MQ}\,\hat{\cH}_{NR}\,\hat{\cH}_{PS} 
\nn\\
 &\quad     +\frac{1}{4}\,f^{MNP}\,f^{QRS}\,\eta_{MQ}\,\eta_{NR}\,\hat{\cH}_{PS} \,,
\end{align}
where $\hat{\cR}$ is the generalized Ricci scalar for $\hat{\cH}_{MN}$ and $f^{MNP}$ is a gauging defined as
\begin{align}
 f_{MNP} \equiv 3\,\eta_{S[M} \, (U^{-1})^Q{}_N\, (U^{-1})^R{}_{P]} \,\partial_Q U^S{}_R \,. 
\end{align}
In our present example \eqref{eq:532-twist}, the non-vanishing component of the gauging is the $R$-flux,
\begin{align}
 f^{678} = R^{678} = m\,. 
\end{align}
Then, the modification of the Lagrangian $\cR_f$ becomes \cite{Catal-Ozer:2017ycb}
\begin{align}
 \cR_f = -3\,m\, \hat{\cH}_{[6}{}^n\,\hat{\cH}_{7|}{}^M\,\partial_n \hat{\cH}_{|8]M} 
       -\frac{m^2}{12}\,\det\hat{\cH}_{ij}\qquad (i,j=6,7,8) \,. 
\end{align}
An important point is that, the consistency of the GDFT \eqref{consistency-GDFT} requires that the untwisted field and the dilaton should satisfy the condition
\begin{align}
 f^{MNP}\,\partial_P = 0 \quad \Rightarrow \quad R^{mnp}\,\partial_p = 0 \,. 
\end{align}
This shows that, similar to the GSE, the $R$-flux-deformed supergravity requires the backgrounds to admit three Killing vectors. 
In other words, we are essentially describing a seven-dimensional supergravity. 

In this manner, we have transformed the DFT solution \eqref{eq:532-soln} into a winding-coordinate-independent solution \eqref{eq:532-soln-deformed} of the deformed supergravity. 
In principle, we can repeat this procedure to all of the ``elementary'' domain-wall solutions in DFT/EFT and obtain various (effectively lower-dimensional) supergravities that are deformed by the locally non-geometric fluxes. 
In the following, we will just consider several examples, without calculating the action nor the equations of motion of the deformed supergravities explicitly.

\subsection{D8 solutions in the Romans massive type IIA supergravity}

Before considering further new examples, let us briefly go back to the well-studied D8-brane solution \eqref{eq:D8-soln}. 
In this case, the R--R 1-form potential $A_1$ includes a linear winding-coordinate dependence,
\begin{align}
 A_1(x) = \hat{A}_1(x^i) + m\,\tilde{x}_8\,\rmd x^8 \,,
\label{eq:massive-IIA-ansatz}
\end{align}
where $\hat{A}_1(x^i)$ does not include the $x^8$ dependence that vanishes in the D8-brane solution \eqref{eq:D8-soln}. 
As it was studied in \cite{Hohm:2011cp}, in such case, the R--R 0-form field strength becomes constant and the equations of motion of DFT reproduce those of the Romans massive IIA supergravity \cite{Romans:1985tz}. 
The modifications of the gauge symmetry and $T$-duality transformation rules can be reproduced from those of DFT by considering the ansatz \eqref{eq:massive-IIA-ansatz}. 
Once the winding-coordinate dependence is absorbed into the mass parameter, or the deformation of the supergravity, the D8-brane solution \eqref{eq:D8-soln} without R--R fields
\begin{align}
 \rmd s^2 = \tau_2^{-1/2} \,\bigl(\rmd x^2_{01\cdots 8} + \tau_2 \,\rmd x_{9}^2\bigr)\,,\qquad 
 \Exp{-2\Phi}= \tau_2^{5/2} \,,
\end{align}
becomes a solution of the Romans massive type IIA supergravity. 
Namely, the winding-coordinate-dependent solution becomes a winding-coordinate-independent solution of the modified supergravity. 

As the R--R field depends on a winding coordinate, for the SC to be satisfied, one might expect that it is necessary to require the existence of an isometry direction. 
However, in this case, as it was shown in \cite{Hohm:2011cp}, by relaxing the strong constraint to the weak constraint, we can formulate the massive IIA supergravity in ten dimensions, rather than 9. 
This may be understood as follows. 
We expect that the locally non-geometric $R$-fluxes will always play the role of the gaugings and the consistency conditions will require conditions like $R^{\cdots m}\,\partial_m =0$\,. 
However, in the special case of the D8-brane, the corresponding $R$-flux (i.e.~the R--R 0-form field strength $F_0$) does not have any index and we cannot write a condition for the derivative. 
In terms of M-theory, a D8-brane is uplifted to the $8^{(1,0)}$-brane (also known as the KK9M- or the M9-brane) and the associated $R$-flux is $R^{i,i}$\,. 
Therefore, in this case, we may need to require a condition $R^{i,i}\,\partial_i=0$\,. 
This consideration is consistent with the fact that the eleven-dimensional uplift of the massive IIA supergravity depend on a certain Killing vector in the eleven dimensions \cite{Bergshoeff:1997ak}. 

\subsection{KK8A and M9 solutions}

Let us consider the solution of the $7_3^{(1,0)}(1\cdots 7,,8)$-brane \eqref{eq:p1(7-p)3-soln}, which is also known as the KK8A-brane. 
At the same time, we consider its eleven-dimensional uplift, the solution of the $8_{12}^{(1,0)}(1234567z,,8)$-brane \eqref{eq:810M-soln}. 
In this case, the linear winding-coordinate dependence is included in $\gamma^{8} = m\,\tilde{x}_8$\,, or in terms of M-theory, $A^8=\tilde{G}^{8\rmM}/\tilde{G}^{\rmM\rmM} = - m\,y_{8\rmM}$\,. 
We thus consider the following twist for the generalized metric in EFT:
\begin{align}
 \cM_{IJ} = \bigl(U^\rmT\,\hat{\cM}\,U\bigr)_{IJ}\,,\qquad 
 U \equiv \Exp{-(m\,y_{8\rmM}-c)\,K_8{}^{\rmM}}\,,
\end{align}
where $K_i{}^j$ is a matrix representation of the $\text{GL}(n)$ generator given in \eqref{eq:Kij-M}. 
Substituting this ansatz into the EFT action or the equations of motion, we obtain a deformed type IIA supergravity, which will effectively be nine-dimensional due to $R_{(3)}^{m,\,m}\,\partial_m=0$\,. 

Since the winding-coordinate dependence has been absorbed into the twist matrix, we obtain the following solution of the deformed type IIA supergravity,
\begin{align}
\begin{split}
 &\rmd \tilde{s}^2 = \tau_2^{1/2}\,\bigl(\rmd x^2_{01\cdots p}+\tau_2\,\rmd x_{9}^2\bigr) + \tau_2^{-3/2}\,\rmd x_8^2\,, 
\\
 &\Exp{-2\tilde{\phi}} = \tau_2^{-1/2} \,,\qquad \gamma^{8} = c \,. 
\end{split}
\end{align}
By translating the dual fields into the conventional fields (recall \eqref{eq:p1(7-p)3-soln-conv}), we obtain
\begin{align}
\begin{split}
 &\rmd s^2 = \Bigl(\frac{c^2+\tau_2^2}{\tau_2}\Bigr)^{1/2}\, \bigl(\rmd x^2_{01\cdots 7}+\tau_2\,\rmd x_{9}^2\bigr)
  + \tau_2^{-1}\,\Bigl(\frac{c^2+\tau_2^2}{\tau_2}\Bigr)^{-1/2}\, \rmd x_8^2\,, 
\\
 &\Exp{-2\Phi}= (c^2+\tau_2^2)^{-3/2}\, \tau_2^{\frac{5}{2}} \,,\quad 
  C_{\bar0\cdots\bar8} = - \frac{c^2+\tau_2^2}{\tau_2}\,,\qquad C_{\bar8} = -\frac{c}{c^2+\tau_2^2} \,. 
\end{split}
\end{align}
This is precisely the KK8A solution given in Eq.~(6.20) of \cite{Eyras:1999at}. 
By choosing $c=0$\,, we obtain the KK8A solution originally obtained in \cite{Meessen:1998qm}, which is not a solution of any type II supergravity, but instead of a deformed ten-dimensional supergravity (which is not Romans' supergravity).

On the other hand, the $8_{12}^{(1,0)}(1234567z,,8)$ solution becomes a solution of the deformed eleven-dimensional supergravity,
\begin{align}
 \rmd \tilde{s}^2 = \tau_2^{1/3}\, \bigl(\rmd x^2_{0\cdots7z} + \tau_2\,\rmd x_{9}^2\bigr) + \tau_2^{-5/3}\,\bigl(\rmd x^2_{8}+c\,\rmd x_z^2\bigr)\,.
\end{align}
By choosing $c=0$\,, the conventional metric becomes
\begin{align}
 \rmd s^2 = \tau_2^{1/3}\, \bigl(\rmd x^2_{0\cdots7z} + \tau_2\,\rmd x_{9}^2\bigr) + \tau_2^{-5/3}\, \rmd x^2_{8} \,,
\end{align}
which is the KK9M solution obtained in \cite{Bergshoeff:1998bs,Meessen:1998qm}. 

Note that if we consider the $T$-dual of the $7_3^{(1,0)}(1\cdots 7,,8)$ solution along the $x^8$-direction, we obtain $7_3(1\cdots 7)$ solution, which is a solution of the undeformed type IIB supergravity. 
On the other hand, if we consider the background of the $(p,q)$-$7(1\cdots 7)$ brane and perform the $T$-duality along the $x^8$-direction, we obtain a solution corresponding to the bound state of the $8_1(1\cdots 8)$-brane and the $7_3^{(1,0)}(1\cdots 7,,8)$-brane. 
This is a solution of a deformed type IIA supergravity. 
The eleven-dimensional uplift corresponds to the bound state of the $8_{12}^{(1,0)}(1\cdots 8,,\text{M})$-brane and the $8_{12}^{(1,0)}(1\cdots 7\text{M},,8)$-brane, and the corresponding background will be a solution of the $\SL(2)$-covariant eleven-dimensional massive supergravity \cite{Meessen:1998qm}. 

\subsection{$6^{(1,1)}_{3}$ solution in a deformed IIB supergravity}

Finally, let us consider the $6^{(1,1)}_{3}(1\cdots 6,7,8)$ solution of \eqref{eq:p1(7-p)3-soln}. 
In this case, the winding-coordinate dependence is contained in $\gamma^{78} = m\,\tilde{x}_8$\,. 
The corresponding twist is
\begin{align}
 \cM_{IJ} = \bigl(U^\rmT\,\hat{\cM}\,U\bigr)_{IJ}\,,\qquad 
 U \equiv \Exp{-(m\,\tilde{x}_{8}-c)\,R^2_{78}}\,,
\end{align}
where the matrix $R^\alpha_{mn}$ can be found in \eqref{eq:R-gamma-pq}. 
In this case, the deformed supergravity will effectively be eight dimensional since the $6^{(1,1)}_{3}$-brane requires two isometry directions or the $R$-flux contains two antisymmetric indices. 

By absorbing the winding-coordinates into the twist matrix and choosing $c=0$\,, the solution \eqref{eq:p1(7-p)3-soln-conv} reduces to a purely gravitational solution,
\begin{align}
 \rmd s^2 = \tau_2^{1/2} \, \bigl(\rmd x^2_{01\cdots 6}+\tau_2\,\rmd x_{9}^2\bigr) + \tau_2^{-1/2} \, \rmd x^2_{7} + \tau_2^{-3/2} \, \rmd x_8^2\,,\qquad 
 \Exp{-2\Phi}= 1 \,.
\end{align}
This is precisely a solution corresponding to the ``\emph{Unknown} brane (6,2,1)'' obtained in Eq.~(6.9) of \cite{Meessen:1998qm}.

\section*{Acknowledgment}

We appreciate useful discussions during the workshop ``Geometry, Duality and Strings 2018'' at Departamento de F\'isica, Universidad de Murcia. 
We also would like to thank Shozo Uehara and Satoshi Watamura for valuable comments. 
J.J.F.-M. gratefully acknowledges the support of JSPS (Postdoctoral Fellowship) and the Fundaci\'on S\'eneca/Universidad de Murcia (Programa Saavedra Fajardo) and JSPS KAKENHI (Grant-in-Aid for JSPS Fellows) Grant Number JP16F16741.
The work of TK is supported by the Iwanami-Fujukai Foundation. 
The work of YS is supported by the JSPS KAKENHI Grant Number JP 18K13540 and 18H01214, and by the Supporting Program for Interaction based Initiative Team Studies (SPIRITS) from Kyoto University. 
J.J.F.-M. would like to dedicate this work to the memory of Jos\'e M. Rodr\'iguez Arnaldos.

\appendix

\section{Conventions}
\label{app:conventions}

We define the totally antisymmetric delta functions as
\begin{align}
 \delta^{m_1\cdots m_p}_{n_1\cdots n_p} \equiv \delta^{[m_1}_{[n_1}\cdots \delta^{m_p]}_{n_p]} \,,
\end{align}
where the antisymmetrization is defined as
\begin{align}
 A_{[m_1\cdots m_n]} \equiv \frac{1}{n!}\,\bigl(A_{m_1\cdots m_n} \pm \text{permutations}\bigr) \,.
\end{align}
We also define the antisymmetrized metric as
\begin{align}
 G_{m_1\cdots m_p,\,n_1\cdots n_p} \equiv G_{m_1r_1}\cdots G_{m_pr_p}\,\delta^{r_1\cdots r_p}_{n_1\cdots n_p}\,. 
\end{align}

The Hodge dual operator is defined as
\begin{align}
\begin{split}
 &(* \alpha_q)_{m_1\cdots m_{p+1-q}} =\frac{1}{q!}\, \varepsilon^{n_1\cdots n_q}{}_{m_1\cdots m_{p+1-q}}\,\alpha_{n_1\cdots n_q} \,,\qquad 
 \rmd^d x = \rmd x^1\wedge\cdots\wedge \rmd x^d \,,
\\
 &* (\rmd \sigma^{m_1}\wedge \cdots \wedge \rmd \sigma^{m_q}) =\frac{1}{(p+1-q)!}\,\varepsilon^{m_1\cdots m_q}{}_{n_1\cdots n_{p+1-q}}\,\rmd x^{n_1}\wedge \cdots \wedge \rmd x^{n_{p+1-q}} \,,
\\
 &\varepsilon^{1\cdots d}=-\frac{1}{\sqrt{-G}}\,,\qquad 
 \varepsilon_{1\cdots d}= \sqrt{-G} \,.
\end{split}
\end{align}

\section{Parameterizations of the generalized metric in EFT}
\label{app:EFT-parameterization}

In this appendix, we review the parameterization of the generalized metric in $E_{n(n)}$ EFT ($n\leq 7$). 
We follow the convention used in \cite{Sakatani:2017nfr}. 

\subsection{M-theory parameterization}

When we consider M-theory, we can parameterize the generalized metric $\cM_{IJ}$ in terms of the conventional supergravity fields $G_{ij}$, $A_{i_1i_2i_3}$, and $A_{i_1\cdots i_6}$ as follows \cite{Berman:2011jh} (see \cite{Duff:1990hn,Berman:2010is,Berman:2011pe} for earlier works):
\begin{align}
\begin{split}
 &\cM_{IJ} = (L_6^\rmT\,L_3^\rmT\,\hat{\cM}\,L_3\,L_6)_{IJ} \,,\quad 
 L_3 =\Exp{\frac{1}{3!}\,R^{i_1i_2i_3}\,A_{i_1i_2i_3}}\,,\quad L_6 = \Exp{\frac{1}{6!}\,R^{i_1\cdots i_6}\,A_{i_1\cdots i_6}}\,, 
\label{eq:M-conv}
\\
 &\hat{\cM} \equiv \abs{G}^{\frac{1}{d-2}}\,
 \begin{pmatrix}
 G_{ij} & 0 & 0 & 0 \\
 0 & G^{i_1i_2,\,j_1j_2} & 0 & 0 \\
 0 & 0 & G^{i_1\cdots i_5,\,j_1\cdots j_5} & 0 \\
 0 & 0 & 0 & G^{i_1\cdots i_7,\,j_1\cdots j_7}\,G^{ij}
 \end{pmatrix} , \quad 
 \abs{G}\equiv \det(G_{ij})\,,
\\
 &\text{with}\qquad \Exp{A}\equiv I+\sum_{n=1}^\infty\frac{1}{n!}\,A^n\,,\quad 
 I\equiv{\tiny
 \begin{pmatrix}
 \delta^i_j & 0 & 0 & 0 \\
 0 & \delta_{i_1i_2}^{j_1j_2} & 0 & 0 \\
 0 & 0 & \delta_{i_1\cdots i_5}^{j_1\cdots j_5} & 0\\
 0 & 0 & 0 & \delta_{i_1\cdots i_7}^{j_1\cdots j_7}\,\delta^j_i
 \end{pmatrix}} ,
\end{split}
\end{align}
where $d\equiv 11-n$ is the dimension of the external space and we have introduced the matrix representations of the $E_{n(n)}$ generators $\{K_{k_1}{}^{k_2},\,R_{k_1k_2k_3},\,R^{k_1k_2k_3},\,R_{k_1\cdots k_6},\,R^{k_1\cdots k_6}\}$ as
\begin{align}
 &(K_{k_1}{}^{k_2})^I{}_J \equiv {\arraycolsep=0.2mm {\scriptsize
 \begin{pmatrix}
 \delta_{k_1}^i \delta_j^{k_2} & 0 & 0 & 0 \\
 0 & -\frac{\bdelta_{i_1i_2}^{k_2l} \bdelta_{k_1l}^{j_1j_2}}{\sqrt{2!\,2!}} & 0 & 0 \\
 0 & 0 & -\frac{\bdelta_{i_1\cdots i_5}^{k_2l_1\cdots l_4} \bdelta_{k_1l_1\cdots l_4}^{j_1\cdots j_5}}{4!\sqrt{5!\,5!}} & 0 \\
 0 & 0 & 0 & -\frac{\frac{1}{6!}\bdelta_{i_1\cdots i_7}^{k_2l_1\cdots l_6} \bdelta_{k_1l_1\cdots l_6}^{j_1\cdots j_7}\delta_i^j +\bdelta_{i_1\cdots i_7}^{j_1\cdots j_7} \delta_{i}^{k_2}\delta_{k_1}^j}{\sqrt{7!\,7!}}
 \end{pmatrix} + \frac{\delta_{k_1}^{k_2}}{9-n}}}\,\delta^I_J 
\label{eq:Kij-M}\,, 
\\
 &(R_{k_1k_2k_3})^I{}_J \equiv {\scriptsize
 \begin{pmatrix}
 0 & -\frac{\bdelta^{i j_1j_2}_{k_1k_2k_3}}{\sqrt{2!}} & 0 & 0 \\
 0 & 0 & \frac{\bdelta_{i_1i_2 k_1k_2k_3}^{j_1\cdots j_5}}{\sqrt{2!\,5!}} & 0 \\
 0 & 0 & 0 & \frac{\bdelta_{i_1\cdots i_5 l_1l_2}^{j_1\cdots j_7}\,\bdelta^{l_1l_2j}_{k_1k_2k_3}}{2!\sqrt{5!\,7!}} \\
 0 & 0 & 0 & 0
 \end{pmatrix} } , 
\\
 &(R^{k_1k_2k_3})^I{}_J \equiv {\scriptsize
 \begin{pmatrix}
 0 & 0 & 0 & 0 \\
 -\frac{\bdelta_{i_1i_2 j}^{k_1k_2k_3}}{\sqrt{2!}} & 0 & 0 & 0 \\
 0 & \frac{\bdelta^{j_1j_2 k_1k_2k_3}_{i_1\cdots i_5}}{\sqrt{2!\,5!}} & 0 & 0 \\
 0 & 0 & \frac{\bdelta^{j_1\cdots j_5 l_1l_2}_{i_1\cdots i_7}\,\bdelta_{l_1l_2i}^{k_1k_2k_3}}{2!\sqrt{5!\,7!}} & 0
 \end{pmatrix} } , 
\\
 &(R_{k_1\cdots k_6})^I{}_J \equiv {\scriptsize
 \begin{pmatrix}
 0 & 0 & \frac{\bdelta^{j_1\cdots j_5 i}_{k_1\cdots k_6}}{\sqrt{5!}} & 0 \\
 0 & 0 & 0 & \frac{\bdelta_{i_1i_2 l_1\cdots l_5}^{j_1\cdots j_7}\,\bdelta^{l_1\cdots l_5j}_{k_1\cdots k_6}}{5!\sqrt{2!\,7!}} \\
 0 & 0 & 0 & 0 \\
 0 & 0 & 0 & 0
 \end{pmatrix} } ,
\\
 &(R^{k_1\cdots k_6})^I{}_J \equiv {\scriptsize
 \begin{pmatrix}
 0 & 0 & 0 & 0 \\
 0 & 0 & 0 & 0 \\
 \frac{\bdelta_{i_1\cdots i_5 j}^{k_1\cdots k_6}}{\sqrt{5!}} & 0 & 0 & 0\\
 0 & \frac{\bdelta^{j_1j_2 l_1\cdots l_5}_{i_1\cdots i_7}\,\bdelta_{l_1\cdots l_5i}^{k_1\cdots k_6}}{5!\sqrt{2!\,7!}} & 0 & 0
 \end{pmatrix} } ,
\end{align}
where $\bdelta_{i_1\cdots i_p}^{j_1\cdots j_p}\equiv p!\,\delta_{i_1\cdots i_p}^{j_1\cdots j_p}$\,. 
If we consider the $E_{8(8)}$ EFT, the generalized metric additionally contains the dual graviton $A_{i_1\cdots i_8,\,j}$ \cite{Godazgar:2013rja} although the explicit parameterization in our convention is not determined yet.

We can also parameterize the same $\cM_{IJ}$ in terms of the dual fields, $\tilde{G}_{ij}$, $\Omega^{i_1i_2i_3}$, and $\Omega^{i_1\cdots i_6}$ as follows \cite{Lee:2016qwn} (see \cite{Malek:2012pw,Malek:2013sp} for earlier results in $\SL(5)$ EFT):
\begin{align}
\begin{split}
 \cM_{IJ} &= (\tilde{L}_6^\rmT\,\tilde{L}_3^\rmT\,\tilde{\cM}\,\tilde{L}_3\,\tilde{L}_6)_{IJ} \,,\quad 
 \tilde{L}_3 \equiv\Exp{\frac{1}{3!}\,R_{i_1i_2i_3}\,\Omega^{l_1l_2l_3}}\,,\quad \tilde{L}_6 \equiv \Exp{\frac{1}{6!}\,R_{i_1\cdots i_6}\,\Omega^{i_1\cdots i_6}}\,,
\label{eq:M-dual}
\\
 \tilde{\cM} &\equiv \abs{\tilde{G}}^{\frac{1}{d-2}}\,
 \begin{pmatrix}
 \tilde{G}_{ij} & 0 & 0 & 0 \\
 0 & \tilde{G}^{i_1i_2,\,j_1j_2} & 0 & 0 \\
 0 & 0 & \tilde{G}^{i_1\cdots i_5,\,j_1\cdots j_5} & 0 \\
 0 & 0 & 0 & \tilde{G}^{i_1\cdots i_7,\,j_1\cdots j_7}\,\tilde{G}^{ij}
 \end{pmatrix}\,, \quad 
 \abs{\tilde{G}}\equiv \det(\tilde{G}_{ij}) \,.
\end{split}
\end{align}
This is called the non-geometric parameterization. 

Similar to the identification in DFT \eqref{eq:conv-dual-DFT}, by comparing the two parameterizations \eqref{eq:M-conv} and \eqref{eq:M-dual} as \cite{Malek:2012pw,Malek:2013sp,Lee:2016qwn}
\begin{align}
 \cM_{IJ}^{\text{\tiny(M)}} = \cM_{IJ}^{\text{\tiny(M, non-geometric)}}\,,
\end{align}
we can in principle express the dual fields $(\tilde{G}_{ij},\,\Omega^{i_1i_2i_3},\,\Omega^{i_1\cdots i_6})$ in terms of the conventional fields $(G_{ij},\, A_{i_1i_2i_3},\, A_{i_1\cdots i_6})$. 

\subsection{Type IIB parameterization}
\label{app:IIB-parameterization}

When we consider type IIB theory, we parameterize the generalized metric as \cite{Tumanov:2014pfa,Lee:2016qwn} (see also \cite{Blair:2014zba} for the case of $\SL(5)$ EFT)
\begin{align}
\begin{split}
 &\cM_{IJ} = (\sfL_6^\rmT\,\sfL_4^\rmT\,\sfL_2^\rmT\,\hat{\sfM}\,\sfL_2\,\sfL_4\,\sfL_6)_{IJ} \,, \qquad 
 \abs{g^{\rmE}} \equiv \det(g^{\rmE}_{mn})\,,
\label{eq:IIB-conv}
\\
 &\hat{\sfM}\equiv \abs{g^{\rmE}}^{\frac{1}{d-2}} {\arraycolsep=0.5mm {\scriptsize\left(\begin{array}{ccccc} 
 g^{\rmE}_{mn} & 0 & 0 & 0 & 0 \\
 0 & m_{\alpha\beta} \,g_{\rmE}^{mn} & 0 & 0 & 0 \\
 0 & 0 & g_{\rmE}^{m_1m_2m_3,\,n_1n_2n_3} & 0 & 0 \\
 0 & 0 & 0 & m_{\alpha\beta} \,g_{\rmE}^{m_1\cdots m_5,\,n_1\cdots n_5} & 0 \\
 0 & 0 & 0 & 0 & g_{\rmE}^{m_1\cdots m_6,\,n_1\cdots n_6}\,g_{\rmE}^{mn}
 \end{array}\right)}} \,,
\\
 &\sfL_2\equiv \Exp{\frac{1}{2!}\,R^{p_1p_2}_\gamma\,B_{p_1p_2}^\gamma} \,, \qquad
  \sfL_4\equiv \Exp{\frac{1}{4!}\,R^{p_1\cdots p_4}\,D_{p_1\cdots p_4}} \,, \qquad
  \sfL_6\equiv \Exp{\frac{1}{6!}\,R_\gamma^{p_1\cdots p_6} \,D_{p_1\cdots p_6}^\gamma} \,,
\\
 &\text{with}\qquad \Exp{A}\equiv \sfI+\sum_{n=1}^\infty\frac{1}{n!}\,A^n\,,\quad 
 \sfI\equiv {\arraycolsep=0.0mm {\tiny\left(\begin{array}{ccccc} 
 \delta^m_n & 0 & 0 & 0 & 0 \\
 0 & \delta^\alpha_\beta\,\delta_m^n & 0 & 0 & 0 \\
 0 & 0 & \delta_{m_1m_2m_3}^{n_1n_2n_3} & 0 & 0 \\
 0 & 0 & 0 & \delta^\alpha_\beta \,\delta_{m_1\cdots m_5}^{n_1\cdots n_5} & 0 \\
 0 & 0 & 0 & 0 & \delta_{m_1\cdots m_6}^{n_1\cdots n_6}\,\delta_m^n 
 \end{array}\right)}}\,,
\end{split}
\end{align}
where we introduced the type IIB parameterizations of the $E_{n(n)}$ generators as
\begin{align}
 &(K_{p_1}{}^{p_2})^I{}_J \equiv {\arraycolsep=0.2mm{\scriptsize
 \begin{pmatrix}
 \delta_{p_1}^{m} \delta_{n}^{p_2} & 0 & 0 & 0 \\
 0 & - \bdelta_{m}^{p_2q} \bdelta_{p_1q}^{n}\delta^\alpha_\beta & 0 & 0 \\
 0 & 0 & -\frac{\bdelta_{m_1m_2m_3}^{p_2q_1q_2} \bdelta_{p_1q_1q_2}^{n_1n_2n_3}\delta^\alpha_\beta}{2!\sqrt{3!\,3!}} & 0 \\
 0 & 0 & 0 & -\frac{\frac{1}{5!}\bdelta_{m_1\cdots m_6}^{p_2q_1\cdots q_5} \bdelta_{p_1q_1\cdots q_5}^{n_1\cdots n_6}\delta_m^n +\bdelta_{m_1\cdots m_6}^{n_1\cdots n_6} \delta_{m}^{p_2}\delta_{p_1}^n}{\sqrt{6!\,6!}}
 \end{pmatrix} + \frac{\delta_{p_1}^{p_2}}{9-n}}}\,\delta^I_J \,, 
\\
 &(R_{\gamma\delta})^I{}_J \equiv {\scriptsize
 \begin{pmatrix}
 0 & 0 & 0 & 0 \\
 0 & - \delta^\beta_{(\gamma}\,\epsilon_{\delta)\alpha}\,\delta_{m}^{n} & 0 & 0 \\
 0 & 0 & - \delta^\beta_{(\gamma}\,\epsilon_{\delta)\alpha}\, \delta_{m_1\cdots m_5}^{n_1\cdots n_5} & 0 \\
 0 & 0 & 0 & 0
 \end{pmatrix}} , 
\\
 &(R_\gamma^{p_1p_2})^I{}_J\equiv {\arraycolsep=0.5mm {\scriptsize\left(\begin{array}{ccccc} 
 0 & 0 & 0 & 0 & 0 \\
 \delta^\alpha_\gamma\,\bdelta_{mn}^{p_1p_2} & 0 & 0 & 0 & 0 \\
 0 & \frac{\epsilon_{\beta\gamma} \,\bdelta_{m_1m_2m_3}^{n p_1p_2}}{\sqrt{3!}} & 0 & 0 & 0 \\
 0 & 0 & \frac{\delta^\alpha_\gamma\,\bdelta_{m_1\cdots m_5}^{n_1n_2n_3p_1p_2}}{\sqrt{3!\,5!}} & 0 & 0 \\
 0 & 0 & 0 & -\frac{2\,\epsilon_{\beta\gamma} \,\bdelta_{m_1\cdots m_6}^{n_1\cdots n_5[p_1}\,\delta_{m}^{p_2]}}{\sqrt{5!\,6!}} & ~0~
 \end{array}\right)}} ,
\\
 &(R^\gamma_{p_1p_2})^I{}_J\equiv {\arraycolsep=0.5mm {\scriptsize\left(\begin{array}{ccccc} 
 ~0~ & \delta_\beta^\gamma\,\bdelta^{nm}_{p_1p_2} & 0 & 0 & 0 \\
 0 & 0 & \frac{\epsilon^{\alpha\gamma} \,\bdelta^{n_1n_2n_3}_{m p_1p_2}}{\sqrt{3!}} & 0 & 0 \\
 0 & 0 & 0 & \frac{\delta_\beta^\gamma\,\bdelta^{m_1\cdots m_5}_{n_1n_2n_3p_1p_2}}{\sqrt{3!\,5!}} & 0 \\
 0 & 0 & 0 & 0 & -\frac{2\,\epsilon^{\alpha\gamma} \,\bdelta^{n_1\cdots n_6}_{m_1\cdots m_5[p_1}\,\delta^{n}_{p_2]}}{\sqrt{5!\,6!}} \\
 0 & 0 & 0 & 0 & 0 
 \end{array}\right)}} ,
\label{eq:R-gamma-pq}
\\
 &(R^{p_1\cdots p_4})^I{}_J \equiv {\arraycolsep=0.5mm {\scriptsize\left(\begin{array}{ccccc} 
 0 & 0 & 0 & 0 & 0 \\
 0 & 0 & 0 & 0 & 0 \\
 \frac{\bdelta_{m_1m_2m_3 n}^{p_1\cdots p_4}}{\sqrt{3!}} & 0 & 0 & 0 & 0 \\
 0 & -\frac{\delta^\alpha_\beta \,\bdelta_{m_1\cdots m_5}^{np_1\cdots p_4}}{\sqrt{5!}} & 0 & 0 & 0 \\
 0 & 0 & -\frac{4\,\bdelta_{m_1\cdots m_6}^{n_1n_2n_3[p_1p_2p_3}\,\delta_{m}^{p_4]}}{\sqrt{3!\,6!}} & ~0~ & ~0~
 \end{array}\right)}} , 
\\
 &(R_{p_1\cdots p_4})^I{}_J \equiv {\arraycolsep=0.5mm {\scriptsize\left(\begin{array}{ccccc} 
 ~0~ & ~0~ & \frac{\bdelta^{n_1n_2n_3 m}_{p_1\cdots p_4}}{\sqrt{3!}} & 0 & 0 \\
 0 & 0 & 0 & -\frac{\delta_\beta^\alpha \,\bdelta^{n_1\cdots n_5}_{mp_1\cdots p_4}}{\sqrt{5!}} & 0 \\
 0 & 0 & 0 & 0 & -\frac{4\,\bdelta^{n_1\cdots n_6}_{m_1m_2m_3[p_1p_2p_3}\,\delta^{n}_{p_4]}}{\sqrt{3!\,6!}} \\
 0 & 0 & 0 & 0 & 0 \\
 0 & 0 & 0 & 0 & 0 
 \end{array}\right)}} , 
\\
 &(R_\gamma^{p_1\cdots p_6})^I{}_J \equiv {\arraycolsep=0.5mm {\scriptsize\left(\begin{array}{ccccc} 
 0 & 0 & ~0~ & ~0~ & ~0~ \\
 0 & 0 & 0 & 0 & 0 \\
 0 & 0 & 0 & 0 & 0 \\
 \frac{\delta^\alpha_\gamma\,\bdelta_{m_1\cdots m_5 n}^{p_1\cdots p_6}}{\sqrt{5!}} & 0 & 0 & 0 & 0 \\
 0 & \frac{6\,\epsilon_{\beta\gamma}\,\bdelta_{m_1\cdots m_6}^{n[p_1\cdots p_5}\, \delta_{m}^{p_6]}}{\sqrt{6!}} & 0 & 0 & 0
 \end{array}\right)}} ,
\\
 &(R^\gamma_{p_1\cdots p_6})^I{}_J \equiv {\arraycolsep=0.5mm {\scriptsize\left(\begin{array}{ccccc} 
 ~0~ & ~0~ & ~0~ & \frac{\delta_\beta^\gamma\,\bdelta^{n_1\cdots n_5 m}_{p_1\cdots p_6}}{\sqrt{5!}} & 0 \\
 0 & 0 & 0 & 0 & \frac{6\,\epsilon^{\alpha\gamma}\,\bdelta^{n_1\cdots n_6}_{m[p_1\cdots p_5}\, \delta^{n}_{p_6]}}{\sqrt{6!}} \\
 0 & 0 & 0 & 0 & 0 \\
 0 & 0 & 0 & 0 & 0 \\
 0 & 0 & 0 & 0 & 0
 \end{array}\right)}} .
\end{align}
where $\bdelta_{m_1\cdots m_p}^{n_1\cdots n_p}\equiv p!\,\delta_{m_1\cdots m_p}^{n_1\cdots n_p}$\,. 

The introduced $\SL(2)$-covariant fields $(g^{\rmE}_{mn},\,m_{\alpha\beta},\,B_{mn}^\gamma,\,D_{m_1\cdots m_4},\,D_{m_1\cdots m_6}^\gamma)$ are related to the standard type IIB supergravity fields as follows. 
The metric $g^{\rmE}_{mn}$ is the standard Einstein-frame metric and $\Phi$ is the standard dilaton and the string-frame metric is defined as $g_{mn} \equiv \Exp{\frac{1}{2}\,\Phi} g^{\rmE}_{mn}$\,. 
Other fields are further parameterized as
\begin{align}
\begin{split}
 &\bigl(m_{\alpha\beta}\bigr) \equiv \Exp{\Phi}\,\begin{pmatrix}
 \Exp{-2\Phi} + (C_0)^2 & C_0 \\
 C_0 & 1
 \end{pmatrix}\,,\quad B^\alpha_{mn} \equiv \begin{pmatrix} B_{mn}\\ -C_{mn} \end{pmatrix} \,,
\\
 &D_{m_1\cdots m_4} =C_{m_1\cdots m_4} - 3\,B_{[m_1m_2}\, C_{m_3m_4]}\,,\qquad
 D^\alpha_{m_1\cdots m_6} \equiv \begin{pmatrix} D^1_{m_1\cdots m_6} \\ D^2_{m_1\cdots m_6} \end{pmatrix} \,,
\\
 &D^1_{m_1\cdots m_6}\equiv C_{m_1\cdots m_6}-15\,D_{[m_1\cdots m_4}\,B_{m_5m_6]} - 15\, B_{[m_1m_2}\,B_{m_3m_4}\, C_{m_5m_6]} \,,
\\
 &D^2_{m_1\cdots m_6}\equiv -D_{m_1\cdots m_6}+ 15\,D_{[m_1\cdots m_4}\,C_{m_5m_6]} +30\,C_{[m_1m_2}\,C_{m_3m_4}\,B_{m_5m_6]} \,.
\end{split}
\end{align}
Then, $\Phi$, $B_{mn}$, $C_{m_1\cdots m_{2n}}$, and $D_{m_1\cdots m_6}$ are the standard dilaton, the $B$-field, the R--R potentials, and the dual potential of the $B$-field.

We can also provide the non-geometric parameterization as \cite{Lee:2016qwn}
\begin{align}
\begin{split}
 &\cM_{IJ} = (\tilde{\sfL}_6^\rmT\,\tilde{\sfL}_4^\rmT\,\tilde{\sfL}_2^\rmT\, \tilde{\sfM}\, \tilde{\sfL}_2\,\tilde{\sfL}_4\,\tilde{\sfL}_6)_{IJ} \,, \qquad 
 \abs{\tilde{g}^{\rmE}} \equiv \det(\tilde{g}^{\rmE}_{mn})\,,
\label{eq:IIB-dual}
\\
 &\tilde{\sfM} = \abs{\tilde{g}^{\rmE}}^{\frac{1}{d-2}}{\arraycolsep=0.5mm {\scriptsize\left(\begin{array}{ccccc} 
 \tilde{g}^{\rmE}_{mn} & 0 & 0 & 0 & 0 \\
 0 & \tilde{m}_{\alpha\beta} \,\tilde{g}_{\rmE}^{mn} & 0 & 0 & 0 \\
 0 & 0 & \tilde{g}_{\rmE}^{m_1m_2m_3,\,n_1n_2n_3} & 0 & 0 \\
 0 & 0 & 0 & \tilde{m}_{\alpha\beta} \,\tilde{g}_{\rmE}^{m_1\cdots m_5,\,n_1\cdots n_5} & 0 \\
 0 & 0 & 0 & 0 & \tilde{g}_{\rmE}^{m_1\cdots m_6,\,n_1\cdots n_6}\,\tilde{g}_{\rmE}^{mn} \end{array}\right)}} ,
\\
 &\tilde{\sfL}_2 = \Exp{\frac{1}{2!}\,R^\gamma_{p_1p_2}\,\beta_\gamma^{p_1p_2}} \,, \qquad
  \tilde{\sfL}_4 = \Exp{\frac{1}{4!}\,R_{p_1\cdots p_4}\,\eta^{p_1\cdots p_4}} \,, \qquad
  \tilde{\sfL}_6 = \Exp{\frac{1}{6!}\,R^\gamma_{p_1\cdots p_6}\,\eta^{p_1\cdots p_6}_\gamma} \,.
\end{split}
\end{align}
By using the dual metric $g^{\rmE}_{mn}$ and the dual dilaton $\Phi$\,, we define the dual string-frame metric as $\tilde{g}_{mn} \equiv \Exp{\frac{1}{2}\,\tilde{\phi}} \tilde{g}^{\rmE}_{mn}$\,. 
Again, we further parameterize other fields as
\begin{align}
\begin{split}
 &\bigl(\tilde{m}_{\alpha\beta}\bigr)= \Exp{\tilde{\phi}}\,\begin{pmatrix}
 1 & \gamma \\
 \gamma & \Exp{-2\tilde{\phi}} + \gamma^2
 \end{pmatrix}\,, \quad \beta_\alpha^{mn} \equiv \begin{pmatrix} \beta^{mn}\\ -\gamma^{mn} \end{pmatrix} \,,
\\
 &\eta^{m_1\cdots m_4} \equiv \gamma^{m_1\cdots m_4} + 3\,\beta^{[m_1m_2}\, \gamma^{m_3m_4]} \,,\qquad
 \eta_\alpha^{m_1\cdots m_6} \equiv \begin{pmatrix} \eta_1^{m_1\cdots m_6} \\ \eta_2^{m_1\cdots m_6} \end{pmatrix} \,,
\\
 &\eta_1^{m_1\cdots m_6} \equiv \gamma^{m_1\cdots m_6}+15\,\eta^{[m_1\cdots m_4}\,\beta^{m_5m_6]} - 15\, \beta^{[m_1m_2}\,\beta^{m_3m_4}\, \gamma^{m_5m_6]} \,,
\\
 &\eta_2^{m_1\cdots m_6} \equiv -\beta^{m_1\cdots m_6} - 15\,\eta^{[m_1\cdots m_4}\,\gamma^{m_5m_6]} + 30\,\gamma^{[m_1m_2}\,\gamma^{m_3m_4}\,\beta^{m_5m_6]} \,.
\end{split}
\end{align}
In this paper, we use the fields $(\tilde{g}_{mn},\,\tilde{\phi},\,\beta^{mn},\,\gamma^{m_1\cdots m_{2n}},\,\beta^{m_1\cdots m_6})$ to describe supergravity solutions of exotic branes in type IIB theory. 

Again, by comparing the two parameterizations \eqref{eq:IIB-conv} and \eqref{eq:IIB-dual} as
\begin{align}
 \cM_{IJ}^{\text{\tiny(IIB)}} = \cM_{IJ}^{\text{\tiny(IIB, non-geometric)}}\,,
\end{align}
we can in principle determine the dual fields in terms of the conventional supergravity fields.

\section{Contents of the $p$-brane multiplets}
\label{app:multiplets}

In this appendix, we provide a list of branes contained in the $p$-brane multiplets. 

\subsection{``Elementary'' branes}

We first provide a list of ``elementary'' branes that are connected to the standard branes. 

\begin{table}[H]
 \begin{center}
  \newcommand{\parboxM}[1]{\parbox{0.15\textwidth}{\vspace{.2\baselineskip}\raggedright #1\vspace{.2\baselineskip}}}
  \newcommand{\parboxII}[1]{\parbox{0.3\textwidth}{\vspace{.2\baselineskip}\raggedright #1\vspace{.2\baselineskip}}}
{\scriptsize
\begin{tabular}{|c||c||l|l|l|}
 \hline
 $d$ & $D$ & \multicolumn{1}{|c|}{M} & \multicolumn{1}{|c|}{IIA} & \multicolumn{1}{|c|}{IIB} \\ \hline\hline
  9 & $1$ & \parboxM{
 $2_3$~({\bf 1})
}
   & \parboxII{
 $1_0$~({\bf 1})
}
   & \parboxII{
 $\text{P}$~({\bf 1})
} 
\\ \cline{2-5}
 & $2$ & \parboxM{
 $\text{P}$~({\bf 2})
}
   & \parboxII{
 $\text{P}$~({\bf 1}),
 $0_1$~({\bf 1})
}
   & \parboxII{
 $1_0$~({\bf 1}),
 $1_1$~({\bf 1})
} 
\\ \hline
  8 & $6$ & \parboxM{
 $\text{P}$~({\bf 3}),
 $2_3$~({\bf 3})
} 
   & \parboxII{
 $\text{P}$~({\bf 2}),
 $1_0$~({\bf 2}),
 $0_1$~({\bf 1}),
 $2_1$~({\bf 1})
}
   & \parboxII{
 $\text{P}$~({\bf 2}),
 $1_0$~({\bf 2}),
 $1_1$~({\bf 2})
} 
\\ \hline
  7 & $10$ & \parboxM{
 $\text{P}$~({\bf 4}),
 $2_3$~({\bf 6})
} 
   & \parboxII{
 $\text{P}$~({\bf 3}),
 $1_0$~({\bf 3}),
 $0_1$~({\bf 1}),
 $2_1$~({\bf 3})
}
   & \parboxII{
 $\text{P}$~({\bf 3}),
 $1_0$~({\bf 3}),
 $1_1$~({\bf 3}),
 $3_1$~({\bf 1})
} 
\\ \hline
  6 & $16$ & \parboxM{
 $\text{P}$~({\bf 5}),
 $2_3$~({\bf 10}),
 $5_6$~({\bf 1})
}
   & \parboxII{
 $\text{P}$~({\bf 4}),
 $1_0$~({\bf 4}),
 $0_1$~({\bf 1}),
 $2_1$~({\bf 6}),
 $4_1$~({\bf 1})
}
   & \parboxII{
 $\text{P}$~({\bf 4}),
 $1_0$~({\bf 4}),
 $1_1$~({\bf 4}),
 $3_1$~({\bf 4})
} 
\\ \hline
  5 & $27$ & \parboxM{
 $\text{P}$~({\bf 6}),
 $2_3$~({\bf 15}),
 $5_6$~({\bf 6})
}
   & \parboxII{
 $\text{P}$~({\bf 5}),
 $1_0$~({\bf 5}),
 $0_1$~({\bf 1}),
 $2_1$~({\bf 10}),
 $4_1$~({\bf 5}),
 $5_2$~({\bf 1})
}
   & \parboxII{
 $\text{P}$~({\bf 5}),
 $1_0$~({\bf 5}),
 $1_1$~({\bf 5}),
 $3_1$~({\bf 10}),
 $5_1$~({\bf 1}),
 $5_2$~({\bf 1})
}
\\
\hline
 \end{tabular}
}
 \end{center}
\caption{Branes in the particle multiplet.}
\label{tab:BPS-0-brane}
\end{table}

\begin{table}[H]
 \begin{center}
  \newcommand{\parboxM}[1]{\parbox{0.15\textwidth}{\vspace{.2\baselineskip}\raggedright #1\vspace{.2\baselineskip}}}
  \newcommand{\parboxII}[1]{\parbox{0.3\textwidth}{\vspace{.2\baselineskip}\raggedright #1\vspace{.2\baselineskip}}}
{\scriptsize
 \begin{tabular}{|c||c||l|l|l|}
 \hline
 $d$ & $D$ & \multicolumn{1}{|c|}{M} & \multicolumn{1}{|c|}{IIA} & \multicolumn{1}{|c|}{IIB} \\ \hline\hline
 9 & $2$ & \parboxM{
 $2_3$~({\bf 2})
}
   & \parboxII{
 $1_0$~({\bf 1}),
 $2_1$~({\bf 1})
}
   & \parboxII{
 $1_0$~({\bf 1}),
 $1_1$~({\bf 1})
} 
\\ \hline
 8 & $3$ & \parboxM{
 $2_3$~({\bf 3})
}
   & \parboxII{
 $1_0$~({\bf 1}),
 $2_1$~({\bf 2})
}
   & \parboxII{
 $1_0$~({\bf 1}),
 $1_1$~({\bf 1}),
 $3_1$~({\bf 1})
} 
\\ \hline
 7 & $5$ & \parboxM{
 $2_3$~({\bf 4}),
 $5_6$~({\bf 1})
}
   & \parboxII{
 $1_0$~({\bf 1}),
 $2_1$~({\bf 3}),
 $4_1$~({\bf 1})
}
   & \parboxII{
 $1_0$~({\bf 1}),
 $1_1$~({\bf 1}),
 $3_1$~({\bf 3})
} 
\\ \hline
 6 & $10$ & \parboxM{
 $2_3$~({\bf 5}),
 $5_6$~({\bf 5})
}
   & \parboxII{
 $1_0$~({\bf 1}),
 $2_1$~({\bf 4}),
 $4_1$~({\bf 4}),
 $5_2$~({\bf 1})
}
   & \parboxII{
 $1_0$~({\bf 1}),
 $1_1$~({\bf 1}),
 $3_1$~({\bf 6}),
 $5_1$~({\bf 1}),
 $5_2$~({\bf 1})
} 
\\ \hline
 5 & $27$ & \parboxM{
 $2_3$~({\bf 6}),
 $5_6$~({\bf 15}),
 $6_{9}^{1}$~({\bf 6})}
   & \parboxII{
 $1_0$~({\bf 1}),
 $2_1$~({\bf 5}),
 $4_1$~({\bf 10}),
 $6_1$~({\bf 1}),
 $5_2$~({\bf 5}),
 $5_{2}^{1}$~({\bf 5})
}
   & \parboxII{
 $1_0$~({\bf 1}),
 $1_1$~({\bf 1}),
 $3_1$~({\bf 10}),
 $5_1$~({\bf 5}),
 $5_2$~({\bf 5}),
 $5_{2}^{1}$~({\bf 5})
}
\\
\hline
 \end{tabular}
}
 \end{center}
\caption{Branes in the string multiplet.}
\label{tab:BPS-1-brane}
\end{table}

\begin{table}[H]
 \begin{center}
  \newcommand{\parboxM}[1]{\parbox{0.15\textwidth}{\vspace{.2\baselineskip}\raggedright #1\vspace{.2\baselineskip}}}
  \newcommand{\parboxII}[1]{\parbox{0.3\textwidth}{\vspace{.2\baselineskip}\raggedright #1\vspace{.2\baselineskip}}}
{\tiny
 \begin{tabular}{|c||c||l|l|l|}
 \hline
 $d$ & $D$ & \multicolumn{1}{|c|}{M} & \multicolumn{1}{|c|}{IIA} & \multicolumn{1}{|c|}{IIB} \\ \hline\hline
  9 & $1$ & \parboxM{
 $2_3$~({\bf 1})
}
   & \parboxII{
 $2_1$~({\bf 1})
}
   & \parboxII{
 $3_1$~({\bf 1})
} 
\\ \hline
  8 & $2$ & \parboxM{
 $2_3$~({\bf 1}),
 $5_6$~({\bf 1})
} 
   & \parboxII{
 $2_1$~({\bf 1}),
 $4_1$~({\bf 1})
}
   & \parboxII{
 $3_1$~({\bf 2})
} 
\\ \hline
  7 & $5$ & \parboxM{
 $2_3$~({\bf 1}),
 $5_6$~({\bf 4})
}
   & \parboxII{
 $2_1$~({\bf 1}),
 $4_1$~({\bf 3}),
 $5_2$~({\bf 1})
}
   & \parboxII{
 $3_1$~({\bf 3}),
 $5_1$~({\bf 1}),
 $5_2$~({\bf 1})
} 
\\ \hline
  6 & $16$ & \parboxM{
 $2_3$~({\bf 1}),
 $5_6$~({\bf 10}),
 $6_{9}^{1}$~({\bf 5})
}
   & \parboxII{
 $2_1$~({\bf 1}),
 $4_1$~({\bf 6}),
 $6_1$~({\bf 1}),
 $5_2$~({\bf 4}),
 $5_{2}^{1}$~({\bf 4})
}
   & \parboxII{
 $3_1$~({\bf 4}),
 $5_1$~({\bf 4}),
 $5_2$~({\bf 4}),
 $5_{2}^{1}$~({\bf 4})
} 
\\ \hline
  5 & $72$ & \parboxM{
 $2_3$~({\bf 1}),
 $5_6$~({\bf 20}),
 $6_{9}^{1}$~({\bf 30}),
 $\Def{5_{12}^{3}}$~({\bf 20}),
 $\Def{2_{15}^{6}}$~({\bf 1})
}
   & \parboxII{
 $2_1$~({\bf 1}),
 $4_1$~({\bf 10}),
 $6_1$~({\bf 5}),
 $5_2$~({\bf 10}),
 $5_{2}^{1}$~({\bf 20}),
 $\Def{5_{2}^{2}}$~({\bf 10}),
 $\Def{6_{3}^{1}}$~({\bf 5}),
 $\Def{4_{3}^{3}}$~({\bf 10}),
 $\Def{2_{3}^{5}}$~({\bf 1})
}
   & \parboxII{
 $3_1$~({\bf 5}),
 $5_1$~({\bf 10}),
 $\Def{7_1}$~({\bf 1}),
 $5_2$~({\bf 10}),
 $5_{2}^{1}$~({\bf 20}),
 $\Def{5_{2}^{2}}$~({\bf 10}),
 $\Def{7_3}$~({\bf 1}),
 $\Def{5_{3}^{2}}$~({\bf 10}),
 $\Def{3_{3}^{4}}$~({\bf 5})
}
\\
\hline
 \end{tabular}
}
 \end{center}
\caption{Branes in the membrane multiplet.}
\label{tab:BPS-2-brane}
\end{table}

\begin{table}[H]
 \begin{center}
  \newcommand{\parboxM}[1]{\parbox{0.15\textwidth}{\vspace{.2\baselineskip}\raggedright #1\vspace{.2\baselineskip}}}
  \newcommand{\parboxII}[1]{\parbox{0.3\textwidth}{\vspace{.2\baselineskip}\raggedright #1\vspace{.2\baselineskip}}}
{\scriptsize
 \begin{tabular}{|c||c||l|l|l|}
 \hline
 $d$ & $D$ & \multicolumn{1}{|c|}{M} & \multicolumn{1}{|c|}{IIA} & \multicolumn{1}{|c|}{IIB} \\ \hline\hline
  9 & $1$ & \parboxM{
 $5_6$~({\bf 1})
}
   & \parboxII{
 $4_1$~({\bf 1})
}
   & \parboxII{
 $3_1$~({\bf 1})
} 
\\ \hline
  8 & $3$ & \parboxM{
 $5_6$~({\bf 3})
} 
   & \parboxII{
 $4_1$~({\bf 2}),
 $5_2$~({\bf 1})
}
   & \parboxII{
 $3_1$~({\bf 1}),
 $5_1$~({\bf 1}),
 $5_2$~({\bf 1})
} 
\\ \hline
  7 & $10$ & \parboxM{
 $5_6$~({\bf 6}),
 $6_{9}^{1}$~({\bf 4})
}
   & \parboxII{
 $4_1$~({\bf 3}),
 $6_1$~({\bf 1}),
 $5_2$~({\bf 3}),
 $5_{2}^{1}$~({\bf 3})
}
   & \parboxII{
 $3_1$~({\bf 1}),
 $5_1$~({\bf 3}),
 $5_2$~({\bf 3}),
 $5_{2}^{1}$~({\bf 3})
} 
\\ \hline
  6 & $40$ & \parboxM{
 $5_6$~({\bf 10}),
 $6_{9}^{1}$~({\bf 20}),
 $\Def{5_{12}^{3}}$~({\bf 10})
}
   & \parboxII{
 $4_1$~({\bf 4}),
 $6_1$~({\bf 4}),
 $5_2$~({\bf 6}),
 $5_{2}^{1}$~({\bf 12}),
 $\Def{5_{2}^{2}}$~({\bf 6}),
 $\Def{6_{3}^{1}}$~({\bf 4}),
 $\Def{4_{3}^{3}}$~({\bf 4})
}
   & \parboxII{
 $3_1$~({\bf 1}),
 $5_1$~({\bf 6}),
 $\Def{7_1}$~({\bf 1}),
 $5_2$~({\bf 6}),
 $5_{2}^{1}$~({\bf 12}),
 $\Def{5_{2}^{2}}$~({\bf 6}),
 $\Def{7_3}$~({\bf 1}),
 $\Def{5_{3}^{2}}$~({\bf 6}),
 $\Def{3_{3}^{4}}$~({\bf 1})
} 
\\
\hline
 \end{tabular}
}
 \end{center}
\caption{Branes in the $3$-brane multiplet.}
\label{tab:BPS-3-brane}
\end{table}

\begin{table}[H]
 \begin{center}
  \newcommand{\parboxM}[1]{\parbox{0.15\textwidth}{\vspace{.2\baselineskip}\raggedright #1\vspace{.2\baselineskip}}}
  \newcommand{\parboxII}[1]{\parbox{0.3\textwidth}{\vspace{.2\baselineskip}\raggedright #1\vspace{.2\baselineskip}}}
{\scriptsize
 \begin{tabular}{|c||c||l|l|l|}
 \hline
 $d$ & $D$ & \multicolumn{1}{|c|}{M} & \multicolumn{1}{|c|}{IIA} & \multicolumn{1}{|c|}{IIB} \\ \hline\hline
  9 & $2$ & \parboxM{
 $5_6$~({\bf 2})
}
   & \parboxII{
 $4_1$~({\bf 1}),
 $5_2$~({\bf 1})
}
   & \parboxII{
 $5_1$~({\bf 1}),
 $5_2$~({\bf 1})
} 
\\ \hline
  8 & $6$ & \parboxM{
 $5_6$~({\bf 3}),
 $6_{9}^{1}$~({\bf 3})
} 
   & \parboxII{
 $4_1$~({\bf 1}),
 $6_1$~({\bf 1}),
 $5_2$~({\bf 2}),
 $5_{2}^{1}$~({\bf 2})
}
   & \parboxII{
 $5_1$~({\bf 2}),
 $5_2$~({\bf 2}),
 $5_{2}^{1}$~({\bf 2})
} 
\\ \hline
  7 & $20$ & \parboxM{
 $5_6$~({\bf 4}),
 $6_{9}^{1}$~({\bf 12}),
 $\Def{5_{12}^{3}}$~({\bf 4})
}
   & \parboxII{
 $4_1$~({\bf 1}),
 $6_1$~({\bf 3}),
 $5_2$~({\bf 3}),
 $5_{2}^{1}$~({\bf 6}),
 $\Def{5_{2}^{2}}$~({\bf 3}),
 $\Def{6_{3}^{1}}$~({\bf 3}),
 $\Def{4_{3}^{3}}$~({\bf 1})
}
   & \parboxII{
 $5_1$~({\bf 3}),
 $\Def{7_1}$~({\bf 1}),
 $5_2$~({\bf 3}),
 $5_{2}^{1}$~({\bf 6}),
 $\Def{5_{2}^{2}}$~({\bf 3}),
 $\Def{7_3}$~({\bf 1}),
 $\Def{5_{3}^{2}}$~({\bf 3})
} 
\\
\end{tabular}
}
 \end{center}
\caption{Branes in the $4$-brane multiplet.}
\label{tab:BPS-4-brane}
\end{table}

\begin{table}[H]
 \begin{center}
  \newcommand{\parboxM}[1]{\parbox{0.15\textwidth}{\vspace{.2\baselineskip}\raggedright #1\vspace{.2\baselineskip}}}
  \newcommand{\parboxII}[1]{\parbox{0.3\textwidth}{\vspace{.2\baselineskip}\raggedright #1\vspace{.2\baselineskip}}}
{\scriptsize
 \begin{tabular}{|c||c||l|l|l|}
 \hline
 $d$ & $D$ & \multicolumn{1}{|c|}{M} & \multicolumn{1}{|c|}{IIA} & \multicolumn{1}{|c|}{IIB} \\ \hline\hline
  9 & $1$ & \parboxM{
 $5_6$~({\bf 1})
}
   & \parboxII{
 $5_2$~({\bf 1})
}
   & \parboxII{
 $5_{2}^{1}$~({\bf 1})
} 
\\ \cline{2-5}
 & $2$ & \parboxM{
 $6_{9}^{1}$~({\bf 2})
}
   & \parboxII{
 $6_1$~({\bf 1}),
 $5_{2}^{1}$~({\bf 1})
}
   & \parboxII{
 $5_1$~({\bf 1}),
 $5_2$~({\bf 1})
} 
\\ \hline
  8 & $2$ & \parboxM{
 $5_6$~({\bf 1}),
 $\Def{5_{12}^{3}}$~({\bf 1})
} 
   & \parboxII{
 $5_2$~({\bf 1}),
 $\Def{5_{2}^{2}}$~({\bf 1})
}
   & \parboxII{
 $5_{2}^{1}$~({\bf 2})
} 
\\ \cline{2-5}
 & $6$ & \parboxM{
 $6_{9}^{1}$~({\bf 6})
}
   & \parboxII{
 $6_1$~({\bf 2}),
 $5_{2}^{1}$~({\bf 2}),
 $\Def{6_{3}^{1}}$~({\bf 2})
}
   & \parboxII{
 $5_1$~({\bf 1}),
 $\Def{7_1}$~({\bf 1}),
 $5_2$~({\bf 1}),
 $\Def{5_{2}^{2}}$~({\bf 1}),
 $\Def{7_3}$~({\bf 1}),
 $\Def{5_{3}^{2}}$~({\bf 1})
} 
\\ \hline
  7 & $5$ & \parboxM{
 $5_6$~({\bf 1}),
 $\Def{5_{12}^{3}}$~({\bf 4})
}
   & \parboxII{
 $5_2$~({\bf 1}),
 $\Def{5_{2}^{2}}$~({\bf 3}),
 $\DoW{5_{4}^{3}}$~({\bf 1})
}
   & \parboxII{
 $5_{2}^{1}$~({\bf 3}),
 $\DoW{5_{2}^{3}}$~({\bf 1}),
 $\DoW{5_{4}^{3}}$~({\bf 1})
} 
\\ \cline{2-5}
 & $20$ & \parboxM{
 $6_{9}^{1}$~({\bf 12}),
 $\DoW{8_{12}^{(1,0)}}$~({\bf 4}),
 $\DoW{5_{15}^{(1,3)}}$~({\bf 4})
}
   & \parboxII{
 $6_1$~({\bf 3}),
 $\DoW{8_1}$~({\bf 1}),
 $5_{2}^{1}$~({\bf 3}),
 $\DoW{5_{2}^{3}}$~({\bf 1}),
 $\Def{6_{3}^{1}}$~({\bf 6}),
 $\DoW{7_{3}^{(1,0)}}$~({\bf 3}),
 $\DoW{5_{3}^{(1,2)}}$~({\bf 3})
}
   & \parboxII{
 $5_1$~({\bf 1}),
 $\Def{7_1}$~({\bf 3}),
 $5_2$~({\bf 1}),
 $\Def{5_{2}^{2}}$~({\bf 3}),
 $\Def{7_3}$~({\bf 3}),
 $\Def{5_{3}^{2}}$~({\bf 3}),
 $\DoW{6_{3}^{(1,1)}}$~({\bf 6})
} 
\\
\end{tabular}
}
 \end{center}
\caption{Branes in the $5$-brane multiplet.}
\label{tab:BPS-5-brane}
\end{table}

\begin{table}[H]
 \begin{center}
  \newcommand{\parboxM}[1]{\parbox{0.15\textwidth}{\vspace{.2\baselineskip}\raggedright #1\vspace{.2\baselineskip}}}
  \newcommand{\parboxII}[1]{\parbox{0.3\textwidth}{\vspace{.2\baselineskip}\raggedright #1\vspace{.2\baselineskip}}}
{\scriptsize
 \begin{tabular}{|c||c||l|l|l|}
 \hline
 $d$ & $D$ & \multicolumn{1}{|c|}{M} & \multicolumn{1}{|c|}{IIA} & \multicolumn{1}{|c|}{IIB} \\ \hline\hline
  9 & $2$ & \parboxM{
 $6_{9}^{1}$~({\bf 2})
}
   & \parboxII{
 $6_1$~({\bf 1}),
 $\Def{6_{3}^{1}}$~({\bf 1})
}
   & \parboxII{
 $\Def{7_1}$~({\bf 1}),
 $\Def{7_3}$~({\bf 1})
} 
\\ \hline
  8 & $6$ & \parboxM{
 $6_{9}^{1}$~({\bf 3}),
 $\DoW{8_{12}^{(1,0)}}$~({\bf 3})
} 
   & \parboxII{
 $6_1$~({\bf 1}),
 $\DoW{8_1}$~({\bf 1}),
 $\Def{6_{3}^{1}}$~({\bf 2}),
 $\DoW{7_{3}^{(1,0)}}$~({\bf 2})
}
   & \parboxII{
 $\Def{7_1}$~({\bf 2}),
 $\Def{7_3}$~({\bf 2}),
 $\DoW{6_{3}^{(1,1)}}$~({\bf 2})
}
\\ \hline
  7 & $20$ & \parboxM{
 $6_{9}^{1}$~({\bf 4}),
 $\DoW{8_{12}^{(1,0)}}$~({\bf 12}),
 $\SpF{6_{18}^{(3,1)}}$~({\bf 4})
}
   & \parboxII{
 $6_1$~({\bf 1}),
 $\DoW{8_1}$~({\bf 3}),
 $\Def{6_{3}^{1}}$~({\bf 3}),
 $\DoW{7_{3}^{(1,0)}}$~({\bf 6}),
 $\SpF{6_{3}^{(2,1)}}$~({\bf 3}),
 $\SpF{8_{4}^{(1,0)}}$~({\bf 3}),
 $\SpF{6_{4}^{(3,0)}}$~({\bf 1})
}
   & \parboxII{
 $\Def{7_1}$~({\bf 3}),
 $\SpF{9_1}$~({\bf 1}),
 $\Def{7_3}$~({\bf 3}),
 $\DoW{6_{3}^{(1,1)}}$~({\bf 6}),
 $\SpF{7_{3}^{(2,0)}}$~({\bf 3}),
 $\SpF{9_4}$~({\bf 1}),
 $\SpF{7_{4}^{(2,0)}}$~({\bf 3})
} 
\\ \hline
 \end{tabular}
}
 \end{center}
\caption{Branes in the $6$-brane multiplet.}
\label{tab:BPS-6-brane}
\end{table}

\begin{table}[H]
 \begin{center}
  \newcommand{\parboxM}[1]{\parbox{0.15\textwidth}{\vspace{.2\baselineskip}\raggedright #1\vspace{.2\baselineskip}}}
  \newcommand{\parboxII}[1]{\parbox{0.3\textwidth}{\vspace{.2\baselineskip}\raggedright #1\vspace{.2\baselineskip}}}
{\scriptsize
 \begin{tabular}{|c||c||l|l|l|}
 \hline
 $d$ & $D$ & \multicolumn{1}{|c|}{M} & \multicolumn{1}{|c|}{IIA} & \multicolumn{1}{|c|}{IIB} \\ \hline\hline
  9 & $2$ & \parboxM{
 $\DoW{8_{12}^{(1,0)}}$~({\bf 2})
}
   & \parboxII{
 $\DoW{8_1}$~({\bf 1}),
 $\DoW{7_{3}^{(1,0)}}$~({\bf 1})
}
   & \parboxII{
 $\Def{7_1}$~({\bf 1}),
 $\Def{7_3}$~({\bf 1})
}
\\ \hline
  8 & $6$ & \parboxM{
 $\DoW{8_{12}^{(1,0)}}$~({\bf 6})
}
   & \parboxII{
 $\DoW{8_1}$~({\bf 2}),
 $\DoW{7_{3}^{(1,0)}}$~({\bf 2}),
 $\SpF{8_{4}^{(1,0)}}$~({\bf 2})
}
   & \parboxII{
 $\Def{7_1}$~({\bf 1}),
 $\SpF{9_1}$~({\bf 1}),
 $\Def{7_3}$~({\bf 1}),
 $\SpF{7_{3}^{(2,0)}}$~({\bf 1}),
 $\SpF{9_4}$~({\bf 1}),
 $\SpF{7_{4}^{(2,0)}}$~({\bf 1})
} 
\\ \hline
 \end{tabular}
}
 \end{center}
\caption{Branes in the $7$-brane multiplet.}
\label{tab:BPS-7-brane}
\end{table}

\begin{table}[H]
 \begin{center}
  \newcommand{\parboxM}[1]{\parbox{0.15\textwidth}{\vspace{.2\baselineskip}\raggedright #1\vspace{.2\baselineskip}}}
  \newcommand{\parboxII}[1]{\parbox{0.3\textwidth}{\vspace{.2\baselineskip}\raggedright #1\vspace{.2\baselineskip}}}
{\scriptsize
 \begin{tabular}{|c||c||l|l|l|}
 \hline
 $d$ & $D$ & \multicolumn{1}{|c|}{M} & \multicolumn{1}{|c|}{IIA} & \multicolumn{1}{|c|}{IIB} \\ \hline\hline
  9 & $2$ & \parboxM{
 $\DoW{8_{12}^{(1,0)}}$~({\bf 2})
}
   & \parboxII{
 $\DoW{8_1}$~({\bf 1}),
 $\SpF{8_{4}^{(1,0)}}$~({\bf 1})
}
   & \parboxII{
 $\SpF{9_1}$~({\bf 1}),
 $\SpF{9_4}$~({\bf 1})
} 
\\ \hline
 \end{tabular}
}
 \end{center}
\caption{Branes in the $8$-brane multiplet.}
\label{tab:BPS-8-brane}
\end{table}

\subsection{Missing states}

We here provide a list of the missing states in each multiplet.
\begin{table}[H]
 \begin{center}
  \newcommand{\parboxM}[1]{\parbox{0.15\textwidth}{\vspace{.2\baselineskip}\raggedright #1\vspace{.2\baselineskip}}}
  \newcommand{\parboxII}[1]{\parbox{0.25\textwidth}{\vspace{.2\baselineskip}\raggedright #1\vspace{.2\baselineskip}}}
{\tiny
 \begin{tabular}{|c||c|c||l|l|l|}
 \hline
 $d$ & $D$ & $N$ & \multicolumn{1}{|c|}{M} & \multicolumn{1}{|c|}{IIA} & \multicolumn{1}{|c|}{IIB} \\ \hline\hline
  4 & $7\subset 133$ & 7 & \parboxM{
 $8_9$~({\bf 1})
}
   & \parboxII{
 $7_2$~({\bf 1})
}
   & \parboxII{
 $7_2$~({\bf 1})
} 
\\ \hline\hline
  3 & $1680\subset 3875$ & 7 & \parboxM{
 $8_9$~({\bf 8}),
 $7_{12}^{2}$~({\bf 28}),
 $4_{15}^{5}$~({\bf 56}),
 $2_{18}^{(1,6)}$~({\bf 56}),
 $1_{21}^{(3,5)}$~({\bf 56}),
 $1_{24}^{(6,2)}$~({\bf 28}),
 $1_{27}^{(1,7,0)}$~({\bf 8})
}
   & \parboxII{
 $7_2$~({\bf 7}),
 $7_{2}^{1}$~({\bf 7}),
 $8_3$~({\bf 1}),
 $6_{3}^{2}$~({\bf 21}),
 $4_{3}^{4}$~({\bf 35}),
 $2_{3}^{6}$~({\bf 7}),
 $3_{4}^{5}$~({\bf 21}),
 $2_{4}^{(1,5)}$~({\bf 42}),
 $1_{4}^{(2,5)}$~({\bf 21}),
 $1_{5}^{(1,6)}$~({\bf 7}),
 $1_{5}^{(3,4)}$~({\bf 35}),
 $1_{5}^{(5,2)}$~({\bf 21}),
 $1_{5}^{(7,0)}$~({\bf 1}),
 $1_{6}^{(6,1)}$~({\bf 7}),
 $1_{6}^{(1,6,0)}$~({\bf 7})
}
   & \parboxII{
 $7_2$~({\bf 7}),
 $7_{2}^{1}$~({\bf 7}),
 $7_{3}^{1}$~({\bf 7}),
 $5_{3}^{3}$~({\bf 35}),
 $3_{3}^{5}$~({\bf 21}),
 $1_{3}^{7}$~({\bf 1}),
 $3_{4}^{5}$~({\bf 21}),
 $2_{4}^{(1,5)}$~({\bf 42}),
 $1_{4}^{(2,5)}$~({\bf 21}),
 $1_{5}^{7}$~({\bf 1}),
 $1_{5}^{(2,5)}$~({\bf 21}),
 $1_{5}^{(4,3)}$~({\bf 35}),
 $1_{5}^{(6,1)}$~({\bf 7}),
 $1_{6}^{(6,1)}$~({\bf 7}),
 $1_{6}^{(1,6,0)}$~({\bf 7})
} 
\\ \cline{2-6}
  & $35\subset 3875$ & 35 & \parboxM{
 $1_{18}^{8}$~({\bf 1})
}
   & \parboxII{
 $1_{4}^{7}$~({\bf 1})
}
   & \parboxII{
 $1_{4}^{7}$~({\bf 1})
} 
\\ \hline
\end{tabular}
}
\end{center}
\caption{Missing states in the string multiplet.}
\label{tab:missing-1-brane}
\end{table}

\begin{table}[H]
 \begin{center}
  \newcommand{\parboxM}[1]{\parbox{0.15\textwidth}{\vspace{.2\baselineskip}\raggedright #1\vspace{.2\baselineskip}}}
  \newcommand{\parboxII}[1]{\parbox{0.25\textwidth}{\vspace{.2\baselineskip}\raggedright #1\vspace{.2\baselineskip}}}
{\tiny
 \begin{tabular}{|c||c|c||l|l|l|}
 \hline
 $d$ & $D$ & $N$ & \multicolumn{1}{|c|}{M} & \multicolumn{1}{|c|}{IIA} & \multicolumn{1}{|c|}{IIB} \\ \hline\hline
  5 & $6\subset 78$ & 6 & \parboxM{
 $8_9$~({\bf 1})
}
   & \parboxII{
 $7_2$~({\bf 1})
}
   & \parboxII{
 $7_2$~({\bf 1})
} 
\\ \hline\hline
  4 & $336\subset 912$ & 6 & \parboxM{
 $8_9$~({\bf 7}),
 $7_{12}^{2}$~({\bf 21}),
 $4_{15}^{5}$~({\bf 21}),
 $2_{18}^{(1,6)}$~({\bf 7})
}
   & \parboxII{
 $7_2$~({\bf 6}),
 $7_{2}^{1}$~({\bf 6}),
 $8_3$~({\bf 1}),
 $6_{3}^{2}$~({\bf 15}),
 $4_{3}^{4}$~({\bf 15}),
 $2_{3}^{6}$~({\bf 1}),
 $3_{4}^{5}$~({\bf 6}),
 $2_{4}^{(1,5)}$~({\bf 6})
}
   & \parboxII{
 $7_2$~({\bf 6}),
 $7_{2}^{1}$~({\bf 6}),
 $7_{3}^{1}$~({\bf 6}),
 $5_{3}^{3}$~({\bf 20}),
 $3_{3}^{5}$~({\bf 6}),
 $3_{4}^{5}$~({\bf 6}),
 $2_{4}^{(1,5)}$~({\bf 6})
} 
\\
\end{tabular}
}
\end{center}
\caption{Missing states in the membrane multiplet.}
\label{tab:missing-2-brane}
\end{table}

\begin{table}[H]
 \begin{center}
  \newcommand{\parboxM}[1]{\parbox{0.15\textwidth}{\vspace{.2\baselineskip}\raggedright #1\vspace{.2\baselineskip}}}
  \newcommand{\parboxII}[1]{\parbox{0.25\textwidth}{\vspace{.2\baselineskip}\raggedright #1\vspace{.2\baselineskip}}}
{\scriptsize
 \begin{tabular}{|c||c|c||l|l|l|}
 \hline
 $d$ & $D$ & $N$ & \multicolumn{1}{|c|}{M} & \multicolumn{1}{|c|}{IIA} & \multicolumn{1}{|c|}{IIB} \\ \hline\hline
  6 & $5\subset 45$ & 5 & \parboxM{
 $8_9$~({\bf 1})
}
   & \parboxII{
 $7_2$~({\bf 1})
}
   & \parboxII{
 $7_2$~({\bf 1})
} 
\\ \hline\hline
  5 & $135\subset 351$ & 5 & \parboxM{
 $8_9$~({\bf 6}),
 $7_{12}^{2}$~({\bf 15}),
 $4_{15}^{5}$~({\bf 6})
}
   & \parboxII{
 $7_2$~({\bf 5}),
 $7_{2}^{1}$~({\bf 5}),
 $8_3$~({\bf 1}),
 $6_{3}^{2}$~({\bf 10}),
 $4_{3}^{4}$~({\bf 5}),
 $3_{4}^{5}$~({\bf 1})
}
   & \parboxII{
 $7_2$~({\bf 5}),
 $7_{2}^{1}$~({\bf 5}),
 $7_{3}^{1}$~({\bf 5}),
 $5_{3}^{3}$~({\bf 10}),
 $3_{3}^{5}$~({\bf 1}),
 $3_{4}^{5}$~({\bf 1})
} 
\\
\end{tabular}
}
\end{center}
\caption{Missing states in the 3-brane multiplet.}
\label{tab:missing-3-brane}
\end{table}

\begin{table}[H]
 \begin{center}
  \newcommand{\parboxM}[1]{\parbox{0.15\textwidth}{\vspace{.2\baselineskip}\raggedright #1\vspace{.2\baselineskip}}}
  \newcommand{\parboxII}[1]{\parbox{0.25\textwidth}{\vspace{.2\baselineskip}\raggedright #1\vspace{.2\baselineskip}}}
{\scriptsize
 \begin{tabular}{|c||c|c||l|l|l|}
 \hline
 $d$ & $D$ & $N$ & \multicolumn{1}{|c|}{M} & \multicolumn{1}{|c|}{IIA} & \multicolumn{1}{|c|}{IIB} \\ \hline\hline
  7 & $4\subset 24$ & 4 & \parboxM{
 $8_9$~({\bf 1})
}
   & \parboxII{
 $7_2$~({\bf 1})
}
   & \parboxII{
 $7_2$~({\bf 1})
} 
\\ \hline\hline
  6 & $64\subset 144$ & 4 & \parboxM{
 $8_9$~({\bf 5})
 $7_{12}^{2}$~({\bf 10})
 $4_{15}^{5}$~({\bf 1})
}
   & \parboxII{
 $7_2$~({\bf 4}),
 $7_{2}^{1}$~({\bf 4}),
 $8_3$~({\bf 1}),
 $6_{3}^{2}$~({\bf 6}),
 $4_{3}^{4}$~({\bf 1})
}
   & \parboxII{
 $7_2$~({\bf 4}),
 $7_{2}^{1}$~({\bf 4}),
 $7_{3}^{1}$~({\bf 4}),
 $5_{3}^{3}$~({\bf 4})
} 
\\ \hline\hline
  5 & $864\subset 1728$ & 4 & \parboxM{
 $8_9$~({\bf 15}),
 $7_{12}^{2}$~({\bf 60}),
 $4_{15}^{5}$~({\bf 6}),
 $7_{15}^{(1,2)}$~({\bf 60}),
 $5_{18}^{(2,3)}$~({\bf 60}),
 $4_{21}^{(4,2)}$~({\bf 15})
}
   & \parboxII{
 $7_2$~({\bf 10}),
 $8_3$~({\bf 5}),
 $7_{2}^{1}$~({\bf 20}),
 $7_{2}^{2}$~({\bf 10}),
 $6_{3}^{2}$~({\bf 30}),
 $4_{3}^{4}$~({\bf 5}),
 $7_{3}^{(1,1)}$~({\bf 20}),
 $5_{3}^{(1,3)}$~({\bf 20}),
 $7_{4}^{2}$~({\bf 10}),
 $6_{4}^{(1,2)}$~({\bf 30}),
 $5_{4}^{(2,2)}$~({\bf 30}),
 $4_{4}^{(3,2)}$~({\bf 10}),
 $4_{5}^{5}$~({\bf 1}),
 $4_{5}^{(2,3)}$~({\bf 10}),
 $4_{5}^{(4,1)}$~({\bf 5})
}
   & \parboxII{
 $7_2$~({\bf 10}),
 $7_{2}^{1}$~({\bf 20}),
 $7_{2}^{2}$~({\bf 10}),
 $7_{3}^{1}$~({\bf 20}),
 $5_{3}^{3}$~({\bf 20}),
 $8_{3}^{(1,0)}$~({\bf 5}),
 $6_{3}^{(1,2)}$~({\bf 30}),
 $4_{3}^{(1,4)}$~({\bf 5}),
 $7_{4}^{2}$~({\bf 10}),
 $6_{4}^{(1,2)}$~({\bf 30}),
 $5_{4}^{(2,2)}$~({\bf 30}),
 $4_{4}^{(3,2)}$~({\bf 10}),
 $4_{5}^{(1,4)}$~({\bf 5}),
 $4_{5}^{(3,2)}$~({\bf 10}),
 $4_{5}^{(5,0)}$~({\bf 1})
} 
\\
\end{tabular}
}
\end{center}
\caption{Missing states in the 4-brane multiplet.}
\label{tab:missing-4-brane}
\end{table}

\begin{table}[H]
 \begin{center}
  \newcommand{\parboxM}[1]{\parbox{0.15\textwidth}{\vspace{.2\baselineskip}\raggedright #1\vspace{.2\baselineskip}}}
  \newcommand{\parboxII}[1]{\parbox{0.25\textwidth}{\vspace{.2\baselineskip}\raggedright #1\vspace{.2\baselineskip}}}
{\scriptsize
 \begin{tabular}{|c||c|c||l|l|l|}
 \hline
 $d$ & $D$ & $N$ & \multicolumn{1}{|c|}{M} & \multicolumn{1}{|c|}{IIA} & \multicolumn{1}{|c|}{IIB} \\ \hline\hline
  8 & $2\subset 8$ & 2 & \parboxM{
 $8_9$~({\bf 1})
}
   & \parboxII{
 $7_2$~({\bf 1})
}
   & \parboxII{
 $7_2$~({\bf 1})
} 
\\ \cline{2-6}
  & $1\subset 3$ & 1 & \parboxM{
 $8_9$~({\bf 1})
}
   & \parboxII{
 $7_2$~({\bf 1})
}
   & \parboxII{
 $7_2$~({\bf 1})
} 
\\ \hline\hline
  7 & $20\subset 40$ & 2 & \parboxM{
 $8_9$~({\bf 4}),
 $7_{12}^{2}$~({\bf 6})
}
   & \parboxII{
 $7_2$~({\bf 3}),
 $7_{2}^{1}$~({\bf 3}),
 $8_3$~({\bf 1}),
 $6_{3}^{2}$~({\bf 3})
}
   & \parboxII{
 $7_2$~({\bf 3}),
 $7_{2}^{1}$~({\bf 3}),
 $7_{3}^{1}$~({\bf 3}),
 $5_{3}^{3}$~({\bf 1})
} 
\\ \cline{2-6}
  & $10\subset 15$ & 1 & \parboxM{
 $8_9$~({\bf 4}),
 $7_{12}^{2}$~({\bf 6})
}
   & \parboxII{
 $7_2$~({\bf 3}),
 $7_{2}^{1}$~({\bf 3}),
 $8_3$~({\bf 1}),
 $6_{3}^{2}$~({\bf 3})
}
   & \parboxII{
 $7_2$~({\bf 3}),
 $7_{2}^{1}$~({\bf 3}),
 $7_{3}^{1}$~({\bf 3}),
 $5_{3}^{3}$~({\bf 1})
} 
\\ \hline\hline
  6 & $160\subset 320$ & 2 & \parboxM{
 $8_9$~({\bf 10})
 $7_{12}^{2}$~({\bf 30})
 $7_{15}^{(1,2)}$~({\bf 30})
 $5_{18}^{(2,3)}$~({\bf 10})
}
   & \parboxII{
 $7_2$~({\bf 6}),
 $7_{2}^{1}$~({\bf 12}),
 $7_{2}^{2}$~({\bf 6}),
 $8_3$~({\bf 4}),
 $6_{3}^{2}$~({\bf 12}),
 $7_{3}^{(1,1)}$~({\bf 12}),
 $5_{3}^{(1,3)}$~({\bf 4}),
 $7_{4}^{2}$~({\bf 6}),
 $6_{4}^{(1,2)}$~({\bf 12}),
 $5_{4}^{(2,2)}$~({\bf 6})
}
   & \parboxII{
 $7_2$~({\bf 6}),
 $7_{2}^{1}$~({\bf 12}),
 $7_{2}^{2}$~({\bf 6}),
 $7_{3}^{1}$~({\bf 12}),
 $5_{3}^{3}$~({\bf 4}),
 $8_{3}^{(1,0)}$~({\bf 4}),
 $6_{3}^{(1,2)}$~({\bf 12}),
 $7_{4}^{2}$~({\bf 6}),
 $6_{4}^{(1,2)}$~({\bf 12}),
 $5_{4}^{(2,2)}$~({\bf 6})
} 
\\
\end{tabular}
}
\end{center}
\caption{Missing states in the 5-brane multiplet.}
\label{tab:missing-5-brane}
\end{table}

\begin{table}[H]
 \begin{center}
  \newcommand{\parboxM}[1]{\parbox{0.15\textwidth}{\vspace{.2\baselineskip}\raggedright #1\vspace{.2\baselineskip}}}
  \newcommand{\parboxII}[1]{\parbox{0.25\textwidth}{\vspace{.2\baselineskip}\raggedright #1\vspace{.2\baselineskip}}}
{\scriptsize
 \begin{tabular}{|c||c|c||l|l|l|}
 \hline
 $d$ & $D$ & $N$ & \multicolumn{1}{|c|}{M} & \multicolumn{1}{|c|}{IIA} & \multicolumn{1}{|c|}{IIB} \\ \hline\hline
  8 & $6\subset 12$ & 1 & \parboxM{
 $8_9$~({\bf 3}),
 $7_{12}^{2}$~({\bf 3})
}
   & \parboxII{
 $7_2$~({\bf 2}),
 $8_3$~({\bf 1}),
 $7_{2}^{1}$~({\bf 2}),
 $6_{3}^{2}$~({\bf 1})
}
   & \parboxII{
 $7_2$~({\bf 2}),
 $7_{2}^{1}$~({\bf 2}),
 $7_{3}^{1}$~({\bf 2})
} 
\\ \hline\hline
  7 & $30\subset 70$ & 1 & \parboxM{
 $8_9$~({\bf 6}),
 $7_{12}^{2}$~({\bf 12}),
 $7_{15}^{(1,2)}$~({\bf 12})
}
   & \parboxII{
 $7_2$~({\bf 3}),
 $7_{2}^{1}$~({\bf 6}),
 $7_{2}^{2}$~({\bf 3}),
 $8_3$~({\bf 3}),
 $6_{3}^{2}$~({\bf 3}),
 $7_{3}^{(1,1)}$~({\bf 6}),
 $7_{4}^{2}$~({\bf 3}),
 $6_{4}^{(1,2)}$~({\bf 3})
}
   & \parboxII{
 $7_2$~({\bf 3}),
 $7_{2}^{1}$~({\bf 6}),
 $7_{2}^{2}$~({\bf 3}),
 $7_{3}^{1}$~({\bf 6}),
 $8_{3}^{(1,0)}$~({\bf 3}),
 $6_{3}^{(1,2)}$~({\bf 3}),
 $7_{4}^{2}$~({\bf 3}),
 $6_{4}^{(1,2)}$~({\bf 3})
} 
\\ \cline{2-6}
  & $20\subset 70$ & 4 & \parboxM{
 $9_{12}^{1}$~({\bf 4}),
 $6_{15}^{4}$~({\bf 1})
}
   & \parboxII{
 $9_2$~({\bf 1}),
 $8_{3}^{1}$~({\bf 3}),
 $6_{3}^{3}$~({\bf 1})
}
   & \parboxII{
 $9_2$~({\bf 1}),
 $9_3$~({\bf 1}),
 $7_{3}^{2}$~({\bf 3})
} 
\\ \hline
\end{tabular}
}
\end{center}
\caption{Missing states in the 6-brane multiplet.}
\label{tab:missing-6-brane}
\end{table}

\begin{table}[H]
 \begin{center}
  \newcommand{\parboxM}[1]{\parbox{0.15\textwidth}{\vspace{.2\baselineskip}\raggedright #1\vspace{.2\baselineskip}}}
  \newcommand{\parboxII}[1]{\parbox{0.25\textwidth}{\vspace{.2\baselineskip}\raggedright #1\vspace{.2\baselineskip}}}
{\scriptsize
 \begin{tabular}{|c||c|c||l|l|l|}
 \hline
 $d$ & $D$ & $N$ & \multicolumn{1}{|c|}{M} & \multicolumn{1}{|c|}{IIA} & \multicolumn{1}{|c|}{IIB} \\ \hline\hline
  8 & $3\subset 15$ & 1 & \parboxM{
 $7_{12}^{2}$~({\bf 3})
}
   & \parboxII{
 $7_{2}^{1}$~({\bf 2}),
 $7_{4}^{2}$~({\bf 1})
}
   & \parboxII{
 $7_2$~({\bf 1}),
 $7_{2}^{2}$~({\bf 1}),
 $7_{4}^{2}$~({\bf 1})
} 
\\ \cline{2-6}
  & $6\subset 15$ & 2 & \parboxM{
 $9_{12}^{1}$~({\bf 3})
}
   & \parboxII{
 $9_2$~({\bf 1}),
 $8_{3}^{1}$~({\bf 2})
}
   & \parboxII{
 $9_2$~({\bf 1}),
 $9_3$~({\bf 1}),
 $7_{3}^{2}$~({\bf 1})
} 
\\ \hline
 \end{tabular}
}
\end{center}
\caption{Missing states in the 7-brane multiplet.}
\label{tab:missing-7-brane}
\end{table}

\section{Counting of mixed-symmetry potentials}
\label{app:counting-mixed-symmetry}

In a series of work on the mixed-symmetry potentials \cite{Bergshoeff:2011mh,Bergshoeff:2011ee,Bergshoeff:2012ex,Bergshoeff:2017gpw}, the number of supersymmetric branes that couple to the mixed-symmetry potentials has been counted up to $\alpha=-7$\,. 
We reproduce the same results by counting the number of branes contained in Tables \ref{tab:BPS-0-brane}--\ref{tab:BPS-8-brane}. 
Our results include all of the ``elementary'' exotic branes in $d\geq 3$ dimensions, i.e.~up to $\alpha=-11$\,. 

For convenience, we consider several examples to elucidate how to reproduce the following tables. 
Let us consider the string multiplet ($p=1$) \eqref{tab:BPS-1-brane} in $d=4$\,. 
In type IIA theory, the number of D-branes are $2_1$ ({\bf 6}), $4_1$ ({\bf 20}), and $6_1$ ({\bf 6}). 
In total, there are 32 D-branes, which is consistent with Table \ref{tab:D-brane-number}. 
Even if we count the number of type IIB D-branes, the result is the same. 
As another example, let us consider the membrane multiplet ($p=2$) \eqref{tab:BPS-2-brane} in $d=3$\,. 
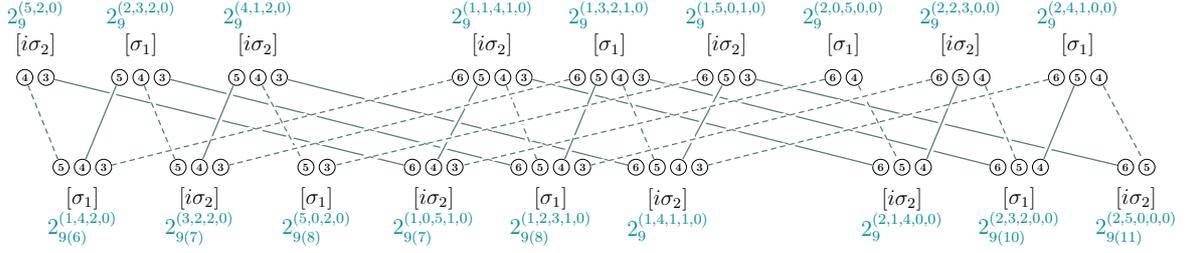
The number of the E$^{(9;5)}$-branes in type IIB theory are 
 $2_{9}^{(1,4,2,0)}$ ({\bf 105}),
 $2_{9}^{(3,2,2,0)}$ ({\bf 210}),
 $2_{9}^{(5,0,2,0)}$ ({\bf 21}),
 $2_{9}^{(1,0,5,1,0)}$ ({\bf 42}),
 $2_{9}^{(1,2,3,1,0)}$ ({\bf 420}),
 $2_{9}^{(1,4,1,1,0)}$ ({\bf 210}),
 $2_{9}^{(2,1,4,0,0)}$ ({\bf 105}),
 $2_{9}^{(2,3,2,0,0)}$ ({\bf 210}), and 
 $2_{9}^{(2,5,0,0,0)}$ ({\bf 21}). 
In total, there are 1344 E$^{(9;5)}$-branes, which are classified in Table \ref{tab:K5-brane-number}. 
Repeating a similar argument, we obtain the set of Tables \ref{tab:Fbranes}--\ref{tab:Efinalbranes}. 

\begin{table}[H]
\begin{center}
{\scriptsize
 \begin{tabular}{|c||c|c|c|c|c|c|c|}\hline
 $p\backslash d$ & 9 & 8 & 7 & 6 & 5 & 4 & 3 \\ \hline\hline
 0 &  2 &  4 &  6 &  8 & 10 & 12 & 14 \\ \hline
 1 &  1 &  1 &  1 &  1 &  1 &  1 &  1 \\ \hline
\end{tabular}
}
\end{center}
\caption{Number of F-branes in the $p$-brane multiplet in $d$-dimensions \cite{Bergshoeff:2011mh}.}
\label{tab:Fbranes}
\end{table}
\begin{table}[H]
\begin{center}
{\scriptsize
 \begin{tabular}{|c||c|c|c|c|c|c|c|}\hline
 $p\backslash d$ & 9 & 8 & 7 & 6 & 5 & 4 & 3 \\ \hline\hline
 0 &  1 &  2 &  4 &  8 & 16 & 32 & 64 \\ \hline
 1 &  1 &  2 &  4 &  8 & 16 & 32 & 64 \\ \hline
 2 &  1 &  2 &  4 &  8 & 16 & 32 & 64 \\ \hline
 3 &  1 &  2 &  4 &  8 & 16 & 32 &    \\ \hline
 4 &  1 &  2 &  4 &  8 & 16 &    &    \\ \hline
 5 &  1 &  2 &  4 &  8 &    &    &    \\ \hline
 6 &  1 &  2 &  4 &    &    &    &    \\ \hline
 7 &  1 &  2 &    &    &    &    &    \\ \hline
 8 &  1 &    &    &    &    &    &    \\ \hline
\end{tabular}
}
\end{center}
\caption{Number of D-branes in the $p$-brane multiplet in $d$-dimensions \cite{Bergshoeff:2011mh}.}
\label{tab:D-brane-number}
\end{table}
\begin{table}[H]
\begin{center}
{\scriptsize
 \begin{tabular}{|c||c|c|c|c|c|c|c|}\hline
 $p\backslash d$ & 9 & 8 & 7 & 6 & 5 & 4 & 3 \\ \hline\hline
 0 &     &     &     &     &  1 &  12 &  84 \\ \hline
 1 &     &     &     &  1  & 10 &  60 & 280 \\ \hline
 2 &     &     &  1  &  8  & 40 & 160 & 560 \\ \hline
 3 &     &  1  &  6  & 24  & 80 & 240 &     \\ \hline
 4 &  1  &  4  & 12  & 32  & 80 &     &     \\ \hline
 5 & 1+1 & 2+2 & 4+4 & 8+8 &    &     &     \\ \hline
\end{tabular}
}
\end{center}
\caption{Number of S-branes in the $p$-brane multiplet in $d$-dimensions \cite{Bergshoeff:2011mh}.}
\end{table}
\begin{table}[H]
\begin{center}
{\scriptsize
 \begin{tabular}{|c||c|c|c|c|c|c|c|}\hline
 $p\backslash d$ & 9 & 8 & 7 & 6 & 5 & 4 & 3 \\ \hline\hline
 0 &   &   &    &    &     &     &   64 \\ \hline
 1 &   &   &    &    &     &  32 &  448 \\ \hline
 2 &   &   &    &    &  16 & 192 & 1344 \\ \hline
 3 &   &   &    &  8 &  80 & 480 &      \\ \hline
 4 &   &   &  4 & 32 & 160 &     &      \\ \hline
 5 &   & 2 & 12 & 48 &     &     &      \\ \hline
 6 & 1 & 4 & 12 &    &     &     &      \\ \hline
 7 & 1 & 2 &    &    &     &     &      \\ \hline
\end{tabular}
}
\end{center}
\caption{Number of E-branes in the $p$-brane multiplet in $d$-dimensions \cite{Bergshoeff:2011ee}.}
\end{table}
\begin{table}[H]
\begin{center}
{\scriptsize
 \begin{tabular}{|c||c|c|c|c|c|c|c|}\hline
 $p\backslash d$ & 9 & 8 & 7 & 6 & 5 & 4 & 3 \\ \hline\hline
 0 &   &   &    &    &     &    & 14 \\ \hline
 1 &   &   &    &    &     &  1 & 14 \\ \hline
\end{tabular}
}
\end{center}
\caption{Number of E$^{(4;6)}$-branes in the $p$-brane multiplet in $d$-dimensions \cite{Bergshoeff:2017gpw}.}
\end{table}
\begin{table}[H]
\begin{center}
{\scriptsize
 \begin{tabular}{|c||c|c|c|c|c|c|c|}\hline
 $p\backslash d$ & 9 & 8 & 7 & 6 & 5 & 4 & 3 \\ \hline\hline
 1 &   &   &    &    &    &     &  560 \\ \hline
 2 &   &   &    &    &    & 160 & 2240 \\ \hline
 3 &   &   &    &    & 40 & 480 &      \\ \hline
 4 &   &   &    &  8 & 80 &     &      \\ \hline
 5 &   &   &  1 &  8 &    &     &      \\ \hline
\end{tabular}
}
\end{center}
\caption{Number of E$^{(4;3)}$-branes in the $p$-brane multiplet in $d$-dimensions \cite{Bergshoeff:2017gpw}.}
\end{table}
\begin{table}[H]
\begin{center}
{\scriptsize
 \begin{tabular}{|c||c|c|c|c|c|c|c|}\hline
 $p\backslash d$ & 9 & 8 & 7 & 6 & 5 & 4 & 3 \\ \hline\hline
 2 &   &   &   &   &    &    & 64 \\ \hline
 3 &   &   &   &   &    & 32 &    \\ \hline
 4 &   &   &   &   & 16 &    &    \\ \hline
 5 &   &   &   & 8 &    &    &    \\ \hline
 6 &   &   & 4 &   &    &    &    \\ \hline
 7 &   & 2 &   &   &    &    &    \\ \hline
 8 & 1 &   &   &   &    &    &    \\ \hline
\end{tabular}
}
\end{center}
\caption{Number of E$^{(4;0)}$-branes in the $p$-brane multiplet in $d$-dimensions \cite{Bergshoeff:2012ex}.}
\end{table}
\begin{table}[H]
\begin{center}
{\scriptsize
 \begin{tabular}{|c||c|c|c|c|c|c|c|}\hline
 $p\backslash d$ & 9 & 8 & 7 & 6 & 5 & 4 & 3 \\ \hline\hline
 1 &   &   &   &   &    &    & 448 \\ \hline
 2 &   &   &   &   &    & 32 & 448 \\ \hline
\end{tabular}
}
\end{center}
\caption{Number of E$^{(5;6)}$-branes in the $p$-brane multiplet in $d$-dimensions \cite{Bergshoeff:2017gpw}.}
\end{table}
\begin{table}[H]
\begin{center}
{\scriptsize
 \begin{tabular}{|c||c|c|c|c|c|c|c|}\hline
 $p\backslash d$ & 9 & 8 & 7 & 6 & 5 & 4 & 3 \\ \hline\hline
 2 &   &   &   &   &    &     & 2240 \\ \hline
 3 &   &   &   &   &    & 480 &      \\ \hline
 4 &   &   &   &   & 80 &     &      \\ \hline
 5 &   &   &   & 8 &    &     &      \\ \hline
\end{tabular}
}
\end{center}
\caption{Number of E$^{(5;4)}$-branes in the $p$-brane multiplet in $d$-dimensions \cite{Bergshoeff:2017gpw}.}
\end{table}
\begin{table}[H]
\begin{center}
{\scriptsize
 \begin{tabular}{|c||c|c|c|c|c|c|c|}\hline
 $p\backslash d$ & 9 & 8 & 7 & 6 & 5 & 4 & 3 \\ \hline\hline
 1 &   &   &   &   &    &     & 280 \\ \hline
\end{tabular}
}
\end{center}
\caption{Number of E$^{(6;4)}$-branes in the $p$-brane multiplet in $d$-dimensions.}
\end{table}
\begin{table}[H]
\begin{center}
{\scriptsize
 \begin{tabular}{|c||c|c|c|c|c|c|c|}\hline
 $p\backslash d$ & 9 & 8 & 7 & 6 & 5 & 4 & 3 \\ \hline\hline
 2 &   &   &   &   &    &     & 3360 \\ \hline
 3 &   &   &   &   &    & 240 &      \\ \hline
\end{tabular}
}
\end{center}
\caption{Number of E$^{(6;2)}$-branes in the $p$-brane multiplet in $d$-dimensions \cite{Bergshoeff:2017gpw}.}
\end{table}
\begin{table}[H]
\begin{center}
{\scriptsize
 \begin{tabular}{|c||c|c|c|c|c|c|c|}\hline
 $p\backslash d$ & 9 & 8 & 7 & 6 & 5 & 4 & 3 \\ \hline\hline
 1 &   &   &   &  &   &   & 64 \\ \hline
\end{tabular}
}
\end{center}
\caption{Number of E$^{(7;7)}$-branes in the $p$-brane multiplet in $d$-dimensions.}
\end{table}
\begin{table}[H]
\begin{center}
{\scriptsize
 \begin{tabular}{|c||c|c|c|c|c|c|c|}\hline
 $p\backslash d$ & 9 & 8 & 7 & 6 & 5 & 4 & 3 \\ \hline\hline
 2 &   &   &   &  &   &   & 2240 \\ \hline
\end{tabular}
}
\end{center}
\caption{Number of E$^{(7;4)}$-branes in the $p$-brane multiplet in $d$-dimensions.}
\end{table}
\begin{table}[H]
\begin{center}
{\scriptsize
 \begin{tabular}{|c||c|c|c|c|c|c|c|}\hline
 $p\backslash d$ & 9 & 8 & 7 & 6 & 5 & 4 & 3 \\ \hline\hline
 2 &   &   &   &  &   &    & 448 \\ \hline
 3 &   &   &   &  &   & 32 &     \\ \hline
\end{tabular}
}
\end{center}
\caption{Number of E$_{(7;6)}$-branes in the $p$-brane multiplet in $d$-dimensions \cite{Bergshoeff:2017gpw}.}
\end{table}
\begin{table}[H]
\begin{center}
{\scriptsize
 \begin{tabular}{|c||c|c|c|c|c|c|c|}\hline
 $p\backslash d$ & 9 & 8 & 7 & 6 & 5 & 4 & 3 \\ \hline\hline
 1 &   &   &   &  &   &    & 1 \\ \hline
\end{tabular}
}
\end{center}
\caption{Number of E$^{(8;7)}$-branes in the $p$-brane multiplet in $d$-dimensions.}
\end{table}
\begin{table}[H]
\begin{center}
{\scriptsize
 \begin{tabular}{|c||c|c|c|c|c|c|c|}\hline
 $p\backslash d$ & 9 & 8 & 7 & 6 & 5 & 4 & 3 \\ \hline\hline
 2 &   &   &   &  &   &    & 64 \\ \hline
\end{tabular}
}
\end{center}
\caption{Number of E$^{(8;0)}$-branes in the $p$-brane multiplet in $d$-dimensions.}
\end{table}
\begin{table}[H]
\begin{center}
{\scriptsize
 \begin{tabular}{|c||c|c|c|c|c|c|c|}\hline
 $p\backslash d$ & 9 & 8 & 7 & 6 & 5 & 4 & 3 \\ \hline\hline
 2 &   &   &   &  &   &    & 2240 \\ \hline
\end{tabular}
}
\end{center}
\caption{Number of E$^{(8;3)}$-branes in the $p$-brane multiplet in $d$-dimensions.}
\end{table}
\begin{table}[H]
\begin{center}
{\scriptsize
 \begin{tabular}{|c||c|c|c|c|c|c|c|}\hline
 $p\backslash d$ & 9 & 8 & 7 & 6 & 5 & 4 & 3 \\ \hline\hline
 2 &   &   &   &  &   &    & 1344 \\ \hline
\end{tabular}
}
\end{center}
\caption{Number of E$^{(9;5)}$-branes in the $p$-brane multiplet in $d$-dimensions.}
\label{tab:K5-brane-number}
\end{table}
\begin{table}[H]
\begin{center}
{\scriptsize
 \begin{tabular}{|c||c|c|c|c|c|c|c|}\hline
 $p\backslash d$ & 9 & 8 & 7 & 6 & 5 & 4 & 3 \\ \hline\hline
 2 &   &   &   &  &   &    & 560 \\ \hline
\end{tabular}
}
\end{center}
\caption{Number of E$^{(10;3)}$-branes in the $p$-brane multiplet in $d$-dimensions.}
\end{table}
\begin{table}[H]
\begin{center}
{\scriptsize
 \begin{tabular}{|c||c|c|c|c|c|c|c|}\hline
 $p\backslash d$ & 9 & 8 & 7 & 6 & 5 & 4 & 3 \\ \hline\hline
 2 &   &   &   &  &   &    & 64 \\ \hline
\end{tabular}
}
\end{center}
\caption{Number of E$^{(11;7)}$-branes in the $p$-brane multiplet in $d$-dimensions.}
\label{tab:Efinalbranes}
\end{table}


\end{document}